\documentclass[twoside,openright,titlepage,numbers=noenddot,headinclude,
                footinclude=true,cleardoublepage=empty,abstractoff, 
                BCOR=5mm,paper=letter,fontsize=11pt,
                ngerman,american,
                ]{scrreprt}
                
\PassOptionsToPackage{eulerchapternumbers,listings,drafting,%
				 pdfspacing,
				 subfig,beramono,eulermath,parts}{classicthesis}										

\usepackage{ifthen}
\newboolean{enable-backrefs} 
\setboolean{enable-backrefs}{false} 

\newcommand{\myTitle}{\textbf{Open Quantum Systems At Low Temperature}\xspace}

\newcommand{\myName}{Johan Fabian Triana Galvis\xspace}
\newcommand{\myProf}{Prof. Leonardo A. Pach\'on \xspace}

\newcommand{\myFaculty}{Instituto de F\'isica\xspace}
\newcommand{\myDepartment}{Facultad de Ciencas Exactas y Naturales\xspace}
\newcommand{\myUni}{Universidad de Antioquia - UdeA -\xspace}
\newcommand{\myLocation}{Medell\'in, Colombia\xspace}
\newcommand{\myTime}{\today\xspace}

\newcounter{dummy} 
\providecommand{\mLyX}{L\kern-.1667em\lower.25em\hbox{Y}\kern-.125emX\@}
\newcommand{\ie}{i.\,e.}

\newcommand{\eg}{e.\,g.}

\newcommand{\ad}{\hat{a}^{\dagger}}
\newcommand{\an}{\hat{a}}
\newcommand{\ant}{\hat{a}(t)}

\newcommand{\bd}{\hat{b}^{\dagger}}
\newcommand{\bn}{\hat{b}}

\newcommand{\bnt}{\hat{b}(t)}

\newcommand{\OC}{\omega_{\mathrm{C}}}
\newcommand{\OD}{\omega_{\mathrm{D}}}
\newcommand{\OM}{\omega_{\mathrm{m}}}
\newcommand{\Hs}{\hat{H}_{\mathrm{S}}}
\newcommand{\Hb}{\hat{H}_{\mathrm{B}}}
\newcommand{\Hi}{\hat{H}_{\mathrm{SB}}}
\newcommand{\Oc}{\omega_{\mathrm{cav}}}

\usepackage{acronym}
\newacro{nm}[non-markovian]{non-Markovian}
\newacro{ucl}[UCL]{Ullersma-Caldeira-Leggett model}
\newacro{oqs}[OQS]{Open Quantum Systems}

\PassOptionsToPackage{latin9}{inputenc}	
 \usepackage{inputenc}				

 \usepackage{babel}					

\PassOptionsToPackage{square,numbers,sort&compress}{natbib}
 \usepackage{natbib}

\PassOptionsToPackage{fleqn}{amsmath}		
 \usepackage{amsmath}

\usepackage{calligra} 
\usepackage{alltt}
\usepackage{spverbatim}
\usepackage{pbsi}
\usepackage{sqrcaps}
\usepackage{chngcntr} 
\usepackage{multicol}
\usepackage{multirow}
\PassOptionsToPackage{T1}{fontenc} 
	\usepackage{fontenc}     
\usepackage{textcomp} 
\usepackage{scrhack} 
\usepackage{xspace} 
\usepackage{mparhack} 
\usepackage{fixltx2e} 
\PassOptionsToPackage{printonlyused,smaller}{acronym}
	\usepackage{acronym} 

\usepackage{tabularx} 
\usepackage{caption}
\usepackage{subfig}  

\usepackage{listings} 
\PassOptionsToPackage{pdftex,hyperfootnotes=false,pdfpagelabels}{hyperref}
	\usepackage{hyperref}  
\PassOptionsToPackage{pdftex}{graphicx}
	\usepackage{graphicx} 
	\usepackage{rotating}

\newcommand{\backrefnotcitedstring}{\relax}
\newcommand{\backrefcitedsinglestring}[1]{(Cited on page~#1.)}
\newcommand{\backrefcitedmultistring}[1]{(Cited on pages~#1.)}
\ifthenelse{\boolean{enable-backrefs}}%
{%
		\PassOptionsToPackage{hyperpageref}{backref}
		\usepackage{backref} 
		   \renewcommand*{\backref}[1]{}  
		   \renewcommand*{\backrefalt}[4]{
		      \ifcase #1 %
		         \backrefnotcitedstring%
		      \or%
		         \backrefcitedsinglestring{#2}%
		      \else%
		         \backrefcitedmultistring{#2}%
		      \fi}%
}{\relax}    

\hypersetup{%
    colorlinks=true, linktocpage=true, pdfstartpage=3, pdfstartview=FitV,%
    breaklinks=true, pdfpagemode=UseNone, pageanchor=true, pdfpagemode=UseOutlines,%
    plainpages=false, bookmarksnumbered, bookmarksopen=true, bookmarksopenlevel=1,%
    hypertexnames=true, pdfhighlight=/O,
    urlcolor=webbrown, linkcolor=RoyalBlue, citecolor=webgreen, 
    pdftitle={\myTitle},%
    pdfauthor={\textcopyright\ \myName, \myUni, \myFaculty},%
    pdfsubject={},%
    pdfkeywords={},%
    pdfcreator={pdfLaTeX},%
    pdfproducer={LaTeX with hyperref and classicthesis}%
}   

\makeatletter
\@ifpackageloaded{babel}%
    {%
       \addto\extrasamerican{%
				}%
       \addto\extrasngerman{%
				}%
    }{\relax}
\makeatother

\usepackage[dottedtoc]{classicthesis} 
\PassOptionsToPackage{fleqn}{amsmath}		
 \usepackage{amsmath}
\usepackage{amssymb,amscd}
\numberwithin{equation}{chapter}

\areaset[current]{410pt}{750pt} 
\begin{document}
\frenchspacing
\raggedbottom
\selectlanguage{american} 
\pagenumbering{roman}
\pagestyle{plain}
\thispagestyle{empty}
\begin{addmargin}[-1cm]{-3cm}
\begin{center}
  \large  

        \hfill

        \vfill
    \spacedlowsmallcaps{\myName} \\ \medskip                        

    \begingroup
        \color{Maroon}\spacedallcaps{\myTitle}
    \endgroup
    
      \vfill   
\end{center}        
\end{addmargin}

\begin{titlepage}
	\begin{addmargin}[-1cm]{-3cm}
    \begin{center}
        \large  

        \hfill

        \vfill

        \begingroup
            \color{Maroon}\spacedallcaps{\myTitle} \\ \bigskip
        \endgroup

        \spacedlowsmallcaps{\myName}

        \vfill

        \includegraphics[width=8cm]{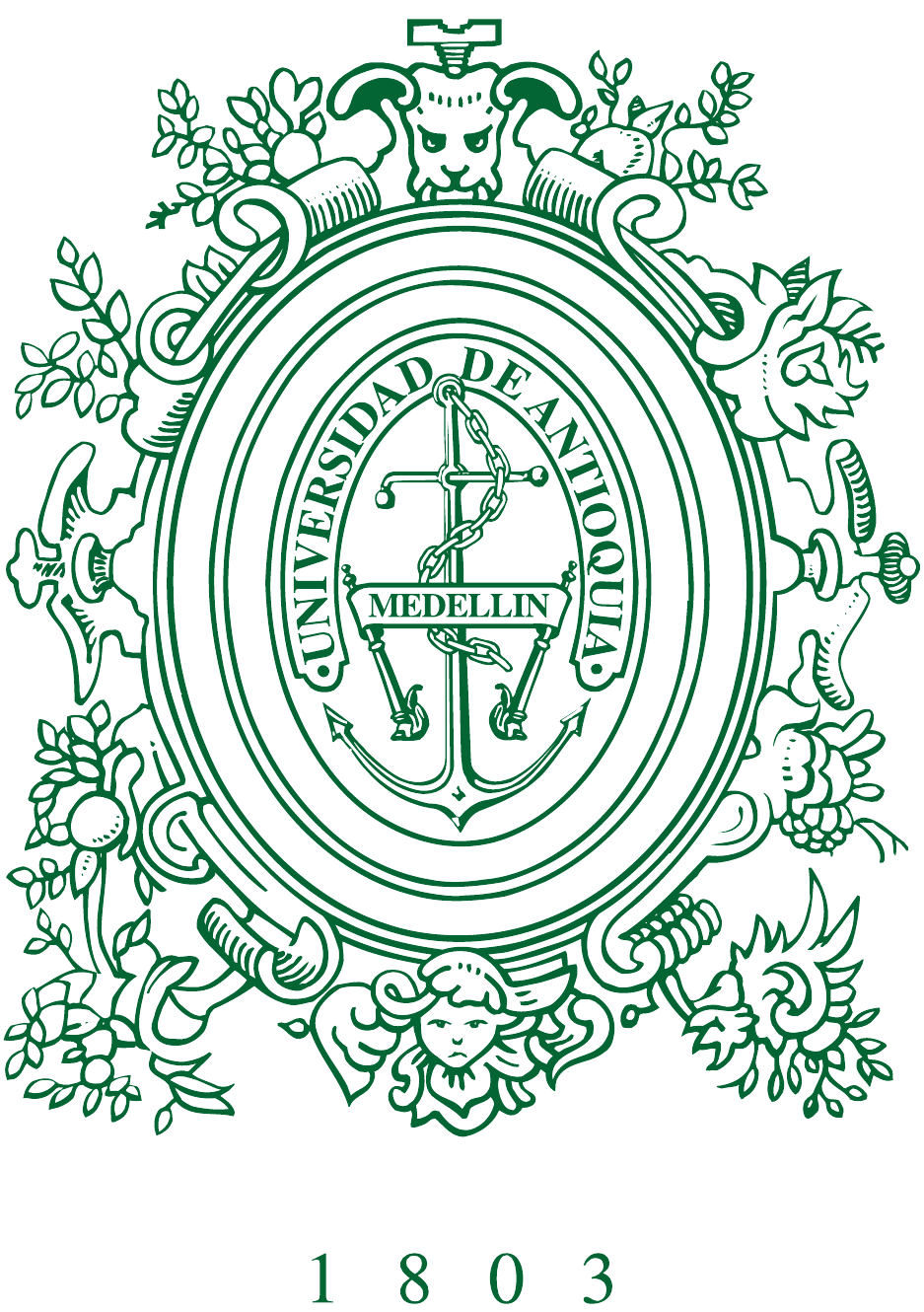} \\ \medskip

        \vfill

        \myFaculty \\
        \myDepartment \\                            
        \myUni \\ \bigskip

        \myTime

        \vfill                      

    \end{center}  
  \end{addmargin}       
\end{titlepage}

\thispagestyle{empty}

\hfill

\vfill

\noindent\myName: \textit{\myTitle,} \\ 
\textcopyright\ \myTime

\bigskip
\noindent\spacedlowsmallcaps{Supervisor}: \\
\myProf \\
\vspace{5mm}
\\
%
%
\medskip
\noindent\spacedlowsmallcaps{Location}: \\
\myLocation
\medskip
%

\cleardoublepage
\thispagestyle{empty}
\refstepcounter{dummy}
\pdfbookmark[1]{Dedication}{Dedication}

\vspace*{3cm}

\begin{center}
    \emph{Ohana} means family. \\
    Family means nobody gets left behind, or forgotten. \\ \medskip
    --- Lilo \& Stitch    
\end{center}

\medskip

\begin{center}
	God doesn't play dice with the world. \\ \medskip
	--- Albert Einstein
\end{center}

\medskip

\begin{center}
    \textit{Dedicated to my grandparents, Gabriel and Flor.} \\ \smallskip
\end{center}

\cleardoublepage
\cleardoublepage
\pdfbookmark[1]{Abstract}{Abstract}
\begingroup
\let\clearpage\relax
\let\cleardoublepage\relax
\let\cleardoublepage\relax

\chapter*{Abstract}
It is known that the origin of the deviations from standard thermodynamics proceed from the strong coupling to the bath. Here, it is shown that these deviations are related to the power spectrum of the bath.
%
%
Specifically, it is shown that the system thermal-equilibrium-state cannot be characterized by the canonical Boltzmann's
distribution in quantum mechanics. 
This is because the uncertainty principle imposed a lower bound of the dispersion of the total energy of the system that is dominated by the spectral density of the bath.
%
However, in the classical case, for a wide class of systems that interact via central forces
with pairwise-self-interacting environment, 
the system thermal equilibrium state is \emph{exactly} characterized by the canonical Boltzmann 
distribution.
%
%
As a consequence of this analysis and taking into account all energy scales in the system and reservoir interaction,
an \emph{effective coupling} to the environment is introduced.
%
Sample computations in different regimes predicted by this effective coupling are shown. 
Specifically, in the strong coupling effective regime, the system exhibits deviations from standard thermodynamics and in the
weak coupling effective regime, quantum features such as stationary 
entanglement are possible at high temperatures. 
%

Moreover, it is known that the spectrum of thermal baths depends on the non-Markovian character. 
Hence, non-Markovian interactions have a important role in the thermal equilibrium state of physical systems. 
For example, in quantum optomechanics is looked up to cool the mechanical system through an auxiliary 
system which generally is a cavity. 
This cooling process takes into account the non-Markovian interaction and as it is shown
here, it is more effective than if we use only the Markovian approximation in the equation of motion for the different modes. 

Finally, we are also interested in the dynamics of the cooling process and how to maintain the minimum phonon number once is achieved. This is acomplished by varying the strength of the coupling between the two resonators along the time and it is found via an optimization process.
\vfill


\endgroup			

\vfill

\cleardoublepage
\pdfbookmark[1]{Publications}{publications}
\chapter*{Publications}

Some ideas and figures have appeared previously in the following publications:

\bigskip

\noindent L. A. Pach\'on, J. F. Triana, D. Zueco, and P. Brumer, Uncertainty principle consequences at thermal equilibrium, arXiv, vol. 1401.1418, 2014.

\noindent J. F. Triana, A. F. Estrada, L. A. Pach\'on, Ultrafast Optimal sideband cooling under non-Markovian evolution, arXiv, vol. 1508.04869, 2015.

\cleardoublepage
\pdfbookmark[1]{Acknowledgments}{acknowledgments}

\begin{flushright}{\slshape    
    We keep moving forward, opening up new doors and \\ 
	doing new things, because we are curious... \\
	and curiosity keeps leading us down new paths.} \\ \medskip
	--- Walt Disney
\end{flushright}

\bigskip

\begingroup
\let\clearpage\relax
\let\cleardoublepage\relax
\let\cleardoublepage\relax
\chapter*{Acknowledgments}

\textit{First of all, I want to thank my family for trusting and supporting me in all the decisions I have made in my life. Specially to my grandparents, Gabriel and Flor, for educating me when I was just a child. There are no words for thanking them. Similarly, to my mother Marlen for her support, aid and comprehension. }

\textit{Second, I thank to my advisor Leonardo A. Pach\'on who I have shared with the best moments in my academic and professional life and who tought me a great passion for science. Likewise, he is an example both personal and academic. I also have to thank -CODI- and GFAM group at Universidad de Antioquia, for giving me the support along the time of the Master.}

\textit{Third, I thank the other professors of GFAM group, Jos\'e Luis Sanz, Boris A. Rodriguez, Alvaro Valdes and Jorge Mahecha, who share with the students aspects of life different from academia. }

\textit{Similarly, I thank the members of GFAM group for the social activities, discussions, moments shared and the working atmosphere at SIU. All this has been to grow up both personal and professional. }

\textit{Last, I thank to all new members of my ``family'' in Medell\'in for making possible the familiar coexistence. }

%

\endgroup

\pagestyle{scrheadings}

\cleardoublepage

\refstepcounter{dummy}
\pdfbookmark[1]{\contentsname}{tableofcontents}
\setcounter{tocdepth}{1} 
\setcounter{secnumdepth}{2} 
\manualmark
\markboth{\spacedlowsmallcaps{\contentsname}}{\spacedlowsmallcaps{\contentsname}}
\tableofcontents 
\automark[section]{chapter}
\renewcommand{\chaptermark}[1]{\markboth{\spacedlowsmallcaps{#1}}{\spacedlowsmallcaps{#1}}}
\renewcommand{\sectionmark}[1]{\markright{\thesection\enspace\spacedlowsmallcaps{#1}}}
\clearpage

\begingroup 
    \let\clearpage\relax
    \let\cleardoublepage\relax
    \let\cleardoublepage\relax
    \refstepcounter{dummy}
    \pdfbookmark[1]{\listfigurename}{lof}
    \listoffigures

    \vspace*{8ex}

%
    
%
       
\endgroup

\cleardoublepage

\pagenumbering{arabic}

\markboth{\spacedlowsmallcaps{introduction}}{\spacedlowsmallcaps{introduction}}

\chapter*{Introduction}\label{ch:introduction}
\addtocontents{toc}{\protect\vspace{\beforebibskip}} 
\addcontentsline{toc}{chapter}{\tocEntry{Introduction}}

The history of thermodynamics has been inexorably related to the development of machines, and undoubtedly was the fulcrum of the 18th century industrial revolution. Currently, we are at the dawn of a new revolution: the nanotechnological revolution. This time, the aim of constructing and designing machines at nanometer-length scales, such as atomic motors \cite{PDH09}, quantum photocells \cite{Scu10}, gyrators \cite{RJ&12,KK&12} or quantum heat engines (heat pumps or refrigerators) \cite{FR07,SC&11}, has brought the need of developing a quantum version of thermodynamics \cite{EK&08,CHT11}. One of the foundational conundra in this emerging field is, to what extent nanomachines can display quantum features and how this quantum behaviour could be used to improve their efficiency. 

At low temperature, phenomenology of quantum systems differs non-trivially from high temperatures \cite{Wei93}, phenomena such as superconductivity \cite{TI&08}, superfluidity \cite{Gri05} or Bose-Einstein condensation \cite{KM11,Mel11} are clear examples. This is because at low temperatures, quantum fluctuations can dominate over thermal fluctuations. Therefore, the low temperature regime is ideal for exploring the quantum behavior of physical systems. 
However, defining the precise meaning of ``low temperature'' is not a trivial task and depends on the characteristics of each system. At the same temperature $T$, different systems may be at temperature regimes completely different, \eg, ambient temperature may be considered as low temperature in the context of electronic transitions \cite{PB11}, but high temperature in the context of mechanical nanoresonators \cite{PDH09,Scu10,FR07,RJ&12}. A similar analysis can also be formulated for the case of driven systems out of thermodynamic equilibrium \cite{GPZ10,Ved10}. A specific topic dedicated to this revolution is the cooling of nanomechanical structures, which have had several results in both theoretical \cite{SN&11,WV&11,MC&12} and experimental aspects \cite{RM&10,OH&10,SC&13}.

According to what is mentioned above, one of the concerns at this moment, both in the cooling of nanomechanical structures as in quantum mechanical systems, is to get theoretical results the closest to physical reality as possible.
An example of these results were initially derived in statistical mechanics, where it is known that the thermal equilibrium state of a system weakly coupled to an environment is typically well characterized by the canonical Boltzmann distribution or Gibbs state \cite{Gre95}. In quantum systems, this result is derived under the assumption that there is a weak coupling between the system and the environment \cite{GL&06,DY&07}, or that the total system consisting of the system plus the environment can be seen as a macroscopic system, as it is assumed by statistical mechanics \cite{PSW06}.
Therefore, bearing in mind that at low temperatures there are deviations from the Gibbs state \cite{GWT84,PT&14}, in this work the effects of non-Markovian interactions are incorporated through of an effective coupling in the thermodynamic equilibrium regime, where systems are capable of reflecting their quantum characteristics or features.

This is possible due to the evidence in different investigations of thermodynamic equilibrium \cite{GPZ10,PB11}, where the effects of non-Markovian interactions have an impact on the effective temperature. 
Hence, because our particular interests in this work, we focus on the modifications of the thermodynamic properties of the system at equilibrium due to non-Markovian interactions. 
Also, once we known how the non-Markovian character acts on the thermodynamic properties of the system at equilibrium, we explain the origin of these deviations from the canonical state or Gibbs state and relate it to the Heisenberg's uncertainty principle.

Taking into account the deviations in the canonical state, we have that these deviations  induce corrections to the partition function, \ie, changes in the thermodynamic properties of the system. 
Thus, by changing some thermodynamic quantities of the system, unusual phenomena are generated at equilibrium such as squeezing and entanglement \cite {GPZ10,PT&14, AW08}. 
The latter is studied in an effective coupling regime described by low effective temperatures and the Markovian and non-Markovian interactions (see Chapter \ref{ch:influence}).

Likewise, the cooling limit, \eg, in nanoresonators, is affected by the change of the thermodynamic quantities of the system, in particular, by the non-Markovian character of the environment. 
Bearing in mind that the cooling limit is evaluated through the cooling factor, or the average phonon occupancy in the resonator, and that the more recent theories on cooling \cite{OH&10,LX&13,MC&12,AA&09b,MC&07} are performed under the assumption that the system behaves in the Markovian regime, this limit is re-evaluated from the point of view of a model that includes the non-Markovian character. This is done motivated by the recent experimental evidence of the non-Markovian dynamics  in mechanical resonators \cite{LL&11,GT&13}.

Since in open quantum systems thermodynamic equilibrium is a consequence of the coupling to the environment, in this work, we study how  quantum fluctuations are able to modify the thermodynamic properties of a system $S$. 
In general, we show following results:
\begin{enumerate}
\item Study and quantify, in detail, how quantum fluctuations induce deviations from the canonical state \cite {GL&06, PSW06}.
\item The role of the effective coupling in the entanglement of mechanical nanoresonators at equilibrium and in the role of the coupling in the production of squeezing at equilibrium \cite{GPZ10,PT&14}.
\item The role of the effective coupling in the theoretical limits of cooling in mechanical resonators \cite{MC&07,LX&13,MC&12}.
\item The optimal coupling between two resonators in the cooling process via optimal control theory in a non-Markovian evolution \cite{TEP15}.
\end{enumerate}

The study, analysis and characterization of the above three situations allow for the discussions  of the thermodynamic consequences of quantum fluctuations, not only from an academic point of view but also from a technological point of view. This combination of interests strengthenes the motivation of the development of this work and contextualizes the relevance of their potential outcomes.



%
%
%
%


\cleardoublepage
\part{Quantum Thermodynamics and Thermal Equilibrium}
\cleardoublepage
\begin{flushright}{\slshape    
    To those who do not know mathematics it is difficult to get across a real feeling \\
    as to the beauty, the deepest beauty, of nature... If you want to learn about nature, \\
    to appreciate nature, it is necessary to understand the language that she speaks in.} \\ \medskip
	--- Richard P. Feynman
\end{flushright}

\bigskip

\begingroup
\let\clearpage\relax
\let\cleardoublepage\relax
\let\cleardoublepage\relax

\chapter{Quantum Thermodynamics and Concepts}\label{ch:thermo}
\counterwithin{figure}{chapter}


A few centuries ago, statistical mechanics developed by James Clerk Maxwell, Ludwig Boltzmann, and Josiah Willard Gibbs was incorporated in thermodynamics which carried to the development of machines in the age of Industrial Revolution. These physicists believed in the existence of atoms and developed the mathematical methods for describing their statistical and thermodynamical properties. However, contrary to the large scales used in the Industrial Revolution and the results gotten by Maxwell, Boltzmann and Gibbs, our age is in a new revolution, the nanotechnological revolution. This has been influenced by technologies designed and constructed in micrometre- or nanometre-length scales \cite{PDH09,Scu10,FR07,SC&11,RJ&12,KK&12}, 
which have brought the need of developing a quantum version of thermodynamics \cite{EK&08,CHT11}, which is still incomplete.

In this work, we attempt to explain questions along this line of work based on the intimate link between quantum thermodynamics and the theory of open quantum systems. In this chapter, we first make a brief summary of what is quantum thermodynamics and which are their limitations and the link with the open quantum systems, one of the main topics in this work.


\section{What is Quantum Thermodynamics?}

To begin with, both thermodynamics and quantum mechanics are two different areas in physics and they are applied to very different kinds of systems. The works of Maxwell, Boltzmann and Gibbs gave rise to thermodynamics and the description of macroscopic systems where the number of particles is around Avogadro's number ($6.022\times10^{23}$). In contrast, quantum mechanics describes the behavior of atoms and molecules, i.e., microscopic particles. Therefore, the regimes where these two theories can be applied are very different. However, these regimes are not impediments to find connections between them. Some of these connections have arisen during the last few years, \eg, the definition of the laws of quantum thermodynamics \cite{BW&13,LAK12,Kos13} and the fundamental limitations of quantum thermodynamics \cite{LPS10,HO13}.

One of the ways to relate both theories can be set as follows: What happens if a microscopic system (quantum system) is put to interact with a thermodynamic system (thermal bath)? The link between them is an area of physics known as Open Quantum Systems (OQS) \cite{Wei93,BP07}.
Then, we can say that  \textit{quantum thermodynamics} is the study of thermodynamic processes within the context of quantum dynamics. Further, currently there is a major ongoing effort to develop a consistent and well defined extension of quantum thermodynamics \cite{EK&08,CHT11,AN00}. However, the majority of these theories are primarily based on Boltzmann's original ideas and are therefore plagued by issues concerning irreversibility, the origin of the Zero Law, the origin of the Second Law, the relation between physics and information, the meaning of ergodicity, etc \cite{BW&13,HO13,GM03}.

One of these issues, the origin of the Zero Law, which within the framework of classical thermodynamics is typically stated follows: If a system $A$ and a system $C$ are each in thermal equilibrium with a system $B$, $A$ is also in equilibrium with $C$. Besides, the thermal equilibrium state of a classical and quantum system weakly coupled to a thermal bath is well characterized by the canonical Boltzmann distribution \cite{PSW06,GL&06}. However, when the system is strongly coupled to a thermal bath, the thermal equilibrium state deviates from the canonical Boltzmann distribution in the classical case \cite{Jar04} as in the quantum case \cite{AN00,GWT84,CTH09,PT&14}. Later, in chapter \ref{ch:thermaleq}, we will show that the origin of the deviations in the quantum regime from the canonical Boltzmann distribution are not simply produced by the strong coupling but are intimately liked with the uncertainty principle. Further, we will explain the consequences of these deviations.

Another matter to consider is that ultracold matter and quantum information processing are closely related to quantum thermodynamics. Cooling mechanical systems unravels their quantum character. As the temperature of the system decreases, degrees of freedom freeze out, leaving a simplified dilute effective Hilbert space \cite{Kos13}. Ultracold quantum systems contributed significantly to our understanding of the basic quantum concepts. In addition, such systems form the basis for emerging quantum technologies \cite{PDH09,Scu10,FR07,SC&11,RJ&12}. To reach sufficiently low temperatures we must focus on the cooling process itself, \ie, quantum refrigeration \cite{LAK12,LPS10}, sideband cooling \cite{MC&07,WN&07,PW09} or another implementations such as ultraefficient cooling of resonators \cite{WV&11}. Later, in chapter \ref{ch:optomechcool} we will show a methodology more efficient than previous ones, \eg, sideband cooling 
\cite{WN&07,MC&07,RD&11,GV&08}.
In the following we make a brief summary of the principal concepts needed along this work. 

\subsection{Open Quantum Systems}
\label{ssec:oqs}

In nature, all systems, either classical or quantum, are continuously exposed to their environments. For example, electrons or atoms interact with the different electromagnetic fields that exist in our everyday life. More specifically, if we go to a particular branch of physics, we find that impurities in a lattice interact with nuclear spins of the atoms in the lattice. Hence, the term \acl{oqs} refers to the description of a quantum system in presence of its environment, where both are interacting. In some cases, the quantum system can be viewed as a distinguishable part of a larger closed quantum system, while the other part of the closed system is the environment. However, there must be some motivation to study them. The importance lies on the interaction between quantum systems and their environment, which in most cases, leads to the quick disappearance of quantum properties, such as entanglement and quantum coherence. 


In order to formally address our study of open quantum systems, we assume that the total Hamiltonian can be explicitly written as
\begin{equation}
\hat{H}=\Hs + \Hb + \Hi,
\label{eq:basicham}
\end{equation}
\noindent where $\Hs$, $\Hb$ and $\Hi$ are the Hamiltonian of the system, the Hamiltonian of the bath and the Hamiltonian of system-bath interaction, respectively. Nevertheless, we need to find the expectation value of the observable quantities to make a description of the quantum thermodynamics. The expectation value of these observables ($\langle\hat{O}\rangle$) is obtained as
\begin{equation}
\left\langle \hat{O} \right\rangle = \mathrm{Tr} \{ \hat{\rho}_{\rm{S}}\hat{O} \},
\end{equation}
\noindent $\hat{\rho}_{\rm{S}}$ being the thermal equilibrium state of the system S. This thermal equilibrium state of the system generally is characterized by the canonical Boltzmann distribution or Gibbs state \cite{GL&06,PSW06}. This, despite the fact that it is well known that when we are in the strong coupling regime, the thermal equilibrium state of the system is different from the Gibbs state \cite{CHT11,CTH09,GSI88}. In this work, we will show what are the most general conditions needed to reach a thermal equilibrium state different from Gibbs state.


\subsection{Quantum Correlations}

Correlations can be established on both quantum physics and classical physics. In classical physics, correlations are usually described in terms of conditional probabilities to get outcomes which depend on other variables. However, one of the most outstanding features of quantum mechanics is its non-local nature, which are in sharp contrast to classical physics. In quantum mechanics, we have correlations between values of measurements performed in spatially separated locations, some that can never occur in systems described by classical physics. The most common type of quantum correlation is entanglement. This is resource to enhance quantum technologies like quantum metrology \cite{GLM06,JM&11,BK&13,MC&11}. Hence, this is the need for preserving it as long as possible from the destructive action of the environment which causes its quick disappearance. Thus, to preserve entanglement for long times, one needs to study its properties in the presence of noise, which is caused by the action of the environment of the systems. 

In this work, as we will show in some of the next chapters, if we consider the most general conditions, \ie, that the coupling to the environment is weak or that environment has infinite degrees of freedom, we reach the thermal equilibrium state characterize by the Gibbs state. However, if we consider an effective coupling, we obtain that the thermal equilibrium state differs from the Gibbs state and one of their consequences is that entanglement can survive at thermal equilibrium and thus attaining longer lifetimes for quantum correlations.

\section{Markovian and Non-Markovian Dynamics}

All investigations on open quantum systems dynamics differentiate between two types of dynamical behaviors, known in literature as Markovian and non-Markovian dynamics. Markovian dynamics has been associated to the semigroup property of the dynamical map describing the system evolution. If one thinks in terms of a microscopic model of system, environment and interaction, the Markovian description for open system needs a particular assumptions, among them are that the system is weakly coupled to the environment and there is not memory effects. Such approaches lead to a master equation in the Lindblad form \cite{BP07}.  However, such approximations always are not always justified and one needs to go beyond perturbation theory.

Hence, currently there is a lot of investigations to study and understand non-Markovian effects of the dynamics in open quantum systems \cite{BP07}. However, these investigations have a large number of problems, not only practical but also fundamental reasons, \eg, ignorance of an accurate measure of non-Markovianity of one process \cite{AB&14}, deficiency of general characterization of the equations of motion, such as that in the Markovian setting thanks to the theory of quantum dynamical semigroups \cite{CK10}. Moreover, the treatment of non-Markovian systems is particularly demanding because the expressions are extremely cumbersome (see. Chapter \ref{ch:osc}). In this case, one cannot consider assumptions such as weak coupling, separation of time scales between the system and the environment and the factorization of the system-environment state \cite{VS&11}. 

Despite the difficulty in the investigations with non-Markovian dynamics, recently, this topic has achieved significant results and important improvements to reach a better understanding of a few of their issues in the theory of open quantum systems. 
Manipulating the environment or performing reservoir engineering is useful for improving quantum devices and thus to extend the persistence of quantum properties in the system \cite{AB&14}. There is evidence that, for certain specific systems, non-Markovian environments has an advantage over Markovian environments for quantum devices \cite{CHP12,VO&11,HRP12,SN&11,TE&09} and further, at sufficiently low temperatures, as in the case of sideband cooling, or for reservoirs with structured spectral mode densities, the dynamics of open quantum systems are non-Markovian \cite{SN&11}.

One of the differences between Markovian and non-Markovian dynamics can be seen in terms of the dynamical maps, \eg, if the initial state of the system and the environment can be factorized, the dynamics of an open quantum system can be described by a trace preserving completely positive (CPT) map \cite{AB&14}, so that the initial state of the system $\hat{\rho}(t_0)$ evolving to the state $\hat{\rho}(t)$ in a time $t$ as
\begin{equation}
\hat{\rho}(t)=\Phi(t,t_0)\hat{\rho}(t_0).
\label{eq:dinmap}
\end{equation}

In order to define Markovianity or non-Markovianity in this configuration, one can assume that the map $\Phi(t,t_0)$ is known and one can look at certain mathematical properties of the map itself. One of this is, \eg, a suitable notion of distinguishability of quantum states \cite{Hay06}, an approach that captures the idea of information flow between the system and the environment. 
Further, there is also a different form in the equations, \eg, Markovian equations generally are local in time while non-Markovian equations are non-local in time, \ie, they are integro-differential equations. 

These two approaches can be understood as follow: if the map $\Phi(t,0)$ can be split according to
\begin{equation}
\Phi(t,0)=\Phi(t,s)\Phi(s,0),
\label{eq:dinmaps}
\end{equation}
where we choose $t_0 = 0$, for any $0\le s \le t$, with $\Phi(t,s)$ itself being a CPT map, we say that the map $\Phi(t,0)$ is divisible, which implies that $\Phi(t,s)$ and $\Phi(s,0)$ are well-defined. At this point, the existence of the map $\Phi(t,s)$ means that the evolution of the physical system is Markovian due to the divisibility implies that memory effects are neglected. 

Similarly, Markovian and non-Markovian dynamics can be defined in terms of the Langevin equation \cite{Ing02}. For example, if we have a constant dissipation over the system we have a local equation or a local contribution and the dynamics is Markovian. 
On the other hand, if we have a dissipation kernel over the system we get a non-local equation or a non-local contribution and the dynamics in non-Markovian.

This work included the non-Markovian behaviour both in the thermal equilibrium state of a physical system in contact to environment and in the cooling process of mechanical resonators. Similarly, we explain the connection and differences between these two approaches.

\section{Heisenberg's Uncertainty Principle}

In classical physics, the behaviour of a physical system can be described by their respectively equations to motion once the initial conditions are given. For example, if the initial parameters of a system are the coordinates $\vec{r}_0$, velocities $\vec{v}_0$, mass $m_0$ and all the forces or external fields that act over the system are known, we can see that the parameters at any time, in classical physics, are deterministic. However, we could ask if these laws or behaviour apply in any physical system, \eg, in the microscopic world.

In quantum mechanics, the systems are represented by a wave function and as the wave functions cannot be localized, the microscopic system or particle cannot be localized in space contrary to a particle in classical physics. 
For example, in the Young's experiment or double-slit experiment it is impossible to determine which slit the electron crosses \cite{Bra00}. Hence, the classical concepts that we known as exact position, exact momentum, and unique path of a particle do not have sense in the quantum domain or at a microscopic scale. 
This leads to the Heisenberg's uncertainty principle, which  states that:  to a given state which corresponds a statistical distribution of the values of an observable $\hat{A}$, a suitable measure of the width of the distribution for this observable is its variance $\Delta\hat{A}$, which corresponds to the standard deviation of the distribution, it is not possible to observe the state of its complementary or canonical conjugate observable $\hat{B}$, at the same time, with the same accuracy than the observable $\hat{A}$ \cite{Zet09}. 
In other words, the observables $\hat{A}$ and $\hat{B}$ do not commute and hence there are quantum correlations between them and their relation, the Heisenberg's uncertainty principle, is given by
\begin{equation}
\Delta\hat{A}\Delta\hat{B}\ge \frac{1}{2}| \langle[\hat{A},\hat{B} ]\rangle |.
\end{equation}

For example, taking the position and momentum as the observables, the Heisenberg's uncertainty principle indicates that, if it is possible to measure the position of a particle accurately, it will not be possible to measure the momentum of the particle with an arbitrary accurately, simultaneously. In other words, we cannot measure one observable without disturbing its canonical conjugate variable, \ie, we cannot measure the position or momentum without disturbing the other observable.
For example, consider a macroscopic object (e.g., a ball rolling in a inclined plane) and a microscopic system (e.g., an electron in an atom). Besides, to calculate the position of the ball, one need simply to observe it and measure with certains measurements instruments. If one of these use light to measure, the light that hits the ball and gets reflected to the detector has not affect the motion of the object. On the other hand, to measure the position of an electron in an atom, one has to use radiation which is high enough to change the path of the electron and hence also change its momentum. Therefore, it impossible to determine the position and the momentum of the electron simultaneously to arbitrary accuracy. 

Despite these facts, Heisenberg's uncertainty principle is not only applied to measure observables. As we will show in the next chapter, the Heisenberg's uncertainty principle or the commutation relations between the Hamiltonian of the system (bath) and the Hamiltonian of system-bath interaction are responsible for the thermal equilibrium state of the system not to be described by the Gibbs state or the canonial Boltzmann distribution. Further, there are some recent investigations that show how Heinserberg's uncertainty principle determines the non locality of quantum mechanics \cite{OW10} and how a violation of the Heinserberg's uncertainty principle implies a violation of the second law of thermodynamics \cite{HW12}.

\endgroup

\cleardoublepage
\begin{flushright}{\slshape    
	In this house, young lady, we obey the laws of thermodynamics!} \\ \medskip
	--- Homer J. Simpson
\end{flushright}

\bigskip

\begingroup
\let\clearpage\relax
\let\cleardoublepage\relax
\let\cleardoublepage\relax

\chapter{Thermal Equilibrium}\label{ch:thermaleq}
\counterwithin{figure}{chapter}

According to the Second Law of Thermodynamics, which establishes the irreversibility of thermodynamic evolutions and processes, there is a certain stationary equilibrium state, into which a thermodynamic system will evolve eventually. Under given constraints, this equilibrium state is stable with respect to perturbations. This equilibrium state is controlled by macroscopic constraints, such as a fixed temperature $T$. Since under this condition the equilibrium state of the system is controlled by the contact to an environment or a heat bath, we have only one specific parameter, the temperature. Therefore, the equilibrium state depends only on this temperature, regardless of its initial state \cite{GMM09}.

In this chapter, we show which are the classical and quantum thermal equilibrium states. Despite that the classical and quantum thermal states will be calculated for the same class of systems, the classical and quantum thermal equilibrium state will turn out to be significantly different for the same class of systems. Hence, we focus now in deriving the different thermal equilibrium states and under which conditions the classical and quantum equilibrium states are different.

\section{Classical Thermal Equilibrium State}

\begin{figure*}[t]
\centering
\includegraphics[width=0.7\textwidth]{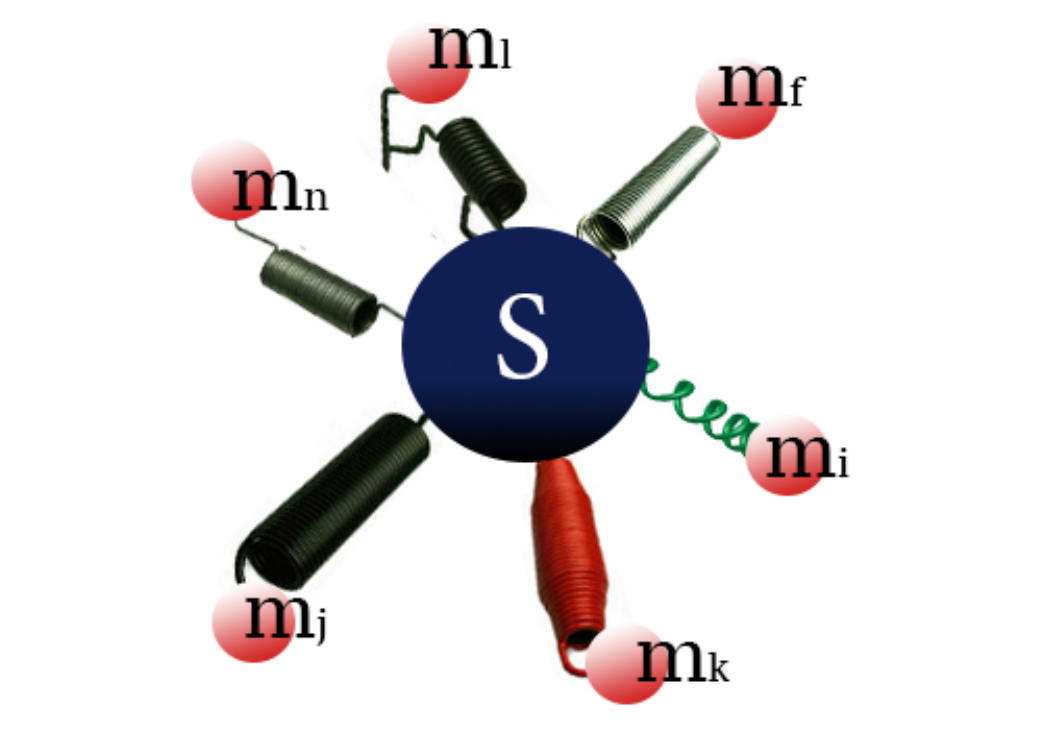}
\caption[Representation of the Ullersma-Caldeira-Leggett model]{  \small Representation of the Ullersma-Caldeira-Leggett model. A system S is coupled to a thermal bath described by a collection of harmonic oscillators (red circles) with different masses and different frequencies.}
\label{fig:caldmodel}
\noindent\rule{\textwidth}{1pt}
\end{figure*}

After reviewing the history of quantum thermodynamics and defining some of the most important concepts of this work, it is appropriate to introduce one model that we are going to use along this part of the work. First at all, as we defined in Section \ref{ssec:oqs}, our model consists of three parts: the Hamiltonian of the system $H_{\mathrm{S}}$, the Hamiltonian of the bath $H_{\mathrm{B}}$ and the Hamiltonian of the system-bath interaction $H_{\mathrm{SB}}$. Second, we use some general models at the beginning that gradually we expand in order to get specific models.

We consider the system as a particle of mass $m$ with potential energy $U_{\rm{S}}(q)$ that depends of the position of the particle $q$, so that the Hamiltonian of the system is given by 
\begin{equation}
H_{\mathrm{S}}=\frac{1}{2m}p^{2}+U_{\mathrm{S}}(q),
\end{equation}
\noindent and consider a bath of $N$ particles interacting via the central force potential $U_{\mathrm{B}}^{\mathrm{i,j}}(\mathfrak{q}_{\mathrm{i}}-\mathfrak{q}_{\mathrm{j}})$ and Hamiltonian 
\begin{equation}
H_{\mathrm{B}}=\sum_{i=1}^{N}\left[\frac{1}{2m_{i}}\mathfrak{p}_{i}^{2}+\sum_{j\neq i}^{N}U_{\mathrm{B}}^{\mathrm{i,j}}(\mathfrak{q}_{\mathrm{i}}-\mathfrak{q}_{\mathrm{j}})\right],
\end{equation}
\noindent where $\mathfrak{q}_{\mathrm{n}}$ is the \textit{nth}-position of the \textit{nth}-particle of the bath. We consider that the system interacts with this bath via the central force potential energy $V_{j}(\mathfrak{q}_{j}-q)$, so that the total Hamiltonian is given by
\begin{equation}
H=H_{\mathrm{S}} (p,q) + H_{\mathrm{B}} (\mathfrak{p,q}) + \sum_{j=1}^{N}V_{j}(\mathfrak{q}_{j}-q).
\label{eq:hamcenfor}
\end{equation}

If we return to the case where the interaction is $H_{\mathrm{SB}}=\sum_{j=1}^{N}V_{j}(\mathfrak{q}_{j}-q)$ and set the interaction between the bath particles to zero, \ie, $U_{\mathrm{B}}^{i,j} = 0$
and consider the second order picture of the system-bath central force interaction, namely,
\begin{equation}
H_{\mathrm{SB}}=\sum_{j=1}^{N}\frac{m_{j}\omega_{j}}{2}(\mathfrak{q}_{j}-q)^{2},
\end{equation} 
this yields a suitable model for dissipative quantum systems which corresponds to the \acl{ucl} \cite{CL81,CL83a}. This model consists of a particle of mass $m$ moving in a potential $U_{\mathrm{S}}(q)$ and the bath is described by a collection of a harmonic oscillators with a bilinear coupling in the position operators of system and bath. The total Hamiltonian of the \acl{ucl} (Fig. \ref{fig:caldmodel}) is then given by
\begin{equation}
H=\frac{p^{2}}{2m} + U_{\mathrm{S}}(q) + \sum_{i=1}^{N}\left( \frac{\mathfrak{p}^{2}}{2m_{i}} + \frac{m_{i}\omega_{i}^{2}}{2}\mathfrak{q}_{i}^{2} \right) - q \sum_{i=1}^{N} c_{i}\mathfrak{q}_{i} + q^{2}\sum_{i=1}^{N}\frac{c_{i}^{2}}{2m_{i}\omega_{i}^{2}}.
\label{eq:hamucl}
\end{equation}
Although this model seems purely academic, the \acl{ucl} can describe very realistic situations. In particular, in the study of the dynamics of appropriate variables of meso or nanoscopic devices which are useful for testing quantum mechanics on the macroscopic level, \eg, through of optomechanical systems.

One of the motivations for using of harmonic oscillators is that they can describe many physical systems. For example, in quantum mechanics, they are the base to describe the fields in quantum optics and the lattices in solid state. Further, there are cases where the bilinear coupling is realistic, \eg, for an environment consisting of a linear electric circuit like the resistor just mentioned or for dipolar coupling to electromagnetic field modes encountered in quantum optics \cite{Ing02}. In addition, harmonic oscillators are used to describe optomechanical systems \cite{MC&07,SC&13} which are used in optomechanical cooling related with the second part of this work. 

Hence, for this part we consider, as we mentioned above, the system as a classical particle of mass $m$ with potential energy $U_{\mathrm{S}}(q)$ which interacts via central forces with pairwise-self-interacting environment. So that the total Hamiltonian, in agreement with Eq. (\ref{eq:hamcenfor}), is given by
\begin{equation}
H=\frac{1}{2m}p^{2}+U_{\mathrm{S}}(q) + \sum_{i=1}^{N}\left[\frac{1}{2m_{i}}\mathfrak{p}_{i}^{2}+\sum_{j\neq i}^{N}U_{\mathrm{B}}^{\mathrm{i,j}}(\mathfrak{q}_{\mathrm{i}}-\mathfrak{q}_{\mathrm{j}})\right] + \sum_{j=1}^{N}V_{j}(\mathfrak{q}_{j}-q).
\label{eq:tothamcenfor}
\end{equation}
In classical statistical mechanics, the thermal equilibrium distribution of the system S is defined by
\begin{equation}
\label{eq:theequstacenfor}
\rho_\mathrm{S}(p,q) = \frac{1}{Z} 
\int
\prod_{\mathfrak{j}}^{N}\mathrm{d}\mathfrak{p}_{\mathfrak{j}} \mathrm{d}\mathfrak{q}_{\mathfrak{j}}
\exp\left[-H(p,q,{\mathfrak{p}}_{\mathfrak{j}},{\mathfrak{q}}_{\mathfrak{j}})
\beta \right],
\end{equation}
\noindent where $Z$ denotes the partition function of the total system and is given by
\begin{equation}
Z = \int \prod_{\mathfrak{j}}^{N}\mathrm{d}\mathfrak{p}_{\mathfrak{j}}
\mathrm{d}\mathfrak{q}_{\mathfrak{j}}
\int \mathrm{d}p \mathrm{d}q \exp(-H \beta),
\label{eq:parfuncenfor}
\end{equation}
\noindent with $\beta = 1/k_{\mathrm{B}} T$ and $T$ being the temperature of the bath.

To derive the classical thermal equilibrium state $\rho_\mathrm{S}$, we replace the Eq. (\ref{eq:parfuncenfor}) into Eq. (\ref{eq:theequstacenfor}) so that
\begin{equation}
\rho_{\mathrm{S}}(q,p)=\frac{\exp\left[-H_{\mathrm{S}}(q,p)\beta\right]\int\,\prod_{\mathfrak{j}}^{N}\mathrm{d}\mathfrak{q}\int\,\mathrm{d}\mathfrak{p} \exp\left\{-\left[\mathrm{H_{B}}(\mathfrak{q},\mathfrak{p}) + \mathrm{H_{SB}}(q,p,\mathfrak{q},\mathfrak{p}) \right]\beta \right\}}{\int\,\mathrm{d}q\int\,\mathrm{d}p \exp\left[-\mathrm{H_{S}}(q,p)\beta\right] \int\,\prod_{\mathfrak{j}}^{N}\mathrm{d}\mathfrak{q}\int\,\mathrm{d}\mathfrak{p} \exp\left\{-\left[\mathrm{H_{B}}(\mathfrak{q},\mathfrak{p}) + \mathrm{H_{SB}}(q,p,\mathfrak{q},\mathfrak{p}) \right]\beta\right\}}.
\label{equ:theequstafull}
\end{equation}
\noindent The integral over $\mathfrak{p}$ in Eq. (\ref{equ:theequstafull}) trivially cancels out with the corresponding contribution in $Z$. Due to the particular dependence of $V(q,\mathfrak{q})$ on $q$ and $\mathfrak{q}$, the integral over $\mathfrak{q}$ can be shifted by $q$ with the net result that whole integral cancels out with the corresponding contribution in $Z$. Thus, in the classical case for a system in contact to a thermal bath via central forces, we have that the thermal equilibrium state is given by
\begin{equation}
\rho_{\mathrm{S}}(q,p)=\frac{1}{Z_{\mathrm{S}}}\exp\left[ -\mathrm{H_{S}}(q,p)\beta \right],
\label{equ:theequstared}
\end{equation}
where $Z_{\mathrm{S}}$ is the partition function of the system and is given by
\begin{equation}
Z_{\mathrm{S}}  = \int \prod_{j}^{N} \mathrm{d}p_j \mathrm{d}q_j \exp(-H_{\mathrm{S}} \beta).
\end{equation}
Therefore, the thermal equilibrium state of a bounded particle in contact to a pairwise-self-interacting thermal bath via central forces, \emph{irrespective of the coupling strength}, is exactly given by the canonical Boltzmann distribution or the Gibbs state.
If the system S is composed, this result remains valid if each constituent of S is coupled to its own independent bath. 
This situation is relevant in, e.g., electronic energy transfer in light-harvesting systems \cite{PB12}.

This result is enlightening because, for a wide class of systems, there is no any physical reason or condition why thermal
equilibrium state is exactly characterized by the Gibbs state and that thermodynamic properties of the system are independent of the coupling to the bath.
%
%
The physical picture that emerges from this result is that \emph{in the long time regime}, any
dissipative mechanism is equally effective in taking the system to thermal equilibrium.
%
%
%
As we show below, quantum dissipative systems are accompanied by decoherence effects, which are produced by the bath-nature, \ie, the spectral density, and the functional form of the system-bath coupling that are capable of inducing a variety of different thermal states.

\section{Quantum Thermal Equilibrium State}

Before deriving the quantum thermal equilibrium state, we need to quantize the Hamiltonian in Eq. (\ref{eq:tothamcenfor}) which is obtained via the standard quantization procedure and is given by
\begin{equation}
\hat{H}=\frac{1}{2m}\hat{p}^{2}+\hat{U}_{\mathrm{S}}(\hat{q}) + \sum_{i=1}^{N}\left[\frac{1}{2m_{i}}\mathfrak{\hat{p}}_{i}^{2}+\sum_{j\neq i}^{N}\hat{U}_{\mathrm{B}}^{\mathrm{i,j}}(\mathfrak{\hat{q}}_{\mathrm{i}}-\mathfrak{\hat{q}}_{\mathrm{j}})\right] + \sum_{j=1}^{N}\hat{V}_{j}(\mathfrak{\hat{q}}_{j}-\hat{q}).
\label{eq:tothamcenforq}
\end{equation}

Based on the general description given in Refs. \cite{CTH09,CHT11,GSI88}, one can easily extend
the classical definition in Eq.~(\ref{eq:theequstacenfor}) to the quantum regime, namely,
\begin{equation}
\label{equ:qtheequsta}
\hat{\rho}_{\mathrm{S}} = \frac{1}{Z}\mathrm{tr}_{\mathrm{B}}
\exp\left\{-\left[ \hat{H}(\hat{p},\hat{q},{\hat{\mathfrak{p}}}_{\mathfrak{j}},{\hat{\mathfrak{q}}}_{\mathfrak{j}})
\right]\beta \right\}.
\end{equation}
The operator character of the various terms in Eq.~(\ref{equ:qtheequsta}) according to the Baker-Campbell-Hausdorff formula and their commutativity relations prevents us from continuing with the procedure followed for the classical case. However, these very same commutative relations allow the immediate formulation of the following set of inequalities:
\begin{subequations}
\label{equ:nonconmbound}
\begin{align}
\label{equ:nonconmbound.a}
[\hat{H}_{\mathrm{S}}, \hat{V} ] \neq 0 \Rightarrow
\Delta \hat{H}_{\mathrm{S}} \Delta \hat{V} &\ge \frac{1}{2}
|\langle[ \hat{H}_{\mathrm{S}},\hat{V}] \rangle|,
\\
\label{equ:nonconmbound.b}
[\hat{V}, \hat{H}_{\mathrm{B}} ] \neq 0 \Rightarrow
\Delta \hat{V} \Delta \hat{H}_{\mathrm{B}} &\ge \frac{1}{2}
|\langle[\hat{V} ,\hat{H}_{\mathrm{B}}] \rangle|,
\end{align}
\end{subequations}
where
$\Delta \hat{O} = \sqrt{\langle \hat{O}^2 \rangle - \langle \hat{O} \rangle^2}$ denotes
the standard deviation of $\hat{O}$, with $\langle \hat{O} \rangle = \mathrm{tr}(\hat{O} \hat{\rho})$,
$\hat{\rho}$ being the thermal equilibrium state of the system S and the bath B.
We should note that Eqs. ~(\ref{equ:nonconmbound}) apply to any system, bath,
and system-bath coupling Hamiltonians, but will prove particularly
noteworthy in the case of Eq.~(\ref{eq:hamcenfor}) where the classical
$\rho_{\rm S}$ is the Gibbs state or the canonical Boltzmann distribution. 
These last results to make sure the deviations from the Gibbs state are presented only in the quantum case.

This is a special case where the thermal equilibrium state differs from the canonical Boltzmann distribution. Moreover, very recently there have been some publications where the thermal equilibrium state differs from the Gibbs state is shown \cite{XL&13,YA&13,RH&14}. 
However, some general implications follow from Eqs.~(\ref{equ:nonconmbound}). 
Specifically, since $|\langle[ \hat{H}_{\mathrm{S}},\hat{V}] \rangle|$ is, to some extent, a measure of the quantum
correlations between the system and the bath, it dictates the lower bound of $\Delta \hat{H}_{\mathrm{S}}\Delta \hat{V}$. The important insight here is that the lower bound is different for each type of interaction since each particular form of $\hat{V}$, constant, linear, quadratic, etc., imposes a different commutation relation. Therefore, the general bounds in Eqs. (\ref{equ:nonconmbound}) can predict different thermal equilibrium states for each type of interaction. For the class of systems studied  above, this is a purely quantum effect.

For example, since $[ \Hs,\hat{V}]= 0$ implies a pure decohering interaction, which can be treated here in the framework of fluctuations without dissipation \cite{FO98}, the equilibrium state is an incoherent mixture of system's eigenstates and is expected to be well characterized by the canonical Boltzmann distribution \cite{DY&07}. 
In this case $[ \Hs,\hat{V}]= 0$, so that $\Delta \hat{H}_{\mathrm{S}} \Delta \hat{V} \ge 0$, meaning that the commutation relation here results in the minimum lower bound on
$\Delta \hat{H}_{\mathrm{S}} \Delta \hat{V}$. 
Note that the same lowest limit is obtained if, as in the classical case, the thermal equilibrium
state of the system $\hat{\rho}_{\mathrm{S}}$ is formally the canonical 
Boltzmann distribution $\hat{\rho}_{\mathrm{S}}^{\mathrm{can}}$.
Specifically, if
\begin{equation}
\hat{\rho} = \hat{\rho}_{\mathrm{S}}^{\mathrm{can}}\otimes \mathrm{tr}_{\mathrm{S}}\hat{\rho}
\end{equation}
then
\begin{equation}
|\langle[ \hat{H}_{\mathrm{S}},\hat{V}] \rangle| = \mathrm{tr}([ \hat{H}_{\mathrm{S}},\hat{V}] \hat{\rho})
=\mathrm{tr}([\hat{\rho}, \hat{H}_{\mathrm{S}}] \hat{V})=0
\end{equation}
since
$[\hat{\rho}_{\mathrm{S}}^{\mathrm{can}},\hat{H}_{\mathrm{S}}]=0$, giving
\begin{equation}
\Delta \hat{H}_{\mathrm{S}} \Delta \hat{V} \ge 0.
\end{equation}
This is just a consequence of the fact that the Boltzmann distribution is the zero-order-in-the-coupling
thermal equilibrium state and therefore, disregards quantum correlations between the system and the
bath.

We thus showed that the quantum equilibrium state described by the canonical Boltzmann distribution does apply for a wide class of systems. However, nature is quantum mechanical. As a consequence, as we have shown here, the uncertainty principle [Eqs. (\ref{equ:nonconmbound})] inhibits the system thermal equilibrium state from being described by this distribution, and the equilibrium state depends explicitly on the system-environment coupling, which induces one distribution for each type of interaction. Further, the uncertainty principle not only inhibits the systems thermal-equilibrium-state from being described by the canonical Boltzmann distribution, it also selects which thermal equilibrium states are physically accessible. This new result about the uncertainty principle gives us new ideas  and generates many views about where the applications of uncertainty principle can go to. For example, lately there have been a few publications about fundamental issues in quantum mechanics and quantum thermodynamics, \eg, nonlocality of quantum mechanics \cite{OW10} and the violation of the Second Law of Thermodynamics which states that the entropy of an isolated system never decreases \cite{HW12}. 

The result shown in Eq. (\ref{equ:nonconmbound}), formulated here for the first time within the framework of quantum thermodynamics, constitutes the cornerstone of the theory of pointer states (the states which are robust against the presence of the environment) \cite{Zur03} and could have deep consequences for an understanding of the thermalization of quantum systems. Below, it shown that the bath spectrum is also related to the lower boundary and thus, Eq.~(\ref{equ:nonconmbound}) will also allow for a clear connection to other fundamental features such as the failure of the Onsager's regression hypothesis in the quantum regime \cite{Tal86,FO96}.

\section{Thermodynamic Perturbation Theory}

As we showed in Eq. (\ref{equ:qtheequsta}), the quantum thermal equilibrium state $\hat{\rho}_{\mathrm{S}}$ differs from the canonical Boltzmann distribution.
However, in principle, we do not have a exact expression for this state. 
Further, some investigations have shown that in the weak coupling regime, the system thermal equilibrium state is characterized by the canonical Boltzmann distribution \cite{PSW06,GL&06}. 
Despite of, we will assume that the system-bath coupling is weak. In the next chapter, we will show that in contrast with the works of Goldstein \cite{GL&06} and Popescu \cite{PSW06}, the system thermal equilibrium state is not the Gibbs state or the canonical Boltzmann distribution despite of being in the weak coupling regime. 
This suggests that there are another parameters and effects that have to be taken into account in the moment calculating the equilibrium state. 
This assumption will allow us to perform the equilibrium calculation in a perturbative order. 
In what follows, it is convenient to split the Hamiltonian interaction $\hat{V} = \Hi$ from the uncoupled Hamiltonian $\hat{H} = \Hs + \Hb$, because, as we mentioned above, we assume that the system-bath interaction is weak. That is, we treat $\hat{V} = \Hi$ as a perturbation.

In order to obtain an approximate expression for $\hat{\rho}$ (the thermal equilibrium state of the complete system), we use the Kubo identity \cite{TK&91}, second we split the Hamiltonian interaction $V = \Hi$ from uncoupled Hamiltonian $\hat{H} = \Hs + \Hb$ and expand to the second order \cite{Zue07}:
\begin{equation}
\hat{\rho} \propto e^{-\beta \hat{H}} \left[ 1 - \int_{0}^{\hbar\beta} \mathrm{d}\sigma\, \hat{V}(-i\hbar\sigma) - \int_{0}^{\hbar\beta}\mathrm{d}\sigma \int_{0}^{\sigma}\mathrm{d}\theta\, \hat{V}(-i\hbar\sigma)\hat{V}(-i\hbar\theta) \right].
\label{eq:rhoper}
\end{equation}
The therm $\hat{V}(-i\hbar\sigma)= e^{\Hs\sigma}\hat{V}e^{-\Hs\sigma}$ is the evolution in imaginary time \cite{Ing02} and here, the expansion is in powers of $\beta V$. 

The result in Eq. (\ref{eq:rhoper}) allows us to calculate the lower bound in Eq. (\ref{equ:nonconmbound.a}), which depends on the two-time correlation of the bath operators and the spectral density of the bath. (See next chapter.)

\endgroup

\cleardoublepage
\begin{flushright}{\slshape    
	Nature isn't classical dammit, and if you want to make a simulation \\
	of nature, you'd better make it quantum mechanical, and by golly \\ 
	it's a wonderful problem because it doesn't look so easy.} \\ \medskip
	--- Richard P. Feynman
\end{flushright}

\bigskip

\begingroup
\let\clearpage\relax
\let\cleardoublepage\relax
\let\cleardoublepage\relax

\chapter{Influence of the Spectrum of the Bath}\label{ch:influence} 
\counterwithin{figure}{chapter}

As we showed in chapter \ref{ch:thermaleq}, the quantum thermal equilibrium state differs from the canonical Boltzmann distribution due to the Heisenberg's uncertainty principle. However, this does not mean that we cannot calculate the quantum thermal equilibrium state. The equilibrium state given in Eq. (\ref{equ:qtheequsta}) would have an influence of the commutation relations given in Eqs. (\ref{equ:nonconmbound}). Hence, despite that the set of inequalities (Eqs. \ref{equ:nonconmbound}) is general, it is not possible to infer the role that standard quantities such as the spectral density or the spectrum of the bath play in establishing the thermodynamic bounds above. 

In this chapter, we discuss the consequences and effects of the low temperatures and the non-Markovian character in the coupling among the system and the bath. These consequences are produced due to the commutation relations among the Hamiltonian of the system (bath) and the interaction depends on the two-time correlation of the bath operators, when we consider the case $|\langle[\Hs,\hat{V}_\mathrm{SB}]\rangle |\neq 0$.

\section{Role of the Spectral Density in the Uncertainty Principle}

To provide a concrete expression for the lower bound in Eq.~(\ref{equ:nonconmbound.a}), we return to the case of 
$\hat{V} = \sum_{\mathfrak{j}}^{{N}} \hat{\mathcal{V}}_{\mathfrak{j}}(\hat{\mathfrak{q}}_{\mathfrak{j}} - \hat{q})$, 
set the interaction between the bath particles to zero, i.e., $U^{\mathfrak{i},\mathfrak{j}}_{\mathrm{B}}=0$ 
and consider the second order picture of the system-bath central force interaction, i.e.,
$\hat{V} \approx \sum^{N}_{\mathfrak{j}} \frac{1}{2}\mathfrak{m}_{\mathfrak{j}} \omega_{\mathfrak{j}}^2
\left(\hat{q}_{\mathfrak{j}} - \hat{q}\right)^2$, which yields to the well-known Ullersma-Caldeira-Leggett
model \cite{CL83b,Ull66} describe in Eq. (\ref{eq:hamucl}). 
The Hamiltonian of \acl{ucl} can be written in general terms as
\begin{equation}
\hat{H}=\Hs + \Hb + \hat{V}_{\mathrm{SB}},
\end{equation}
where $\hat{V}_{\mathrm{SB}}= \hat{S}\otimes \hat{B}$. Here $\hat{B}=\sum_{\mathfrak{j}}^{N} \mathfrak{m}_{\mathfrak{j}} \omega_{\mathfrak{j}}^2 \hat{q}_{\mathfrak{j}}$ and $\hat{S} = \hat{q}$, which act in the Hilbert space of the bath and the system, respectively. Hence, the commutator $[ \hat{H}_{\mathrm{S}},\hat{V}_{\mathrm{SB}}]$, calculated to second order in $\hat{V}_{\mathrm{SB}}\beta$, is then given by
\begin{equation}
[ \hat{H}_{\mathrm{S}},\hat{V}_{\mathrm{SB}}]=[\Hs, \hat{S}\otimes \hat{B}] = [ \Hs,  \hat{S} ]\otimes \hat{B},
\end{equation}
provided by the fact that the commutator among the Hamiltonian of the system and the operator of the bath commute ($[\Hs,\hat{B}]=0$). Now, the  lower bound in Eq.~(\ref{equ:nonconmbound.a}) is given by
\begin{equation}
\left| \langle [ \Hs,\hat{V}_{\mathrm{SB}} ] \rangle \right| = |\mathrm{tr}\left( [ \Hs,  \hat{S} ]\otimes \hat{B} \hat{\rho} \right)|.
\end{equation}

Here we use the result found in Eq. (\ref{eq:rhoper}) where the equilibrium state at first order is given by
\begin{equation}
\hat{\rho}=e^{-\beta\Hs}\left[ 1-\int_{0}^{\hbar\beta}\mathrm{d}\sigma\,V_{\mathrm{SB}}(-i\hbar\sigma) \right],
\end{equation}
therefore
\begin{equation}
\left| \langle [ \Hs,\hat{V}_{\mathrm{SB}} ] \rangle \right| = \left|\mathrm{tr}\left( [ \Hs,  \hat{S} ]\otimes \hat{B} e^{-\beta\Hs}\left[ 1-\int_{0}^{\hbar\beta}\mathrm{d}\sigma\,\hat{S}\otimes\hat{B}(-i\hbar\sigma) \right]  \right)\right|,
\end{equation}
the first term that corresponds to  $\mathrm{tr}\left([ \Hs,  \hat{S} ]\otimes \hat{B} e^{-\beta\Hs}\right)=\mathrm{tr}\left([e^{-\beta\Hs}, \Hs]  \hat{S} \otimes \hat{B} \right)=0$ because the commutator between an operator on a smoth function of it vanishes.
Therefore, after tracing out over the bath, we get that the lower bound in Eq.~(\ref{equ:nonconmbound.a}) is given by
\begin{equation}
\label{equ:thermodynbound}
|\langle[ \hat{H}_{\mathrm{S}},\hat{V}] \rangle| \propto
\mathrm{tr}_{\mathrm{S}}
\left\{
[\hat{H}_{\mathrm{S}}, \hat{S}]\mathrm{e}^{-\hat{H}_{\mathrm{S}}\beta}
\int_0^{\hbar \beta} \mathrm{d}\sigma \hat{S}(-\mathrm{i}\sigma) K(\sigma)
\right\},
\end{equation}
where
$\hbar K(\sigma) = \langle \hat{B}(-\mathrm{i}\sigma)\hat{B}(0) \rangle_{\mathrm{B}}$
denotes the two-time correlation of the bath operators given by  \cite{Ing02}
\begin{equation}
K(\sigma)
= \pi^{-1}\int\mathrm{d}\omega J(\omega) \frac{\cosh\left(\tiny{\frac{1}{2}}\hbar \beta \omega - \mathrm{i}\sigma\right)}
{\sinh\left(\tiny{\frac{1}{2}}\hbar \beta \omega \right)},
\label{eq:noiseke}
\end{equation}
where $J(\omega)$ is the spectral density of the bath given by
\begin{equation}
J(\omega)= \pi \sum_{\mathfrak{j}}^{\infty} \tiny{\frac{1}{2}}\mathfrak{m}_\mathfrak{j}
\omega_\mathfrak{j}^3\delta(\omega-\omega_\mathfrak{j}).
\end{equation}
Note that as long as second order perturbation theory is valid, Eq.~(\ref{equ:thermodynbound})
holds for any $\hat{S}$ and $\hat{B}$ and can be straightforwardly generalized to
the case of $\hat{V}_{\mathrm{SB}}= \sum_{\alpha}\hat{B}_{\alpha}\otimes \hat{S}_{\alpha}$.

The main feature of the quantum thermodynamic bound in
Eq.~(\ref{equ:thermodynbound}) is the presence of the power spectrum of the bath
\begin{equation}
I(\omega,T) = \hbar J(\omega)\coth(\tiny{\frac{1}{2}}\hbar \beta \omega).
\end{equation}
For the most frequently used spectral density, associated with Ohmic dissipation, $J(\omega)=m\gamma \omega$, at high temperatures $\hbar \beta \rightarrow 0$, 
the power spectrum of the bath is flat and given by 
\begin{equation}
I(\omega,T) \approx 2 m \gamma k_{\mathrm{B}} T.
\end{equation}
The Ohmic spectral density is used to define proportionality to frequency merely at low frequencies instead of over the whole frequency range. In fact, in nature the Ohmic spectral density will not diverge for arbitrarily high frequencies. Note that in this high temperature limit, the upper limit of the integral in Eq.~(\ref{equ:thermodynbound}) vanishes, leading to the vanishing of the commutator, even if $[\hat{H}_{\mathrm{S}}, \hat{S}]\neq 0$.
A similar series-expansion analysis leads to the conclusion that the thermal equilibrium
state $\hat{\rho}_{\mathrm{S}}$ formally approaches the canonical Boltzmann distribution 
only when $\hbar \beta \rightarrow 0$.
In other words, in the high temperature limit the quantum correlations between the bath and 
the system disappear and the thermal equilibrium state is described by the canonical 
distribution, irrespective of the coupling strength or the functional form of the spectral 
density $J(\omega)$. 

%
For out-of-equilibrium quantum dynamics, the low temperature condition, finite $\hbar\beta$,
is associated with non-Markovian dynamics \cite{Ing02,HJ07}.
Since at fixed $T$, this non-Markovian character can be modified by the functional form of
the spectral density \cite{PB14}, Eq.~(\ref{equ:thermodynbound}) makes clear that the 
equilibrium system properties depend on the non-Markovian character. 
This means that the quantum equilibrium statistical properties of a system experiencing 
Markovian dynamics (flat spectrum) are expected to differ from those of the same system 
experiencing non-Markovian dynamics (non-flat spectrum), see Fig. \ref{fig:mnmspectrum}, which is in sharp contrast to
the classical case (see Eq.~\ref{equ:theequstared}).
This can be clearly understood in terms of the different thermodynamic lower bounds
resulting from either Markovian or non-Markovian interactions [see Eq.~(\ref{equ:thermodynbound})].

\begin{figure*}[t]
\centering
\includegraphics[width=0.8\textwidth]{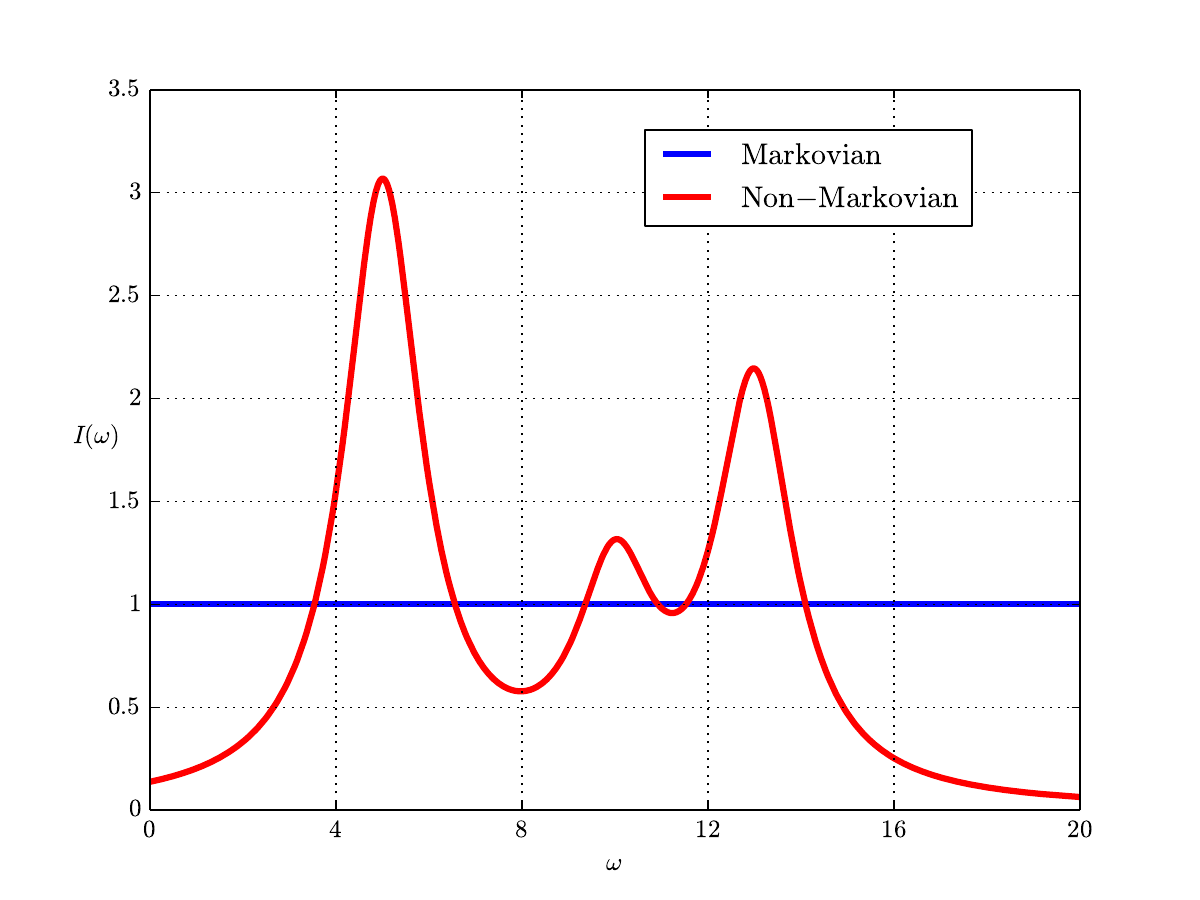}
\caption[Power spectrum for Markovian and Non-Markovian case.]{  \small Power spectrum for Markovian and Non-Markovian case. A flat spectrum (blue line) is associated to Markovian dynamics because in this case the power spectrum is independent of the frequency or the cutoff frequency in the spectral density is $\OD\rightarrow \infty$. In the opposite case, a non-flat spectrum (red line) is associated with non-Markovian dynamics, associated with the  regularized Drude spectral density with a high frequency cutoff $\OD$.
}
\noindent\rule{\textwidth}{1pt}
\label{fig:mnmspectrum}
\end{figure*}

To make a connection with previous studies, the failure Onsager's regression hypothesis 
in quantum mechanics \cite{Tal86,FO96} is discussed next.
In doing so, note that the hypothesis that knowing all mean values suffices to  determine the 
quantum dynamics of the correlation functions is valid only under Markovian dynamics \cite{Tal86,FO96} 
and when correlations between the bath and the system are negligible at equilibrium (in general, 
at any time) \cite{Swa81}.
Based on the fact that formal Markovian dynamics can only be achieved for flat spectra 
(bare Ohmic spectral density with $\hbar \beta \rightarrow 0$ \cite{PB14}), these two 
conditions can be seen as a single one when formulated in terms of Eq.~(\ref{equ:thermodynbound}).
Specifically, Markovian dynamics imply $|\langle[ \hat{H}_{\mathrm{S}},\hat{V}] \rangle| \rightarrow 0$
and, hence, the system-bath correlations vanish. 
This implies that Onsager's regression hypothesis, as well as the Boltzmann distribution, pertains 
exclusively to the classical realm.

To provide some insight into the magnitude and consequences of the fundamental limit found in (Eq. \ref{equ:thermodynbound}) and, in particular, of the role of the spectral density, an effective coupling to the bath is introduced. This effective coupling will be analyzed in two different regimes to show some of its implications and consequences. 

\section{Effective Coupling to the Bath}

It is clear from Eq. (\ref{equ:thermodynbound}) that if $\hbar\beta\rightarrow 0$, then the system bath correlations are 
$|\langle[ \hat{H}_{\mathrm{S}},\hat{V}] \rangle| \rightarrow 0$. However, 
 Eq. (\ref{equ:thermodynbound}) can also vanish if $K(\sigma)$ is sufficiently small.
This result introduces the concept of an effective coupling
to the bath that depends not only on the standard coupling described by the spectral density,
but also on other time and energy scales. 

To be specific, we use the Ullersma-Caldeira-Leggett model, where the equilibrium reduced density operator can be expressed in terms of 
the path integral expression as \cite{FH65,GSI88}
\begin{equation}
\label{equ:CDenMatPI}
\langle q''|\hat{\rho}_{\mathrm{S}}| q'\rangle=\frac{1}{Z} \int_{\bar{q}(0) = q'}^{\bar{q}(\hbar\beta)=q''}
\mathcal{D}\bar{q}\exp\left(-\frac{1}{\hbar}
S^{\mathrm{E}}_{\mathrm{S}}[\bar{q}]\right)\mathcal{F}[\bar{q}],
\end{equation}
where $Z$ is the partition function of the whole system, $S^{\mathrm{E}}[\bar{q}]$ is the Euclidean action of the system S,
which is obtained by inverting the global sign of the potential energy \cite{FH65,GSI88}. 
$\mathcal{F}[\bar{q}]$ is the influence functional that describes the influence of the 
bath B on the system and is given by
\begin{equation}
\label{emu:InfFun}
\mathcal{F}[q]=\exp\left(-\frac{1}{2\hbar}\int\limits_{0}^{\hbar\beta}\mathrm{d}\tau
\int\limits_{0}^{\hbar\beta}\mathrm{d}\sigma k(\tau-\sigma)q(\tau)q(\sigma)\right),
\end{equation}
where $k(\tau)$ condenses the influence of the bath on the system and can be written 
in terms of the Laplace transform, of the damping kernel $\gamma(t)$, as \cite{GSI88,Ing02}
\begin{equation}
\label{equ:CouKer}
k(\tau)=\frac{2m}{\hbar\beta}\sum\limits_{l=1}^{\infty}|\nu_{l}|\, 
\tilde{\gamma}(|\nu_{l}|)\cos(\nu_{l}\tau),
\end{equation}
where $\nu_{l}= l \Omega$, with $\Omega = 2\pi /\hbar\beta$, are the Matsubara frequencies and $\tilde{\gamma}(z)$, the Laplace transform of the damping kernel, defines an effective coupling to the bath which is given by
\begin{equation}
\tilde{\gamma}(z)= \frac{1}{m} \int_0^{\infty} \frac{\mathrm{d}\omega}{\pi} 
\frac{J(\omega)}{\omega}\frac{2z}{\omega^2 + z^2}.
\label{eq:effcouz}
\end{equation}

Note that $\tilde{\gamma}(|\nu_{l}|)$ contains all the information about the 
correlations of the bath operators and therefore defines the influence of the 
bath on the system at thermal equilibrium.

For the subsequent discussion, we adopt one of the most commonly used spectral densities, 
the regularized Drude model with a high frequency cutoff $\omega_{\mathrm{D}}$,
\begin{equation}
J(\omega)   =  m_0 \gamma\omega\,
\frac{\omega_{\mathrm{D}}^2}{\omega^2 + \omega_{\mathrm{D}}^2},
\label{eq:specdend}
\end{equation}
where $\gamma$ is the standard strength coupling constant to the thermal bath
and $\omega_{\mathrm{D}}$  dictates the degree of non-Markovian dynamics.
For this particular case, the effective coupling is given by 
\begin{equation}
\tilde{\gamma}(|\nu_{l}|)=\frac{\gamma}{ 1+ | l | \Omega / \omega_{\mathrm{D}}}.
\label{eq:effcoun}
\end{equation}
Below we analyze the effective weak coupling, $\Omega/ \omega_{{\mathrm{D}}} \gg 1$,
and the effective strong coupling, $\Omega/ \omega_{{\mathrm{D}}} \ll 1$, regimes.

\subsection{Strong effective coupling regime}

To quantify the consequences of non-flat spectra in this regime,   consider as the system a harmonic
oscillator of mass $m_0$ and frequency $\omega_0$ coupled to a thermal bath
 \cite{GWT84,GSI88}.
In particular, we use the \acl{ucl} with the regularized spectral density with a Drude cutoff frequency. We are interested in quantifying, in this regime, the following:

\begin{enumerate}
\item The deviation from the canonical partition function $Z_{\mathrm{can}} = \mathrm{tr}_{\mathrm{S}}\mathrm{e}^{\hat{H}_{\mathrm{S}}\beta}$.
\item The deviation from the canonical von-Neumann entropy $S_{\mathrm{can}}  = \mathrm{tr}_{\mathrm{S}}\left[\hat{\rho}_{\mathrm{can}}\mathrm{ln}(\hat{\rho}_{\mathrm{can}})\right]$.
\item The generation of squeezing in the thermal equilibrium state.
\end{enumerate}

We need to consider the following before continuing with our results. In our calculations we use the spectral density defined in Eq. (\ref{eq:specdend}). However, we know that the effective coupling to the bath is defined in terms of the spectral density as in Eq. (\ref{eq:effcouz}).
The use of different spectral densities changes the functional form of the effective 
coupling (Eq. \ref{eq:effcoun}) and therefore, of the thermal equilibrium properties. 

For the reasons noted above, a change or deviation from the canonical partition function or canonical entropy means that we have deviations from the canonical thermal equilibrium state. Hence, it is relevant to calculate the deviations enunciated previously. Furthermore, deviations not only imply that the thermal equilibrium properties change but, it also can produce the generation of squeezing in this state.

\begin{figure*}[t]
\includegraphics[scale=0.55]{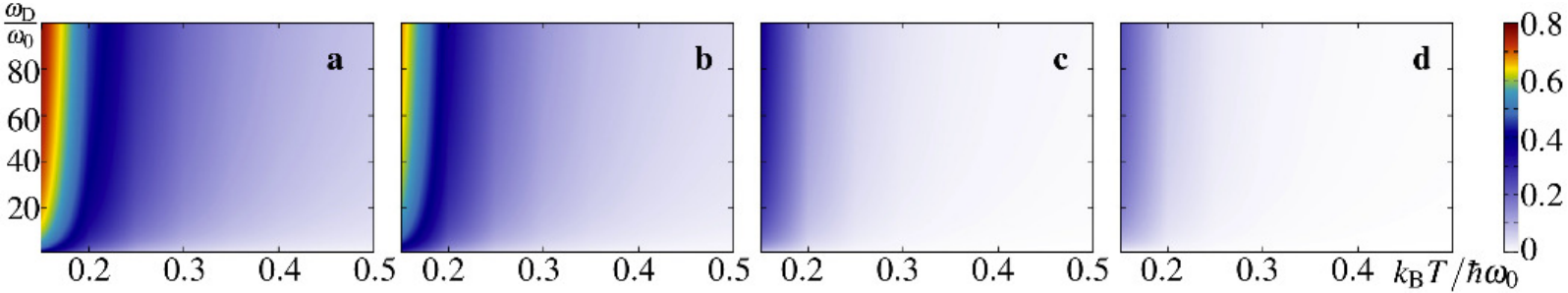}
\caption[$\log(Z/Z_{\mathrm{can}})$ for a harmonic oscillator as a function of the ratios
$k_{\mathrm{B}}T/\hbar \omega_0$ and $\omega_{\mathrm{D}}/\omega_0$.]{  \small $\log(Z/Z_{\mathrm{can}})$ for a harmonic oscillator as a function of the ratios
$k_{\mathrm{B}}T/\hbar \omega_0$ and $\omega_{\mathrm{D}}/\omega_0$.
We compare the partition function for $\gamma = 0.1\omega_0$ (a), $\gamma = 0.05\omega_0$ (b),
$\gamma = 0.01\omega_0$ (c) and $\gamma = 0.005\omega_0$ (d).\\
}
\noindent\rule{\textwidth}{1pt}
\label{fig:partition}
\end{figure*}

As we mentioned above, one of the forms to characterize the deviations from the thermal equilibrium state described by the canonical Boltzmann distribution or Gibbs state is to look for deviations in the canonical partition function ($Z_{\mathrm{can}} = \mathrm{tr}_{\mathrm{S}}\mathrm{e}^{\hat{H}_{\mathrm{S}}\beta}$) and in the canonical entropy ($S_{\mathrm{can}}  = \mathrm{tr}_{\mathrm{S}}\left[\hat{\rho}_{\mathrm{can}}\mathrm{ln}(\hat{\rho}_{\mathrm{can}})\right]$). 
%
We calculate the deviations in the partition function and the entropy as the ratio $Z/Z_{\mathrm{can}}$ and $S/S_{\mathrm{can}}$, respectively, where $Z(S)$ is the partition function (entropy) of the thermal equilibrium state. The partition function for the \acl{ucl} is well known and is given by \cite{GWT84,Ing02}
\begin{equation}
Z=\frac{1}{2\sinh\left( \frac{1}{2}\beta\hbar\omega_{\mathrm{eff}} \right)},
\end{equation}
where the effective frequency is given in terms of the variances for the position $\langle q^{2}\rangle$ and momentum $\langle p^{2}\rangle$ and for the effective coupling described in Eq. (\ref{eq:effcoun}) is given by
\begin{equation}
\omega_{\mathrm{eff}}=\frac{1}{\hbar\beta}\ln\left( \frac{\sqrt{\langle p^{2}\rangle \langle q^{2}\rangle}+\hbar/2}{\sqrt{\langle p^{2}\rangle \langle q^{2}\rangle}-\hbar/2} \right),
\end{equation}
and the variances are \cite{GWT84,GSI88}
\begin{subequations}
\label{eq:variances}
\begin{align}
\label{eq:posvar}\langle q^{2}\rangle &= \frac{1}{m_0\beta} \sum_{n=-\infty}^{\infty} \frac{1}{\omega_{0}^{2}+\nu_{n}^{2}+\gamma\OD|\nu_{n}|/(\OD+|\nu_{n}|)},
\end{align}
 \\
 \begin{align}
 \label{eq:momvar}\langle p^{2}\rangle &= \frac{m_0}{\beta} \sum_{n=-\infty}^{\infty} \frac{\omega_{0}^{2}+\gamma\OD|\nu_{n}|/(\OD+|\nu_{n}|)}{\omega_{0}^{2}+\nu_{n}^{2}+\gamma\OD|\nu_{n}|/(\OD+|\nu_{n}|)} .
 \end{align}
\end{subequations}
Hence, the ratio $Z/Z_{\mathrm{can}}$ is 
\begin{equation}
\frac{Z}{Z_{\mathrm{can}}}=\frac{\sinh\left( \frac{1}{2}\beta\hbar\omega_{\mathrm{can}} \right)}{\sinh\left( \frac{1}{2}\beta\hbar\omega_{\mathrm{eff}} \right)}.
\end{equation}

Here we can clearly see that deviations will depend on the differences between the effective frequency $\omega_{\mathrm{eff}}$ and the canonical frequency $\omega_{\mathrm{can}}=\omega_0$. However, the degree of non-Markovian dynamics $\OD$ would be finally the responsible for deviations since we have used the spectral density given in Eq. (\ref{eq:specdend}) to calculate the effective coupling (Eq. \ref{eq:effcoun}). If the $Z/Z_{\mathrm{can}}\neq 1$, we will have deviations from canonical Boltzmann distribution. As a result, deviations from the canonical result are evident in the partition function $Z$, which are showed below. 
Figure ~\ref{fig:partition} shows the logarithmic of the ratio of $Z$ to the canonical partition
function $Z_{\mathrm{can}}$ as a function of the dimensionless parameters
$k_{\mathrm{B}}T/\hbar \omega_0$ and $\omega_{\mathrm{D}}/\omega_0$ for
(from left to right) $\gamma=0.1\omega_0$, $\gamma=0.05\omega_0$, $\gamma=0.01\omega_0$ and
$\gamma=0.005\omega_0$.
Deviations are observed at low temperatures and for high cutoff frequencies (i.e., in the
effective strong coupling regime).
In the opposite limit, regardless of the coupling parameter $\gamma$, both calculated
partition functions show the same behavior, as expected from the discussion above.

For the von Neumann entropy
\begin{equation}
S  = \mathrm{tr}_{\mathrm{S}}\left[\hat{\rho}_{\mathrm{S}}
\mathrm{ln}(\hat{\rho}_{\mathrm{S}})\right],
\end{equation}
where $\hat{\rho}_{\mathrm{S}}$ is the result in Eq. (\ref{equ:qtheequsta}). The behavior of the ratio $\log(S/S_{\mathrm{can}})$ is  essentially the same as the one
described for the partition function ratio in Fig.~\ref{fig:partition}, and is shown in Figure \ref{fig:Entropy}. Figure ~\ref{fig:Entropy} shows the logarithmic of the ratio of $S$ to the canonical entropy $S_{\mathrm{can}}$ as a function of the dimensionless parameters
$k_{\mathrm{B}}T/\hbar \omega_0$ and $\omega_{\mathrm{D}}/\omega_0$ for
(from left to right) $\gamma=0.5\omega_0$, $\gamma=0.1\omega_0$ and
$\gamma=0.01\omega_0$.

In agreement with Eq. (\ref{eq:effcouz}), the use of different spectral densities changes the functional form of the effective 
coupling and therefore, of the thermal equilibrium properties.
Hence, as long as $\hbar\beta$ remains finite, different spectral densities lead 
to different thermal equilibrium states.

\begin{figure*}[t]
\includegraphics[scale=0.48]{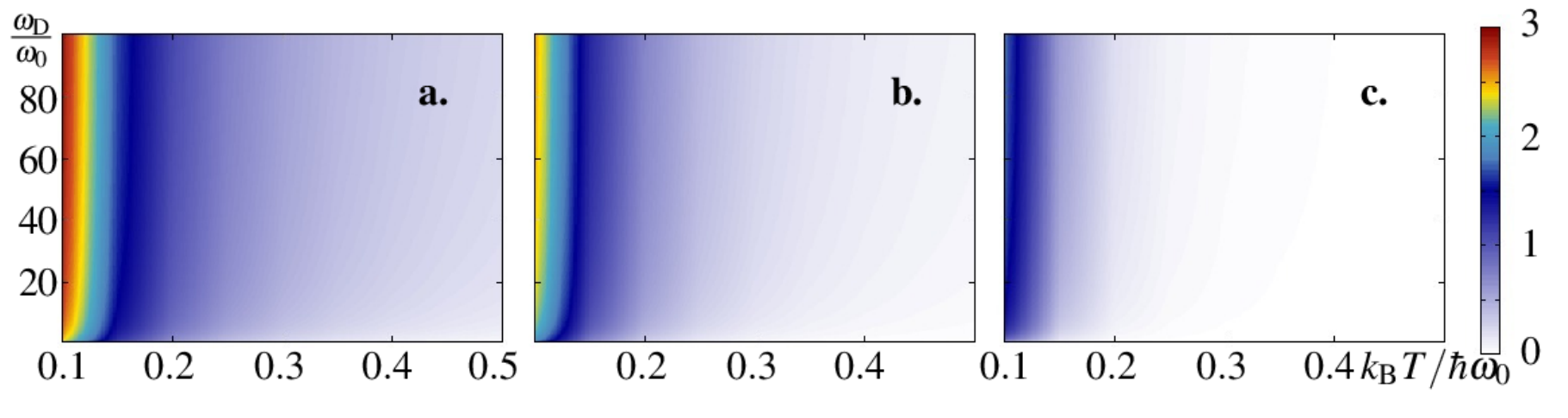}
\caption[$\log(S/S_{\mathrm{can}})$ for a harmonic oscillator as a function of the ratios
$k_{\mathrm{B}}T/\hbar \omega_0$ and $\omega_{\mathrm{D}}/\omega_0$.]{  \small $\log(S/S_{\mathrm{can}})$ for a harmonic oscillator as a function of the ratios
$k_{\mathrm{B}}T/\hbar \omega_0$ and $\omega_{\mathrm{D}}/\omega_0$.
We compare the partition function for $\gamma = 0.5\omega_0$ (a), $\gamma = 0.1\omega_0$ (b) and $\gamma = 0.01\omega_0$ (c).
}
\noindent\rule{\textwidth}{1pt}
\label{fig:Entropy}
\end{figure*}

\subsubsection{Generation of Squeezed Thermal Equilibrium States}

For this case the momentum and position variances given in Eqs. (\ref{eq:variances}) have the following relation \cite{GWT84,GSI88}
\begin{equation}
\langle p^2 \rangle = m_0^2 \omega_0^2 \langle q^2 \rangle + \Delta,
\label{equ:relationp2q2}
\end{equation}
where
\begin{equation}
\langle q^2 \rangle = \langle q^2_{\mathrm{cl}} \rangle +
\frac{2}{m_0 \beta}\sum_{n=1}^{\infty} \frac{1}{\omega_0^2 + \nu_n^2 +  \tilde{\gamma}(|\nu_n|)|\nu_n|}
\end{equation}
and the squeezing parameter
\begin{equation}
\Delta = - \frac{2m_0 \gamma}{\beta} \frac{\partial}{\partial \gamma} \ln Z' 
\end{equation}
with
\begin{equation}
Z'=\frac{1}{\hbar \beta\omega_0}
\prod_{n=1}^{\infty} \frac{\nu_{n}^{2}}{\omega_0^2 + \nu_n^2 +  \tilde{\gamma}(|\nu_n|)|\nu_n|}.
\end{equation}
We recall that for this model, the classical theory predicts
$\langle p^2_{\mathrm{cl}} \rangle = m_0^2 \omega_0^2 \langle q^2_{\mathrm{cl}} \rangle$
and $\langle q^2_{\mathrm{cl}} \rangle = k_{\mathrm{B}}T/m_0 \omega_0^2$, so that
$\Delta_{\mathrm{cl}}=0$.

Now, if we use the different regimes discussed  above, in the effective weak coupling regime $\Omega / \omega_{\mathrm{D}}\gg 1$,
disregarding terms of the order $\omega_0/\omega_{\mathrm{D}}$ and
$\gamma/\omega_{\mathrm{D}}$ the squeezing parameter gives \cite{GWT84}
\begin{equation}
\Delta=\frac{\pi \hbar \gamma m_0 \omega_{\mathrm{D}}}{6 \Omega}.
\end{equation}
Thus $\Delta$ vanishes at high temperatures, and the classical unsqueezed state
is recovered. However, in the other regime, the strong coupling regime $\Omega / \omega_{\mathrm{D}}\ll1$,
the squeezing parameter gives \cite{GWT84}
\begin{equation}
\Delta \approx \frac{\hbar \gamma m_0}{\pi}\, \mathrm{ln} \left(\frac{2\pi  \omega_{\mathrm{D}}}{\Omega}\right),
\end{equation}
meaning that the deviation from the canonical state translates into squeezing
of the equilibrium state.
This feature may be of relevance toward the generation of non-classical states, e.g., in
nano-mechanical resonators.

\subsection{Weak effective coupling regime}

To quantify the consequences of non-flat spectra in this regime, consider as the system two harmonic
oscillators of masses $m_0$ and frequencies $\omega_0$ linearly coupled with coupling constant $c_0$ and each one coupled to an independent thermal bath  \cite{GWT84,GSI88}. In this way our Hamiltonian is given by
\begin{equation}
\hat{H}= \sum_{i=1}^{2}\left(\frac{1}{2m_0}\hat{p}_{i}^{2} + \frac{1}{2}m_0 \omega_0^{2}\hat{q}_{i}^2 \right) - c_0 q_1 q_2 +
\sum_{\mathfrak{j},\alpha}^{\mathfrak{N},2} \left[\frac{\hat{p}_{\mathfrak{j},\alpha}^2}{2m_{\mathfrak{j}}}
+ \frac{m_{\mathfrak{j}} \omega_{\mathfrak{j}}^2}{2}
\left(\hat{q}_{\mathfrak{j},\alpha} - \hat{q}_{\alpha}\right)^2\right],
\label{equ:HamilEntang}
\end{equation}
\noindent with $\alpha=\{1,2\}$. Since the interaction is bilinear in the position operators of systems and baths, we consider linear response of the bath over the influence of the system, which is valid just for the case where the baths are macroscopic. This case yields a weak interaction between the harmonic oscillators and oscillators of each bath \cite{CL83b,Wei93}.

\begin{figure*}[t]
\centering
\includegraphics[width=0.8\textwidth]{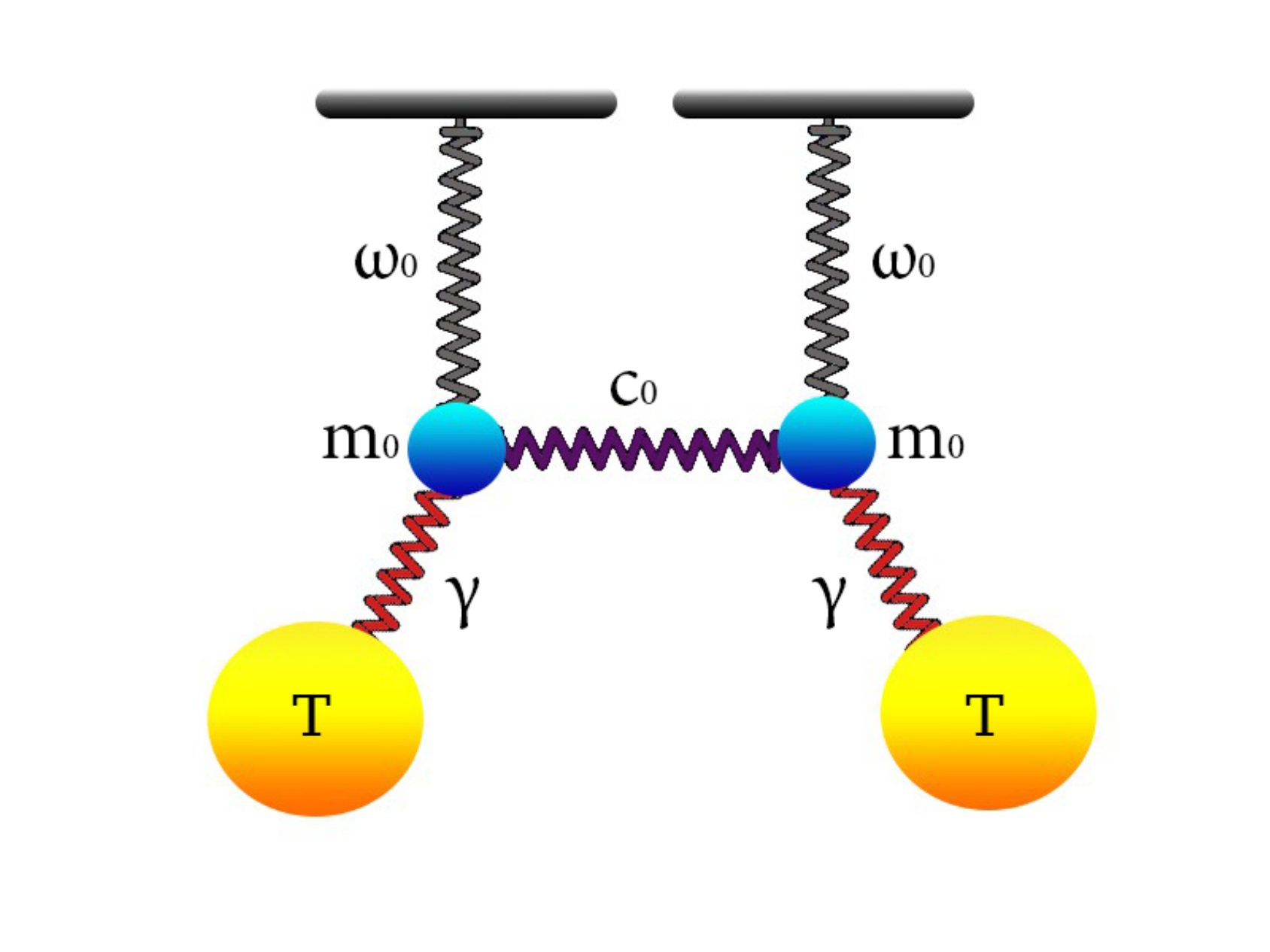}
\caption[Representation of the model for entanglement.]{  \small Representation of the model for entanglement. Here, we have two harmonic oscillators (blue circles) with masses $m_0$ and frequencies $\omega_0$ linearly coupled with coupling constant $c_0$. Likewise, each harmonic oscillator is coupled to a independent thermal bath (yellow circles) with a temperature $T$. $\gamma$ is the dissipation rate of the bath over the system.}
\label{fig:entanmodel}
\noindent\rule{\textwidth}{1pt}

\end{figure*}

The introduction of independent baths for each oscillator ensures that no deviations from
Boltzmann's distribution are present in the classical case. The Hamiltonian used in this regime to calculate the entanglement between two oscillators is described in Eq. (\ref{equ:HamilEntang}). 

%
%
Another consequence or effect of these deviations in the thermal equilibrium state is the capacity to maintain the entanglement between two oscillators at equilibrium, in particular, in the effective weak coupling regime $\Omega / \omega_{\mathrm{D}}\gg 1$. Previous studies predict that at equilibrium, the entanglement between the two harmonic oscillators can survive only when $k_{\mathrm{B}}T/\hbar \omega_0 \ll1$ (see, e.g., Ref.~\cite{GPZ10}).
However, this limit only applies in the Markovian regime and $\gamma\rightarrow 0$.
Thus, based on the discussion above, and supported by the recent observation that non-Markovian
dynamics assists entanglement in the longtime limit \cite{HRP12}, we expect that this
limit needs to be refined in order to account for the non-Markovian character of the interaction
and the finite value of $\gamma$.

\subsubsection{Entanglement measure}

One the most common measures of entanglement is called ``logarithmic negativity'' \cite{HH&09,VW02} which is defined by \cite{GPZ10}
\begin{equation}
E_{N}=-\frac{1}{2}\sum_{i=1}^{4}\log_{2}[\mathrm{min}(1,2|\lambda_{i}|)],
\label{equ:NegLog}
\end{equation}
where $\lambda_{i}$ are the eigenvalues of the symplectic covariance matrix $\sigma$,
\begin{equation}
\sigma=\left(\begin{array}{cccc}
\langle q_{1}^{2} \rangle - \langle q_{1} \rangle \langle q_{1} \rangle&\langle q_{1}q_{2} \rangle - \langle q_{1} \rangle \langle q_{2} \rangle&\langle q_{1}p_{1} \rangle - \langle q_{1} \rangle \langle p_{1} \rangle&\langle q_{1}p_{2} \rangle - \langle q_{1} \rangle\langle p_{2} \rangle\\
\langle q_{2}q_{1} \rangle- \langle q_{2} \rangle\langle q_{1} \rangle&\langle q_{2}^{2} \rangle- \langle q_{2} \rangle\langle q_{2} \rangle&\langle q_{2}p_{1} \rangle- \langle q_{2} \rangle\langle p_{1} \rangle&\langle q_{2}p_{2} \rangle- \langle q_{2} \rangle\langle p_{2} \rangle\\
\langle p_{1}q_{1} \rangle- \langle p_{1} \rangle\langle q_{1} \rangle&\langle p_{1}q_{2} \rangle- \langle p_{1} \rangle\langle q_{2} \rangle&\langle p_{1}^{2} \rangle- \langle p_{1} \rangle\langle p_{1} \rangle&\langle p_{1}p_{2} \rangle- \langle p_{1} \rangle\langle p_{2} \rangle\\
\langle p_{2}q_{1} \rangle- \langle p_{2} \rangle\langle q_{1} \rangle&\langle p_{2}q_{2} \rangle- \langle p_{2} \rangle\langle q_{2} \rangle&\langle p_{2}p_{1} \rangle- \langle p_{2} \rangle\langle p_{1} \rangle&\langle p_{2}^{2} \rangle- \langle p_{2} \rangle\langle p_{2} \rangle\\
\end{array} \right),
\end{equation} 
\ie, they are the eigenvalues of the matrix $-i\Sigma\sigma$, where $\Sigma$ is the symplectic matrix given by
\begin{equation}
\Sigma=\left(\begin{array}{cccc}
0&0&1&0\\
0&0&0&1\\
-1&0&0&0\\
0&-1&0&0\\
\end{array} \right).
\end{equation}

In accordance with the Peres-Horodecki separability criterion \cite{Sim00}, it follows that if the logarithmic negativity is zero, \ie, $E_ {N} = 0 $, then the state of the system of interest is separable, \ie, the oscillators can be described independently, no quantum correlation between them is present.

\subsubsection{Consequences Weak Effective Coupling}

One of the consequences of this effective regime is that we can maintain entanglement among two coupled harmonic oscillators at equilibrium. As we explain above, entanglement is measured by logarithmic negativity. 
Figure~\ref{fig:EntanC0p01} shows the logarithmic negativity (Eq. \ref{equ:NegLog}) for different
values of the damping constant $\gamma$ and different values of the coupling constant $c_0$ as a function
of the dimensionless ratios $k_{\mathrm{B}}T/\hbar \omega_0$ and
$\omega_{\mathrm{D}}/\omega_0$.
As expected, 
\begin{enumerate}
\item The more coupled the oscillators are, the higher the temperature and
the damping rate at which entanglement can survive at equilibrium.
\item The smaller the damping rate (the more isolated the system is), the higher the temperature
at which entanglement can be maintained.
\end{enumerate}
The new feature here is that the more non-Markovian the interaction, the higher the
temperature and the damping rate at which entanglement can be maintained \emph{at equilibrium.}

\begin{figure*}[ht]
\centering
\includegraphics[scale=0.24]{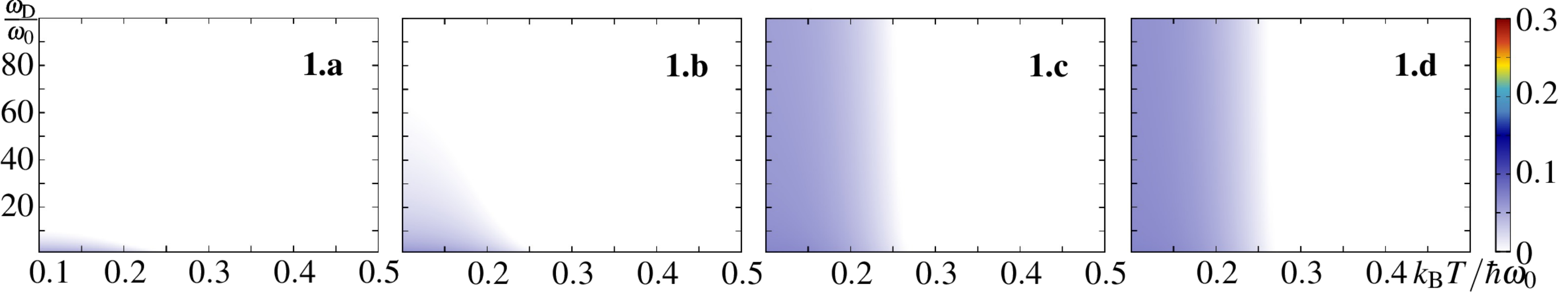}
\\
\includegraphics[scale=0.24]{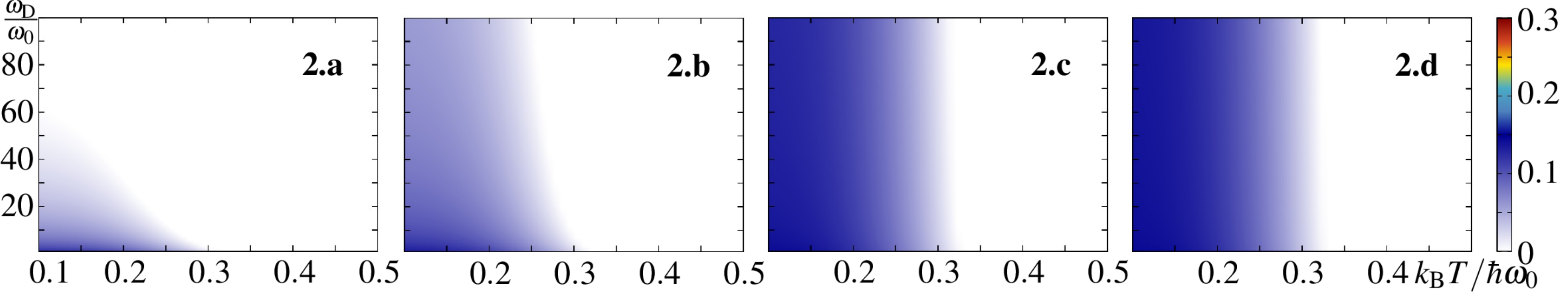}
\\
\includegraphics[scale=0.24]{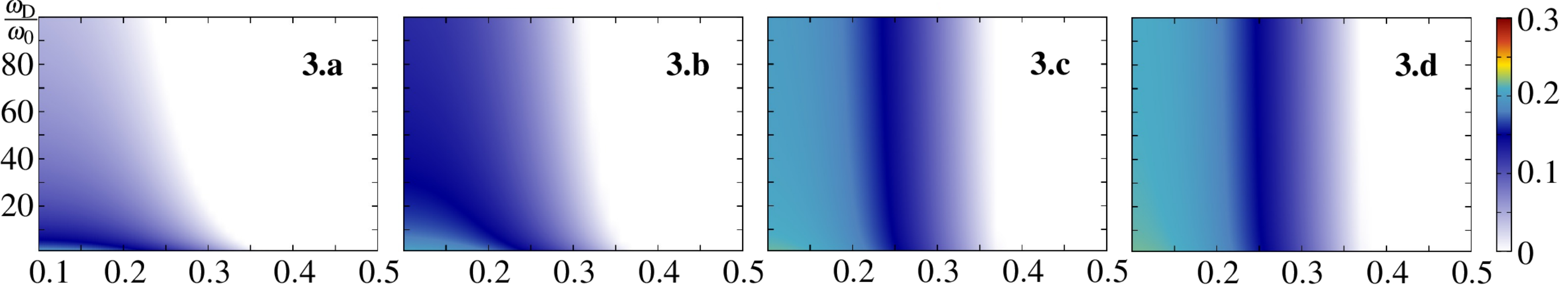}
\\
\includegraphics[scale=0.24]{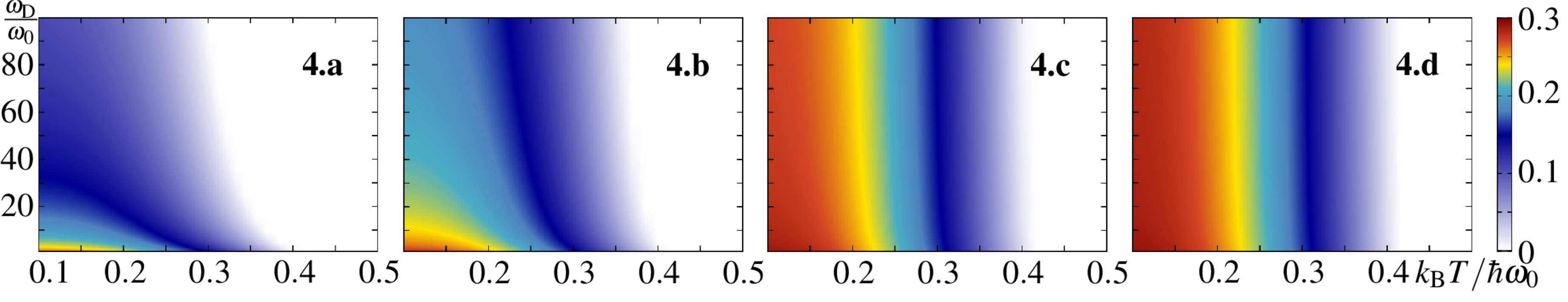}
\caption[Logarithmic negativity in the presence of non-Markovian interactions.]{\small Logarithmic negativity in the presence of non-Markovian interactions
for $c_0=0.05m_0\omega_0^2$ first row,   
$c_0 = 0.10 m_0 \omega_0^2$ second row,
$c_0 = 0.15 m_0 \omega_0^2$ third row and
$c_0 = 0.20 m_0 \omega_0^2$ fourth row.
Parameters are $\gamma = 0.1\omega_0$ (a), $\gamma = 0.05\omega_0$ (b),
$\gamma = 0.01\omega_0$ (c) and $\gamma = 0.005\omega_0$ (d) as a function of the dimensionless
parameters $k_{\mathrm{B}}T/\hbar \omega_0$ and $\omega_{\mathrm{D}}/\omega_0$.
}
\label{fig:EntanC0p01}
\end{figure*}

In this part of the work we showed the importance of the non-Markovian interactions and the low temperature regime, and effects generated by considering these conditions. The next part of the work consists in exploring the physical conditions to reach the effects predated for the low temperatures and the consequences that could emerge in quantum technologies, \eg, in quantum optomechanics.

\endgroup

\cleardoublepage
\part{Quantum Optomechanics}
\cleardoublepage
\begin{flushright}{\slshape    
   Anyone who is not shocked by quantum theory has not understood it.} \\ \medskip
	--- Niels Bohr
\end{flushright}

\bigskip

\begingroup
\let\clearpage\relax
\let\cleardoublepage\relax
\let\cleardoublepage\relax

\chapter{Cavity Optomechanics}\label{ch:cavoptomech} 
\counterwithin{figure}{chapter}

This chapter provides a short introduction to optomechanical and electromechanical systems where the target is knowing, understanding and controlling the interaction between radiation and mechanical nano or micro-resonators. Such interaction is described by the radiation pressure force performed over a resonator, with appropriate parameters for the mechanical quality factor and the optomechanical coupling we can generate proper phenomena of quantum mechanics as entanglement or squeezing \cite{LHM10,KMC14}. In brief, this chapter contains (i) the description of optomechanical and electromechanical systems used in this work, (ii) the Hamiltonian formulation of each one and (iii) the approximations performed based on the experimentally accessible setups.

\section*{Introduction}

As we known, Einstein was one of the people who either helps greatly or little in the construction of quantum mechanics. One of his works was about the statistics of the radiation pressure force fluctuations acting over a moveable mirror \cite{Ein09}, in which he reveals the dual wave-particle nature of radiation. 
Some years after, Frish \cite{Fri33} and Beth \citep{Bet36} were pioneers in their experiments where the photons transfer linear and angular momentum to atoms and macroscopic objects. 
Furthermore, the pressure of radiation has the ability to provide cooling for larger objects and allow them to reach their motional ground state. 
This ability has given rise to many applications between which we can include optical atomic clocks \cite{LWL09,YP10,LSV10}, systematic studies of quantum many-body physics in trapped clouds of atoms \cite{BZ08} and the detection of gravitational waves \cite{BG&73,LPT09,EM&07,AA&09,TZ&08}. 
%

Another ability that quantum optomechanics introduces is the preparation of the mechanical resonator in a well defined quantum state to later control the state in a coherent way and finally measure the state with an accuracy close to the limit imposed by the quantum mechanics. This is important because preparation of the quantum state, coherent control and quantum measurement  are the three steps required to develop any quantum technology. If we can perform these three steps, the quantum regime would be accessible and this would give rise to many applications ranging from fundamental to technological applications \cite{AKM13}. 

As noted above, studying mechanical systems in the quantum regime is a great challenge since we could have different technologies in a few years or decades. 
Hence, the idea of studying this kind of systems is to couple them to an auxiliary system with the objective to perform quantum control over it. The auxiliary systems are picked up in agreement with the characteristics of the resonators or mechanical system among which we can find mass, frequency, quality factor, temperature, etc. The auxiliary systems can take diverse forms, \eg, an electric circuit (quantum electromechanics), optical cavities (quantum optomechanics) or atomic systems (ion trap).

Now, we move to the next section where we are going to describe the different opto-, electro-mechanical systems. Here, the important point is to see the differences and similarities between these systems.

\section{Optomechanical and Electromechanical Systems}\label{sec:optomsyst}

Cavity optomechanics (electromechanics) focuses on the study of the interaction between electromagnetic radiation and mechanical systems. In this kind of cavities, light is confined to a volume separed by two mirrors, in the case of optomechanical systems, or plates in the case of electromechanical systems, where one of these mirrors (plates) oscillates. This effect can be used for different applications, \eg, to obtain quantum effects in a nano-resonator, which has around 10$^{14}$ atoms, \ie, to reach phenomena predicted by quantum mechanics in the regime of mesoscopic objects.

The starting point consists of the formulation of the Hamiltonian that describes the mechanical system coupled to a radiation mode that interacts with a vibrational mode of the system (this vibrational mode can be an oscillating mirror or plate). Before starting the Hamilitonian formulation, we define the proper energies of the different modes. The electromagnetic radiation mode, or optical mode, has an angular frequency $\omega_{\mathrm{cav}}$ and the mechanical mode has an angular frequency $\omega _m$.
This configuration can be represented by two harmonic oscillators one for the cavity and the other for the vibrational mode. Hence, the energies of the optical and mechanical mode are $\hbar\Oc$ and $\hbar\OM$, respectively. The total Hamiltonian should include both the terms that describe the two oscillators and their interaction as well as the terms that describe dissipation and fluctuations. These last terms will be discussed shortly to perform the linear approximation in the Hamiltonian, in each one of the different systems that we will show in this section.

We are going to describe two typical systems. The first system described will be the typical optomechanical system analyzed in a large amount of scientific papers and successfully used in the experiments to date, the Fabry-Perot cavity \cite{MC&07,WN&07,PW09,WV&11,SC&13,AKM13,RD&11,HSK10,VD&12}. In this cavity one the mirrors, the end-mirror is moveable which corresponds to a vibrational mode. The second and the last system is an electromechanical system where the vibrational mode is present in a capacitor coupled to a LC circuit. 

\subsection{Fabry-Perot Cavity}

A Fabry-Perot cavity has many resonant frequencies, however we are interested just in one of them, given by 
\begin{equation}
\omega_n=\frac{n\pi c}{L},
\end{equation}
where $c$ and $L$ are the speed light and the cavity length, respectively. $n$ is an integer that characterizes the normal mode in the cavity \cite{AKM13}. We can see the details of the Fabry-Perot cavity in Fig. \ref{fig:fabper}. Since we are interested only in one of them, we choose one and call it $\Oc$. Experimentally, we should have present another important parameters like the quality factor given by
\begin{equation}
Q_\mathrm{c}=\Oc \tau,
\end{equation}
where $\tau=\kappa^{-1}$ is the photon lifetime inside the cavity and $\kappa$ is the cavity decay rate. Furthermore, the interaction between the electromagnetic radiation and the mechanical systems is produced via radiation pressure force generated by each photon inside the cavity. Each photon, due to reflection that occurs in the cavity, transfer linear momentum $p=2\hbar\omega_\mathrm{cav}/c$. As each single photon takes time $\tau_\mathrm{cav}=2L/c$ in a round trip from one mirror to other, the total momentum transfer of each photon over the cavity is
\begin{equation}
p_{tot}=\frac{\hbar\Oc}{\kappa L}, 
\end{equation}
and the radiation pressure force, considering that there are $n_{\mathrm{ph}}=\langle \ad \an \rangle$ photons inside the cavity, is given by
\begin{equation}
\langle \hat{F}_\mathrm{rad} \rangle = \frac{\hbar\Oc}{L}\langle \ad \an \rangle.
\end{equation}

Now, since we are interested in the most common model used in optomechanics, we use two harmonic oscillators which describe the optical mode ($\omega_{\mathrm{cav}}$) and the mechanical mode ($\OM$). Therefore, the uncouple Hamiltonian is given by
\begin{equation}
\hat{H}_{0}=\hbar\omega_{\mathrm{cav}}\ad\an + \hbar\OM\bd\bn.
\label{eqn:H0}
\end{equation}
\noindent Since we are interested in the case of a cavity with a movable end mirror, the coupling between optical and mechanical mode is parametric, \ie, the cavity resonance frequency is modulated by the mechanical amplitude. Hence, we get in the linear approximation 
\begin{equation}
\omega_{\mathrm{cav}}(x)\approx \omega_{\mathrm{cav}} + x\frac{\partial\omega_{\mathrm{cav}}}{\partial x} + ...
\label{eq:omegafx}
\end{equation}
For the cavity discussed here, a Fabry Perot cavity of length $L$, we have the optical frequency shift per displacement $G=\partial\omega_{\mathrm{cav}}/\partial x=\omega_{\mathrm{cav}}/L$. Therefore, we expand the first term in Eq. (\ref{eqn:H0}), so that
\begin{equation}
\hbar\omega_{\mathrm{cav}}(x)\ad\an\approx\hbar(\Oc + G\hat{x})\ad\an,
\label{eqn:expansionG}
\end{equation}
\noindent where $\hat{x}=x_{\mathrm{ZPF}}(\bn+\bd)$ with $x_{\mathrm{ZPF}}=\sqrt{\hbar/2m_0 \OM}$ the zero-point fluctuation amplitude of the mechanical oscillator. Hence, the Hamiltonian of interaction is given by
\begin{equation}
\hat{H}_{\mathrm{int}}=\hbar g_{0}\ad\an(\bn+\bd),
\label{eqn:Hint}
\end{equation}
\noindent where $g_{0}=Gx_{\mathrm{ZPF}}$, is the vacuum optomechanical coupling strength. Because $g_{0}$ is formed by a part of the optical mode ($G$) and other from the mechanical mode ($x_{\mathrm{ZPF}}$), it quantifies the interaction between radiation (a single photon) with matter (a single phonon). Now, we apply a unitary transformation $\hat{U}=\exp(i\omega_{L}\ad\an t)$ to change the description of the optical mode to a frame rotating at the laser frequency $\omega_{\mathrm{L}}$. This generates a new Hamiltonian of the form
\begin{equation}
\hat{H}=\hbar\Delta\ad\an + \hbar\OM\bd\bn + \hbar g_{0}\ad\an(\bn+\bd)
\label{eqn:hamil}
\end{equation}
\noindent where $\Delta=\Oc-\omega_{\mathrm{d}}$ is the laser detuning. Now, since we are interested in linear coupling, we introduce the linearized approximate description of cavity optomechanics ($\an=\overline{\alpha}+\delta\an$) and therefore the linearized hamiltonian is given by
\begin{eqnarray}
\nonumber \hat{H}&=&\hbar\Delta\ad\an + \hbar\OM\bd\bn + \hbar g_0(\overline{\alpha}+\delta\an)^{\dagger}+(\overline{\alpha}+\delta\an)(\bn+\bd) \\
&=&\hbar\Delta\ad\an + \hbar\OM\bd\bn +\hbar g_{0}(\overline{\alpha}^{2}+\overline{\alpha}(\delta\hat{a}+\delta\hat{a}^{\dagger})+\delta\hat{a}^{\dagger}\delta\hat{a})(\hat{b}^{\dagger}+\hat{b}),
\label{eqn:hamlin}
\end{eqnarray}
\noindent where $\left\langle\an\right\rangle=\overline{\alpha}$ is the average coherent amplitude of the cavity field and $\delta\an$ is the fluctuating term due to vacuum noise.  In Eq. (\ref{eqn:hamlin}) the first term $\overline{\alpha}^{2}=\langle \ad\an \rangle$ can be omitted because it corresponds to the average of the radiation pressure force. The third term $\delta\ad\delta\an$ can be neglected because it is much smaller than the second term for a factor of $\overline{\alpha}$. Therefore, we finally obtain that the linear Hamiltonian of the system  is given by
\begin{equation}
\hat{H}=\hbar\Delta\ad\an + \hbar\OM\bd\bn + \hbar g(\delta\ad+\delta\an)(\bn+\bd),
\label{eqn:hamlint}
\end{equation}
where $g=g_0\overline{\alpha}$ is the optomechanical coupling strength \cite{MC&07,AKM13}.

\begin{figure}[!ht]
\centering
\includegraphics[width=\textwidth]{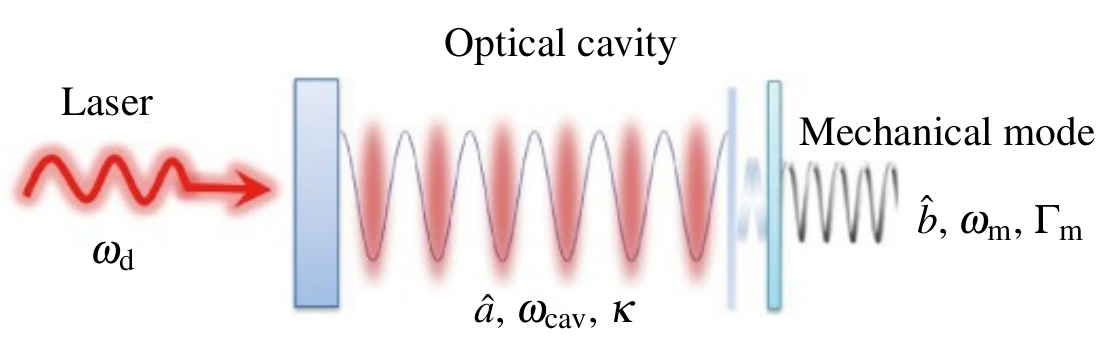}
\caption[Schematic representation of a generic optomechanical system in the optical domain with a laser-driven optical cavity and a vibrating end mirror.]{Schematic representation of a generic optomechanical system in the optical domain with a laser-driven optical cavity and a vibrating end mirror. $\omega_{\mathrm{d}}$, $\omega_{\mathrm{cav}}$ and $\omega_{\mathrm{m}}$ are the frequencies of the laser, cavity and mechanical resonator, respectively. $\kappa$ and $\Gamma_{\mathrm{m}}$ are the dissipation parameters of the cavity and mechanical resonator, respectively. The system consists of two mirrors, one of them is fixed and the end-mirror (mechanical mode) is moveable.}
\label{fig:fabper}
\noindent\rule{\textwidth}{1pt}
\end{figure}

The Hamiltonian described in Eq. (\ref{eqn:hamlint}) is the starting point to define the most important aspects in optomechanical systems. These aspects depend explicitly on the detuning and if the system is the good cavity regime ($\OM>\Gamma_{\mathrm{m}}$). For example, if $\Delta=\OM$ (red-detuned regime) and later to perform the rotating wave approximation, we get that the Hamiltonian of the system is 
\begin{equation}
\hat{H}=\hbar\Delta\ad\an + \hbar\OM\bd\bn + \hbar g(\bn\delta\ad+\bd\delta\an), 
\label{eqn:hamlintt}
\end{equation}
which describes the interaction between a phonon of the mechanical resonator with a photon of the laser that fed the cavity, and in this way, it is possible to cool the resonator. There is a lot of papers about this process which is known as \textit{sideband cooling} \cite{RD&11,AKM13,AA&09,TAS08,MC&07,PW09}. One of the motivations to focus in red-detuned regime is our interest in the non-Markovian cooling of mechanical resonators. We can see how can influence the non-Markovian character in the cooling.

\subsection{Electromechanical System}

In quantum electromechanics there are two common systems used in the last few years, LC circuits and superconductor transmission lines where these systems are coupled to a capacitor with a oscillating plate \cite{AKM13,KK&06,BH&04}. The main idea of the capacitive coupling in these electric systems with the mechanical resonators is to generate a dependence on the capacitance with the displacement of the oscillating plate in the capacitor, so we can change the frequency of the electric normal mode. In other words, to get the same change in the frequency (Eq. \ref{eq:omegafx}) as we did before in the Fabry-Perot cavity with the end-mirror.

Now, we consider a plate of the capacitor in a LC circuit as the mechanical resonator, as we can see in the Fig. \ref{fig:LCcircuit}. The quantized Hamiltonian of the system in terms of the canonical variables for the circuit, $Q$ the charge of the capacitor and $\varphi$ the flux through the inductor, is given by \cite{TS&11}
\begin{equation}
\hat{H}=\frac{\hat{p}^{2}}{2m}+\frac{1}{2}m\omega^{2}\hat{x}^{2}+\frac{\varphi^{2}}{2L}+\frac{Q^{2}}{2C(\hat{x})}+QV(t),
\label{eq:hamLCcir}
\end{equation}
where $\hat{x}$ y $\hat{p}$ defined the displacement and the linear momentum of a mechanical resonator with mass $m$ and frequency $\omega$, and $L$ and $C(\hat{x})$ are the inductance and the capacitance of the circuit, respectively. $V(t)=V_0\cos(\omega_{\mathrm{d}}t)$ is the voltage time-dependent. As we did in the optomechanical system, we consider linear dependence of the capacitance with the displacement of the resonator so that 
\begin{equation}
C(\hat{x})=C_0\left( 1+\frac{\hat{x}}{d_0} \right),
\end{equation}
where $C_0$ and $d_0$ are the capacitance and the separation between the plates in the capacitor when $\hat{x}=0$, respectively. Hence, we get the Hamiltonian
\begin{equation}
\hat{H}=\frac{\hat{p}^{2}}{2m}+\frac{1}{2}m\omega^{2}\hat{x}^{2}+\frac{\varphi^{2}}{2L}+\frac{Q^{2}}{2C_0 d_0}\hat{x}+QV(t).
\label{eq:hamLCcir2}
\end{equation}

\begin{figure}[t]
\centering
\includegraphics[width=0.8\textwidth]{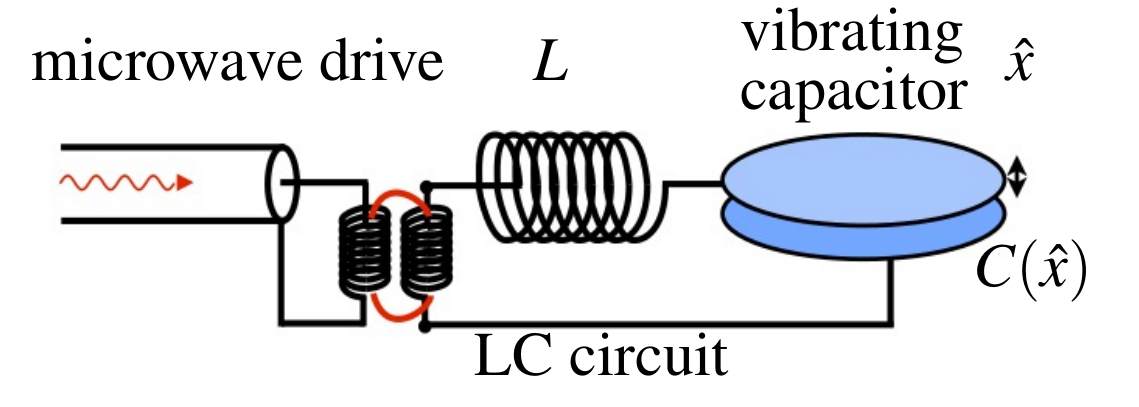}
\caption[Sketch of a generic electromechanical system in the microwave domain with a vibrating capacitor. ]{Sketch of a generic electromechanical system in the microwave domain with a vibrating capacitor. We have depicted a microwave drive entering along a transmission line that is inductively coupled to the LC circuit representing the microwave resonator.}
\label{fig:LCcircuit}
\noindent\rule{\textwidth}{1pt}
\end{figure}

If we define the ladder operators for the LC circuit and the mechanical resonator as
\begin{align}
\nonumber \hat{a}&=\sqrt{\frac{\omega_c L}{2\hbar}}Q+\frac{i}{\sqrt{2\hbar\omega_c L}}\varphi, \\
\hat{b}&=\sqrt{\frac{m \omega_0}{2\hbar}}\hat{q}+\frac{i}{\sqrt{2\hbar m\omega_0}}\hat{p},
\end{align}
where $\omega_c=1/\sqrt{LC_0}$ is the frequency of the circuit normal mode, therefore the Hamiltonian in Eq. (\ref{eq:hamLCcir2}) in terms of the ladder operators is
\begin{equation}
\hat{H}=\hbar\omega_c\ad\an+\hbar\omega_0\bd\bn - \hbar g_0(\bn+\bd)(\an+\ad)^{2} + \hbar\epsilon(e^{i\omega_{\mathrm{d}}t}+e^{-i\omega_{\mathrm{d}}t})(\an+\ad),
\end{equation}
where $g_0$ is the coupling strength between the mechanical resonator and the LC circuit, $\epsilon$ is the coupling strength between the LC circuit and the voltage applied. Following almost the same steps as in the optomechanical case, if we change to an interaction picture in the $\omega_{\mathrm{d}}$ frequency, and later perform the rotating wave approximation \cite{MW11}, the Hamilitonian is
\begin{equation}
\hat{H}=\hbar\Delta\ad\an + \hbar\omega_0\bd\bn - \hbar g_0\ad\an(\bn+\bd) + \hbar\epsilon(\an+\ad),
\label{eq:hamelec}
\end{equation}
where $\Delta=\omega_c - \omega_{\mathrm{d}}$ is the detuning between the voltage frequency and the frequency of the circuit normal mode. Hence, the Hamiltonian for the electromechanical system in Eq. (\ref{eq:hamelec}) has the same form of the Hamiltonian for the optomechanical system in Eq. (\ref{eqn:hamil}). In others words, these two systems are analogous and the different effects as the sideband cooling also can be studied under the electromechanical system.

\endgroup

\cleardoublepage
\begin{flushright}{\slshape    
  Before the discovery of quantum mechanics, the framework of physics \\
  was this: If you tell me how things are now, I can then use the laws of \\
  physics to calculate, and hence predict, how things will be later.} \\ \medskip
	--- Brian Greene
\end{flushright}

\bigskip

\begingroup
\let\clearpage\relax
\let\cleardoublepage\relax
\let\cleardoublepage\relax

\chapter{Optomechanical Cooling}\label{ch:optomechcool} 
\counterwithin{figure}{chapter}

Nowadays there is great effort, both theoretical and experimental, to fabricate and control micro- or nano-mechanical devices which have a numerous promising technological applications \cite{GNQ13}. Further, the nano-mechanical devices could work in the quantum regime and thus can explore different quantum effects as the generation of non-classical states \cite{MVT98} or entanglement \cite{MC&07,MC&12}. Since the potential quantum technologies, or exploration of quantum effects, requires preparing the system, \eg, in the ground state, it is of great interest searching for different schemes that can enhance the effectiveness of cooling in these systems. 
However, the mechanical frequency is in the GHz range for a some nano-mechanical setups \cite{CA&11,EC&09}, even dilution refrigerator temperatures of 20 mK are not sufficient to ensure $k_\mathrm{B}T\ll \hbar\OM$. Thus, we need to include an additional cooling in the mechanical mode.
Many schemes have been proposed recently \cite{MVT98,MC&07,MC&12,WV&11,AKM13,SC&13}, all performed under Markovian processes. Our principal analysis below is performed considering non-Markovian processes because is closer to reality as it has been pointed out by recent experiments in micro-mechanical devices \cite{GT&13}.
Based on results presented above, we emphasize that non-Markovian processes could allow for a minimum 
phonon number lower than the one predicted by Markovian processes. 
We do not only focus in the non-Markovian cooling, but based on optimal control strategies, 
we present a methodology to maintain the minimum phonon number in the resonator for long times.

\section{Markovian Cooling}
%

The purpose of the present section is to develop a quantum theory of optomechanical cooling, which in particular describes the limits for cooling that cannot be obtained from a discussion of the damping rate alone. We will focus in this section on a Fabry-Perot cavity, which was described in the previous chapter. 
Classical theory of an oscillator or resonator at initial temperature $T_{\mathrm{in}}$ exposed to some type of damping, \ie, the damping associate to the laser $\gamma_{\mathrm{opt}}$ predicts that its temperature can be reduced to
\begin{equation}
T_{\mathrm{f}}=T_{\mathrm{in}}\frac{\gamma_{\mathrm{m}}}{\gamma_{\mathrm{m}}+\gamma_{\mathrm{opt}}}.
\end{equation}
However, this classical expression is not consistent at sufficiently low $T_{\mathrm{f}}$, because fluctuations of the radiation pressure force due to photon shot noise can take place to establish a lower bound of the achievable temperature. In the following, we explain the quantum theory that permits to calculate the limits of cooling factor or phonon number to reach the ground state in the resolved sideband regime $\kappa\ll \OM$.
As we mentioned above, we will focus explicitly on radiation pressure cooling in cavity setups, which is conceptually the simplest case.

\subsection{Cooling at Equilibrium}

As we have done in the previous sections, we work in the weak coupling regime, \ie, when $g\le\kappa$. 
The quantum theory of optomechanical cooling \cite{AKM13} is related to earlier approaches for cavity-assisted laser cooling of atomic and molecular motion \cite{HG&98,VC00}. The idea is best explained in a Raman-scattering picture. 
Photons impinging at a frequency red-detuned from the cavity resonance will, 
via the optomechanical interaction, preferentially scatter upwards in energy in 
order to enter the cavity resonance, absorbing a phonon from the oscillator in the process. 
In these processes, phonon absorption  happens at a rate $A^{-}$, which is 
related to Fermi's golden rule that represents the transition rate between system levels $i$ and $f$. 
More precisely, the transition rate  $A^{-}$ happens from  state $n$ to $n-1$ of the phonon. 
The opposite process, where an extra phonon is created happens at a smaller rate $A^{+}$, \ie, the transition rate from phonon state $n-1$ to $n$. 

Given the transition rates ($A^+$ for upward transitions in the mechanical oscillator, $A^-$ for downward transitions), the optomechanical damping rate $\gamma_{\mathrm{opt}}$ is 
\begin{equation}
\gamma_{\mathrm{opt}}=A^- - A^+.
\end{equation}
However, as we mentioned above, our interest is the average phonon number $\overline{n}$, which changes according to the rates $\Gamma_{n\rightarrow n\pm 1}$, leading to the equation of motion for the average phonon number given by
\begin{equation}
\dot{\overline{n}}=(\overline{n}+1)(A^+ + A_{\rm th}^+)-\overline{n}(A^- + A_{\rm th}^-).
\label{equ:dotn}
\end{equation}
Here, $A_{\rm th}^{\pm}$ are the transition rates due to the oscillator's thermal environment, which are given in terms of a mean phonon number $\overline{n}_{\rm th}$ as $A_{\rm}^{+}=\overline{n}_{\rm th}\gamma_{\rm m}$ and $A_{\rm}^{-}=(\overline{n}_{\rm th}+1)\gamma_{\rm m}$. Now, since we are interested in the steady-state final phonon number, it is required that $\dot{\overline{n}} = 0$ in Eq. (\ref{equ:dotn}), hence we get that the minimum phonon number is 
\begin{equation}
\overline{n}_{\rm f}=\frac{A^{+}+\overline{n}_{\rm th}\gamma_{\rm m}}{\gamma_{\rm opt}+\gamma_{\rm m}}.
\end{equation}
In absence of any coupling or a thermal bath over the mechanical oscillator or considering that the dissipation of the mechanical oscillator tends to zero ($\gamma_{\rm m}\rightarrow 0$), this leads to a minimum phonon number given by
\begin{equation}
\overline{n}_{\rm min}=\frac{A^{+}}{\gamma_{\rm opt}}=\frac{A^{+}}{A^{-}-A^{+}}.
\label{equ:nminn}
\end{equation}

As we mentioned above, the rates $A^{\pm}$ can be calculated using Fermi's Golden Rule, 
once the quantum noise spectrum of the radiation pressure force $\hat{F} = h G \ad\an$ is known \cite{CD&10}. According to this, the transition rates $A^{\pm}$ in terms of the noise spectrum are given by
\begin{equation}
A^{\pm}=g_0^{2}S_{\rm NN}(\omega=\mp\OM),
\label{equ:trates}
\end{equation}
where $S_{\rm NN}$ is the quantum noise spectra given by the Fourier transform of phonon number two-time correlation
\begin{equation}
S_{\rm NN}=\int_{-\infty}^{\infty} e^{i\omega t}\langle (\ad\an)(t)(\ad\an)(0) \rangle.
\end{equation}
One can show for a laser-driven cavity, the phonon number spectrum of is \cite{MC&07}
\begin{equation}
S_{\rm NN}=\overline{n}_{\rm cav}\frac{\kappa}{\kappa^{2}/4+(\Delta+\omega)^{2}}.
\label{equ:Snn}
\end{equation} 

Inserting Eq. (\ref{equ:Snn}) in Eq. (\ref{equ:trates}) and this result in Eq. (\ref{equ:nminn}) we obtain that the minimum phonon number is
\begin{equation}
\overline{n}_{\rm min}=\left(\frac{(\kappa/2)^{2}+(\Delta-\OM)^{2}}{(\kappa/2)^{2}+(\Delta+\OM)^{2}}-1\right)^{-1}
\end{equation}
However, in a realistic situation, one can setup the laser detuning $\Delta$ to minimize the minimum phonon number and therefore in the resolved sideband regime $\kappa\ll \OM$, this leads to
\begin{equation}
\overline{n}_{\rm min}=\left( \frac{\kappa}{4\OM} \right)^{2} < 1.
\label{eq:minphonum}
\end{equation}
This result has the important consequence that when $\overline{n}_{\rm min}$ is less than one, it permits the ground state cooling.

In the first part of this work we could see that non-Markovian interactions are the responsible that the thermal equilibrium state of the system to be described by canonical Boltzmann distribution. Hence, the next goal in this work is to find the minimum phonon number in the non-Markovian regime.

\section{Non-Markovian Cooling}

From the linearized Hamiltonian (Eq. \ref{eqn:hamlintt}), we get the equations of motion for mechanical $\bnt$ and optical $\ant$ modes in the non-Markovian regimen 
\begin{align}
\dot{\bn}(t)&=-\dot{\imath}\omega_0\bn-\int\limits_{0}^{t}\mathrm{d}s\,\gamma(t-s)\bn(s) - ig(\ad+\an)- \sqrt{\gamma_{i}}\bn_{in}(t), \\
\dot{\an}(t)&=-i\Delta\an-\int\limits_{0}^{t}\mathrm{d}s\,\kappa(t-s)\an(s) - ig(\bd+\bn) - \sqrt{\kappa_{e}/2}\an_{in}(t)  - \sqrt{\kappa'}\an_{in,i}(t).
\end{align}

In the Fourier domain, the operators for the mechanical and optical modes are found to be
\begin{subequations}
\label{equ:bn}
\begin{align}
\label{equ:bn.a}
\bn(\omega)&=\frac{-\sqrt{\gamma_{i}}\bn_{in}(\omega)}{i(\OM-\omega)+\tilde{\gamma}_{i}(\omega)} - \frac{ig(\an(\omega)+\ad(\omega))}{i(\OM-\omega)+\tilde{\gamma}_{i}(\omega)}, \\
\label{equ:bn.b}
\bd(\omega)&=\frac{-\sqrt{\gamma_{i}}\bd_{in}(\omega)}{-i(\OM+\omega)+\tilde{\gamma}_{i}(\omega)} + \frac{ig(\an(\omega)+\ad(\omega))}{-i(\OM+\omega)+\tilde{\gamma}_{i}(\omega)}, 
\end{align}
\end{subequations}
\begin{subequations}
\label{equ:an}
\begin{align}
\label{equ:an.a}
\an(\omega)&=\frac{-\sqrt{\kappa_{e}/2}\an_{in}(\omega)-\sqrt{\kappa'}\an_{in,i}-ig(\bn(\omega)+\bd(\omega))}{i(\Delta-\omega)+\tilde{\kappa}(\omega)}, \\
\label{equ:an.b}
\ad(\omega)&=\frac{-\sqrt{\kappa_{e}/2}\ad_{in}(\omega)-\sqrt{\kappa'}\ad_{in,i}+ig(\bn(\omega)+\bd(\omega))}{-i(\Delta+\omega)+\tilde{\kappa}(\omega)}.
\end{align}
\end{subequations}

Using the equations (\ref{equ:bn}-\ref{equ:an}), we get the operator for the mechanical fluctuations
\begin{eqnarray}
\nonumber \bn(\omega)&=&\frac{-\sqrt{\gamma_{i}}\bn_{in}(\omega)}{i(\OM-\omega)+\tilde{\gamma}(\omega)}
+\frac{ig}{i(\OM-\omega)+\tilde{\gamma}(\omega)}\bigg[ \frac{\sqrt{\kappa_{e}/2}\an_{in}(\omega)+\sqrt{\kappa'}\an_{in,i}}{i(\Delta-\omega)+\tilde{\kappa}(\omega)}\bigg] \\
&&+\frac{ig}{i(\OM-\omega)+\tilde{\gamma}(\omega)}\bigg[ \frac{\sqrt{\kappa_{e}/2}\ad_{in}(\omega)+\sqrt{\kappa'}\ad_{in,i}}{-i(\Delta+\omega)+\tilde{\kappa}(\omega)} \bigg],
\label{eqn:opb}
\end{eqnarray}
\noindent where $\OM=\omega_0+\delta\omega_{\mathrm{m}}$ and $\tilde{\gamma}=\tilde{\gamma_{i}}+\gamma_{\mathrm{OM}}$. $\delta\omega_{\mathrm{m}}$ and $\gamma_{\mathrm{OM}}$ are the mechanical frequency shift and the optomechanical damping rate respectively, which are given by
\begin{eqnarray}
\delta\omega_{\mathrm{m}}&=&|g|^{2}\mathrm{Im}\bigg[\frac{1}{i(\Delta-\OM)+\tilde{\kappa}(\omega)}-\frac{1}{-i(\Delta+\OM)+\tilde{\kappa}(\omega)}\bigg] \\
\label{eqn:optdam}\gamma_{\mathrm{OM}}&=&2|g|^{2}\mathrm{Re}\bigg[\frac{1}{i(\Delta-\OM)+\tilde{\kappa}(\omega)}-\frac{1}{-i(\Delta+\OM)+\tilde{\kappa}(\omega)}\bigg].
\end{eqnarray}

From the Eq. (\ref{eqn:optdam}), we can see that the maximum optical damping occurs in the red-detuned regime where $\Delta=\OM$ and it is the same regime where occurs the maximum cooling as we see below.

\subsection{Quantum Noise Spectrum}

The noise spectrum is the starting point to find our final goal which is the minimum phonon number with non-Markovian  interactions. This quantity can be obtained with the operator in Eq. (\ref{eqn:opb}) by
\begin{equation}
S_{bb}(\omega)=\int_{-\infty}^{\infty}\mathrm{d}\omega'\,\langle\bd(\omega)\bn(\omega')\rangle.
\end{equation}
\noindent Considering that the optical bath is the vacuum state and the mechanical mode is in contact with a thermal bath of occupancy $n_{b}$, we have that the noise correlations associated with the input fluctuations are given by \cite{SC&13,Lou73}
\begin{align}
\langle \an_{in}(\omega)\ad_{in}(\omega') \rangle &= \delta(\omega+\omega') \\
\langle \ad_{in}(\omega)\an_{in}(\omega') \rangle &= 0 \\
\langle \bn_{in}(\omega)\bd_{in}(\omega') \rangle &= (\overline{n}_{th}+1)\delta(\omega+\omega') \\
\langle \bd_{in}(\omega)\bn_{in}(\omega') \rangle &= \overline{n}_{th}\delta(\omega+\omega'),
\end{align}
\noindent hence, the quantum noise spectrum is given by
\begin{equation}
S_{bb}(\omega)=\frac{\tilde{\gamma}(\omega) n_{f}(\omega)}{[\tilde{\gamma}(\omega)-i(\OM+\omega)[\tilde{\gamma}(-\omega)+i(\OM+\omega)]},
\label{eqn:nospec}
\end{equation}
\noindent where $n_{f}(\omega)$ is defined like the back-action modified phonon number given by
\begin{equation}
n_{f}(\omega)=\frac{\gamma n_{b}}{\tilde{\gamma}(\omega)} + \frac{g^{2}\tilde{\kappa}(\omega)}{\tilde{\gamma}(\omega)} \left[ \frac{1}{[ \tilde{\kappa}(\omega)-i(\Delta-\omega) ] [ \tilde{\kappa}(-\omega)+i(\Delta-\omega) ]} \right].
\label{eqn:nf}
\end{equation}

At this point, in accordance with the definition of noise spectrum which is a real and positive function, in Eq. (\ref{eqn:nospec}) the Fourier transform of the dissipation kernels of mechanical $\tilde{\gamma}(\omega)$ and optical mode $\tilde{\kappa}(\omega)$ should be symmetric. In principle, we have a problem because not all spectral densities work. However, we have a great variety of spectral densities without any problem and we can continue with our calculations.

\subsection{Spectral Densities and Dissipation Kernels}

Two spectral densities with a cutoff frequency used in this work are (\textit{i}) the spectral density with a Drude cutoff frequency given by
\begin{equation}
J_{\mathrm{D}}(\omega)=m\gamma\omega\frac{\OD^{2}}{\omega^{2}+\OD^{2}},
\label{eqn:drudespec}
\end{equation}
\noindent and (\textit{ii}) the Ohmic spectral density with a cutoff frequency $\OC$ given by
\begin{equation}
J_{\mathrm{C}}(\omega)=\kappa\omega\exp\left(-\frac{\omega}{\OC}\right).
\end{equation}

Our goal is finding the Fourier transform of the kernel dissipation which in terms of the spectral density is given by
\begin{align}
\gamma(t)&=\frac{2}{m}\int_{0}^{\infty}\frac{\mathrm{d}\omega}{\pi}\frac{J_{\mathrm{D}}(\omega)}{\omega}\cos(\omega t), \\
\kappa(t)&=\frac{2}{\pi}\int_{0}^{\infty}\mathrm{d}\omega\,\frac{J_{\mathrm{C}}(\omega)}{\omega}\cos(\omega t), 
\end{align}
\noindent therefore, the Fourier transform of dissipation kernels are
\begin{align}
\label{eqn:dkernelD}\tilde{\gamma}(\omega)&=\gamma\frac{\OD^{2}}{\omega^{2}+\OD^{2}}, \\
\nonumber \tilde{\kappa}(\omega)&=\kappa\exp\left(-\frac{\omega}{\OC}\right)\left[\exp\left(\frac{2\omega}{\OC}\right)\Theta(-\omega)+\Theta(\omega)\right], \\
\label{eqn:dkernelC} \tilde{\kappa}(\omega)&=\kappa\exp\left(-\frac{|\omega|}{\OC}\right),
\end{align}
\noindent where $\Theta(\omega)$ is the Heaviside function. In Eqs. (\ref{eqn:dkernelD}) and (\ref{eqn:dkernelC}) we can see that they are symmetric functions, therefore we have that the quantum noise spectrum finally is given by
\begin{equation}
S_{bb}(\omega)=\frac{\tilde{\gamma}(\omega) n_{f}(\omega)}{[\tilde{\gamma}(\omega)]^{2}+(\OM+\omega)^{2}},
\label{eqn:nospecf}
\end{equation}
\noindent and the back-action modified phonon number  $n_{f}(\omega)$ given by
\begin{equation}
n_{f}(\omega)=\frac{\gamma n_{b}}{\tilde{\gamma}(\omega)} + \frac{g^{2}\tilde{\kappa}(\omega)}{\tilde{\gamma}(\omega)} \left[ \frac{1}{[ \tilde{\kappa}(\omega)]^{2}+(\Delta-\omega)^{2}} \right]
\label{eqn:nff}
\end{equation}

As we can see in Eq. (\ref{eqn:nff}), the minimum phonon number depends explicitely on the non-Markovian interactions due to the Fourier transform of dissipation kernels $\tilde{\kappa}$ and $\tilde{\gamma}$ depend on the frequency $\omega$ and the cutoff frequency $\OC$ ($\omega_{\mathrm{D}}$) as it is defined in Eqs. (\ref{eqn:dkernelD}) and (\ref{eqn:dkernelC}).

\subsection{Cooling at Equilibrium}

From equation (\ref{eqn:nff}), we have that the minimum phonon occupation number or the cooling limit is reached when the system is in the red-detuned regime ($\Delta=\OM$) and the $n_{f}(\omega)$ is evaluated in $-\OM$. Hence, 
%
%
%
using the Fourier transform of the dissipation kernels, for Lorentzian spectral density we get 
\begin{align}
\label{equ:dkernell} \tilde{\kappa}_{\rm L}(\omega)&=\kappa\frac{\OC^{2}}{\omega^{2}+\OC^{2}},
\end{align}
\noindent and for exponential spectral density we get
\begin{align}
\label{equ:dkernele} \tilde{\kappa}_{\rm E}(\omega)&=\kappa\exp\left(-\frac{|\omega|}{\OC}\right).
\end{align}

\begin{figure}[t]
\centering
\includegraphics[width=\textwidth]{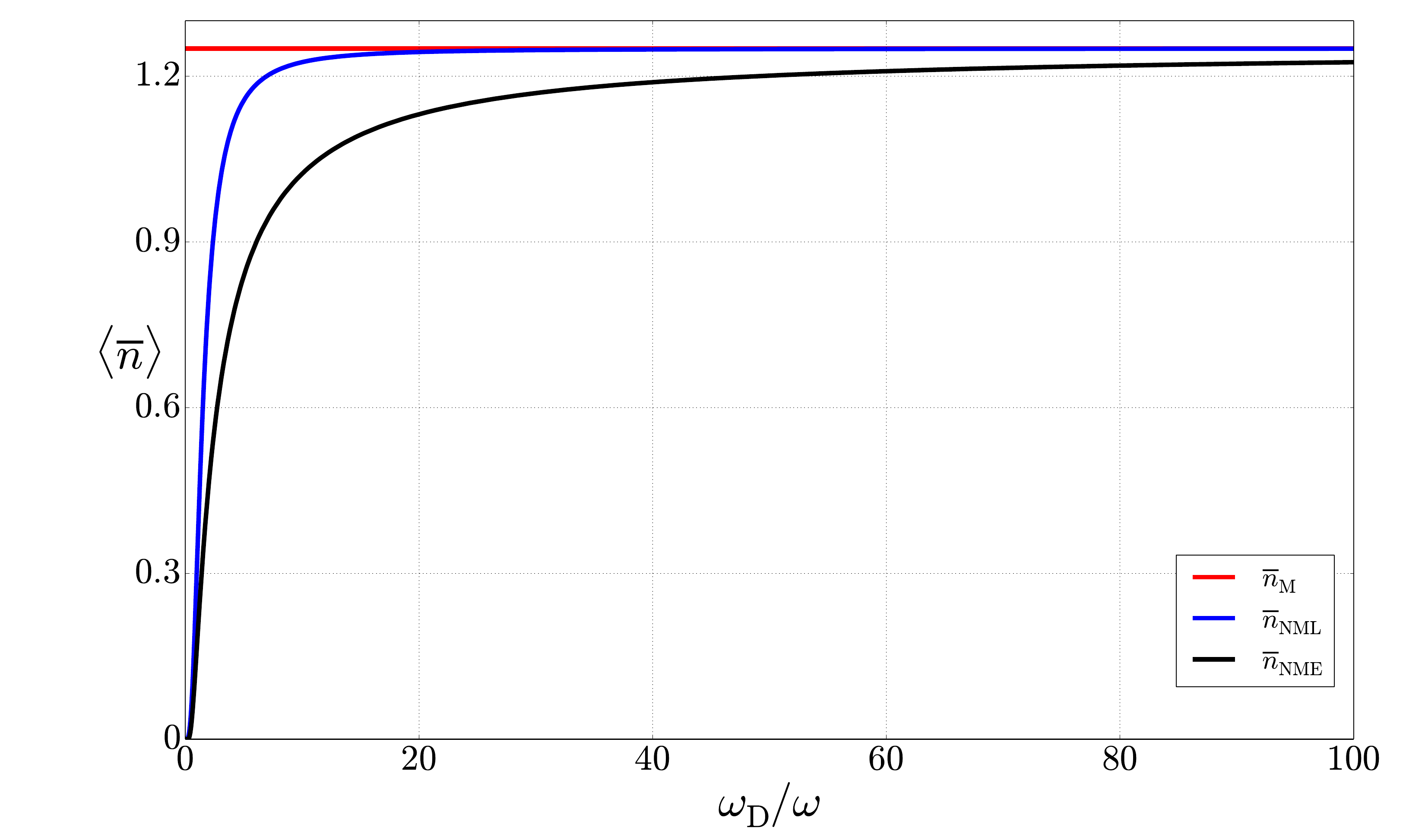}
\caption[Minimum phonon number at equilibrium.]{Minimum phonon number at equilibrium in terms of cutoff frequency $\OD$ as a measure of non-Markovianity of the system. The dissipation parameter is $\kappa=10^{-5}\omega$ The $y$-axis is scale of $10^{-12}$. Here, $n_{\rm M}$, $n_{\rm NML}$ and $n_{\rm NME}$ are the Markovian minimum phonon occupation number , the non-Markovian minimum phonon occupation number with the Lorentzian spectral density  and the non-Markovian minimum phonon occupation number with the exponential spectral density, respectively.}
\label{fig:npromeq}
\noindent\rule{\textwidth}{1pt}
\end{figure}

Taking into account the common approximations to get the minimum phonon occupation number in Eq. (\ref{eqn:nff}), i.e., that $\gamma_{0}\rightarrow0$, $n_{\rm b}=0$, $\Delta=\omega$
\begin{equation}
\label{equ:pona}
n(\omega)=\frac{g^{2}\tilde{\kappa}(\omega)}{\gamma_{\rm OM}(\omega)}\left[ \frac{1}{\tilde{\kappa}(\omega)^2 + 4\omega^{2}} \right].
\end{equation}
From Eq. (\ref{eqn:optdam}), with $\Delta=\omega$ we have
\begin{equation}
\gamma_{\rm OM}=\frac{8g^{2}\tilde{\kappa}(\omega)\omega^{2}}{\tilde{\kappa}(\omega)^{2}[\tilde{\kappa}(\omega)^{2}+4\omega^{2}]}.
\end{equation}
Thus, the minimum phonon occupation number is given by
\begin{eqnarray}
n_{\rm min}(\omega)=\frac{\tilde{\kappa}(\omega)^{2}}{8\omega^{2}}.
\label{equ:nminwd}
\end{eqnarray}
For the cutoff frequency $\OC=\omega$, we get in Eqs. (\ref{equ:dkernell}) and (\ref{equ:dkernele}) 
\begin{align}
\tilde{\kappa}_{\rm L}(\omega)&=\frac{\kappa}{2\pi} \\
\tilde{\kappa}_{\rm E}(\omega)&=\frac{\kappa}{e\pi} ,
\end{align}
therefore, the Markovian minimum phonon occupation number $n_{\rm M}$, the non-Markovian minimum phonon occupation number with the Lorentzian spectral density $n_{\rm NML}$, and the non-Markovian minimum phonon occupation number with the exponential spectral density $n_{\rm NME}$ are given by
\begin{align}
\label{eqn:mpn} n_{\rm M}&=\frac{\kappa^{2}}{8\omega^{2}}, \\
\label{eqn:nmpnl} n_{\rm NML}&=\frac{1}{4\pi^{2}}\left(\frac{\kappa^{2}}{8\omega^{2}} \right)\approx0.025n_{\rm M}, \\
\label{eqn:nmpne} n_{\rm NME}&=\frac{1}{e^{2}\pi^{2}}\left(\frac{\kappa^{2}}{8\omega^{2}} \right)\approx0.0135n_{\rm M}.
\end{align}
Here, for the Markovian minimum phonon occupation $n_{\rm M}$ we consider that in the Fourier transforms of the dissipation kernels, in Eqs. (\ref{equ:dkernell}) and (\ref{equ:dkernele}),  the cutoff frequencies $\OC\rightarrow\infty$. This limit, as we had explained corresponds to the Markovian approximation. 
More precisely, we calculate the minimum phonon number that would be reached with experimental data given by Groeblacher \emph{et. al} \cite{GT&13}. In this case, the three phonon numbers are $n_{\rm M}=6.4\times10^{-3}$, $n_{\rm NML}=1.6\times10^{-3}$ and $n_{\rm NME}=8.7\times10^{-4}$.

Thus, as it was our hypothesis, the quantum thermal equilibrium state that depends explicitely of the non-Markovian interactions help us reaching a lower minimum phonon number (Eqs. \ref{eqn:nmpnl} and \ref{eqn:nmpne}). This time, the minimum phonon number is much lower than the minimum phonon number found in the Markovian approximation (Eq. \ref{eqn:mpn}).

In Fig. \ref{fig:npromeq} it is shown the minimum phonon number obtained from Eq. (\ref{equ:nminwd}) in terms of cutoff frequency $\OC$. In this plot, we can see that as the cutoff frequency increases, the non-Markovian minimum phonon numbers approach to the Markovian minimum phonon number. This agrees with the theory because the higher cutoff frequency $\OC$, the system will be in the Markovian regime. 

%
%
%


\endgroup

\cleardoublepage
\begin{flushright}{\slshape    
    Science cannot solve the ultimate mystery of nature. And that \\ 
    is because, in the last analysis, we ourselves are part of nature \\ 
    and therefore part of the mystery that we are trying to solve.} \\ \medskip
	--- Max Planck
\end{flushright}

\bigskip

\begingroup
\let\clearpage\relax
\let\cleardoublepage\relax
\let\cleardoublepage\relax
\chapter{Optimal Sideband Cooling}\label{ch:osc} 
\counterwithin{figure}{chapter}

Mechanical micro- and nano-resonators cooled to very low temperatures can be used to explore quantum effects such as superposition of states, entanglement at macroscopic scales \cite{KB06} and, as we showed in the first part of this work,  states different from the canonical Boltzmann distribution when further, we consider non-Markovian interactions.
Additionally, when they are coupled to optical systems or superconducting qubits at these low temperatures, they can be used 
as a tool to study fundamental issues such as the quantum-mechanical transition \cite{SR05}. 
%
However, the time scales where cooling takes or would take place are very short, of the order of the resonator period and the temperatures involved in the process are very low, it is then expected that non-Markovian effects of the interaction between the resonator and its surrounding medium dominate the transfer of energy and entropy between them \cite{SN&11,EK&08,AG&04}. 
This makes that in achieving a complete understanding of the cooling process and in designing more robust techniques, the effects of non-Markovian dynamics must be considered in, \eg, all protocols mentioned above \cite{KB06,AA&09b,OH&10,TD&11}. 
To be sure, the role of non-Markovian dynamics in cooling process has been studied in other physical systems like spins with very positive results \cite{EK&08,AG&04}.

In this chapter, we study the effects of the non-Markovian character of the dynamics in the optimal cooling with light pulses \cite{WV&11,MC&12}. 
More specifically, we will use one of the most promising proposal so far \cite{WV&11}, which provides the introduction of optimal light pulses in the technique known as ``sideband cooling'' \cite{TD&11}. Therefore, in this chapter we show a general treatment, and in detail, of the cooling process.

\section{Optimal Control Theory}

Optimal control theory has been well developed for over forty years. With the advances of computer techniques, optimal control is now widely used in multi-disciplinary applications such as biological systems, communication networks and socio-economic systems. As a result, more and more people will greatly benefit by learning how to solve the optimal control problems numerically.

The objective of optimal control theory is to determine the control of signals that will cause a process to satisfy the physical constraints and at the same time minimize some performance criterion. This design is generally done by a trial-and-error process in which various methods of analysis are used iteratively to determine the design parameters of an ``aceptable'' system. Acceptable performance is generally defined in terms of time and frequency domain criteria such as rise time, settling time, peak overshoot and bandwidth \cite{Kir12}. 
Radically different performance criteria must be satisfied, however, by the complex, multiple-input, multiple-output systems required to meet the demands of modern technology, this is a theory based on a combination of variational techniques with high-speed computation. 
Therefore, this technique is especially used for those problems with free final time and nonlinear dynamics.

To get an idea of how optimal control theory works, we first need to have or formulate the problem in a specific way. Problem formulation of an optimal control problem, first of all, requires three specific components:
\begin{enumerate}
\item Mathematical description (or model) of the process to be controlled, \ie, the description of the system in terms of $n$ first-order differential equations, as a state vector of the system $\boldsymbol{x}(t)$ and the control vector $\boldsymbol{u}(t)$. Hence, the state equations can be written as
\begin{equation}
\boldsymbol{\dot{x}}=\boldsymbol{a}(\boldsymbol{x}(t),\boldsymbol{u}(t),t).
\label{eq:stateequ}
\end{equation}
\item A statement of the physical constraints, \ie, the initial (final) conditions of the system constraints $\boldsymbol{x}(t_0)$. 
\item Specification of a criterion to minimize or maximize. At this point, we should specify the condition to minimize (maximize) as
\begin{equation}
J=h(\boldsymbol{x}(t),t)+\int_{0}^{t_f}g(\boldsymbol{x}(t),\boldsymbol{u}(t),t)\,\mathrm{d}t,
\label{eq:condmin}
\end{equation}
\noindent where $h$ and $g$ are scalar functions \cite{Kir12}. 
\end{enumerate}

After doing so, optimal control theory is aimed at solving the following problem. Find a control function $\boldsymbol{u^{\star}}$ which causes the system described by Eq. (\ref{eq:stateequ}) follows the behaviour $\boldsymbol{x^{\star}}$ that minimizes the condition in Eq. (\ref{eq:condmin}). Here $\boldsymbol{u^{\star}}$ is called the\textit{optimal control} and $\boldsymbol{x^{\star}}$ is called the \textit{optimal trajectory}. However, there is an aspect to consider. This is that even if an optimal control exists, it may not be unique. At this point, the optimal control will depend on the initial control, \ie, one should prove different initials parameters for not to fall in a local minimum but to reach the global minimum of the condition to minimize $J$.

As we mentioned at the beginning, optimal control theory has been used in many disciplines with different algorithms or methodologies. Nevertheless, in this work we refer to two previous investigations which are pioneers in the optimal control of open quantum systems and the effects of the driving \cite{SN&11,WV&11}. Inspired by the work of  Wang \textit{et al.}, we address the problem of finding the optimal coupling function between a nano-mechanical resonator and a cavity mode. In doing so, we connect the results in Ref. \cite{WV&11}, in particular, we found that to reach the minimum phonon number predicted by them, the coupling amplitude should be the twice the one predicted by them.
We also find the optimal pulse to maintain the minimum phonon number once it is reached. Later, we thus implement the methodology in the non-Markovian case using the results for variances in \cite{Est13,EP15} to find the minimum phonon number in the non-Markovian case and thus observe the effects of non-Markovian interactions in the sideband cooling.

\section{Markovian Cooling}

One the techniques to reach minimum phonon number is known as sideband cooling. Sideband cooling uses a linear coupling between the resonator and the cavity, which in practice is usually obtained from a ``radiation pressure force'' interaction by strongly driving the cavity (see chapter \ref{ch:optomechcool}). In a recent work \cite{WV&11} it was demonstrated that one can cool significantly better than traditional sideband cooling by using quantum control, based on the ``\textit{The Method of Steepest Descent}'' \cite{Kir12}, (see Appendix \ref{app:algo} for details).

In general, cooling is characterized by the average phonon number, $\langle \hat{n} \rangle=\langle \ad\an \rangle$, which is a second moment of operators $\an$ and $\ad$. Because the dynamics of the resonators are linear (that is, the evolution can be described by a set of linear quantum Langevin equations \cite{CD&10,GZ04}) one can derive a set of equations for the variances and covariances of the ladder operators. 
To describe damping, in Ref. \cite{WV&11} the authors use the Markovian version of the Brownian-motion master equation. However, this set of equations violates the positivity of the trace in the density operator, in others words, the Markovian version of the Brownian-motion master equation cannot be written in the Lindblad form (see chapter 3 in Ref. \cite{BP07}). Moreover, they use the adjoint version of the master equation, which is only valid for time independent Hamiltonians, a condition that obviously is not fulfil in this case. Hence, we assume that results in \cite{WV&11}, shown in Fig. \ref{fig:cooljac}, can be redesigned using the correct equations of motion.

\begin{figure}[t]
\centering
\includegraphics[width=\textwidth]{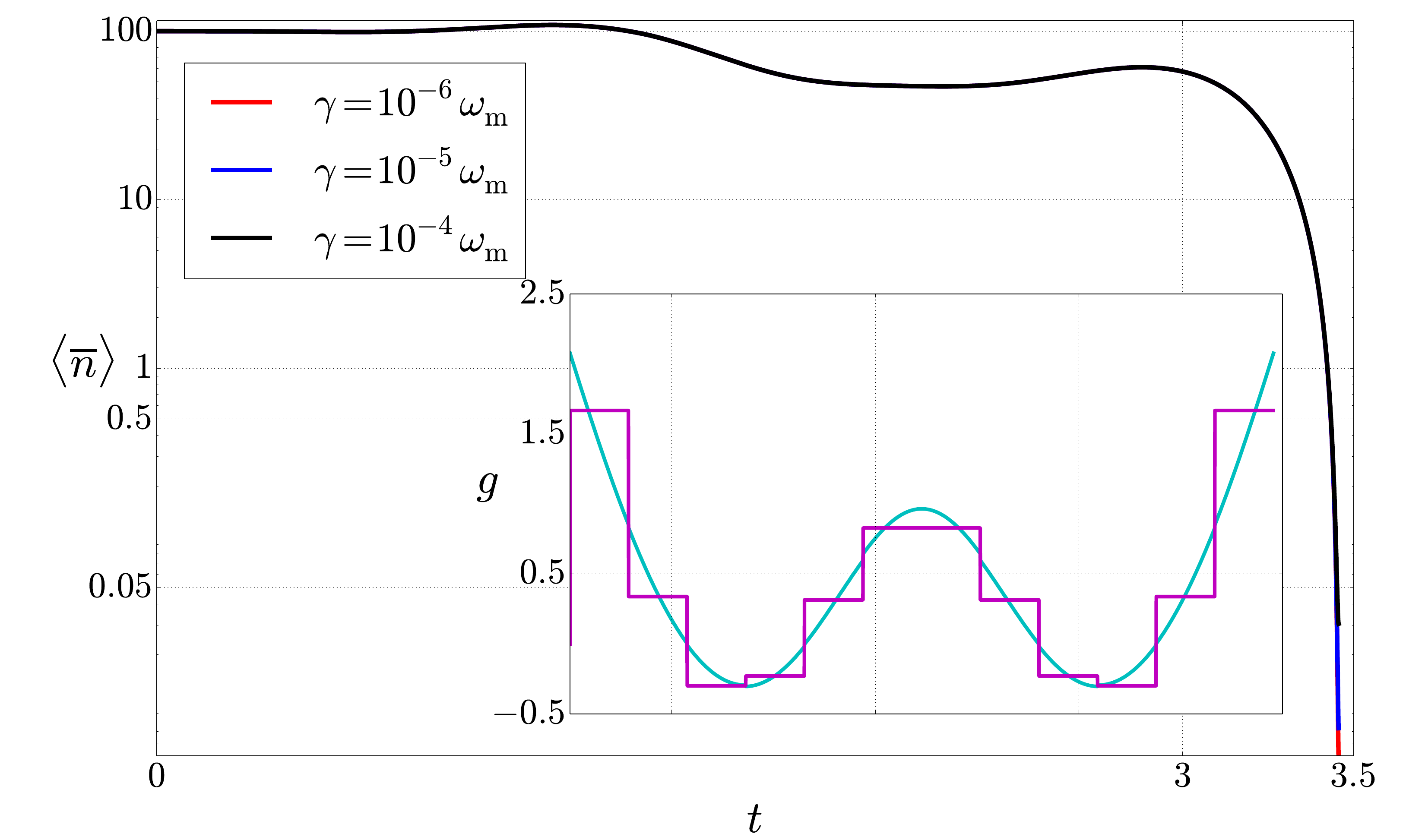}
\caption[Dynamics of the cooling with the Markovian version of the Brownian-motion master equation.]{Dynamics of the cooling with the Markovian version of the Brownian-motion master equation for different parameters of dissipation. The initial parameters are $n_{\mathrm{T}}=100$, $n_{\mathrm{cav}}=0$, $\kappa=2.15\times10^{-4}\omega_0$. Inset: The optimal 12-segment piecewise-constant control pulse for the coupling rate $g(t)$ (magenta line) and the true $g(t)$ function found in the optimization (cyan line). The $y$-axis is in logarithmic scale.}
\label{fig:cooljac}
\noindent\rule{\textwidth}{1pt}
\end{figure}

Hence, we use a master equation in the Linbland form and assume that the commutator between the Liouvillian and the driving force can be neglected. Thus, we derive the correct equations of motion for variances from the position and momentum operators in terms of the ladder operators (calculation of these equations is showed in Appendix \ref{app:equmot}). Additionally, we use the adjoint master equation given by
\begin{eqnarray}
\nonumber \frac{d}{dt}A_{\mathrm{H}}(t)=& &i\omega_0 [\ad\an,A_{\mathrm{H}}(t)] \\
\nonumber &+& \gamma(n_{T}+1)\left\{ \ad A_{\mathrm{H}}(t)\an - \frac{1}{2}\ad\an A_{\mathrm{H}}(t) - \frac{1}{2}A_{\mathrm{H}}(t)\ad\an \right\} \\
\nonumber &+& \gamma n_{T}\left\{ \an A_{\mathrm{H}}(t)\ad - \frac{1}{2}\an\ad A_{\mathrm{H}}(t) - \frac{1}{2}A_{\mathrm{H}}(t)\an\ad \right\} \\
\nonumber &+& \kappa(n_{\mathrm{cav}}+1)\left\{ \bd A_{\mathrm{H}}(t)\bn - \frac{1}{2}\bd\bn A_{\mathrm{H}}(t) - \frac{1}{2}A_{\mathrm{H}}(t)\bd\bn \right\} \\
&+& \kappa n_{\mathrm{cav}}\left\{ \bn A_{\mathrm{H}}(t)\bd - \frac{1}{2}\bn\bd A_{\mathrm{H}}(t) - \frac{1}{2}A_{\mathrm{H}}(t)\bn\bd \right\},
\end{eqnarray}
because this allows for deriving the expressions for dissipation rate in the equations of motion for the expectation values \cite{BP07}. In this expression, $\gamma$ and $\kappa$ accounts for the non-unitary processes in the mechanical mode and optical mode, respectively. $n_{T}$ ($n_{\mathrm{cav}}$) denotes the thermal occupation in the mechanical (optical) mode and the operator $A_{\mathrm{H}}(t)$ will be every variance of the ladder operators. This approximation obeys the positivity of the trace that in contrast to previous works, this had not been considered. 


Fig. \ref{fig:corrcool} show the results for the minimum phonon number $\overline{n}=\langle \ad\an \rangle$ as a function of time for different parameters of dissipation. We note that the initial coupling function before optimization was a constant function $g=0.01$.
As we can see in Fig. \ref{fig:corrcool}, the smaller the dissipation $\gamma$ in the resonator is, the smaller minimum phonon number. 
This result is in agreement with the result obtained for the minimum phonon number in Eq. (\ref{eq:minphonum}), since these results were obtained in the approximation $\gamma\rightarrow0$, \ie, the final minimum phonon number will be obtained when the dissipation rate tends to zero or it is small enough to not affect the phonon number.

Another important result found with the correct equations of motion is a factor of one half in the coupling. As we can see in the inset of Fig. \ref{fig:corrcool}, the limits of the coupling function $g_{\mathrm{opt}}(t)$ are the double of the function coupling found by X. Wang \textit{et al.} \cite{WV&11}. 
Despite the inadequateness of the equation of motion in Ref. \cite{WV&11}, it seems that the only noticeable effect is a scaling factor of the field. Please note that at the experimental level, due to the need of strong coupling, the differences may dictate the feasibility of an experimental validation.
However, a tremendous increase in $g$ was demonstrated recently in an experiment by Teufel \textit{et al}. \cite{TD&11}. This has brought $g$ within a factor of 10 of $\omega_0$, which demonstrated that in the future the increase of the coupling $g$ is feasible. 

\begin{figure}[t]
\centering
\includegraphics[width=\textwidth]{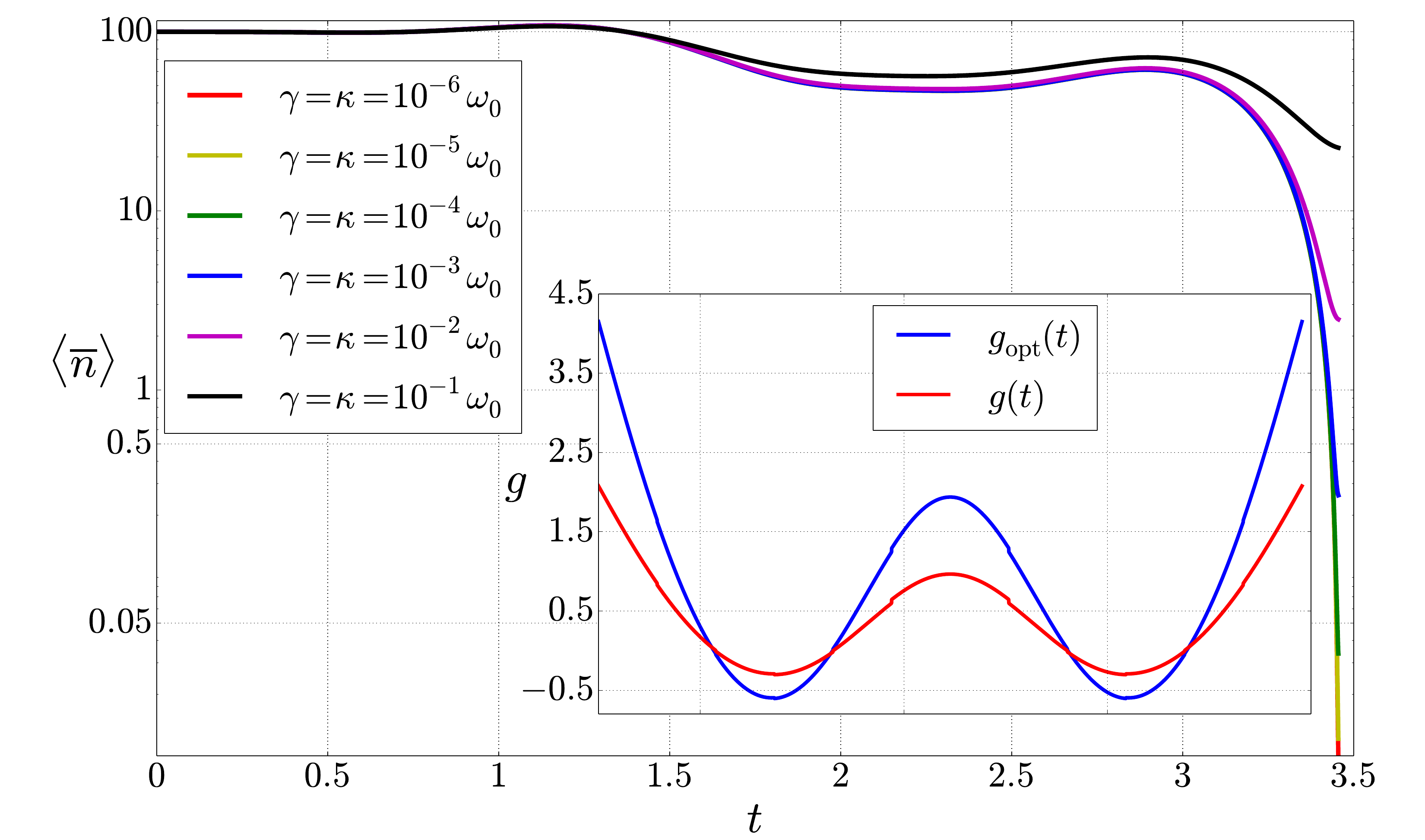}
\caption[Dynamics of the cooling with the correct equations of motion.]{Dynamics of the cooling with the correct equations of motion for different parameters for dissipation. The initial parameters are $n_{\mathrm{T}}=100$, $n_{\mathrm{cav}}=0$. Inset: The optimal control pulse for the coupling rate $g(t)$. The $y$-axis is in logarithmic scale.}
\label{fig:corrcool}
\noindent\rule{\textwidth}{1pt}
\end{figure}

\subsection{Optimal control pulse to maintain the minimum phonon number}

%
Reaching a very low phonon number in a short period of time is a very desirable goal, however, since the resonator is continuously coupled to its environment, keeping that phonon number is a must. 
%
%
Here, we found an optimal pulse to maintain the minimum phonon number which depends on the initial coupling function. 
\begin{figure}[h!]
\centering
a. $\gamma=10^{-6}\omega_0$ \\
\includegraphics[width=0.715\textwidth]{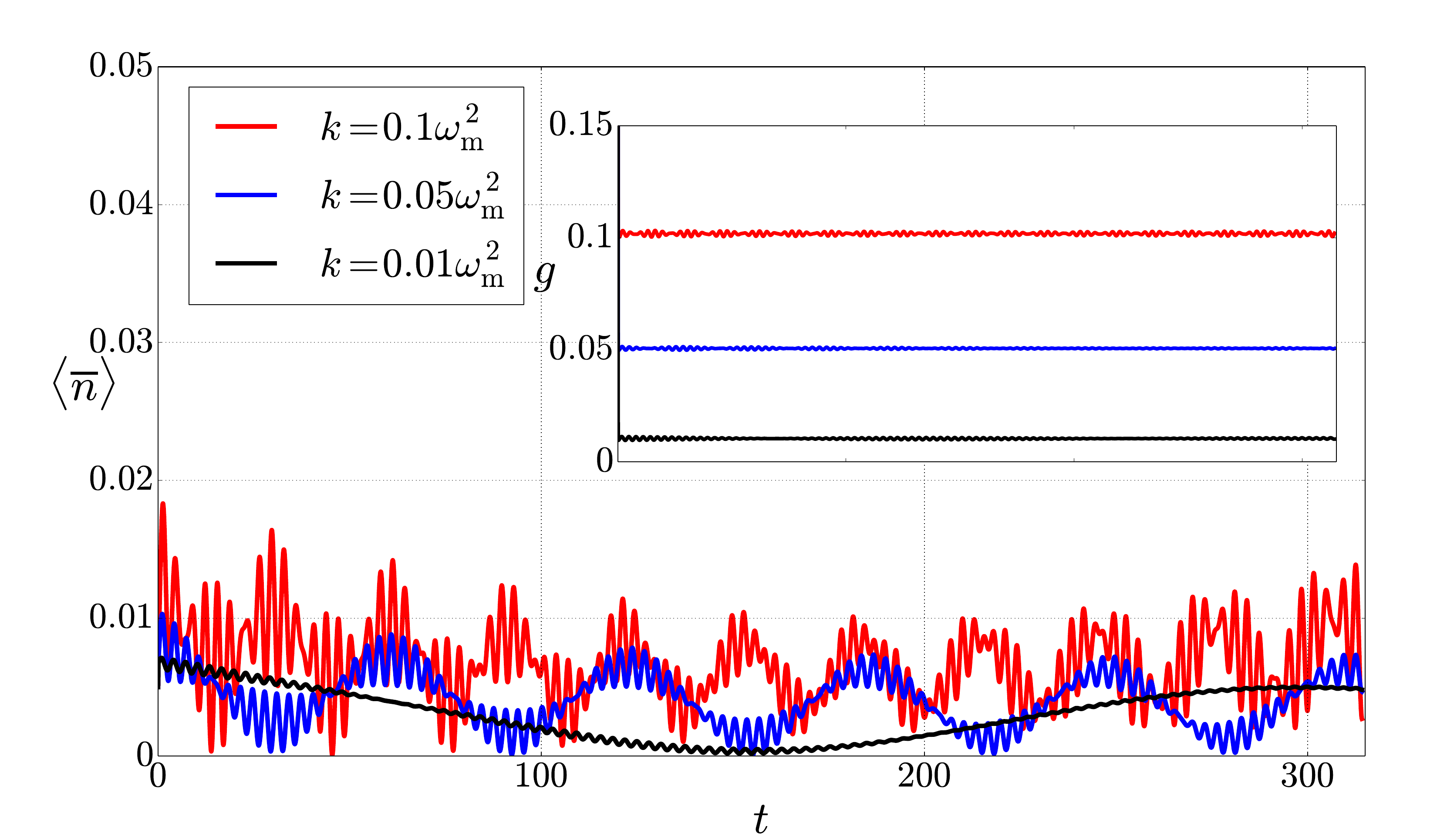}\\
b. $\gamma=10^{-5}\omega_0$ \\
\includegraphics[width=0.715\textwidth]{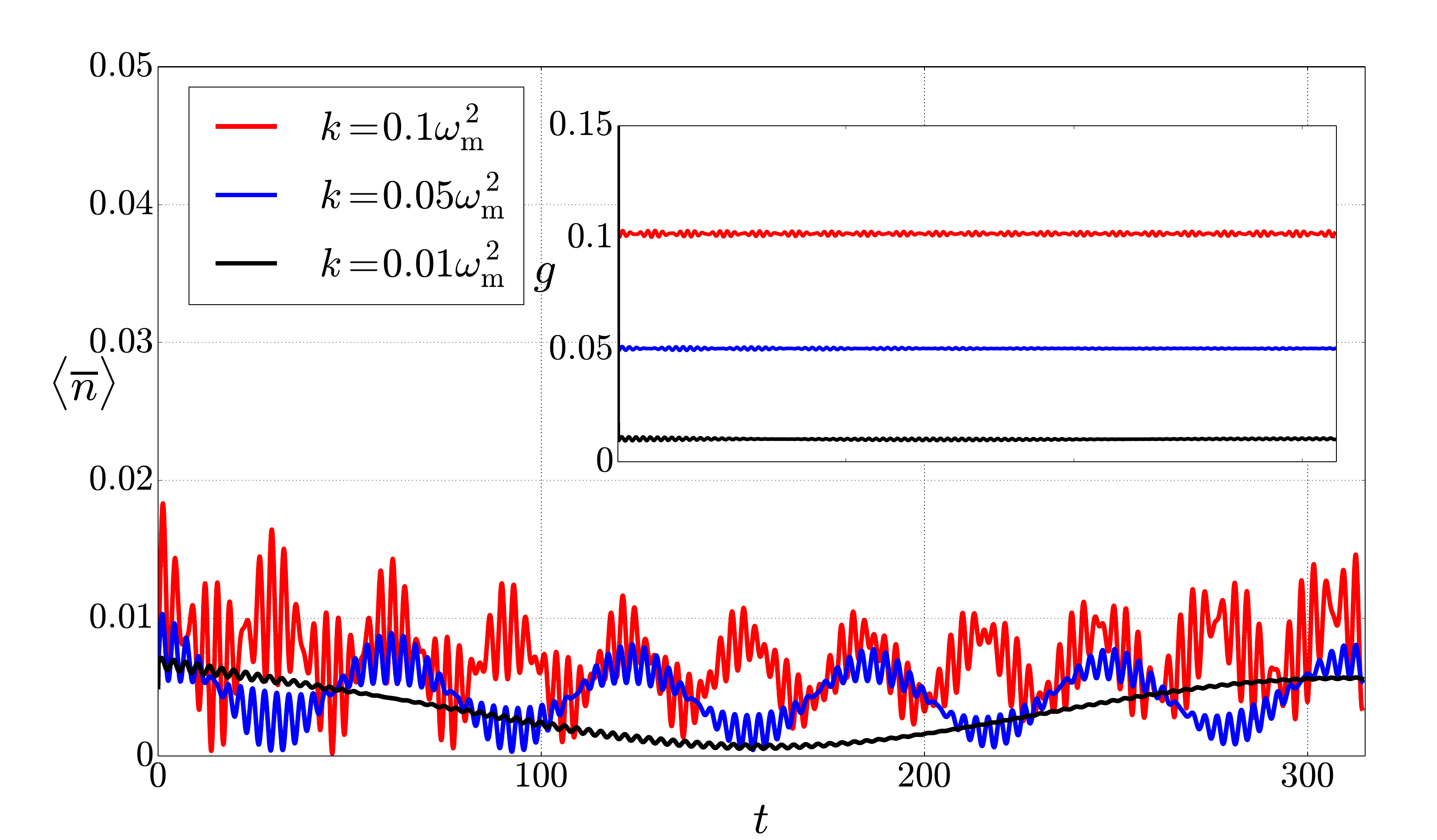}\\
c. $\gamma=10^{-4}\omega_0$ \\
\includegraphics[width=0.715\textwidth]{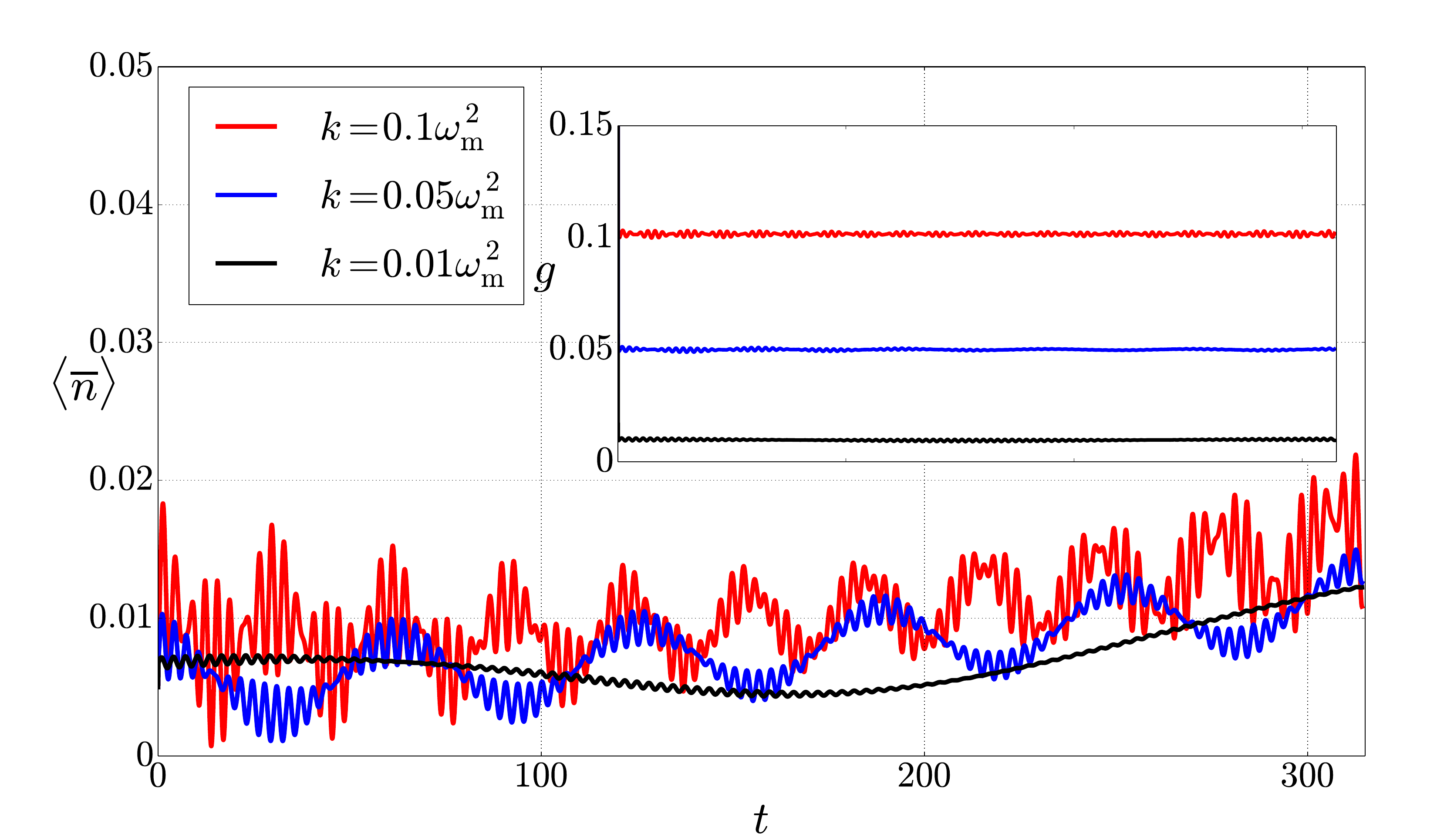}\\
\caption[Phonon number as a function of time during the optimal control protocol aimed at maintaing the minimum phonon number.]{Phonon number as a function of time during the optimal control protocol aimed at maintaing the minimum phonon number with the correct equations of motion for different parameters for dissipation. The initial parameters are $n_{\mathrm{T}}=100$, $n_{\mathrm{cav}}=0$, $\kappa=2.15\times10^{-4}\omega_0$. Inset: The optimal control pulse for the coupling rate $g(t)$.}
\label{fig:coolman}
\noindent\rule{\textwidth}{1pt}
\end{figure}

We found that to maintain the minimum phonon number, the coupling function should be an oscillating function in time over a constant value. We proposed as initial coupling function an exponential decay which ends in a constant value given by
\begin{equation}
g_{\mathrm{m}}(t)=e^{-50 t} + k.
\label{eq:pulseman}
\end{equation}
The decay factor 50 is to reach the factor $k$ as soon as possible, because when the coupling to the bath is strong, the phonon number tends to increase. 
As we can see in Fig. \ref{fig:coolman}, for all the different parameters of dissipation used ($\gamma=10^{-6}\omega_0$, $\gamma=10^{-5}\omega_0$, $\gamma=10^{-4}\omega_0$), the best result is the smallest factor $k$ because this has minor oscillations along the time, \ie, the minimum phonon number keeps almost constant.
Despite of the control protocol, when dissipation increases, the minimum phonon number also increases, as we see in Fig. \ref{fig:coolman} for the dissipation parameter $\gamma=10^{-4}\omega_0$. 
However, this value of dissipation should not worry because in systems that could be used to prove this, \eg, a microwave superconducting cavities, 
the typical frequencies in these resonators are in the GHz regime, operating usually in the millidegrees Kelvin range \cite{MD&08,RZ&10}.
Moreover, the typical dissipation rate in these systems is $\gamma=10^{-4}\omega_0$, or even less. Further, current experiments with nanomechanical resonators have these typical parameters: $\omega=2\pi\times15$ MHz, $m=10^{-17}$ kg, $g\sim10^{-3}\omega^{2}$, and a quality factor $Q\sim20000$, which yields a damping $\gamma=5\times10^{-5}\omega$ \cite{WMC08}.

As we explained above, the pulse found in the inset in Fig. \ref{fig:corrcool} is to reach the minimum phonon number and the pulse found in Fig. \ref{fig:coolman} is to maintain the minimum phonon number. Therefore, it is worth mentioning that the optimal pulse $g_{\mathrm{m}}^{\star}(t)$ obtained from Eq. (\ref{eq:pulseman}) after the optimization process is the continuation of the optimal pulse to cool $g(t)$ showed in Fig. \ref{fig:corrcool}. Thus, it takes into account that the minimum phonon number is reached approximately at $t_{\mathrm{cool}}=1.1\pi/\omega$, we define an optimal pulse as \cite{TEP15}
\begin{equation}
c(t)=\left\{\begin{array}{l c}
g(t) & 0 \le t \le 1.1\pi/\omega \\
g_{\mathrm{m}}^{\star}(t) & t > 1.1\pi/\omega
\end{array}\right.
\end{equation}
which encompasses the entire cooling process, \ie, to reach the minimum phonon number and maintain this number once is obtained. Likewise, with the optimal pulse $c(t)$ we can maintain the phonon number for about 50 periods. 

The next step, as we made at equilibrium, it is to perform the cooling and optimization process including the non-Markovian interactions. As we could see above, the minimum phonon number at equilibrium in lower when we take into account the non-Markovian interactions than when we use the Markovian approximation. We expect to find evidence this phenomena in our out of equilibrium optimal control numerical experiment.



\section{Non-Markovian Cooling}

In this section we use the influence functional theory by Feynman-Vernon \cite{GSI88} which allows for the study of dynamics of open quantum systems without the rotating wave approximation (RWA) and Markovian approximation. This will allow us to perform a description of the system without any approximation, \ie, a full description of the non-Markovian dynamics.

Before starting to explain the case of a dissipative quantum system, it is useful to give the main ideas and notation by studying the unitary time evolution of a system. Henceforth, the state, pure or mixed, of a system S will be described by the density operator $\hat{\rho}_{\mathrm{S}}$. For an isolated system described by the Hamiltonian $\Hs$, the temporal evolution of the density operator is given in terms of the time evolution operator 
\begin{equation}
\hat{U}_{\mathrm{S}}(t'',t')=\hat{T}\exp\left( -\frac{i}{\hbar}\int_{t'}^{t''}\mathrm{d}s\,\Hs(s) \right), 
\end{equation}
and its adjoint operator $\hat{U}_{\mathrm{S}}^{\dagger}$ through the relation
\begin{equation}
\hat{\rho}_{\mathrm{S}}(t)=\hat{U}_{\mathrm{S}}^{\dagger}(t)\hat{\rho}_{\mathrm{S}}(0)\hat{U}_{\mathrm{S}}(t),
\end{equation}
where $\hat{T}$ is the time ordering operator. In the position representation, the density operator is given by
\begin{equation}
\rho_{\mathrm{S}}(q''_{+},q''_{-},t)=\int\mathrm{d}q'_+ \mathrm{d}q'_- J(q''_+,q''_-,t'';q'_+,q'_-,t')\rho_{\mathrm{S}}(q'_+,q'_-,t'),
\label{eq:denmatsyss}
\end{equation}
where $J(q''_+,q''_-,t'';q'_+,q'_-,t')=U(q''_+,q'_+,t)U^{*}(q''_-,q'_-,t)$ is the propagator of the density operator $\rho_{\mathrm{S}}(q_+,q_-)=\langle q_+|\hat{\rho}_{\mathrm{S}}|q_-\rangle$ and $U(q''_\pm,q'_\pm,t)=\langle q''_\pm|\hat{U}(t'',t')|q'_\pm\rangle$. In order to with Feynman-Vernon theory \cite{FH12}, the time evolution operator can be expressed by a path integral as
\begin{equation}
U(q''_\pm,q'_\pm,t)=\int\mathfrak{D}q_{\pm}\exp\left[ \frac{i}{\hbar}S_{\mathrm{S}}(q_{\pm},t)\right],
\end{equation}
where $S_{\mathrm{S}}$ is the action associated to the path $q(t')$. 

Now, we are going to consider the case of a dissipative system for the \acl{ucl} in Eq. (\ref{eq:hamucl}). In this class of systems one is usually not interested in the full dynamics, but so it reduced dynamics of the system of interest. Before tracing over the degrees of freedom of the bath, it is necessary to specify the initial density operator of the system and the bath. For simplicity, we restrict to the case where the density operator of the system and bath are factorized \cite{CL83b}. Under this approximation, the initial density operator may be expressed as
\begin{equation}
\hat{\rho}(0)=\hat{\rho}_{\mathrm{S}}(0)\frac{e^{-\beta\Hb}}{Z_{\mathrm{B}}},
\label{eq:initialstat}
\end{equation}
\ie, as the product between the initial density matrix of the system $\hat{\rho}_{\mathrm{S}}(0)$ and the initial density matrix of the bath $\hat{\rho}_{\mathrm{B}}(0)=\frac{e^{-\beta\Hb}}{Z_{\mathrm{B}}}$ at inverse temperature $\beta=1/k_{\mathrm{B}T}$. In this expression, $Z_{\mathrm{B}}$ denotes the partition function of the bath.

Dynamics of the reduced density matrix of the system S is obtained tracing over the degrees of freedom of the bath. This can perform analytically if the bath is described by a collection of harmonic oscillators with masses $m_{j}$ and frequencies $\omega_{j}$ and with coordinates coupled bilinearly to the coordinates of the system S.
It is convenient to mention that the microscopic details of the bath and its coupling to the system emerge in the reduced dynamics of the system through of the spectral density of the bath given by
\begin{equation}
J(\omega)=\pi\sum_{j=1}^{\infty} \frac{c_{j}^{2}}{2m_{j}\omega_{j}}\delta(\omega-\omega_{j}),
\end{equation}
where $c_{j}$ is the system-bath coupling. Similarly, we introduced the noise kernel $K(z)$ induced by the bath given in Eq. (\ref{eq:noiseke}) and the friction kernel as \cite{Ing02}
\begin{equation}
\eta(t)=\frac{2}{\pi}\int_{0}^{\infty}\mathrm{d}\omega\frac{J(\omega)}{\omega}\cos(\omega t).
\end{equation}

After tracing over the bath, we get the temporal evolution of the reduced density matrix of the system S. If we assume the initial conditions in Eq. (\ref{eq:initialstat}), we get the same form that in Eq. (\ref{eq:denmatsyss}). $J(q''_+,q''_-,t'';q'_+,q'_-,t')$ is the propagation function which is determined by a path integral as
\begin{equation}
J(q''_+,q''_-,t'';q'_+,q'_-,t')=\frac{1}{Z}\int\mathfrak{D}q_+\mathfrak{D}q_-\,\exp\left\{\frac{i}{\hbar}[ S_{\mathrm{S}}(q_+) - S_{\mathrm{S}}(q_-) ]\right\}\mathfrak{F}(q_+,q_-).
\label{eq:funcpro}
\end{equation}
Here, $Z$ is the partition function as the ratio between the partition function of the system-bath $Z_{\mathrm{S+B}}$ and the bath $Z_{\mathrm{B}}$. The influence of the bath over the system in Eq. (\ref{eq:funcpro}) is given by
\begin{equation}
\mathfrak{F}(q_+,q_-)=\exp\left[ -\frac{1}{\hbar} \Phi(q_+,q_-) \right],
\end{equation}
with
\begin{align}
\nonumber \Phi(q_+,q_-)=&\int_{0}^{t}\mathrm{d}s\int_{0}^{s}\mathrm{d}u\, K(s-u)[q_+(s)-q_-(s)][q_+(u)-q_-(u)] \\
&+\frac{i}{2}\int_{0}^{t}\mathrm{d}s\int_{0}^{s}\mathrm{d}u\,\eta(s-u)[q_+(s)-q_-(s)][\dot{q}_+(u)-\dot{q}_-(u)] \\
\nonumber &+\frac{i}{2}[q'_+ + q'_-]\int_{0}^{t}\mathrm{d}s\,[q_+(s)-q_-(s)].
\end{align}
Hence, non-Markovian character that we want to take into account in the dynamics is expressed through the double temporal integrals which included all the history in the evolution of the system. 

Now, since in the cooling we need to calculate the phonon number, we calculate the mean value as
\begin{equation}
\langle \hat{n}(t) \rangle = \mathrm{tr}[\hat{\rho}_{\mathrm{S}}\hat{n}] = \int\mathrm{d}q\mathrm{d}\tilde{q}\langle q | \hat{\rho}_{\mathrm{S}} | \tilde{q} \rangle \langle \tilde{q} | \hat{n} | q \rangle.
\end{equation}

Our goal is to calculate the minimum phonon number of the resonator $\langle \hat{n} \rangle(t_{f})$ in the shortest possible time $t_{f}$. In terms of the second moments of the position $\langle \hat{q}^{2} \rangle$ and momentum $\langle \hat{p}^{2} \rangle$, the minimum phonon number is given by
\begin{equation}
\langle \hat{n} \rangle(t) = \frac{1}{2\hbar\omega}\left( \frac{\langle \hat{p}^{2} \rangle(t)}{m} + m\omega^{2}\langle \hat{q}^{2}\rangle(t) \right) - \frac{1}{2},
\label{eq:npromnm}
\end{equation}
$\langle \hat{p}^{2}\rangle(t)$, $\langle \hat{q}^{2}\rangle(t)$ can be found in Appendix \ref{app:variances}.

The results of the comparison between the dynamics in the Markovian approximation found in Fig. \ref{fig:corrcool}, the Markovian approximation with a high cutoff frequency $\OD=10\OM$ and the non-Markovian $\OD=\OM$ are shown in Fig. \ref{fig:coolnonm}. Here the time at which the minimum phonon number in the resonator is obtained is $t_{f}=t_{\mathrm{cool}}=1.1\pi/\omega_{\mathrm{m}}$. The insets of Fig. \ref{fig:coolnonm} show a magnification in the final time thus allowing us a  better comparison between the three cooling scenarios. In these three processes we used the coupling function $g_{\mathrm{opt}}(t)$ found in Fig. \ref{fig:corrcool} and the minimum phonon numbers of each one is shown in the following table:

\begin{center}
\begin{tabular}{c||c||c||c}
\multirow{2}{*}{Dissipation rate} & \multicolumn{3}{c}{$\langle \hat{n}(t_{f}) \rangle$} \\
\cline{2-4}
 & Markovian & Markovian $\omega_{\mathrm{D}}$ & Non-Markovian \\
 \hline \hline
$\kappa=\gamma=10^{-6}\omega_0$ & 9.03$\times10^{-3}$ & 8.96$\times10^{-3}$ & 8.86$\times10^{-3}$ \\
\hline
$\kappa=\gamma=10^{-5}\omega_0$ & 1.04$\times10^{-2}$ & 1.03$\times10^{-2}$ & 1.02$\times10^{-2}$ \\
\hline
$\kappa=\gamma=10^{-4}\omega_0$ & 3.28$\times10^{-2}$ & 3.30$\times10^{-2}$ & 2.39$\times10^{-2}$ \\
\hline
$\kappa=\gamma=10^{-3}\omega_0$ & 2.61$\times10^{-1}$ & 2.62$\times10^{-1}$ & 1.61$\times10^{-1}$ \\
\hline
$\kappa=\gamma=10^{-2}\omega_0$ & 2.45 & 2.50 & 1.52 \\
\hline
$\kappa=\gamma=10^{-1}\omega_0$ & 21.12 & 24.77 & 14.64 
\end{tabular}
\captionof{table}{Minimum phonon number at time $t=1.1\pi/\omega_{0}$ using the coupling function $g(t)$.}
\label{tab:mpn}
\end{center}

We see in the Table \ref{tab:mpn} that for non-Markovian dynamics a smaller phonon number is obtained and when dissipation increases, the results are most noticeable. However, the coupling function $g(t)$ used is these three processes was found through an optimization process in the Markovian approximation. Hence, we also perform the optimization process described in Appendix \ref{app:algo} to reach the optimal coupling function in non-Markovian dynamics. However, this optimization process has a big degree of complexity because  we have around 40 first order equations, where their solutions are used to find the second moments of the position and momentum. This implies that we have to derive 40 auxiliary first order equations to optimize and find the optimal coupling function.  
Moreover, to find the ``initial'' values of the auxiliary variables we need to calculate the derivatives of the second moments (see Appendix \ref{app:variances}), which are expressions too long (see Appendix \ref{app:algo} for details).



\begin{figure*}[h!]
\centering
a. $\gamma=10^{-6}\omega_0$ \\
\includegraphics[width=0.78\textwidth]{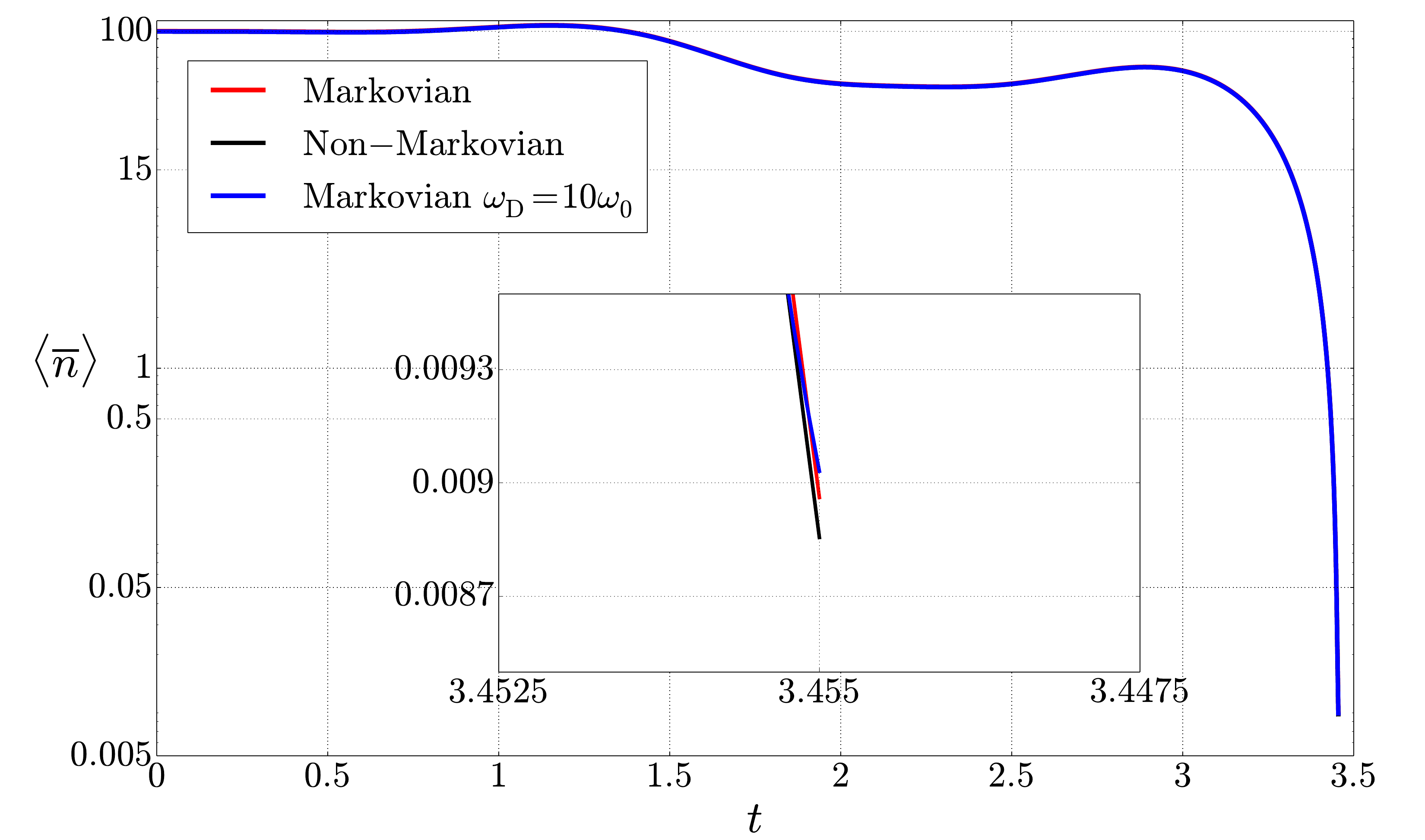}\\
b. $\gamma=10^{-5}\omega_0$ \\
\includegraphics[width=0.78\textwidth]{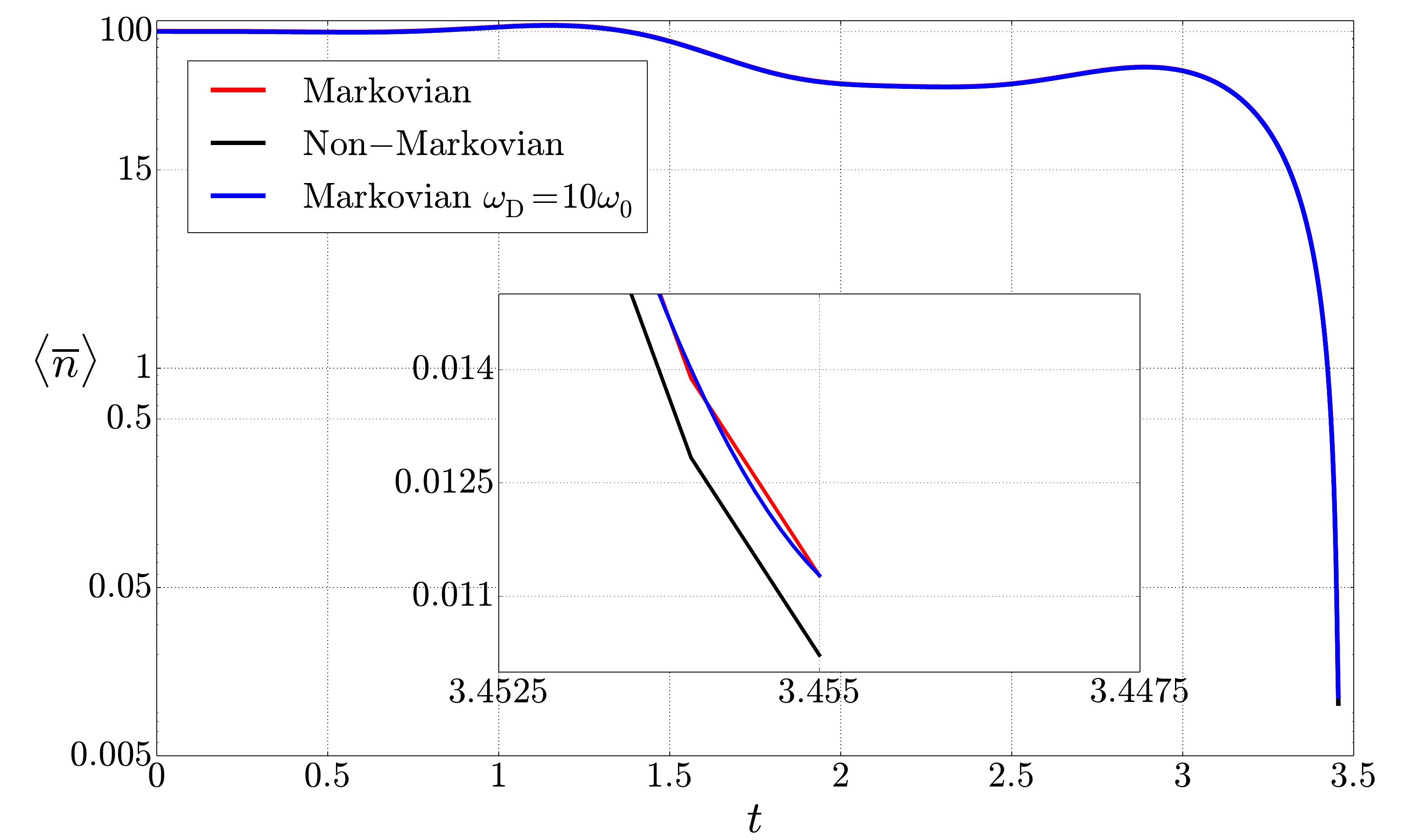}\\
c. $\gamma=10^{-4}\omega_0$ \\
\includegraphics[width=0.78\textwidth]{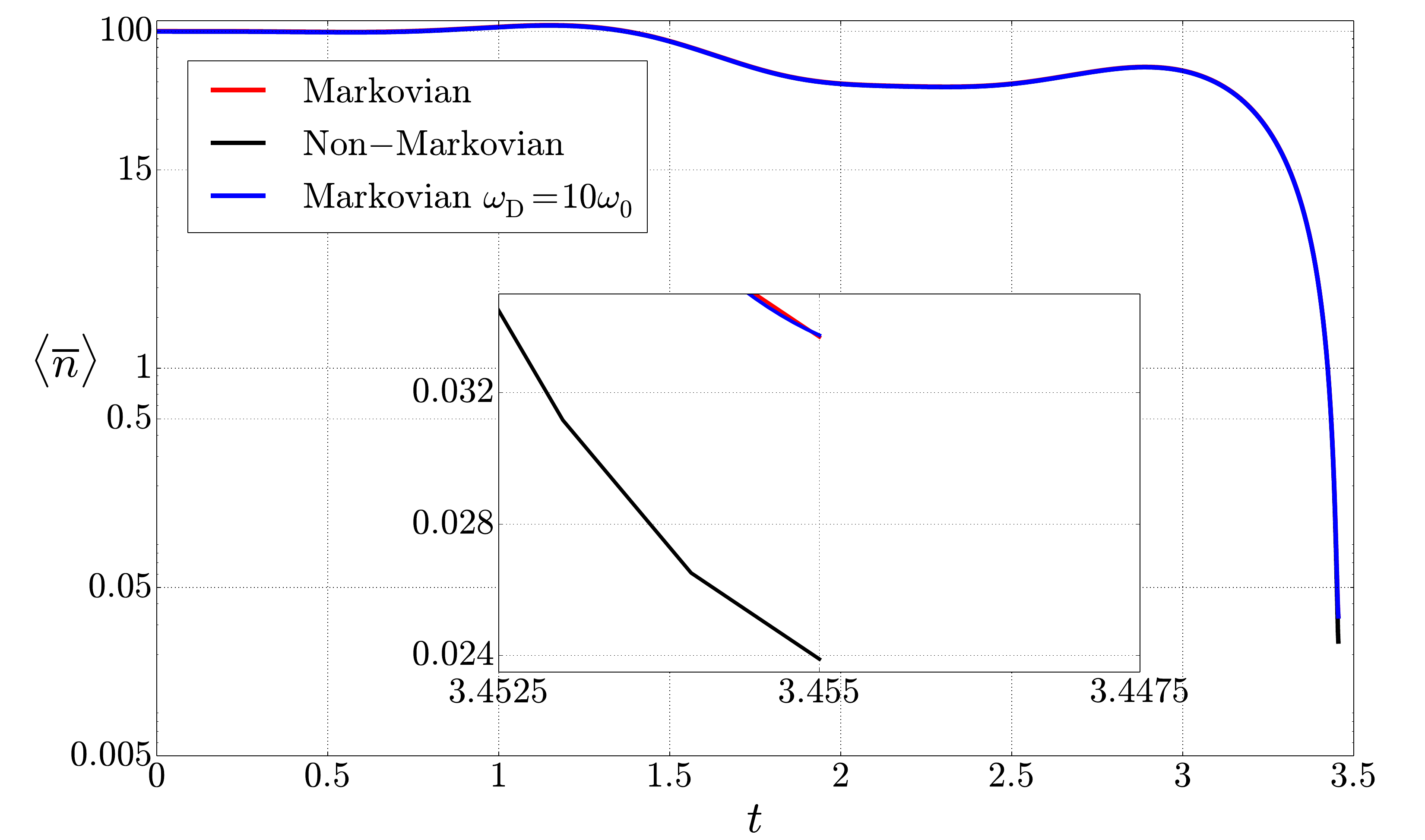}\\
\noindent\rule{\textwidth}{1pt}
\end{figure*}
\begin{figure}[h!]
\centering
d. $\gamma=10^{-3}\omega_0$ \\
\includegraphics[width=0.7278\textwidth]{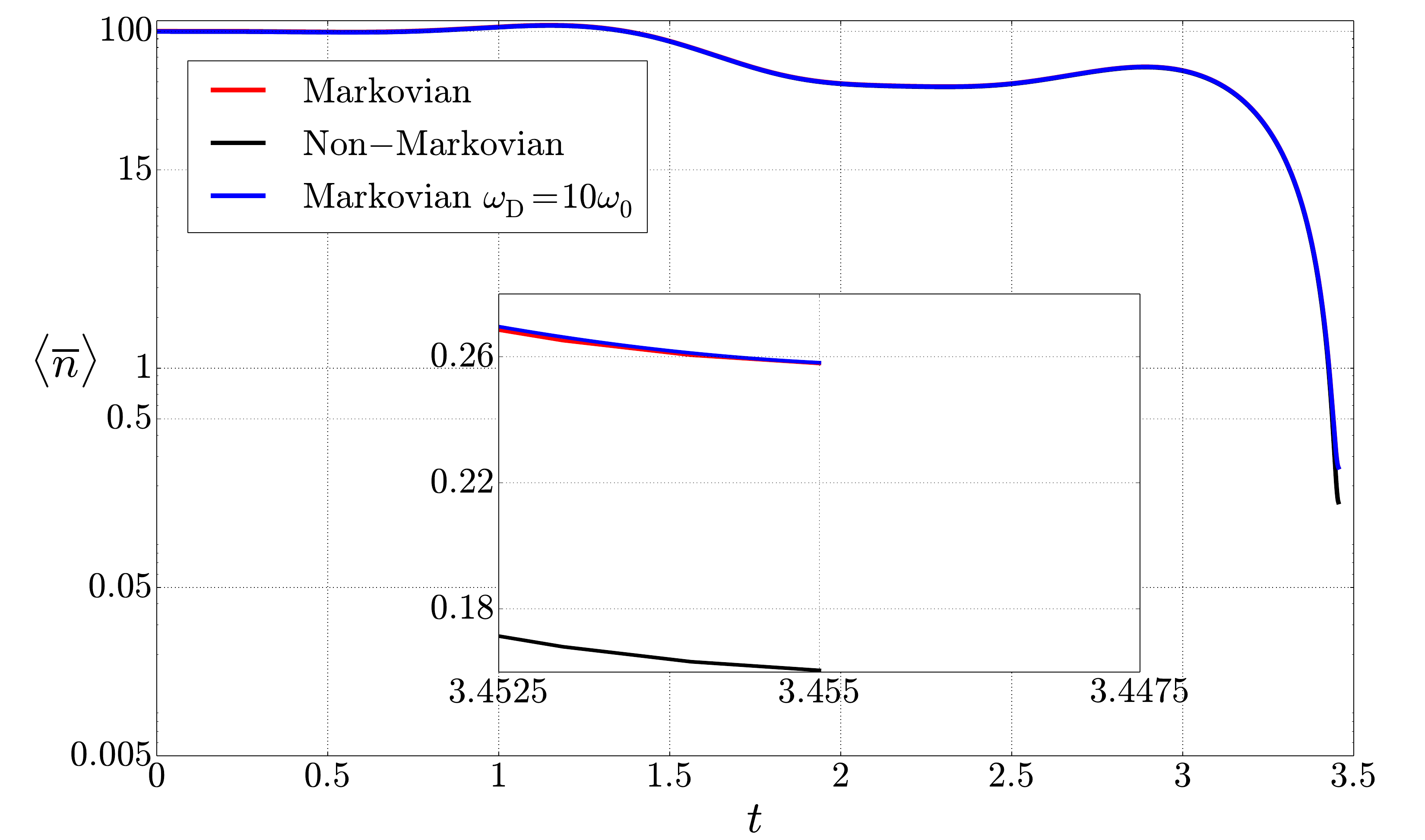}\\
e. $\gamma=10^{-2}\omega_0$ \\
\includegraphics[width=0.7278\textwidth]{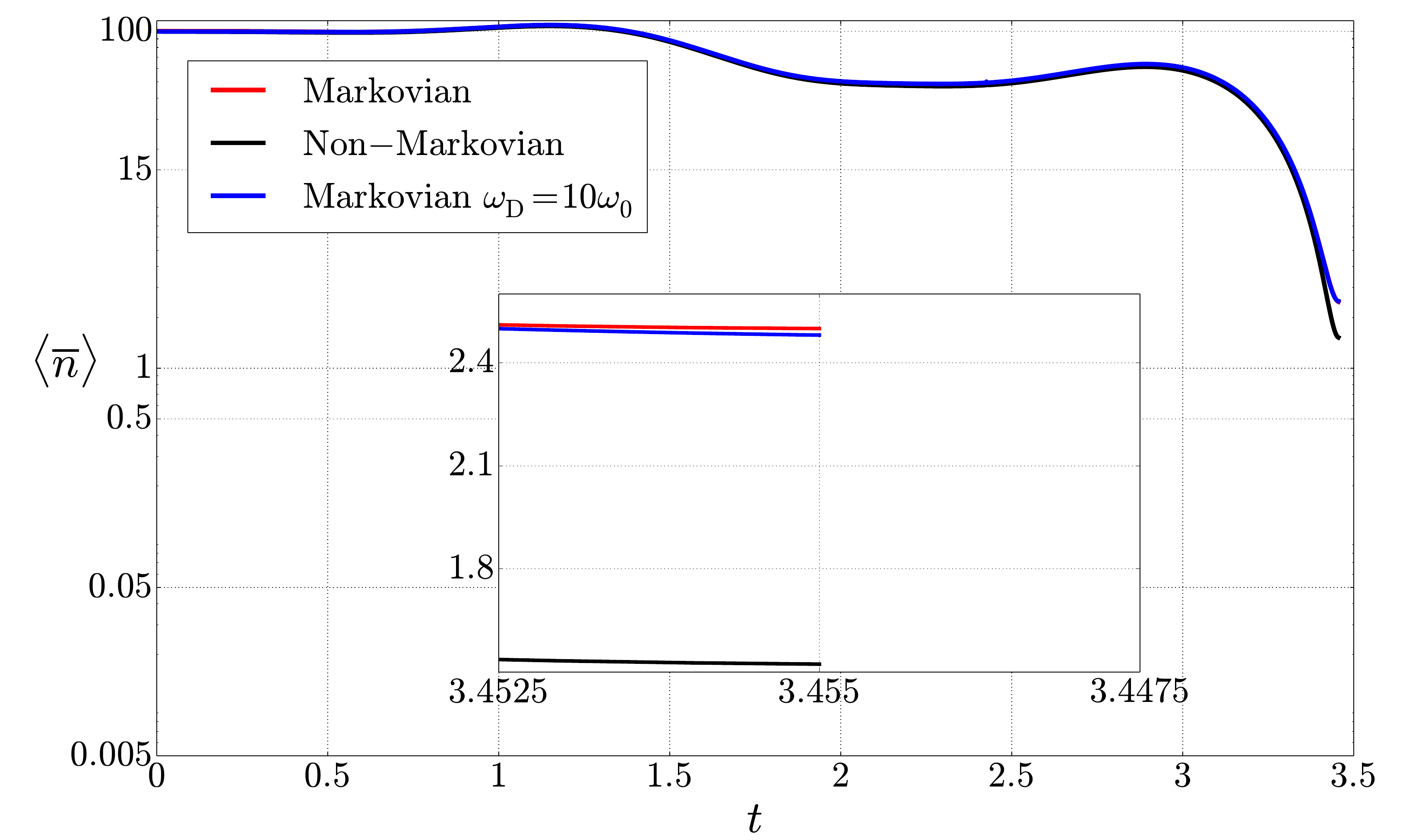}\\
f. $\gamma=10^{-1}\omega_0$ \\
\includegraphics[width=0.7278\textwidth]{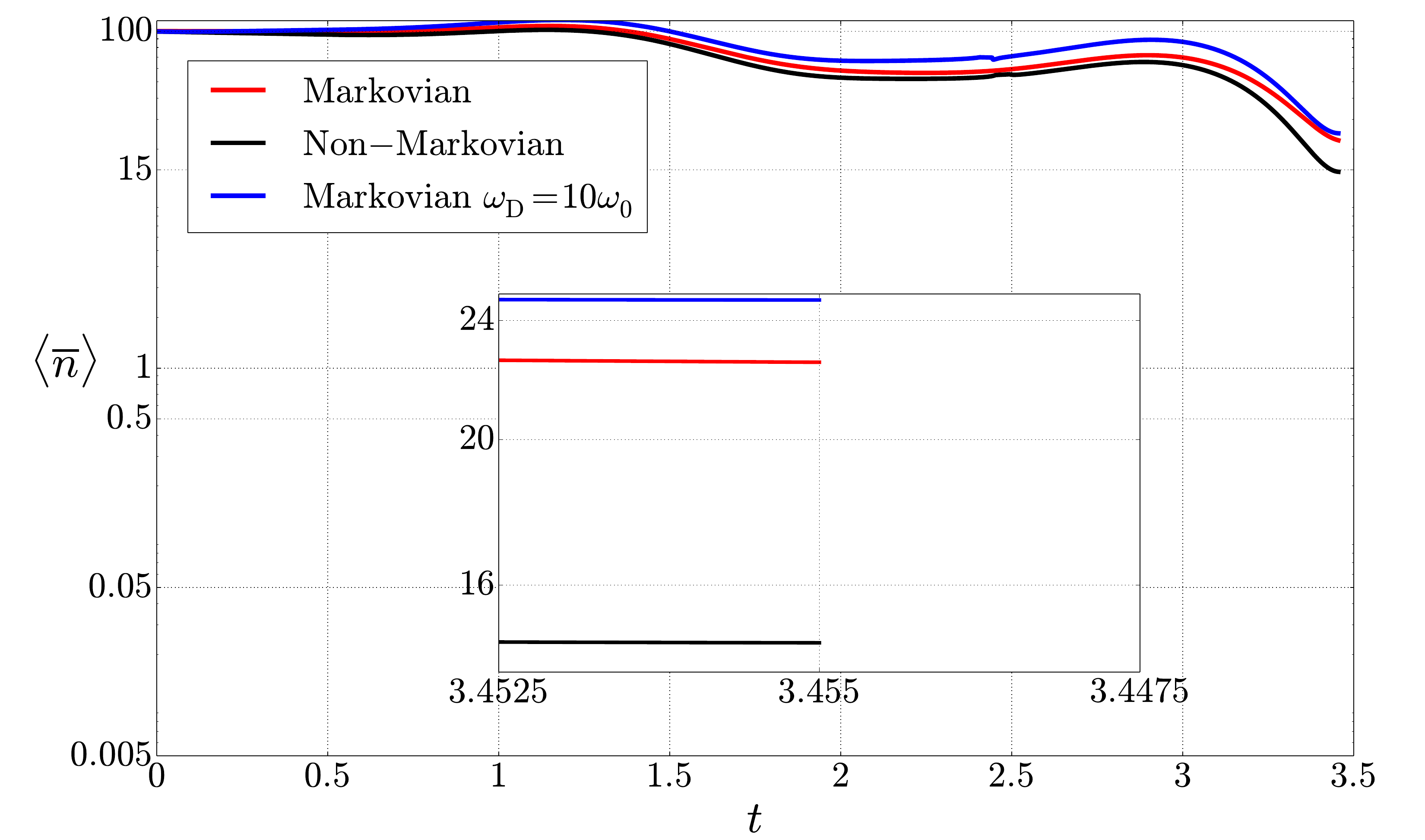}\\
\caption[Comparison of the dynamics of the cooling between the three processes, Markovian, non-Markovian and Markovian with $\OD=10\OM$.]{Comparison of the dynamics of the cooling between the three processes, Markovian, non-Markovian and Markovian with $\OD=10\OM$, for different parameters for dissipation. The initial parameters are $n_{\mathrm{T}}=100$, $n_{\mathrm{cav}}=0$. Insets: Magnification in the final time $t_{\mathrm{cool}}=1.1\pi/\omega_{0}$}
\label{fig:coolnonm}
\noindent\rule{\textwidth}{1pt}
\end{figure}


The optimal coupling function in the non-Markovian dynamics is shown in the inset of Fig. \ref{fig:ncoolnmop}. As we can see, the optimal coupling $g_{NM}(t)$ has the same behaviour that the coupling function $g(t)$ found in the Markovian case and both are almost equal. Despite of a minimum variation in the coupling function, we get a significant change in the minimum phonon number.

The minimum phonon number found after optimization for different values of dissipation is shown in Table \ref{tab:mpno}. There, we can see that when the dissipation increases, the minimum phonon number does not have a significant decrease.

\begin{center}
\begin{tabular}{c||c||c||c}
\multirow{2}{*}{Dissipation rate} & \multicolumn{2}{c||}{$\langle \hat{n}(t_{f}) \rangle$} & \multirow{2}{*}{Percentage}\\
\cline{2-3}
 & Non-Markovian & Non-Markovian Optimized & \\
 \hline \hline
$\kappa=\gamma=10^{-6}\omega_0$ & 8.86$\times10^{-3}$ & 3.43$\times10^{-3}$ & 61.29 \% \\
\hline
$\kappa=\gamma=10^{-5}\omega_0$ & 1.02$\times10^{-2}$ & 4.79$\times10^{-3}$ & 53.04 \% \\
\hline
$\kappa=\gamma=10^{-4}\omega_0$ & 2.39$\times10^{-2}$ & 1.83$\times10^{-2}$ & 23.43 \% \\
\hline
$\kappa=\gamma=10^{-3}\omega_0$ & 1.61$\times10^{-1}$ & 1.53$\times10^{-1}$ & 4.97 \% \\
\hline
$\kappa=\gamma=10^{-2}\omega_0$ & 1.52 & 1.50 & 1.32 \% \\
\hline
$\kappa=\gamma=10^{-1}\omega_0$ & 14.64 & 13.52 & 0.68 \%
\end{tabular}
\captionof{table}{Comparison of minimum phonon number in the non-Markovian regimen before and after optimization at time $t=1.1\pi/\omega_{0}$}
\label{tab:mpno}
\end{center}

With these last results, as we mentioned in the Chapter \ref{ch:influence}, we showed that non-Markovian character plays an important role both the characterization of the thermal equilibrium state and in sideband cooling.
This suggests that the non-Markovian character should be included in order to reach more realistic results in many different kind of systems. Another key result here is the robustness of the optimization process in the non-Markovian case \cite{TEP15}.

\begin{figure}[t]
\centering
\includegraphics[width=\textwidth]{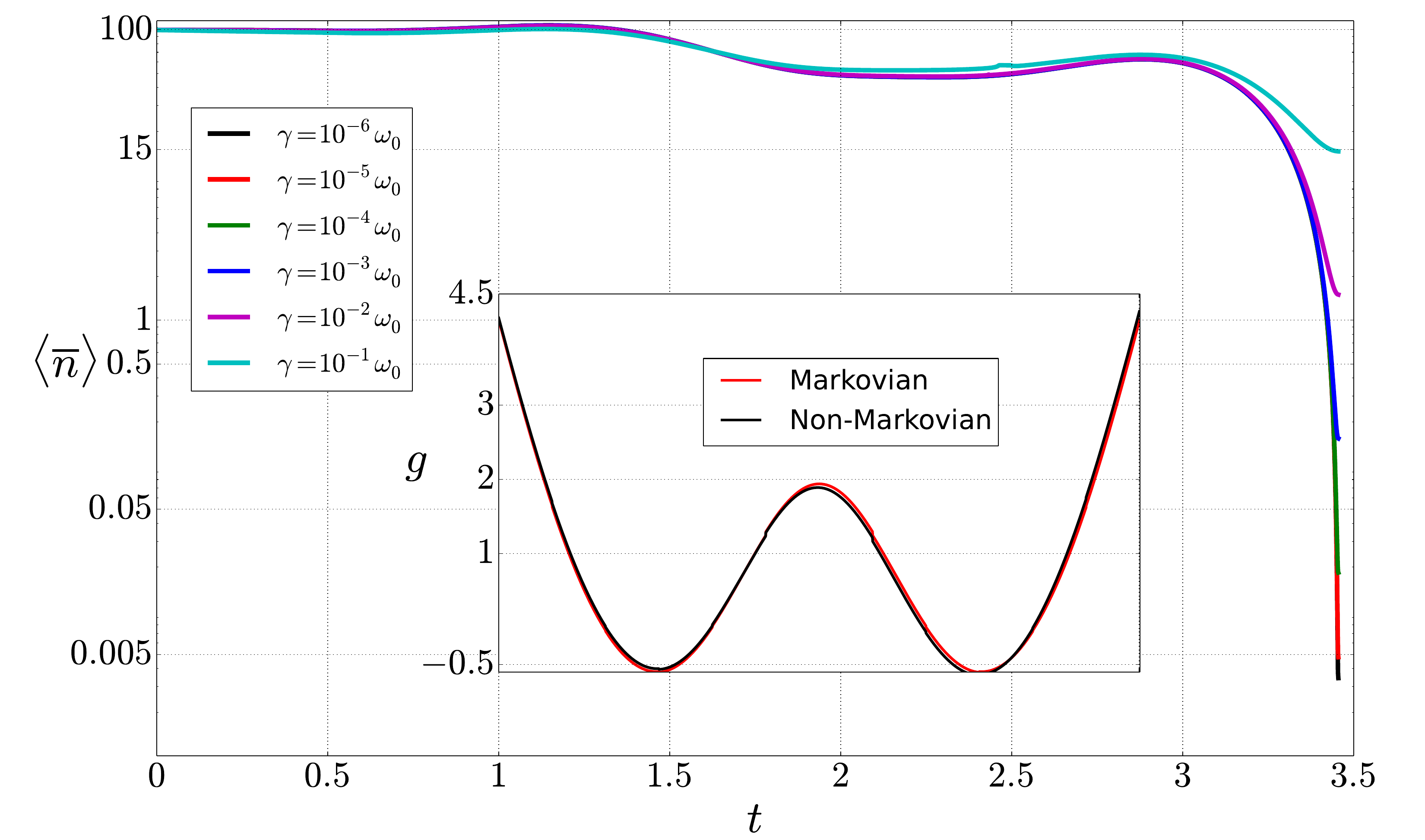}
\caption[Dynamics of the cooling with the optimal coupling function.]{Dynamics of the cooling with the optimal coupling function found later to the optimization process. The initial parameters are $n_{\mathrm{T}}=100$, $n_{\mathrm{cav}}=0$. Inset: The optimal control pulse for the coupling rate $g_{NM}(t)$ compare with the optimal control pulse for the coupling rate $g(t)$ shown in Fig. \ref{fig:corrcool}. The $y$-axis is in logarithmic scale.}
\label{fig:ncoolnmop}
\noindent\rule{\textwidth}{1pt}
\end{figure}

\endgroup

\cleardoublepage
\chapter*{Conclusions and Disscusions}\label{ch:conclu}
\addtocontents{toc}{\protect\vspace{\beforebibskip}} 
\addcontentsline{toc}{chapter}{\tocEntry{Conclusions}}

We elaborate on the role of the Heisenberg uncertainty principle [Eq.~(\ref{equ:thermodynbound})]
in the quantum thermal-equilibrium-state. It inhibits the system thermal equilibrium state to be 
characterized by the Gibbs state.
%
By contrast, for a wide class of systems, in the classical case the thermal-equilibrium-state
is exactly characterized by the Gibbs state. Further, in quantum mechanics, the Gibbs state is only 
recovered in the classical-high $T$ limit ($\hbar \beta \rightarrow 0$).
Nervertheless, the low-$T$ regime has some implications in quantum thermodynamics, because 
the non-Markovian effects should be taken into account. 
Among these implications, we find the failure for the Onsager hypothesis, or the difficulty in defining the specific heat 
\cite{IHT09}, and temperature \cite{FGA12}.
Specifically,  the  high  temperature  regime defined by $\hbar \beta  \rightarrow  0$,  modified
by  an appropriate effective coupling,  emerges  as  the main condition for the vanishing of
these deviations.

For the quantum version of the wide class of classical systems discussed above (for which the
thermal equilibrium state exactly corresponds to the Boltzmann distribution), the dependence
of thermal equilibrium state on the spectral density is clearly a pure quantum effect.
This feature can be explored as a quantum resource, e.g., in the one-photon phase control of
biochemical and biophysical systems \cite{PB13,PB13b,PYB13} or in understanding the coherent extent
of excitation with incoherent light in biological systems \cite{PB13}.
A phase-space formulation of this situation for non-harmonic systems, based on the Wigner 
representation of quantum mechanics and semiclassical approximations \cite{PID10,DGP10}, 
is currently under development.

Although  the  examples presented in Chapter \ref{ch:influence} are performed in the second order 
approximation of the potential, the methodology used in this work is general and could be used in non-linear 
cases and with non-Gaussian statistics, which are relevant, e.g., in nuclear physics \cite{WS&13}.
The  results  presented  show the role of non-Markovian effects and its relevance in the
thermal equilibrium state, which contribute to protect quantum features such as entanglement and squeezing.
Further,  this effects may be important or may shed light on the role in the derivation of fundamental limits  in  
areas  such  as  quantum metrology \cite{CHP12} and quantum speed limits \cite{DL13}.

Likewise, if the quantum thermal equilibrium state differs from the Gibbs state and depends implicitly on the 
non-Markovian interactions, this new equilibrium state helps us reaching a minimum phonon number 
(Eqs.~\ref{eqn:nmpnl} and \ref{eqn:nmpnl}) much lower than the minimum phonon number found in the 
Markovian approximation (Eq. \ref{eqn:mpn}). 
Similarly, when non-Markovian dynamics is considered in the optimal cooling of resonators by sideband 
cooling, the cooling process is more effective than in the Markovian approximation despite of using optimal 
control theory in order to find the optimal coupling function between the Fabry-Perot cavity and the resonator. 
This is compared with results found in previous works about sideband cooling \cite{WV&11,MC&07,MC&12,LX&13}, 
all these made in Markovian approximation.
Note that the spectral density can be extracted experimentally by means of spectroscopic techniques \cite{PB14b}.

\appendix
\cleardoublepage
\part{Appendix}
\counterwithin{figure}{chapter}

%

\chapter{Optimization Algorithm} \label{app:algo}

The algorithm used to solve the optimization problem was ``\textit{the method of steepest descent}'' whose algorithm consists of 4 steps:

\begin{enumerate}
\item Subdivide the interval $[ t_0,t_f ]$ into $N$ equal subintervals and assume an initial piecewise-constant control $g^{(0)}(t) = g{(0)}(t_k)$, $t \in [t_k, t_{k+1} ]$ $k = 0,1,...,N-1$.

\item Apply the assumed control $g^{(i)}$ to integrate the state equations from $t_0$ to $t_f$ with initial conditions $\mathbf{x}(t_0) = \mathbf{x}_0$ and store the state trajectory $\mathbf{x}^{(i)}$.

\item Apply $g^{(i)}$ and $\mathbf{x}^{(i)}$ to integrate costate equations backward, \ie, from $[ t_f,t_0 ]$. The ``initial value'' $\mathbf{p}^{(i)}(t_f)$ is obtained by:
\begin{equation}
\mathbf{p}^{(i)}(t_f)=\frac{\partial h}{\partial\mathbf{x}}(\mathbf{x}^{(i)}(t_f)).
\label{eq:auxequ}
\end{equation}
Evaluate $\partial H^{(i)}(t)/\partial g$, $t \in [t_0,t_f]$ and store this vector.

\item If
\begin{equation}
\label{eq:conditional} \left|\left| \frac{\partial H^{(i)}}{\partial g} \right|\right| \le \epsilon,
\end{equation}
\begin{equation}
\left|\left| \frac{\partial H^{(i)}}{\partial g} \right|\right|^{2} \equiv \int_{t_0}^{t_f} \left[ \left|\left| \frac{\partial H^{(i)}}{\partial g} \right|\right| \right]^{T}\left[ \left|\left| \frac{\partial H^{(i)}}{\partial g} \right|\right| \right]\, \mathrm{d}t
\end{equation}
then stop the iterative procedure. Here $\epsilon$ is a preselected small positive constant used as a tolerance. If Eq. (\ref{eq:conditional}) is not satisfied, adjust the piecewise-constant control function by:
\begin{equation}
g^{(i+1)}(t_k)=g^{(i)}-\tau\frac{\partial H^{(i)}}{\partial g}(t_k), \hspace{1cm} k=0,1,...,N-1.
\end{equation}
\end{enumerate}

Replace $g^{(i)}$ by $g^{(i+1)}$ and return to step 2. Here, $\tau$ is the step size.

In this algorithm we should take into account that the condition to minimize is given by
\begin{equation}
J=h(\mathbf{x}(t),t)+\int_{t_0}^{t_f}V(\mathbf{x}(t),\mathbf{g}(t),t)\,\mathrm{d}t
\end{equation}
and
\begin{equation}
H=V(\mathbf{x}(t),\mathbf{g}(t),t) + \lambda\mathbf{x}.
\end{equation}

\chapter{Equations of motion for the variances in the Markovian case}\label{app:equmot}

For the resonator and cavity the position $\hat{q}_i$ and momentum $\hat{p}_i$ operators are given by
\begin{equation}
\hat{q}_{\mathrm{m}}=\frac{1}{\sqrt{2}}(\an+\ad) \hspace{1cm} \hat{p}_{\mathrm{m}}=-\frac{i}{\sqrt{2}}(\an-\ad),
\end{equation}
\begin{equation}
\hat{q}_{\mathrm{cav}}=\frac{1}{\sqrt{2}}(\bn+\bd) \hspace{1cm} \hat{p}_{\mathrm{cav}}=-\frac{i}{\sqrt{2}}(\bn-\bd).
\end{equation} 
where the operators $\hat{q}_{\mathrm{m}}$ ($\hat{q}_{\mathrm{cav}}$) and $\hat{p}_{\mathrm{m}}$ ($\hat{p}_{\mathrm{cav}}$) correspond to the position and momentum operators of the mechanical (optical modes).

The Hamiltonian of the system in terms of the position and momentum is 
\begin{equation}
\hat{H}_{\mathrm{S}}=\sum_{i=1}^{2}\left(\frac{1}{2m_i}\hat{p}^{2}_{i} + \frac{1}{2}m_{i}\omega_{i}\hat{q}^{2}_{i} \right) + c(t)\hat{q}_{1}\hat{q}_{2},
\end{equation}
where $m_{i}$ and $\omega_{i}$ are the masses and frequencies of the modes.

The second moments are calculated by
\begin{eqnarray}
\nonumber \dot{ \hat{O}_{i} \hat{O}_{j}}\,\, (t) &=& \frac{\mathrm{d}\hat{O}_{i}(t)}{\mathrm{d}t} \hat{O}_{j}(t) + \hat{O}_{i}(t)\frac{\mathrm{d}\hat{O}_{j}(t)}{\mathrm{d}t} \\
&=& \frac{1}{\sqrt{2}}\left(\dot{\hat{q}}_{i}(t)+i\dot{\hat{p}}_{i}(t)\right)\hat{O}_{j}(t) + \frac{1}{\sqrt{2}} \hat{O}_{i}(t) \left(\dot{\hat{q}}_{j}(t)+i\dot{\hat{p}}_{i}(t)\right),
\end{eqnarray}
where $\dot{\hat{O}}_{i}$ are calculated in the Heisenberg picture. After performing all the calculations and including the terms of dissipation, we get the following equations of motion for the variances  $\phantom{\dot{\hat{}}_{i}}$

\begin{align}
&\langle\dot{{\hat{a} \hat{a}}}\rangle (t)=-2i\OM\langle\an\an\rangle + i c(t)(\langle\an\bn\rangle+\langle\an\bd\rangle) - \gamma \langle\an\an\rangle \\
\nonumber &\langle\dot{\an\ad}\rangle (t)=-\frac{1}{2}i c(t)\left(\langle\bd\an\rangle-\langle\bd\ad\rangle+\langle\bn\an\rangle-\langle\bn\ad\rangle\right) \\
&\hspace*{18mm}- \gamma(n_{\mathrm{th}}+1) \langle\ad\an\rangle + \gamma n_{\mathrm{th}} \langle\an\ad\rangle \\
\nonumber &\langle\dot{\an\bn}\rangle (t)=-2i\OM\langle\an\bn\rangle+\frac{1}{2}i c(t)\left(\langle\bn\bn\rangle+\langle\an\an\rangle+\langle\bd\bn\rangle+\langle\an\ad\rangle\right) \\
&\hspace*{18mm}- \frac{1}{2}\gamma\langle\an\bn\rangle -\frac{1}{2} \kappa \langle\bn\an\rangle \\
&\langle\dot{\an\bd}\rangle (t)=\frac{1}{2}i c(t)\left(\langle\bn\bd\rangle-\langle\an\ad\rangle+\langle\bd\bd\rangle-\langle\an\an\rangle\right)  - \frac{1}{2}\gamma\langle\an\bd\rangle -\frac{1}{2} \kappa \langle\bd\an\rangle
\end{align}

\begin{align}
\nonumber&\langle\dot{\bn\an}\rangle (t)=-2i\OM\langle\bn\an\rangle+\frac{1}{2}i c(t)\left(\langle\bn\bn\rangle+\langle\an\an\rangle+\langle\bd\bn\rangle+\langle\an\ad\rangle\right) \\
&\hspace*{18mm}- \frac{1}{2}\gamma\langle\an\bn\rangle -\frac{1}{2} \kappa \langle\bn\an\rangle, \\
\nonumber &\langle\dot{\bn\ad}\rangle (t)=-\frac{1}{2}i c(t)\left(\langle\bd\bn\rangle-\langle\ad\an\rangle+\langle\bn\bn\rangle-\langle\ad\ad\rangle\right) \\
&\hspace*{18mm}- \frac{1}{2}\gamma \langle\bn\ad\rangle - \frac{1}{2}\kappa \langle\bn\ad\rangle, \\
&\langle\dot{\bn\bn}\rangle (t)=-2i\OM\langle\bn\bn\rangle + i c(t)(\langle\bn\an\rangle+\langle\bn\ad\rangle) - \kappa \langle\bn\bn\rangle, \\
\nonumber&\langle\dot{\bn\bd}\rangle (t)=-\frac{1}{2}i c(t)\left(\langle\bn\ad\rangle-\langle\bd\ad\rangle+\langle\bn\an\rangle-\langle\bd\an\rangle\right) \\
&\hspace*{18mm}- \kappa(n_{\mathrm{cav}}+1) \langle\bd\bn\rangle + \kappa n_{\mathrm{cav}} \langle\bn\bd\rangle, \\
\nonumber &\langle\dot{\ad\an}\rangle (t)=\frac{1}{2}i c(t)\left(\langle\bn\ad\rangle+\langle\bd\ad\rangle-\langle\bn\an\rangle-\langle\bd\an\rangle\right) \\
&\hspace*{18mm}- \gamma(n_{\mathrm{th}}+1) \langle\ad\an\rangle + \gamma n_{\mathrm{th}} \langle\an\ad\rangle, \\
&\langle\dot{\ad\ad}\rangle (t)=2i\OM\langle\ad\ad\rangle + i c(t)(\langle\bd\ad\rangle+\langle\bn\ad\rangle) - \gamma \langle\ad\ad\rangle, \\
\nonumber&\langle\dot{\ad\bn}\rangle (t)=-\frac{1}{2}i c(t)\left(\langle\bd\bn\rangle-\langle\ad\an\rangle+\langle\bn\bn\rangle-\langle\ad\ad\rangle\right) \\
&\hspace*{18mm}- \gamma \langle\ad\bn\rangle + \kappa \langle\ad\bn\rangle, \\
\nonumber&\langle\dot{\ad\bd}\rangle (t)=2i\OM\langle\ad\bd\rangle-\frac{1}{2}i c(t)\left(\langle\ad\ad\rangle+\langle\bd\bd\rangle+\langle\bn\bd\rangle+\langle\ad\an\rangle\right) \\
&\hspace*{18mm}- \frac{1}{2}\gamma\langle\ad\bd\rangle -\frac{1}{2} \kappa \langle\bd\ad\rangle, \\
&\langle\dot{\bd\an}\rangle (t)=\frac{1}{2}i c(t)\left(\langle\bn\bd\rangle-\langle\an\ad\rangle+\langle\bd\bd\rangle-\langle\an\an\rangle\right)  - \frac{1}{2}\gamma\langle\bd\an\rangle -\frac{1}{2} \kappa \langle\an\bd\rangle, \\
\nonumber&\langle\dot{\bd\ad}\rangle (t)=2i\OM\langle\bd\ad\rangle-\frac{1}{2}i c(t)\left(\langle\ad\ad\rangle+\langle\bd\bd\rangle+\langle\bn\bd\rangle+\langle\ad\an\rangle\right) \\
&\hspace*{18mm}- \frac{1}{2}\gamma\langle\bd\ad\rangle -\frac{1}{2} \kappa \langle\ad\bd\rangle, \\
\nonumber &\langle\dot{\bd\bn}\rangle (t)=\frac{1}{2}i c(t)\left(\langle\bd\an\rangle+\langle\bd\ad\rangle-\langle\bn\an\rangle-\langle\bn\ad\rangle\right) \\
&\hspace*{18mm}- \kappa(n_{\mathrm{cav}}+1) \langle\bd\bn\rangle + \gamma n_{\mathrm{cav}} \langle\bn\bd\rangle, \\
&\langle\dot{\bd\bd}\rangle (t)=2i\OM\langle\bd\bd\rangle + i c(t)(\langle\bd\ad\rangle+\langle\bd\an\rangle) - \kappa \langle\bd\bd\rangle. \\
\end{align}

\chapter{Variances and Optimization Equations}\label{app:variances}

\section{Variance for position operator}

The initial conditions for the thermal states of the mechanical (1) and optical (2) modes are
%
\begin{eqnarray}
\langle q_{1} \rangle (0) &=& 0 \\
\langle q_{2} \rangle (0) &=& 0 \\
\langle p_{1} \rangle (0) &=& 0 \\
\langle p_{2} \rangle (0) &=& 0 \\
\langle q_{1}q_{1} \rangle (0) &=& \frac{1}{2}\coth\left(\frac{1}{2}\beta_{1}\omega_{1} \right) \\
%
\langle q_{2}q_{2} \rangle (0) &=& \frac{1}{2m\omega_{2}}\coth\left(\frac{1}{2}\beta_{2} \omega_{2} \right) \\
\langle p_{1}p_{1} \rangle (0) &=& \frac{1}{2}\coth\left(\frac{1}{2}\beta_{1} \right) \\
\langle p_{2}p_{2} \rangle (0) &=& \frac{m\omega_{2}}{2}\coth\left(\frac{1}{2}\beta_{2}\omega_{2} \right) \\
\langle q_{1}p_{1} \rangle (0) &=& 0 \\
\langle q_{2}p_{2} \rangle (0) &=& 0 
\end{eqnarray}

The second moments of the mechanical resonator $\langle q_{1}q_{1} \rangle(t)=$sq1q1[t] and $\langle p_{1}p_{1} \rangle$=sp1p1[t] are given by

\begin{spverbatim}
sq1q1[t_]:=(sp2p2i (U2p[0] U4[t]-U2[t] U4p[0])^2)/(4 (2 U2[t] V2[t]-2 U4[t] V4[t])^2 (((-U2p[0] V2[t]+U4p[0] V4[t]) (-m U2[t] V2p[0]+m U4[t] V4p[0]))/(2 U2[t] V2[t]-2 U4[t] V4[t])^2-(m (U2p[0] U4[t]-U2[t] U4p[0]) (V2p[0] V4[t]-V2[t] V4p[0]))/(2 U2[t] V2[t]-2 U4[t] V4[t])^2)^2)+(vep2i^2 (U2p[0] U4[t]-U2[t] U4p[0])^2)/(4 (2 U2[t] V2[t]-2 U4[t] V4[t])^2 (((-U2p[0] V2[t]+U4p[0] V4[t]) (-m U2[t] V2p[0]+m U4[t] V4p[0]))/(2 U2[t] V2[t]-2 U4[t] V4[t])^2-(m (U2p[0] U4[t]-U2[t] U4p[0]) (V2p[0] V4[t]-V2[t] V4p[0]))/(2 U2[t] V2[t]-2 U4[t] V4[t])^2)^2)+((U2p[0] U4[t]-U2[t] U4p[0])^2 (u22[t] (u4[t] v1[t]-u3[t] v2[t])^2+u2[t]^2 (u44[t] v1[t]^2-v2[t] (-(u33[t]+v11[t]) v2[t]+v1[t] (u34[t]+u43[t]+v12[t]+v21[t]))+v1[t]^2 v22[t])+u24[t] u3[t] u4[t] v1[t] v4[t]-u23[t] u4[t]^2 v1[t] v4[t]-u32[t] u4[t]^2 v1[t] v4[t]+u3[t] u4[t] u42[t] v1[t] v4[t]-u4[t]^2 v1[t] v14[t] v4[t]-u24[t] u3[t]^2 v2[t] v4[t]+u23[t] u3[t] u4[t] v2[t] v4[t]+u3[t] u32[t] u4[t] v2[t] v4[t]-u3[t]^2 u42[t] v2[t] v4[t]+u3[t] u4[t] v14[t] v2[t] v4[t]+u3[t] u4[t] v1[t] v24[t] v4[t]-u3[t]^2 v2[t] v24[t] v4[t]-u3[t] u34[t] u4[t] v4[t]^2+u33[t] u4[t]^2 v4[t]^2-u3[t] u4[t] u43[t] v4[t]^2+u3[t]^2 u44[t] v4[t]^2+u4[t]^2 v11[t] v4[t]^2-u3[t] u4[t] v12[t] v4[t]^2-u3[t] u4[t] v21[t] v4[t]^2+u3[t]^2 v22[t] v4[t]^2-u4[t]^2 v1[t] v4[t] v41[t]+u3[t] u4[t] v2[t] v4[t] v41[t]+u3[t] u4[t] v1[t] v4[t] v42[t]-u3[t]^2 v2[t] v4[t] v42[t]+u2[t] (u24[t] v1[t] (-u4[t] v1[t]+u3[t] v2[t])+u4[t] (-u42[t] v1[t]^2+u23[t] v1[t] v2[t]+u32[t] v1[t] v2[t]+v1[t] v14[t] v2[t]-v1[t]^2 v24[t]+u34[t] v1[t] v4[t]+u43[t] v1[t] v4[t]+v1[t] v12[t] v4[t]-2 u33[t] v2[t] v4[t]-2 v11[t] v2[t] v4[t]+v1[t] v21[t] v4[t]+v1[t] v2[t] v41[t]-v1[t]^2 v42[t])+u3[t] (u42[t] v1[t] v2[t]-u23[t] v2[t]^2-u32[t] v2[t]^2-v14[t] v2[t]^2+v1[t] v2[t] v24[t]-2 u44[t] v1[t] v4[t]+u34[t] v2[t] v4[t]+u43[t] v2[t] v4[t]+v12[t] v2[t] v4[t]+v2[t] v21[t] v4[t]-2 v1[t] v22[t] v4[t]-v2[t]^2 v41[t]+v1[t] v2[t] v42[t]))+(u4[t] v1[t]-u3[t] v2[t])^2 v44[t]))/(2 (u2[t] v2[t]-u4[t] v4[t])^2 (2 U2[t] V2[t]-2 U4[t] V4[t])^2 (((-U2p[0] V2[t]+U4p[0] V4[t]) (-m U2[t] V2p[0]+m U4[t] V4p[0]))/(2 U2[t] V2[t]-2 U4[t] V4[t])^2-(m (U2p[0] U4[t]-U2[t] U4p[0]) (V2p[0] V4[t]-V2[t] V4p[0]))/(2 U2[t] V2[t]-2 U4[t] V4[t])^2)^2)-(vep1i vep2i (U2p[0] U4[t]-U2[t] U4p[0]) (-m U2[t] V2p[0]+m U4[t] V4p[0]))/(2 (2 U2[t] V2[t]-2 U4[t] V4[t])^2 (((-U2p[0] V2[t]+U4p[0] V4[t]) (-m U2[t] V2p[0]+m U4[t] V4p[0]))/(2 U2[t] V2[t]-2 U4[t] V4[t])^2-(m (U2p[0] U4[t]-U2[t] U4p[0]) (V2p[0] V4[t]-V2[t] V4p[0]))/(2 U2[t] V2[t]-2 U4[t] V4[t])^2)^2)-(vep2i veq1i (U2p[0] U4[t]-U2[t] U4p[0]) (U2[t] (-U1p[0] V2[t]+U4p[0] V3[t])+U4[t] (-U2p[0] V3[t]+U1p[0] V4[t])+U1[t] (U2p[0] V2[t]-U4p[0] V4[t])) (-m U2[t] V2p[0]+m U4[t] V4p[0]))/((2 U2[t] V2[t]-2 U4[t] V4[t])^3 (((-U2p[0] V2[t]+U4p[0] V4[t]) (-m U2[t] V2p[0]+m U4[t] V4p[0]))/(2 U2[t] V2[t]-2 U4[t] V4[t])^2-(m (U2p[0] U4[t]-U2[t] U4p[0]) (V2p[0] V4[t]-V2[t] V4p[0]))/(2 U2[t] V2[t]-2 U4[t] V4[t])^2)^2)-(sq2p2i (U2p[0] U4[t]-U2[t] U4p[0]) (U2[t] (U4p[0] V1[t]-U3p[0] V2[t])+U4[t] (-U2p[0] V1[t]+U3p[0] V4[t])+U3[t] (U2p[0] V2[t]-U4p[0] V4[t])) (-m U2[t] V2p[0]+m U4[t] V4p[0]))/((2 U2[t] V2[t]-2 U4[t] V4[t])^3 (((-U2p[0] V2[t]+U4p[0] V4[t]) (-m U2[t] V2p[0]+m U4[t] V4p[0]))/(2 U2[t] V2[t]-2 U4[t] V4[t])^2-(m (U2p[0] U4[t]-U2[t] U4p[0]) (V2p[0] V4[t]-V2[t] V4p[0]))/(2 U2[t] V2[t]-2 U4[t] V4[t])^2)^2)-(vep2i veq2i (U2p[0] U4[t]-U2[t] U4p[0]) (U2[t] (U4p[0] V1[t]-U3p[0] V2[t])+U4[t] (-U2p[0] V1[t]+U3p[0] V4[t])+U3[t] (U2p[0] V2[t]-U4p[0] V4[t])) (-m U2[t] V2p[0]+m U4[t] V4p[0]))/((2 U2[t] V2[t]-2 U4[t] V4[t])^3 (((-U2p[0] V2[t]+U4p[0] V4[t]) (-m U2[t] V2p[0]+m U4[t] V4p[0]))/(2 U2[t] V2[t]-2 U4[t] V4[t])^2-(m (U2p[0] U4[t]-U2[t] U4p[0]) (V2p[0] V4[t]-V2[t] V4p[0]))/(2 U2[t] V2[t]-2 U4[t] V4[t])^2)^2)-((U2p[0] U4[t]-U2[t] U4p[0]) (u12[t] u2[t] u4[t] v1[t] v2[t]+u2[t] u21[t] u4[t] v1[t] v2[t]-u2[t]^2 u41[t] v1[t] v2[t]+u13[t] u2[t]^2 v2[t]^2-u12[t] u2[t] u3[t] v2[t]^2-u2[t] u21[t] u3[t] v2[t]^2+u2[t]^2 u31[t] v2[t]^2+u2[t]^2 v13[t] v2[t]^2-u2[t]^2 v1[t] v2[t] v23[t]-2 u2[t] u24[t] u4[t] v1[t] v3[t]+2 u22[t] u4[t]^2 v1[t] v3[t]-2 u2[t] u4[t] u42[t] v1[t] v3[t]+2 u2[t]^2 u44[t] v1[t] v3[t]+u2[t] u24[t] u3[t] v2[t] v3[t]-u2[t]^2 u34[t] v2[t] v3[t]+u2[t] u23[t] u4[t] v2[t] v3[t]-2 u22[t] u3[t] u4[t] v2[t] v3[t]+u2[t] u32[t] u4[t] v2[t] v3[t]+u2[t] u3[t] u42[t] v2[t] v3[t]-u2[t]^2 u43[t] v2[t] v3[t]-u2[t]^2 v12[t] v2[t] v3[t]+u2[t] u4[t] v14[t] v2[t] v3[t]-u2[t]^2 v2[t] v21[t] v3[t]+2 u2[t]^2 v1[t] v22[t] v3[t]-2 u2[t] u4[t] v1[t] v24[t] v3[t]+u2[t] u3[t] v2[t] v24[t] v3[t]+u2[t]^2 v2[t]^2 v31[t]-u2[t]^2 v1[t] v2[t] v32[t]+u2[t] u4[t] v1[t] v2[t] v34[t]-u2[t] u3[t] v2[t]^2 v34[t]-u12[t] u4[t]^2 v1[t] v4[t]-u21[t] u4[t]^2 v1[t] v4[t]+u2[t] u4[t] u41[t] v1[t] v4[t]-2 u13[t] u2[t] u4[t] v2[t] v4[t]+u12[t] u3[t] u4[t] v2[t] v4[t]+u21[t] u3[t] u4[t] v2[t] v4[t]-2 u2[t] u31[t] u4[t] v2[t] v4[t]+u2[t] u3[t] u41[t] v2[t] v4[t]-2 u2[t] u4[t] v13[t] v2[t] v4[t]+u2[t] u4[t] v1[t] v23[t] v4[t]+u2[t] u3[t] v2[t] v23[t] v4[t]+u24[t] u3[t] u4[t] v3[t] v4[t]+u2[t] u34[t] u4[t] v3[t] v4[t]-u23[t] u4[t]^2 v3[t] v4[t]-u32[t] u4[t]^2 v3[t] v4[t]+u3[t] u4[t] u42[t] v3[t] v4[t]+u2[t] u4[t] u43[t] v3[t] v4[t]-2 u2[t] u3[t] u44[t] v3[t] v4[t]+u2[t] u4[t] v12[t] v3[t] v4[t]-u4[t]^2 v14[t] v3[t] v4[t]+u2[t] u4[t] v21[t] v3[t] v4[t]-2 u2[t] u3[t] v22[t] v3[t] v4[t]+u3[t] u4[t] v24[t] v3[t] v4[t]-2 u2[t] u4[t] v2[t] v31[t] v4[t]+u2[t] u4[t] v1[t] v32[t] v4[t]+u2[t] u3[t] v2[t] v32[t] v4[t]-u4[t]^2 v1[t] v34[t] v4[t]+u3[t] u4[t] v2[t] v34[t] v4[t]+u13[t] u4[t]^2 v4[t]^2+u31[t] u4[t]^2 v4[t]^2-u3[t] u4[t] u41[t] v4[t]^2+u4[t]^2 v13[t] v4[t]^2-u3[t] u4[t] v23[t] v4[t]^2+u4[t]^2 v31[t] v4[t]^2-u3[t] u4[t] v32[t] v4[t]^2-u14[t] (u2[t] v1[t]-u3[t] v4[t]) (u2[t] v2[t]-u4[t] v4[t])+u2[t] u4[t] v2[t] v3[t] v41[t]-u4[t]^2 v3[t] v4[t] v41[t]-2 u2[t] u4[t] v1[t] v3[t] v42[t]+u2[t] u3[t] v2[t] v3[t] v42[t]+u3[t] u4[t] v3[t] v4[t] v42[t]+u2[t] u4[t] v1[t] v2[t] v43[t]-u2[t] u3[t] v2[t]^2 v43[t]-u4[t]^2 v1[t] v4[t] v43[t]+u3[t] u4[t] v2[t] v4[t] v43[t]+2 u4[t] (u4[t] v1[t]-u3[t] v2[t]) v3[t] v44[t]+u1[t] (2 u22[t] v2[t] (-u4[t] v1[t]+u3[t] v2[t])+u2[t] (u24[t] v1[t] v2[t]+u42[t] v1[t] v2[t]-u23[t] v2[t]^2-u32[t] v2[t]^2-v14[t] v2[t]^2+v1[t] v2[t] v24[t]-2 u44[t] v1[t] v4[t]+u34[t] v2[t] v4[t]+u43[t] v2[t] v4[t]+v12[t] v2[t] v4[t]+v2[t] v21[t] v4[t]-2 v1[t] v22[t] v4[t]-v2[t]^2 v41[t]+v1[t] v2[t] v42[t])+v4[t] (u24[t] (u4[t] v1[t]-2 u3[t] v2[t])+u4[t] (-(u34[t]+u43[t]+v12[t]+v21[t]) v4[t]+v2[t] (u23[t]+u32[t]+v14[t]+v41[t])+v1[t] (u42[t]+v24[t]+v42[t]))-2 u3[t] (-(u44[t]+v22[t]) v4[t]+v2[t] (u42[t]+v24[t]+v42[t])))+2 v2[t] (-u4[t] v1[t]+u3[t] v2[t]) v44[t])) (-m U2[t] V2p[0]+m U4[t] V4p[0]))/(2 (u2[t] v2[t]-u4[t] v4[t])^2 (2 U2[t] V2[t]-2 U4[t] V4[t])^2 (((-U2p[0] V2[t]+U4p[0] V4[t]) (-m U2[t] V2p[0]+m U4[t] V4p[0]))/(2 U2[t] V2[t]-2 U4[t] V4[t])^2-(m (U2p[0] U4[t]-U2[t] U4p[0]) (V2p[0] V4[t]-V2[t] V4p[0]))/(2 U2[t] V2[t]-2 U4[t] V4[t])^2)^2)+(sp1p1i (-m U2[t] V2p[0]+m U4[t] V4p[0])^2)/(4 (2 U2[t] V2[t]-2 U4[t] V4[t])^2 (((-U2p[0] V2[t]+U4p[0] V4[t]) (-m U2[t] V2p[0]+m U4[t] V4p[0]))/(2 U2[t] V2[t]-2 U4[t] V4[t])^2-(m (U2p[0] U4[t]-U2[t] U4p[0]) (V2p[0] V4[t]-V2[t] V4p[0]))/(2 U2[t] V2[t]-2 U4[t] V4[t])^2)^2)+(vep1i^2 (-m U2[t] V2p[0]+m U4[t] V4p[0])^2)/(4 (2 U2[t] V2[t]-2 U4[t] V4[t])^2 (((-U2p[0] V2[t]+U4p[0] V4[t]) (-m U2[t] V2p[0]+m U4[t] V4p[0]))/(2 U2[t] V2[t]-2 U4[t] V4[t])^2-(m (U2p[0] U4[t]-U2[t] U4p[0]) (V2p[0] V4[t]-V2[t] V4p[0]))/(2 U2[t] V2[t]-2 U4[t] V4[t])^2)^2)+(sq1p1i (U2[t] (-U1p[0] V2[t]+U4p[0] V3[t])+U4[t] (-U2p[0] V3[t]+U1p[0] V4[t])+U1[t] (U2p[0] V2[t]-U4p[0] V4[t])) (-m U2[t] V2p[0]+m U4[t] V4p[0])^2)/((2 U2[t] V2[t]-2 U4[t] V4[t])^3 (((-U2p[0] V2[t]+U4p[0] V4[t]) (-m U2[t] V2p[0]+m U4[t] V4p[0]))/(2 U2[t] V2[t]-2 U4[t] V4[t])^2-(m (U2p[0] U4[t]-U2[t] U4p[0]) (V2p[0] V4[t]-V2[t] V4p[0]))/(2 U2[t] V2[t]-2 U4[t] V4[t])^2)^2)+(vep1i veq1i (U2[t] (-U1p[0] V2[t]+U4p[0] V3[t])+U4[t] (-U2p[0] V3[t]+U1p[0] V4[t])+U1[t] (U2p[0] V2[t]-U4p[0] V4[t])) (-m U2[t] V2p[0]+m U4[t] V4p[0])^2)/((2 U2[t] V2[t]-2 U4[t] V4[t])^3 (((-U2p[0] V2[t]+U4p[0] V4[t]) (-m U2[t] V2p[0]+m U4[t] V4p[0]))/(2 U2[t] V2[t]-2 U4[t] V4[t])^2-(m (U2p[0] U4[t]-U2[t] U4p[0]) (V2p[0] V4[t]-V2[t] V4p[0]))/(2 U2[t] V2[t]-2 U4[t] V4[t])^2)^2)+(sq1q1i (U2[t] (-U1p[0] V2[t]+U4p[0] V3[t])+U4[t] (-U2p[0] V3[t]+U1p[0] V4[t])+U1[t] (U2p[0] V2[t]-U4p[0] V4[t]))^2 (-m U2[t] V2p[0]+m U4[t] V4p[0])^2)/((2 U2[t] V2[t]-2 U4[t] V4[t])^4 (((-U2p[0] V2[t]+U4p[0] V4[t]) (-m U2[t] V2p[0]+m U4[t] V4p[0]))/(2 U2[t] V2[t]-2 U4[t] V4[t])^2-(m (U2p[0] U4[t]-U2[t] U4p[0]) (V2p[0] V4[t]-V2[t] V4p[0]))/(2 U2[t] V2[t]-2 U4[t] V4[t])^2)^2)+(veq1i^2 (U2[t] (-U1p[0] V2[t]+U4p[0] V3[t])+U4[t] (-U2p[0] V3[t]+U1p[0] V4[t])+U1[t] (U2p[0] V2[t]-U4p[0] V4[t]))^2 (-m U2[t] V2p[0]+m U4[t] V4p[0])^2)/((2 U2[t] V2[t]-2 U4[t] V4[t])^4 (((-U2p[0] V2[t]+U4p[0] V4[t]) (-m U2[t] V2p[0]+m U4[t] V4p[0]))/(2 U2[t] V2[t]-2 U4[t] V4[t])^2-(m (U2p[0] U4[t]-U2[t] U4p[0]) (V2p[0] V4[t]-V2[t] V4p[0]))/(2 U2[t] V2[t]-2 U4[t] V4[t])^2)^2)+(vep1i veq2i (U2[t] (U4p[0] V1[t]-U3p[0] V2[t])+U4[t] (-U2p[0] V1[t]+U3p[0] V4[t])+U3[t] (U2p[0] V2[t]-U4p[0] V4[t])) (-m U2[t] V2p[0]+m U4[t] V4p[0])^2)/((2 U2[t] V2[t]-2 U4[t] V4[t])^3 (((-U2p[0] V2[t]+U4p[0] V4[t]) (-m U2[t] V2p[0]+m U4[t] V4p[0]))/(2 U2[t] V2[t]-2 U4[t] V4[t])^2-(m (U2p[0] U4[t]-U2[t] U4p[0]) (V2p[0] V4[t]-V2[t] V4p[0]))/(2 U2[t] V2[t]-2 U4[t] V4[t])^2)^2)+(2 veq1i veq2i (U2[t] (-U1p[0] V2[t]+U4p[0] V3[t])+U4[t] (-U2p[0] V3[t]+U1p[0] V4[t])+U1[t] (U2p[0] V2[t]-U4p[0] V4[t])) (U2[t] (U4p[0] V1[t]-U3p[0] V2[t])+U4[t] (-U2p[0] V1[t]+U3p[0] V4[t])+U3[t] (U2p[0] V2[t]-U4p[0] V4[t])) (-m U2[t] V2p[0]+m U4[t] V4p[0])^2)/((2 U2[t] V2[t]-2 U4[t] V4[t])^4 (((-U2p[0] V2[t]+U4p[0] V4[t]) (-m U2[t] V2p[0]+m U4[t] V4p[0]))/(2 U2[t] V2[t]-2 U4[t] V4[t])^2-(m (U2p[0] U4[t]-U2[t] U4p[0]) (V2p[0] V4[t]-V2[t] V4p[0]))/(2 U2[t] V2[t]-2 U4[t] V4[t])^2)^2)+(sq2q2i (U2[t] (U4p[0] V1[t]-U3p[0] V2[t])+U4[t] (-U2p[0] V1[t]+U3p[0] V4[t])+U3[t] (U2p[0] V2[t]-U4p[0] V4[t]))^2 (-m U2[t] V2p[0]+m U4[t] V4p[0])^2)/((2 U2[t] V2[t]-2 U4[t] V4[t])^4 (((-U2p[0] V2[t]+U4p[0] V4[t]) (-m U2[t] V2p[0]+m U4[t] V4p[0]))/(2 U2[t] V2[t]-2 U4[t] V4[t])^2-(m (U2p[0] U4[t]-U2[t] U4p[0]) (V2p[0] V4[t]-V2[t] V4p[0]))/(2 U2[t] V2[t]-2 U4[t] V4[t])^2)^2)+(veq2i^2 (U2[t] (U4p[0] V1[t]-U3p[0] V2[t])+U4[t] (-U2p[0] V1[t]+U3p[0] V4[t])+U3[t] (U2p[0] V2[t]-U4p[0] V4[t]))^2 (-m U2[t] V2p[0]+m U4[t] V4p[0])^2)/((2 U2[t] V2[t]-2 U4[t] V4[t])^4 (((-U2p[0] V2[t]+U4p[0] V4[t]) (-m U2[t] V2p[0]+m U4[t] V4p[0]))/(2 U2[t] V2[t]-2 U4[t] V4[t])^2-(m (U2p[0] U4[t]-U2[t] U4p[0]) (V2p[0] V4[t]-V2[t] V4p[0]))/(2 U2[t] V2[t]-2 U4[t] V4[t])^2)^2)+((-u14[t] u2[t]^2 v2[t] v3[t]+u12[t] u2[t] u4[t] v2[t] v3[t]+u2[t] u21[t] u4[t] v2[t] v3[t]-u2[t]^2 u41[t] v2[t] v3[t]-u2[t]^2 v2[t] v23[t] v3[t]-u2[t] u24[t] u4[t] v3[t]^2+u22[t] u4[t]^2 v3[t]^2-u2[t] u4[t] u42[t] v3[t]^2+u2[t]^2 u44[t] v3[t]^2+u2[t]^2 v22[t] v3[t]^2-u2[t] u4[t] v24[t] v3[t]^2-u2[t]^2 v2[t] v3[t] v32[t]+u2[t]^2 v2[t]^2 v33[t]+u2[t] u4[t] v2[t] v3[t] v34[t]+u14[t] u2[t] u4[t] v3[t] v4[t]-u12[t] u4[t]^2 v3[t] v4[t]-u21[t] u4[t]^2 v3[t] v4[t]+u2[t] u4[t] u41[t] v3[t] v4[t]+u2[t] u4[t] v23[t] v3[t] v4[t]+u2[t] u4[t] v3[t] v32[t] v4[t]-2 u2[t] u4[t] v2[t] v33[t] v4[t]-u4[t]^2 v3[t] v34[t] v4[t]+u4[t]^2 v33[t] v4[t]^2+u11[t] (u2[t] v2[t]-u4[t] v4[t])^2-u2[t] u4[t] v3[t]^2 v42[t]+u2[t] u4[t] v2[t] v3[t] v43[t]-u4[t]^2 v3[t] v4[t] v43[t]+u4[t]^2 v3[t]^2 v44[t]+u1[t]^2 (u22[t] v2[t]^2-v4[t] (-(u44[t]+v22[t]) v4[t]+v2[t] (u24[t]+u42[t]+v24[t]+v42[t]))+v2[t]^2 v44[t])+u1[t] (u12[t] v2[t] (-u2[t] v2[t]+u4[t] v4[t])+u2[t] (-u21[t] v2[t]^2+u24[t] v2[t] v3[t]+u42[t] v2[t] v3[t]+v2[t] v24[t] v3[t]-v2[t]^2 v34[t]+u14[t] v2[t] v4[t]+u41[t] v2[t] v4[t]+v2[t] v23[t] v4[t]-2 u44[t] v3[t] v4[t]-2 v22[t] v3[t] v4[t]+v2[t] v32[t] v4[t]+v2[t] v3[t] v42[t]-v2[t]^2 v43[t])+u4[t] (-2 u22[t] v2[t] v3[t]+v4[t] (-(u14[t]+u41[t]+v23[t]+v32[t]) v4[t]+v3[t] (u24[t]+u42[t]+v24[t]+v42[t])+v2[t] (u21[t]+v34[t]+v43[t]))-2 v2[t] v3[t] v44[t]))) (-m U2[t] V2p[0]+m U4[t] V4p[0])^2)/(2 (u2[t] v2[t]-u4[t] v4[t])^2 (2 U2[t] V2[t]-2 U4[t] V4[t])^2 (((-U2p[0] V2[t]+U4p[0] V4[t]) (-m U2[t] V2p[0]+m U4[t] V4p[0]))/(2 U2[t] V2[t]-2 U4[t] V4[t])^2-(m (U2p[0] U4[t]-U2[t] U4p[0]) (V2p[0] V4[t]-V2[t] V4p[0]))/(2 U2[t] V2[t]-2 U4[t] V4[t])^2)^2)+(m sq2p2i (U2p[0] U4[t]-U2[t] U4p[0])^2 (U2[t] (-V1p[0] V2[t]+V1[t] V2p[0])+U4[t] (V1p[0] V4[t]-V1[t] V4p[0])+U3[t] (-V2p[0] V4[t]+V2[t] V4p[0])))/((2 U2[t] V2[t]-2 U4[t] V4[t])^3 (((-U2p[0] V2[t]+U4p[0] V4[t]) (-m U2[t] V2p[0]+m U4[t] V4p[0]))/(2 U2[t] V2[t]-2 U4[t] V4[t])^2-(m (U2p[0] U4[t]-U2[t] U4p[0]) (V2p[0] V4[t]-V2[t] V4p[0]))/(2 U2[t] V2[t]-2 U4[t] V4[t])^2)^2)+(m vep2i veq2i (U2p[0] U4[t]-U2[t] U4p[0])^2 (U2[t] (-V1p[0] V2[t]+V1[t] V2p[0])+U4[t] (V1p[0] V4[t]-V1[t] V4p[0])+U3[t] (-V2p[0] V4[t]+V2[t] V4p[0])))/((2 U2[t] V2[t]-2 U4[t] V4[t])^3 (((-U2p[0] V2[t]+U4p[0] V4[t]) (-m U2[t] V2p[0]+m U4[t] V4p[0]))/(2 U2[t] V2[t]-2 U4[t] V4[t])^2-(m (U2p[0] U4[t]-U2[t] U4p[0]) (V2p[0] V4[t]-V2[t] V4p[0]))/(2 U2[t] V2[t]-2 U4[t] V4[t])^2)^2)-(m vep1i veq2i (U2p[0] U4[t]-U2[t] U4p[0]) (-m U2[t] V2p[0]+m U4[t] V4p[0]) (U2[t] (-V1p[0] V2[t]+V1[t] V2p[0])+U4[t] (V1p[0] V4[t]-V1[t] V4p[0])+U3[t] (-V2p[0] V4[t]+V2[t] V4p[0])))/((2 U2[t] V2[t]-2 U4[t] V4[t])^3 (((-U2p[0] V2[t]+U4p[0] V4[t]) (-m U2[t] V2p[0]+m U4[t] V4p[0]))/(2 U2[t] V2[t]-2 U4[t] V4[t])^2-(m (U2p[0] U4[t]-U2[t] U4p[0]) (V2p[0] V4[t]-V2[t] V4p[0]))/(2 U2[t] V2[t]-2 U4[t] V4[t])^2)^2)-(2 m veq1i veq2i (U2p[0] U4[t]-U2[t] U4p[0]) (U2[t] (-U1p[0] V2[t]+U4p[0] V3[t])+U4[t] (-U2p[0] V3[t]+U1p[0] V4[t])+U1[t] (U2p[0] V2[t]-U4p[0] V4[t])) (-m U2[t] V2p[0]+m U4[t] V4p[0]) (U2[t] (-V1p[0] V2[t]+V1[t] V2p[0])+U4[t] (V1p[0] V4[t]-V1[t] V4p[0])+U3[t] (-V2p[0] V4[t]+V2[t] V4p[0])))/((2 U2[t] V2[t]-2 U4[t] V4[t])^4 (((-U2p[0] V2[t]+U4p[0] V4[t]) (-m U2[t] V2p[0]+m U4[t] V4p[0]))/(2 U2[t] V2[t]-2 U4[t] V4[t])^2-(m (U2p[0] U4[t]-U2[t] U4p[0]) (V2p[0] V4[t]-V2[t] V4p[0]))/(2 U2[t] V2[t]-2 U4[t] V4[t])^2)^2)-(2 m sq2q2i (U2p[0] U4[t]-U2[t] U4p[0]) (U2[t] (U4p[0] V1[t]-U3p[0] V2[t])+U4[t] (-U2p[0] V1[t]+U3p[0] V4[t])+U3[t] (U2p[0] V2[t]-U4p[0] V4[t])) (-m U2[t] V2p[0]+m U4[t] V4p[0]) (U2[t] (-V1p[0] V2[t]+V1[t] V2p[0])+U4[t] (V1p[0] V4[t]-V1[t] V4p[0])+U3[t] (-V2p[0] V4[t]+V2[t] V4p[0])))/((2 U2[t] V2[t]-2 U4[t] V4[t])^4 (((-U2p[0] V2[t]+U4p[0] V4[t]) (-m U2[t] V2p[0]+m U4[t] V4p[0]))/(2 U2[t] V2[t]-2 U4[t] V4[t])^2-(m (U2p[0] U4[t]-U2[t] U4p[0]) (V2p[0] V4[t]-V2[t] V4p[0]))/(2 U2[t] V2[t]-2 U4[t] V4[t])^2)^2)-(2 m veq2i^2 (U2p[0] U4[t]-U2[t] U4p[0]) (U2[t] (U4p[0] V1[t]-U3p[0] V2[t])+U4[t] (-U2p[0] V1[t]+U3p[0] V4[t])+U3[t] (U2p[0] V2[t]-U4p[0] V4[t])) (-m U2[t] V2p[0]+m U4[t] V4p[0]) (U2[t] (-V1p[0] V2[t]+V1[t] V2p[0])+U4[t] (V1p[0] V4[t]-V1[t] V4p[0])+U3[t] (-V2p[0] V4[t]+V2[t] V4p[0])))/((2 U2[t] V2[t]-2 U4[t] V4[t])^4 (((-U2p[0] V2[t]+U4p[0] V4[t]) (-m U2[t] V2p[0]+m U4[t] V4p[0]))/(2 U2[t] V2[t]-2 U4[t] V4[t])^2-(m (U2p[0] U4[t]-U2[t] U4p[0]) (V2p[0] V4[t]-V2[t] V4p[0]))/(2 U2[t] V2[t]-2 U4[t] V4[t])^2)^2)+(m^2 sq2q2i (U2p[0] U4[t]-U2[t] U4p[0])^2 (U2[t] (-V1p[0] V2[t]+V1[t] V2p[0])+U4[t] (V1p[0] V4[t]-V1[t] V4p[0])+U3[t] (-V2p[0] V4[t]+V2[t] V4p[0]))^2)/((2 U2[t] V2[t]-2 U4[t] V4[t])^4 (((-U2p[0] V2[t]+U4p[0] V4[t]) (-m U2[t] V2p[0]+m U4[t] V4p[0]))/(2 U2[t] V2[t]-2 U4[t] V4[t])^2-(m (U2p[0] U4[t]-U2[t] U4p[0]) (V2p[0] V4[t]-V2[t] V4p[0]))/(2 U2[t] V2[t]-2 U4[t] V4[t])^2)^2)+(m^2 veq2i^2 (U2p[0] U4[t]-U2[t] U4p[0])^2 (U2[t] (-V1p[0] V2[t]+V1[t] V2p[0])+U4[t] (V1p[0] V4[t]-V1[t] V4p[0])+U3[t] (-V2p[0] V4[t]+V2[t] V4p[0]))^2)/((2 U2[t] V2[t]-2 U4[t] V4[t])^4 (((-U2p[0] V2[t]+U4p[0] V4[t]) (-m U2[t] V2p[0]+m U4[t] V4p[0]))/(2 U2[t] V2[t]-2 U4[t] V4[t])^2-(m (U2p[0] U4[t]-U2[t] U4p[0]) (V2p[0] V4[t]-V2[t] V4p[0]))/(2 U2[t] V2[t]-2 U4[t] V4[t])^2)^2)+(m vep2i veq1i (U2p[0] U4[t]-U2[t] U4p[0])^2 (U2[t] (V2p[0] V3[t]-V2[t] V3p[0])+U1[t] (-V2p[0] V4[t]+V2[t] V4p[0])+U4[t] (V3p[0] V4[t]-V3[t] V4p[0])))/((2 U2[t] V2[t]-2 U4[t] V4[t])^3 (((-U2p[0] V2[t]+U4p[0] V4[t]) (-m U2[t] V2p[0]+m U4[t] V4p[0]))/(2 U2[t] V2[t]-2 U4[t] V4[t])^2-(m (U2p[0] U4[t]-U2[t] U4p[0]) (V2p[0] V4[t]-V2[t] V4p[0]))/(2 U2[t] V2[t]-2 U4[t] V4[t])^2)^2)-(m sq1p1i (U2p[0] U4[t]-U2[t] U4p[0]) (-m U2[t] V2p[0]+m U4[t] V4p[0]) (U2[t] (V2p[0] V3[t]-V2[t] V3p[0])+U1[t] (-V2p[0] V4[t]+V2[t] V4p[0])+U4[t] (V3p[0] V4[t]-V3[t] V4p[0])))/((2 U2[t] V2[t]-2 U4[t] V4[t])^3 (((-U2p[0] V2[t]+U4p[0] V4[t]) (-m U2[t] V2p[0]+m U4[t] V4p[0]))/(2 U2[t] V2[t]-2 U4[t] V4[t])^2-(m (U2p[0] U4[t]-U2[t] U4p[0]) (V2p[0] V4[t]-V2[t] V4p[0]))/(2 U2[t] V2[t]-2 U4[t] V4[t])^2)^2)-(m vep1i veq1i (U2p[0] U4[t]-U2[t] U4p[0]) (-m U2[t] V2p[0]+m U4[t] V4p[0]) (U2[t] (V2p[0] V3[t]-V2[t] V3p[0])+U1[t] (-V2p[0] V4[t]+V2[t] V4p[0])+U4[t] (V3p[0] V4[t]-V3[t] V4p[0])))/((2 U2[t] V2[t]-2 U4[t] V4[t])^3 (((-U2p[0] V2[t]+U4p[0] V4[t]) (-m U2[t] V2p[0]+m U4[t] V4p[0]))/(2 U2[t] V2[t]-2 U4[t] V4[t])^2-(m (U2p[0] U4[t]-U2[t] U4p[0]) (V2p[0] V4[t]-V2[t] V4p[0]))/(2 U2[t] V2[t]-2 U4[t] V4[t])^2)^2)-(2 m sq1q1i (U2p[0] U4[t]-U2[t] U4p[0]) (U2[t] (-U1p[0] V2[t]+U4p[0] V3[t])+U4[t] (-U2p[0] V3[t]+U1p[0] V4[t])+U1[t] (U2p[0] V2[t]-U4p[0] V4[t])) (-m U2[t] V2p[0]+m U4[t] V4p[0]) (U2[t] (V2p[0] V3[t]-V2[t] V3p[0])+U1[t] (-V2p[0] V4[t]+V2[t] V4p[0])+U4[t] (V3p[0] V4[t]-V3[t] V4p[0])))/((2 U2[t] V2[t]-2 U4[t] V4[t])^4 (((-U2p[0] V2[t]+U4p[0] V4[t]) (-m U2[t] V2p[0]+m U4[t] V4p[0]))/(2 U2[t] V2[t]-2 U4[t] V4[t])^2-(m (U2p[0] U4[t]-U2[t] U4p[0]) (V2p[0] V4[t]-V2[t] V4p[0]))/(2 U2[t] V2[t]-2 U4[t] V4[t])^2)^2)-(2 m veq1i^2 (U2p[0] U4[t]-U2[t] U4p[0]) (U2[t] (-U1p[0] V2[t]+U4p[0] V3[t])+U4[t] (-U2p[0] V3[t]+U1p[0] V4[t])+U1[t] (U2p[0] V2[t]-U4p[0] V4[t])) (-m U2[t] V2p[0]+m U4[t] V4p[0]) (U2[t] (V2p[0] V3[t]-V2[t] V3p[0])+U1[t] (-V2p[0] V4[t]+V2[t] V4p[0])+U4[t] (V3p[0] V4[t]-V3[t] V4p[0])))/((2 U2[t] V2[t]-2 U4[t] V4[t])^4 (((-U2p[0] V2[t]+U4p[0] V4[t]) (-m U2[t] V2p[0]+m U4[t] V4p[0]))/(2 U2[t] V2[t]-2 U4[t] V4[t])^2-(m (U2p[0] U4[t]-U2[t] U4p[0]) (V2p[0] V4[t]-V2[t] V4p[0]))/(2 U2[t] V2[t]-2 U4[t] V4[t])^2)^2)-(2 m veq1i veq2i (U2p[0] U4[t]-U2[t] U4p[0]) (U2[t] (U4p[0] V1[t]-U3p[0] V2[t])+U4[t] (-U2p[0] V1[t]+U3p[0] V4[t])+U3[t] (U2p[0] V2[t]-U4p[0] V4[t])) (-m U2[t] V2p[0]+m U4[t] V4p[0]) (U2[t] (V2p[0] V3[t]-V2[t] V3p[0])+U1[t] (-V2p[0] V4[t]+V2[t] V4p[0])+U4[t] (V3p[0] V4[t]-V3[t] V4p[0])))/((2 U2[t] V2[t]-2 U4[t] V4[t])^4 (((-U2p[0] V2[t]+U4p[0] V4[t]) (-m U2[t] V2p[0]+m U4[t] V4p[0]))/(2 U2[t] V2[t]-2 U4[t] V4[t])^2-(m (U2p[0] U4[t]-U2[t] U4p[0]) (V2p[0] V4[t]-V2[t] V4p[0]))/(2 U2[t] V2[t]-2 U4[t] V4[t])^2)^2)+(2 m^2 veq1i veq2i (U2p[0] U4[t]-U2[t] U4p[0])^2 (U2[t] (-V1p[0] V2[t]+V1[t] V2p[0])+U4[t] (V1p[0] V4[t]-V1[t] V4p[0])+U3[t] (-V2p[0] V4[t]+V2[t] V4p[0])) (U2[t] (V2p[0] V3[t]-V2[t] V3p[0])+U1[t] (-V2p[0] V4[t]+V2[t] V4p[0])+U4[t] (V3p[0] V4[t]-V3[t] V4p[0])))/((2 U2[t] V2[t]-2 U4[t] V4[t])^4 (((-U2p[0] V2[t]+U4p[0] V4[t]) (-m U2[t] V2p[0]+m U4[t] V4p[0]))/(2 U2[t] V2[t]-2 U4[t] V4[t])^2-(m (U2p[0] U4[t]-U2[t] U4p[0]) (V2p[0] V4[t]-V2[t] V4p[0]))/(2 U2[t] V2[t]-2 U4[t] V4[t])^2)^2)+(m^2 sq1q1i (U2p[0] U4[t]-U2[t] U4p[0])^2 (U2[t] (V2p[0] V3[t]-V2[t] V3p[0])+U1[t] (-V2p[0] V4[t]+V2[t] V4p[0])+U4[t] (V3p[0] V4[t]-V3[t] V4p[0]))^2)/((2 U2[t] V2[t]-2 U4[t] V4[t])^4 (((-U2p[0] V2[t]+U4p[0] V4[t]) (-m U2[t] V2p[0]+m U4[t] V4p[0]))/(2 U2[t] V2[t]-2 U4[t] V4[t])^2-(m (U2p[0] U4[t]-U2[t] U4p[0]) (V2p[0] V4[t]-V2[t] V4p[0]))/(2 U2[t] V2[t]-2 U4[t] V4[t])^2)^2)+(m^2 veq1i^2 (U2p[0] U4[t]-U2[t] U4p[0])^2 (U2[t] (V2p[0] V3[t]-V2[t] V3p[0])+U1[t] (-V2p[0] V4[t]+V2[t] V4p[0])+U4[t] (V3p[0] V4[t]-V3[t] V4p[0]))^2)/((2 U2[t] V2[t]-2 U4[t] V4[t])^4 (((-U2p[0] V2[t]+U4p[0] V4[t]) (-m U2[t] V2p[0]+m U4[t] V4p[0]))/(2 U2[t] V2[t]-2 U4[t] V4[t])^2-(m (U2p[0] U4[t]-U2[t] U4p[0]) (V2p[0] V4[t]-V2[t] V4p[0]))/(2 U2[t] V2[t]-2 U4[t] V4[t])^2)^2);
\end{spverbatim}

\section{Variance for momentum operator}

\begin{spverbatim}
sp1p1[t_]:=(4 sq1q1i (U1p[t] U2[t] V2[t]-U1[t] U2p[t] V2[t]+U2p[t] U4[t] V3[t]-U2[t] U4p[t] V3[t]-U1p[t] U4[t] V4[t]+U1[t] U4p[t] V4[t])^2)/(2 U2[t] V2[t]-2 U4[t] V4[t])^2+(4 veq1i^2 (U1p[t] U2[t] V2[t]-U1[t] U2p[t] V2[t]+U2p[t] U4[t] V3[t]-U2[t] U4p[t] V3[t]-U1p[t] U4[t] V4[t]+U1[t] U4p[t] V4[t])^2)/(2 U2[t] V2[t]-2 U4[t] V4[t])^2+(8 veq1i veq2i (U1p[t] U2[t] V2[t]-U1[t] U2p[t] V2[t]+U2p[t] U4[t] V3[t]-U2[t] U4p[t] V3[t]-U1p[t] U4[t] V4[t]+U1[t] U4p[t] V4[t]) (U2p[t] U4[t] V1[t]-U2[t] U4p[t] V1[t]-U2p[t] U3[t] V2[t]+U2[t] U3p[t] V2[t]-U3p[t] U4[t] V4[t]+U3[t] U4p[t] V4[t]))/(2 U2[t] V2[t]-2 U4[t] V4[t])^2+(4 sq2q2i (U2p[t] U4[t] V1[t]-U2[t] U4p[t] V1[t]-U2p[t] U3[t] V2[t]+U2[t] U3p[t] V2[t]-U3p[t] U4[t] V4[t]+U3[t] U4p[t] V4[t])^2)/(2 U2[t] V2[t]-2 U4[t] V4[t])^2+(4 veq2i^2 (U2p[t] U4[t] V1[t]-U2[t] U4p[t] V1[t]-U2p[t] U3[t] V2[t]+U2[t] U3p[t] V2[t]-U3p[t] U4[t] V4[t]+U3[t] U4p[t] V4[t])^2)/(2 U2[t] V2[t]-2 U4[t] V4[t])^2+(2 (u22[t] v2[t]^2-v4[t] (-(u44[t]+v22[t]) v4[t]+v2[t] (u24[t]+u42[t]+v24[t]+v42[t]))+v2[t]^2 v44[t]))/(u2[t] v2[t]-u4[t] v4[t])^2+(sp2p2i (-U2p[t] U4[t]+U2[t] U4p[t])^2)/(-m U2[t] V2p[0]+m U4[t] V4p[0])^2+(vep2i^2 (-U2p[t] U4[t]+U2[t] U4p[t])^2)/(-m U2[t] V2p[0]+m U4[t] V4p[0])^2+(2 (-U2p[t] U4[t]+U2[t] U4p[t])^2 (u22[t] (u4[t] v1[t]-u3[t] v2[t])^2+u2[t]^2 (u44[t] v1[t]^2-v2[t] (-(u33[t]+v11[t]) v2[t]+v1[t] (u34[t]+u43[t]+v12[t]+v21[t]))+v1[t]^2 v22[t])+u24[t] u3[t] u4[t] v1[t] v4[t]-u23[t] u4[t]^2 v1[t] v4[t]-u32[t] u4[t]^2 v1[t] v4[t]+u3[t] u4[t] u42[t] v1[t] v4[t]-u4[t]^2 v1[t] v14[t] v4[t]-u24[t] u3[t]^2 v2[t] v4[t]+u23[t] u3[t] u4[t] v2[t] v4[t]+u3[t] u32[t] u4[t] v2[t] v4[t]-u3[t]^2 u42[t] v2[t] v4[t]+u3[t] u4[t] v14[t] v2[t] v4[t]+u3[t] u4[t] v1[t] v24[t] v4[t]-u3[t]^2 v2[t] v24[t] v4[t]-u3[t] u34[t] u4[t] v4[t]^2+u33[t] u4[t]^2 v4[t]^2-u3[t] u4[t] u43[t] v4[t]^2+u3[t]^2 u44[t] v4[t]^2+u4[t]^2 v11[t] v4[t]^2-u3[t] u4[t] v12[t] v4[t]^2-u3[t] u4[t] v21[t] v4[t]^2+u3[t]^2 v22[t] v4[t]^2-u4[t]^2 v1[t] v4[t] v41[t]+u3[t] u4[t] v2[t] v4[t] v41[t]+u3[t] u4[t] v1[t] v4[t] v42[t]-u3[t]^2 v2[t] v4[t] v42[t]+u2[t] (u24[t] v1[t] (-u4[t] v1[t]+u3[t] v2[t])+u4[t] (-u42[t] v1[t]^2+u23[t] v1[t] v2[t]+u32[t] v1[t] v2[t]+v1[t] v14[t] v2[t]-v1[t]^2 v24[t]+u34[t] v1[t] v4[t]+u43[t] v1[t] v4[t]+v1[t] v12[t] v4[t]-2 u33[t] v2[t] v4[t]-2 v11[t] v2[t] v4[t]+v1[t] v21[t] v4[t]+v1[t] v2[t] v41[t]-v1[t]^2 v42[t])+u3[t] (u42[t] v1[t] v2[t]-u23[t] v2[t]^2-u32[t] v2[t]^2-v14[t] v2[t]^2+v1[t] v2[t] v24[t]-2 u44[t] v1[t] v4[t]+u34[t] v2[t] v4[t]+u43[t] v2[t] v4[t]+v12[t] v2[t] v4[t]+v2[t] v21[t] v4[t]-2 v1[t] v22[t] v4[t]-v2[t]^2 v41[t]+v1[t] v2[t] v42[t]))+(u4[t] v1[t]-u3[t] v2[t])^2 v44[t]))/((u2[t] v2[t]-u4[t] v4[t])^2 (-m U2[t] V2p[0]+m U4[t] V4p[0])^2)-(4 vep2i veq1i (-U2p[t] U4[t]+U2[t] U4p[t]) (U1p[t] U2[t] V2[t]-U1[t] U2p[t] V2[t]+U2p[t] U4[t] V3[t]-U2[t] U4p[t] V3[t]-U1p[t] U4[t] V4[t]+U1[t] U4p[t] V4[t]))/((2 U2[t] V2[t]-2 U4[t] V4[t]) (-m U2[t] V2p[0]+m U4[t] V4p[0]))-(4 sq2p2i (-U2p[t] U4[t]+U2[t] U4p[t]) (U2p[t] U4[t] V1[t]-U2[t] U4p[t] V1[t]-U2p[t] U3[t] V2[t]+U2[t] U3p[t] V2[t]-U3p[t] U4[t] V4[t]+U3[t] U4p[t] V4[t]))/((2 U2[t] V2[t]-2 U4[t] V4[t]) (-m U2[t] V2p[0]+m U4[t] V4p[0]))-(4 vep2i veq2i (-U2p[t] U4[t]+U2[t] U4p[t]) (U2p[t] U4[t] V1[t]-U2[t] U4p[t] V1[t]-U2p[t] U3[t] V2[t]+U2[t] U3p[t] V2[t]-U3p[t] U4[t] V4[t]+U3[t] U4p[t] V4[t]))/((2 U2[t] V2[t]-2 U4[t] V4[t]) (-m U2[t] V2p[0]+m U4[t] V4p[0]))+(2 (-U2p[t] U4[t]+U2[t] U4p[t]) (2 u22[t] v2[t] (-u4[t] v1[t]+u3[t] v2[t])+u2[t] (u24[t] v1[t] v2[t]+u42[t] v1[t] v2[t]-u23[t] v2[t]^2-u32[t] v2[t]^2-v14[t] v2[t]^2+v1[t] v2[t] v24[t]-2 u44[t] v1[t] v4[t]+u34[t] v2[t] v4[t]+u43[t] v2[t] v4[t]+v12[t] v2[t] v4[t]+v2[t] v21[t] v4[t]-2 v1[t] v22[t] v4[t]-v2[t]^2 v41[t]+v1[t] v2[t] v42[t])+v4[t] (u24[t] (u4[t] v1[t]-2 u3[t] v2[t])+u4[t] (-(u34[t]+u43[t]+v12[t]+v21[t]) v4[t]+v2[t] (u23[t]+u32[t]+v14[t]+v41[t])+v1[t] (u42[t]+v24[t]+v42[t]))-2 u3[t] (-(u44[t]+v22[t]) v4[t]+v2[t] (u42[t]+v24[t]+v42[t])))+2 v2[t] (-u4[t] v1[t]+u3[t] v2[t]) v44[t]))/((u2[t] v2[t]-u4[t] v4[t])^2 (-m U2[t] V2p[0]+m U4[t] V4p[0]))+(2 vep1i vep2i (-U2p[t] U4[t]+U2[t] U4p[t]) (U2p[t] V2[t]-U4p[t] V4[t]))/((2 U2[t] V2[t]-2 U4[t] V4[t])^2 (((-U2p[0] V2[t]+U4p[0] V4[t]) (-m U2[t] V2p[0]+m U4[t] V4p[0]))/(2 U2[t] V2[t]-2 U4[t] V4[t])^2-(m (U2p[0] U4[t]-U2[t] U4p[0]) (V2p[0] V4[t]-V2[t] V4p[0]))/(2 U2[t] V2[t]-2 U4[t] V4[t])^2))+(4 vep2i veq1i (U2p[0] U4[t]-U2[t] U4p[0]) (U2p[t] V2[t]-U4p[t] V4[t]) (U1p[t] U2[t] V2[t]-U1[t] U2p[t] V2[t]+U2p[t] U4[t] V3[t]-U2[t] U4p[t] V3[t]-U1p[t] U4[t] V4[t]+U1[t] U4p[t] V4[t]))/((2 U2[t] V2[t]-2 U4[t] V4[t])^3 (((-U2p[0] V2[t]+U4p[0] V4[t]) (-m U2[t] V2p[0]+m U4[t] V4p[0]))/(2 U2[t] V2[t]-2 U4[t] V4[t])^2-(m (U2p[0] U4[t]-U2[t] U4p[0]) (V2p[0] V4[t]-V2[t] V4p[0]))/(2 U2[t] V2[t]-2 U4[t] V4[t])^2))+(4 sq2p2i (U2p[0] U4[t]-U2[t] U4p[0]) (U2p[t] V2[t]-U4p[t] V4[t]) (U2p[t] U4[t] V1[t]-U2[t] U4p[t] V1[t]-U2p[t] U3[t] V2[t]+U2[t] U3p[t] V2[t]-U3p[t] U4[t] V4[t]+U3[t] U4p[t] V4[t]))/((2 U2[t] V2[t]-2 U4[t] V4[t])^3 (((-U2p[0] V2[t]+U4p[0] V4[t]) (-m U2[t] V2p[0]+m U4[t] V4p[0]))/(2 U2[t] V2[t]-2 U4[t] V4[t])^2-(m (U2p[0] U4[t]-U2[t] U4p[0]) (V2p[0] V4[t]-V2[t] V4p[0]))/(2 U2[t] V2[t]-2 U4[t] V4[t])^2))+(4 vep2i veq2i (U2p[0] U4[t]-U2[t] U4p[0]) (U2p[t] V2[t]-U4p[t] V4[t]) (U2p[t] U4[t] V1[t]-U2[t] U4p[t] V1[t]-U2p[t] U3[t] V2[t]+U2[t] U3p[t] V2[t]-U3p[t] U4[t] V4[t]+U3[t] U4p[t] V4[t]))/((2 U2[t] V2[t]-2 U4[t] V4[t])^3 (((-U2p[0] V2[t]+U4p[0] V4[t]) (-m U2[t] V2p[0]+m U4[t] V4p[0]))/(2 U2[t] V2[t]-2 U4[t] V4[t])^2-(m (U2p[0] U4[t]-U2[t] U4p[0]) (V2p[0] V4[t]-V2[t] V4p[0]))/(2 U2[t] V2[t]-2 U4[t] V4[t])^2))+(4 vep2i veq1i (-U2p[t] U4[t]+U2[t] U4p[t]) (U2p[t] V2[t]-U4p[t] V4[t]) (U2[t] (-U1p[0] V2[t]+U4p[0] V3[t])+U4[t] (-U2p[0] V3[t]+U1p[0] V4[t])+U1[t] (U2p[0] V2[t]-U4p[0] V4[t])))/((2 U2[t] V2[t]-2 U4[t] V4[t])^3 (((-U2p[0] V2[t]+U4p[0] V4[t]) (-m U2[t] V2p[0]+m U4[t] V4p[0]))/(2 U2[t] V2[t]-2 U4[t] V4[t])^2-(m (U2p[0] U4[t]-U2[t] U4p[0]) (V2p[0] V4[t]-V2[t] V4p[0]))/(2 U2[t] V2[t]-2 U4[t] V4[t])^2))+(4 sq2p2i (-U2p[t] U4[t]+U2[t] U4p[t]) (U2p[t] V2[t]-U4p[t] V4[t]) (U2[t] (U4p[0] V1[t]-U3p[0] V2[t])+U4[t] (-U2p[0] V1[t]+U3p[0] V4[t])+U3[t] (U2p[0] V2[t]-U4p[0] V4[t])))/((2 U2[t] V2[t]-2 U4[t] V4[t])^3 (((-U2p[0] V2[t]+U4p[0] V4[t]) (-m U2[t] V2p[0]+m U4[t] V4p[0]))/(2 U2[t] V2[t]-2 U4[t] V4[t])^2-(m (U2p[0] U4[t]-U2[t] U4p[0]) (V2p[0] V4[t]-V2[t] V4p[0]))/(2 U2[t] V2[t]-2 U4[t] V4[t])^2))+(4 vep2i veq2i (-U2p[t] U4[t]+U2[t] U4p[t]) (U2p[t] V2[t]-U4p[t] V4[t]) (U2[t] (U4p[0] V1[t]-U3p[0] V2[t])+U4[t] (-U2p[0] V1[t]+U3p[0] V4[t])+U3[t] (U2p[0] V2[t]-U4p[0] V4[t])))/((2 U2[t] V2[t]-2 U4[t] V4[t])^3 (((-U2p[0] V2[t]+U4p[0] V4[t]) (-m U2[t] V2p[0]+m U4[t] V4p[0]))/(2 U2[t] V2[t]-2 U4[t] V4[t])^2-(m (U2p[0] U4[t]-U2[t] U4p[0]) (V2p[0] V4[t]-V2[t] V4p[0]))/(2 U2[t] V2[t]-2 U4[t] V4[t])^2))-(2 (U2p[0] U4[t]-U2[t] U4p[0]) (U2p[t] V2[t]-U4p[t] V4[t]) (2 u22[t] v2[t] (-u4[t] v1[t]+u3[t] v2[t])+u2[t] (u24[t] v1[t] v2[t]+u42[t] v1[t] v2[t]-u23[t] v2[t]^2-u32[t] v2[t]^2-v14[t] v2[t]^2+v1[t] v2[t] v24[t]-2 u44[t] v1[t] v4[t]+u34[t] v2[t] v4[t]+u43[t] v2[t] v4[t]+v12[t] v2[t] v4[t]+v2[t] v21[t] v4[t]-2 v1[t] v22[t] v4[t]-v2[t]^2 v41[t]+v1[t] v2[t] v42[t])+v4[t] (u24[t] (u4[t] v1[t]-2 u3[t] v2[t])+u4[t] (-(u34[t]+u43[t]+v12[t]+v21[t]) v4[t]+v2[t] (u23[t]+u32[t]+v14[t]+v41[t])+v1[t] (u42[t]+v24[t]+v42[t]))-2 u3[t] (-(u44[t]+v22[t]) v4[t]+v2[t] (u42[t]+v24[t]+v42[t])))+2 v2[t] (-u4[t] v1[t]+u3[t] v2[t]) v44[t]))/((u2[t] v2[t]-u4[t] v4[t])^2 (2 U2[t] V2[t]-2 U4[t] V4[t])^2 (((-U2p[0] V2[t]+U4p[0] V4[t]) (-m U2[t] V2p[0]+m U4[t] V4p[0]))/(2 U2[t] V2[t]-2 U4[t] V4[t])^2-(m (U2p[0] U4[t]-U2[t] U4p[0]) (V2p[0] V4[t]-V2[t] V4p[0]))/(2 U2[t] V2[t]-2 U4[t] V4[t])^2))+(2 (-U2p[t] U4[t]+U2[t] U4p[t]) (U2p[t] V2[t]-U4p[t] V4[t]) (u12[t] u2[t] u4[t] v1[t] v2[t]+u2[t] u21[t] u4[t] v1[t] v2[t]-u2[t]^2 u41[t] v1[t] v2[t]+u13[t] u2[t]^2 v2[t]^2-u12[t] u2[t] u3[t] v2[t]^2-u2[t] u21[t] u3[t] v2[t]^2+u2[t]^2 u31[t] v2[t]^2+u2[t]^2 v13[t] v2[t]^2-u2[t]^2 v1[t] v2[t] v23[t]-2 u2[t] u24[t] u4[t] v1[t] v3[t]+2 u22[t] u4[t]^2 v1[t] v3[t]-2 u2[t] u4[t] u42[t] v1[t] v3[t]+2 u2[t]^2 u44[t] v1[t] v3[t]+u2[t] u24[t] u3[t] v2[t] v3[t]-u2[t]^2 u34[t] v2[t] v3[t]+u2[t] u23[t] u4[t] v2[t] v3[t]-2 u22[t] u3[t] u4[t] v2[t] v3[t]+u2[t] u32[t] u4[t] v2[t] v3[t]+u2[t] u3[t] u42[t] v2[t] v3[t]-u2[t]^2 u43[t] v2[t] v3[t]-u2[t]^2 v12[t] v2[t] v3[t]+u2[t] u4[t] v14[t] v2[t] v3[t]-u2[t]^2 v2[t] v21[t] v3[t]+2 u2[t]^2 v1[t] v22[t] v3[t]-2 u2[t] u4[t] v1[t] v24[t] v3[t]+u2[t] u3[t] v2[t] v24[t] v3[t]+u2[t]^2 v2[t]^2 v31[t]-u2[t]^2 v1[t] v2[t] v32[t]+u2[t] u4[t] v1[t] v2[t] v34[t]-u2[t] u3[t] v2[t]^2 v34[t]-u12[t] u4[t]^2 v1[t] v4[t]-u21[t] u4[t]^2 v1[t] v4[t]+u2[t] u4[t] u41[t] v1[t] v4[t]-2 u13[t] u2[t] u4[t] v2[t] v4[t]+u12[t] u3[t] u4[t] v2[t] v4[t]+u21[t] u3[t] u4[t] v2[t] v4[t]-2 u2[t] u31[t] u4[t] v2[t] v4[t]+u2[t] u3[t] u41[t] v2[t] v4[t]-2 u2[t] u4[t] v13[t] v2[t] v4[t]+u2[t] u4[t] v1[t] v23[t] v4[t]+u2[t] u3[t] v2[t] v23[t] v4[t]+u24[t] u3[t] u4[t] v3[t] v4[t]+u2[t] u34[t] u4[t] v3[t] v4[t]-u23[t] u4[t]^2 v3[t] v4[t]-u32[t] u4[t]^2 v3[t] v4[t]+u3[t] u4[t] u42[t] v3[t] v4[t]+u2[t] u4[t] u43[t] v3[t] v4[t]-2 u2[t] u3[t] u44[t] v3[t] v4[t]+u2[t] u4[t] v12[t] v3[t] v4[t]-u4[t]^2 v14[t] v3[t] v4[t]+u2[t] u4[t] v21[t] v3[t] v4[t]-2 u2[t] u3[t] v22[t] v3[t] v4[t]+u3[t] u4[t] v24[t] v3[t] v4[t]-2 u2[t] u4[t] v2[t] v31[t] v4[t]+u2[t] u4[t] v1[t] v32[t] v4[t]+u2[t] u3[t] v2[t] v32[t] v4[t]-u4[t]^2 v1[t] v34[t] v4[t]+u3[t] u4[t] v2[t] v34[t] v4[t]+u13[t] u4[t]^2 v4[t]^2+u31[t] u4[t]^2 v4[t]^2-u3[t] u4[t] u41[t] v4[t]^2+u4[t]^2 v13[t] v4[t]^2-u3[t] u4[t] v23[t] v4[t]^2+u4[t]^2 v31[t] v4[t]^2-u3[t] u4[t] v32[t] v4[t]^2-u14[t] (u2[t] v1[t]-u3[t] v4[t]) (u2[t] v2[t]-u4[t] v4[t])+u2[t] u4[t] v2[t] v3[t] v41[t]-u4[t]^2 v3[t] v4[t] v41[t]-2 u2[t] u4[t] v1[t] v3[t] v42[t]+u2[t] u3[t] v2[t] v3[t] v42[t]+u3[t] u4[t] v3[t] v4[t] v42[t]+u2[t] u4[t] v1[t] v2[t] v43[t]-u2[t] u3[t] v2[t]^2 v43[t]-u4[t]^2 v1[t] v4[t] v43[t]+u3[t] u4[t] v2[t] v4[t] v43[t]+2 u4[t] (u4[t] v1[t]-u3[t] v2[t]) v3[t] v44[t]+u1[t] (2 u22[t] v2[t] (-u4[t] v1[t]+u3[t] v2[t])+u2[t] (u24[t] v1[t] v2[t]+u42[t] v1[t] v2[t]-u23[t] v2[t]^2-u32[t] v2[t]^2-v14[t] v2[t]^2+v1[t] v2[t] v24[t]-2 u44[t] v1[t] v4[t]+u34[t] v2[t] v4[t]+u43[t] v2[t] v4[t]+v12[t] v2[t] v4[t]+v2[t] v21[t] v4[t]-2 v1[t] v22[t] v4[t]-v2[t]^2 v41[t]+v1[t] v2[t] v42[t])+v4[t] (u24[t] (u4[t] v1[t]-2 u3[t] v2[t])+u4[t] (-(u34[t]+u43[t]+v12[t]+v21[t]) v4[t]+v2[t] (u23[t]+u32[t]+v14[t]+v41[t])+v1[t] (u42[t]+v24[t]+v42[t]))-2 u3[t] (-(u44[t]+v22[t]) v4[t]+v2[t] (u42[t]+v24[t]+v42[t])))+2 v2[t] (-u4[t] v1[t]+u3[t] v2[t]) v44[t])))/((u2[t] v2[t]-u4[t] v4[t])^2 (2 U2[t] V2[t]-2 U4[t] V4[t])^2 (((-U2p[0] V2[t]+U4p[0] V4[t]) (-m U2[t] V2p[0]+m U4[t] V4p[0]))/(2 U2[t] V2[t]-2 U4[t] V4[t])^2-(m (U2p[0] U4[t]-U2[t] U4p[0]) (V2p[0] V4[t]-V2[t] V4p[0]))/(2 U2[t] V2[t]-2 U4[t] V4[t])^2))-(2 sp2p2i (U2p[0] U4[t]-U2[t] U4p[0]) (-U2p[t] U4[t]+U2[t] U4p[t]) (U2p[t] V2[t]-U4p[t] V4[t]))/((2 U2[t] V2[t]-2 U4[t] V4[t])^2 (-m U2[t] V2p[0]+m U4[t] V4p[0]) (((-U2p[0] V2[t]+U4p[0] V4[t]) (-m U2[t] V2p[0]+m U4[t] V4p[0]))/(2 U2[t] V2[t]-2 U4[t] V4[t])^2-(m (U2p[0] U4[t]-U2[t] U4p[0]) (V2p[0] V4[t]-V2[t] V4p[0]))/(2 U2[t] V2[t]-2 U4[t] V4[t])^2))-(2 vep2i^2 (U2p[0] U4[t]-U2[t] U4p[0]) (-U2p[t] U4[t]+U2[t] U4p[t]) (U2p[t] V2[t]-U4p[t] V4[t]))/((2 U2[t] V2[t]-2 U4[t] V4[t])^2 (-m U2[t] V2p[0]+m U4[t] V4p[0]) (((-U2p[0] V2[t]+U4p[0] V4[t]) (-m U2[t] V2p[0]+m U4[t] V4p[0]))/(2 U2[t] V2[t]-2 U4[t] V4[t])^2-(m (U2p[0] U4[t]-U2[t] U4p[0]) (V2p[0] V4[t]-V2[t] V4p[0]))/(2 U2[t] V2[t]-2 U4[t] V4[t])^2))-(4 (U2p[0] U4[t]-U2[t] U4p[0]) (-U2p[t] U4[t]+U2[t] U4p[t]) (U2p[t] V2[t]-U4p[t] V4[t]) (u22[t] (u4[t] v1[t]-u3[t] v2[t])^2+u2[t]^2 (u44[t] v1[t]^2-v2[t] (-(u33[t]+v11[t]) v2[t]+v1[t] (u34[t]+u43[t]+v12[t]+v21[t]))+v1[t]^2 v22[t])+u24[t] u3[t] u4[t] v1[t] v4[t]-u23[t] u4[t]^2 v1[t] v4[t]-u32[t] u4[t]^2 v1[t] v4[t]+u3[t] u4[t] u42[t] v1[t] v4[t]-u4[t]^2 v1[t] v14[t] v4[t]-u24[t] u3[t]^2 v2[t] v4[t]+u23[t] u3[t] u4[t] v2[t] v4[t]+u3[t] u32[t] u4[t] v2[t] v4[t]-u3[t]^2 u42[t] v2[t] v4[t]+u3[t] u4[t] v14[t] v2[t] v4[t]+u3[t] u4[t] v1[t] v24[t] v4[t]-u3[t]^2 v2[t] v24[t] v4[t]-u3[t] u34[t] u4[t] v4[t]^2+u33[t] u4[t]^2 v4[t]^2-u3[t] u4[t] u43[t] v4[t]^2+u3[t]^2 u44[t] v4[t]^2+u4[t]^2 v11[t] v4[t]^2-u3[t] u4[t] v12[t] v4[t]^2-u3[t] u4[t] v21[t] v4[t]^2+u3[t]^2 v22[t] v4[t]^2-u4[t]^2 v1[t] v4[t] v41[t]+u3[t] u4[t] v2[t] v4[t] v41[t]+u3[t] u4[t] v1[t] v4[t] v42[t]-u3[t]^2 v2[t] v4[t] v42[t]+u2[t] (u24[t] v1[t] (-u4[t] v1[t]+u3[t] v2[t])+u4[t] (-u42[t] v1[t]^2+u23[t] v1[t] v2[t]+u32[t] v1[t] v2[t]+v1[t] v14[t] v2[t]-v1[t]^2 v24[t]+u34[t] v1[t] v4[t]+u43[t] v1[t] v4[t]+v1[t] v12[t] v4[t]-2 u33[t] v2[t] v4[t]-2 v11[t] v2[t] v4[t]+v1[t] v21[t] v4[t]+v1[t] v2[t] v41[t]-v1[t]^2 v42[t])+u3[t] (u42[t] v1[t] v2[t]-u23[t] v2[t]^2-u32[t] v2[t]^2-v14[t] v2[t]^2+v1[t] v2[t] v24[t]-2 u44[t] v1[t] v4[t]+u34[t] v2[t] v4[t]+u43[t] v2[t] v4[t]+v12[t] v2[t] v4[t]+v2[t] v21[t] v4[t]-2 v1[t] v22[t] v4[t]-v2[t]^2 v41[t]+v1[t] v2[t] v42[t]))+(u4[t] v1[t]-u3[t] v2[t])^2 v44[t]))/((u2[t] v2[t]-u4[t] v4[t])^2 (2 U2[t] V2[t]-2 U4[t] V4[t])^2 (-m U2[t] V2p[0]+m U4[t] V4p[0]) (((-U2p[0] V2[t]+U4p[0] V4[t]) (-m U2[t] V2p[0]+m U4[t] V4p[0]))/(2 U2[t] V2[t]-2 U4[t] V4[t])^2-(m (U2p[0] U4[t]-U2[t] U4p[0]) (V2p[0] V4[t]-V2[t] V4p[0]))/(2 U2[t] V2[t]-2 U4[t] V4[t])^2))-(4 sq1p1i (U2p[t] V2[t]-U4p[t] V4[t]) (U1p[t] U2[t] V2[t]-U1[t] U2p[t] V2[t]+U2p[t] U4[t] V3[t]-U2[t] U4p[t] V3[t]-U1p[t] U4[t] V4[t]+U1[t] U4p[t] V4[t]) (-m U2[t] V2p[0]+m U4[t] V4p[0]))/((2 U2[t] V2[t]-2 U4[t] V4[t])^3 (((-U2p[0] V2[t]+U4p[0] V4[t]) (-m U2[t] V2p[0]+m U4[t] V4p[0]))/(2 U2[t] V2[t]-2 U4[t] V4[t])^2-(m (U2p[0] U4[t]-U2[t] U4p[0]) (V2p[0] V4[t]-V2[t] V4p[0]))/(2 U2[t] V2[t]-2 U4[t] V4[t])^2))-(4 vep1i veq1i (U2p[t] V2[t]-U4p[t] V4[t]) (U1p[t] U2[t] V2[t]-U1[t] U2p[t] V2[t]+U2p[t] U4[t] V3[t]-U2[t] U4p[t] V3[t]-U1p[t] U4[t] V4[t]+U1[t] U4p[t] V4[t]) (-m U2[t] V2p[0]+m U4[t] V4p[0]))/((2 U2[t] V2[t]-2 U4[t] V4[t])^3 (((-U2p[0] V2[t]+U4p[0] V4[t]) (-m U2[t] V2p[0]+m U4[t] V4p[0]))/(2 U2[t] V2[t]-2 U4[t] V4[t])^2-(m (U2p[0] U4[t]-U2[t] U4p[0]) (V2p[0] V4[t]-V2[t] V4p[0]))/(2 U2[t] V2[t]-2 U4[t] V4[t])^2))-(4 vep1i veq2i (U2p[t] V2[t]-U4p[t] V4[t]) (U2p[t] U4[t] V1[t]-U2[t] U4p[t] V1[t]-U2p[t] U3[t] V2[t]+U2[t] U3p[t] V2[t]-U3p[t] U4[t] V4[t]+U3[t] U4p[t] V4[t]) (-m U2[t] V2p[0]+m U4[t] V4p[0]))/((2 U2[t] V2[t]-2 U4[t] V4[t])^3 (((-U2p[0] V2[t]+U4p[0] V4[t]) (-m U2[t] V2p[0]+m U4[t] V4p[0]))/(2 U2[t] V2[t]-2 U4[t] V4[t])^2-(m (U2p[0] U4[t]-U2[t] U4p[0]) (V2p[0] V4[t]-V2[t] V4p[0]))/(2 U2[t] V2[t]-2 U4[t] V4[t])^2))-(8 sq1q1i (U2p[t] V2[t]-U4p[t] V4[t]) (U1p[t] U2[t] V2[t]-U1[t] U2p[t] V2[t]+U2p[t] U4[t] V3[t]-U2[t] U4p[t] V3[t]-U1p[t] U4[t] V4[t]+U1[t] U4p[t] V4[t]) (U2[t] (-U1p[0] V2[t]+U4p[0] V3[t])+U4[t] (-U2p[0] V3[t]+U1p[0] V4[t])+U1[t] (U2p[0] V2[t]-U4p[0] V4[t])) (-m U2[t] V2p[0]+m U4[t] V4p[0]))/((2 U2[t] V2[t]-2 U4[t] V4[t])^4 (((-U2p[0] V2[t]+U4p[0] V4[t]) (-m U2[t] V2p[0]+m U4[t] V4p[0]))/(2 U2[t] V2[t]-2 U4[t] V4[t])^2-(m (U2p[0] U4[t]-U2[t] U4p[0]) (V2p[0] V4[t]-V2[t] V4p[0]))/(2 U2[t] V2[t]-2 U4[t] V4[t])^2))-(8 veq1i^2 (U2p[t] V2[t]-U4p[t] V4[t]) (U1p[t] U2[t] V2[t]-U1[t] U2p[t] V2[t]+U2p[t] U4[t] V3[t]-U2[t] U4p[t] V3[t]-U1p[t] U4[t] V4[t]+U1[t] U4p[t] V4[t]) (U2[t] (-U1p[0] V2[t]+U4p[0] V3[t])+U4[t] (-U2p[0] V3[t]+U1p[0] V4[t])+U1[t] (U2p[0] V2[t]-U4p[0] V4[t])) (-m U2[t] V2p[0]+m U4[t] V4p[0]))/((2 U2[t] V2[t]-2 U4[t] V4[t])^4 (((-U2p[0] V2[t]+U4p[0] V4[t]) (-m U2[t] V2p[0]+m U4[t] V4p[0]))/(2 U2[t] V2[t]-2 U4[t] V4[t])^2-(m (U2p[0] U4[t]-U2[t] U4p[0]) (V2p[0] V4[t]-V2[t] V4p[0]))/(2 U2[t] V2[t]-2 U4[t] V4[t])^2))-(8 veq1i veq2i (U2p[t] V2[t]-U4p[t] V4[t]) (U2p[t] U4[t] V1[t]-U2[t] U4p[t] V1[t]-U2p[t] U3[t] V2[t]+U2[t] U3p[t] V2[t]-U3p[t] U4[t] V4[t]+U3[t] U4p[t] V4[t]) (U2[t] (-U1p[0] V2[t]+U4p[0] V3[t])+U4[t] (-U2p[0] V3[t]+U1p[0] V4[t])+U1[t] (U2p[0] V2[t]-U4p[0] V4[t])) (-m U2[t] V2p[0]+m U4[t] V4p[0]))/((2 U2[t] V2[t]-2 U4[t] V4[t])^4 (((-U2p[0] V2[t]+U4p[0] V4[t]) (-m U2[t] V2p[0]+m U4[t] V4p[0]))/(2 U2[t] V2[t]-2 U4[t] V4[t])^2-(m (U2p[0] U4[t]-U2[t] U4p[0]) (V2p[0] V4[t]-V2[t] V4p[0]))/(2 U2[t] V2[t]-2 U4[t] V4[t])^2))-(8 veq1i veq2i (U2p[t] V2[t]-U4p[t] V4[t]) (U1p[t] U2[t] V2[t]-U1[t] U2p[t] V2[t]+U2p[t] U4[t] V3[t]-U2[t] U4p[t] V3[t]-U1p[t] U4[t] V4[t]+U1[t] U4p[t] V4[t]) (U2[t] (U4p[0] V1[t]-U3p[0] V2[t])+U4[t] (-U2p[0] V1[t]+U3p[0] V4[t])+U3[t] (U2p[0] V2[t]-U4p[0] V4[t])) (-m U2[t] V2p[0]+m U4[t] V4p[0]))/((2 U2[t] V2[t]-2 U4[t] V4[t])^4 (((-U2p[0] V2[t]+U4p[0] V4[t]) (-m U2[t] V2p[0]+m U4[t] V4p[0]))/(2 U2[t] V2[t]-2 U4[t] V4[t])^2-(m (U2p[0] U4[t]-U2[t] U4p[0]) (V2p[0] V4[t]-V2[t] V4p[0]))/(2 U2[t] V2[t]-2 U4[t] V4[t])^2))-(8 sq2q2i (U2p[t] V2[t]-U4p[t] V4[t]) (U2p[t] U4[t] V1[t]-U2[t] U4p[t] V1[t]-U2p[t] U3[t] V2[t]+U2[t] U3p[t] V2[t]-U3p[t] U4[t] V4[t]+U3[t] U4p[t] V4[t]) (U2[t] (U4p[0] V1[t]-U3p[0] V2[t])+U4[t] (-U2p[0] V1[t]+U3p[0] V4[t])+U3[t] (U2p[0] V2[t]-U4p[0] V4[t])) (-m U2[t] V2p[0]+m U4[t] V4p[0]))/((2 U2[t] V2[t]-2 U4[t] V4[t])^4 (((-U2p[0] V2[t]+U4p[0] V4[t]) (-m U2[t] V2p[0]+m U4[t] V4p[0]))/(2 U2[t] V2[t]-2 U4[t] V4[t])^2-(m (U2p[0] U4[t]-U2[t] U4p[0]) (V2p[0] V4[t]-V2[t] V4p[0]))/(2 U2[t] V2[t]-2 U4[t] V4[t])^2))-(8 veq2i^2 (U2p[t] V2[t]-U4p[t] V4[t]) (U2p[t] U4[t] V1[t]-U2[t] U4p[t] V1[t]-U2p[t] U3[t] V2[t]+U2[t] U3p[t] V2[t]-U3p[t] U4[t] V4[t]+U3[t] U4p[t] V4[t]) (U2[t] (U4p[0] V1[t]-U3p[0] V2[t])+U4[t] (-U2p[0] V1[t]+U3p[0] V4[t])+U3[t] (U2p[0] V2[t]-U4p[0] V4[t])) (-m U2[t] V2p[0]+m U4[t] V4p[0]))/((2 U2[t] V2[t]-2 U4[t] V4[t])^4 (((-U2p[0] V2[t]+U4p[0] V4[t]) (-m U2[t] V2p[0]+m U4[t] V4p[0]))/(2 U2[t] V2[t]-2 U4[t] V4[t])^2-(m (U2p[0] U4[t]-U2[t] U4p[0]) (V2p[0] V4[t]-V2[t] V4p[0]))/(2 U2[t] V2[t]-2 U4[t] V4[t])^2))-(2 (U2p[t] V2[t]-U4p[t] V4[t]) (-2 u1[t] u22[t] v2[t]^2+2 u22[t] u4[t] v2[t] v3[t]+2 u1[t] u24[t] v2[t] v4[t]-u21[t] u4[t] v2[t] v4[t]+2 u1[t] u42[t] v2[t] v4[t]+2 u1[t] v2[t] v24[t] v4[t]-u24[t] u4[t] v3[t] v4[t]-u4[t] u42[t] v3[t] v4[t]-u4[t] v24[t] v3[t] v4[t]-u4[t] v2[t] v34[t] v4[t]+u14[t] u4[t] v4[t]^2+u4[t] u41[t] v4[t]^2-2 u1[t] u44[t] v4[t]^2-2 u1[t] v22[t] v4[t]^2+u4[t] v23[t] v4[t]^2+u4[t] v32[t] v4[t]^2+u12[t] v2[t] (u2[t] v2[t]-u4[t] v4[t])+2 u1[t] v2[t] v4[t] v42[t]-u4[t] v3[t] v4[t] v42[t]-u4[t] v2[t] v4[t] v43[t]+u2[t] (u21[t] v2[t]^2-u24[t] v2[t] v3[t]-u42[t] v2[t] v3[t]-v2[t] v24[t] v3[t]+v2[t]^2 v34[t]-u14[t] v2[t] v4[t]-u41[t] v2[t] v4[t]-v2[t] v23[t] v4[t]+2 u44[t] v3[t] v4[t]+2 v22[t] v3[t] v4[t]-v2[t] v32[t] v4[t]-v2[t] v3[t] v42[t]+v2[t]^2 v43[t])+2 v2[t] (-u1[t] v2[t]+u4[t] v3[t]) v44[t]) (-m U2[t] V2p[0]+m U4[t] V4p[0]))/((u2[t] v2[t]-u4[t] v4[t])^2 (2 U2[t] V2[t]-2 U4[t] V4[t])^2 (((-U2p[0] V2[t]+U4p[0] V4[t]) (-m U2[t] V2p[0]+m U4[t] V4p[0]))/(2 U2[t] V2[t]-2 U4[t] V4[t])^2-(m (U2p[0] U4[t]-U2[t] U4p[0]) (V2p[0] V4[t]-V2[t] V4p[0]))/(2 U2[t] V2[t]-2 U4[t] V4[t])^2))+(4 m sq1p1i (-U2p[t] U4[t]+U2[t] U4p[t]) (U1p[t] U2[t] V2[t]-U1[t] U2p[t] V2[t]+U2p[t] U4[t] V3[t]-U2[t] U4p[t] V3[t]-U1p[t] U4[t] V4[t]+U1[t] U4p[t] V4[t]) (V2p[0] V4[t]-V2[t] V4p[0]))/((2 U2[t] V2[t]-2 U4[t] V4[t])^3 (((-U2p[0] V2[t]+U4p[0] V4[t]) (-m U2[t] V2p[0]+m U4[t] V4p[0]))/(2 U2[t] V2[t]-2 U4[t] V4[t])^2-(m (U2p[0] U4[t]-U2[t] U4p[0]) (V2p[0] V4[t]-V2[t] V4p[0]))/(2 U2[t] V2[t]-2 U4[t] V4[t])^2))+(4 m vep1i veq1i (-U2p[t] U4[t]+U2[t] U4p[t]) (U1p[t] U2[t] V2[t]-U1[t] U2p[t] V2[t]+U2p[t] U4[t] V3[t]-U2[t] U4p[t] V3[t]-U1p[t] U4[t] V4[t]+U1[t] U4p[t] V4[t]) (V2p[0] V4[t]-V2[t] V4p[0]))/((2 U2[t] V2[t]-2 U4[t] V4[t])^3 (((-U2p[0] V2[t]+U4p[0] V4[t]) (-m U2[t] V2p[0]+m U4[t] V4p[0]))/(2 U2[t] V2[t]-2 U4[t] V4[t])^2-(m (U2p[0] U4[t]-U2[t] U4p[0]) (V2p[0] V4[t]-V2[t] V4p[0]))/(2 U2[t] V2[t]-2 U4[t] V4[t])^2))+(4 m vep1i veq2i (-U2p[t] U4[t]+U2[t] U4p[t]) (U2p[t] U4[t] V1[t]-U2[t] U4p[t] V1[t]-U2p[t] U3[t] V2[t]+U2[t] U3p[t] V2[t]-U3p[t] U4[t] V4[t]+U3[t] U4p[t] V4[t]) (V2p[0] V4[t]-V2[t] V4p[0]))/((2 U2[t] V2[t]-2 U4[t] V4[t])^3 (((-U2p[0] V2[t]+U4p[0] V4[t]) (-m U2[t] V2p[0]+m U4[t] V4p[0]))/(2 U2[t] V2[t]-2 U4[t] V4[t])^2-(m (U2p[0] U4[t]-U2[t] U4p[0]) (V2p[0] V4[t]-V2[t] V4p[0]))/(2 U2[t] V2[t]-2 U4[t] V4[t])^2))+(8 m sq1q1i (-U2p[t] U4[t]+U2[t] U4p[t]) (U1p[t] U2[t] V2[t]-U1[t] U2p[t] V2[t]+U2p[t] U4[t] V3[t]-U2[t] U4p[t] V3[t]-U1p[t] U4[t] V4[t]+U1[t] U4p[t] V4[t]) (U2[t] (-U1p[0] V2[t]+U4p[0] V3[t])+U4[t] (-U2p[0] V3[t]+U1p[0] V4[t])+U1[t] (U2p[0] V2[t]-U4p[0] V4[t])) (V2p[0] V4[t]-V2[t] V4p[0]))/((2 U2[t] V2[t]-2 U4[t] V4[t])^4 (((-U2p[0] V2[t]+U4p[0] V4[t]) (-m U2[t] V2p[0]+m U4[t] V4p[0]))/(2 U2[t] V2[t]-2 U4[t] V4[t])^2-(m (U2p[0] U4[t]-U2[t] U4p[0]) (V2p[0] V4[t]-V2[t] V4p[0]))/(2 U2[t] V2[t]-2 U4[t] V4[t])^2))+(8 m veq1i^2 (-U2p[t] U4[t]+U2[t] U4p[t]) (U1p[t] U2[t] V2[t]-U1[t] U2p[t] V2[t]+U2p[t] U4[t] V3[t]-U2[t] U4p[t] V3[t]-U1p[t] U4[t] V4[t]+U1[t] U4p[t] V4[t]) (U2[t] (-U1p[0] V2[t]+U4p[0] V3[t])+U4[t] (-U2p[0] V3[t]+U1p[0] V4[t])+U1[t] (U2p[0] V2[t]-U4p[0] V4[t])) (V2p[0] V4[t]-V2[t] V4p[0]))/((2 U2[t] V2[t]-2 U4[t] V4[t])^4 (((-U2p[0] V2[t]+U4p[0] V4[t]) (-m U2[t] V2p[0]+m U4[t] V4p[0]))/(2 U2[t] V2[t]-2 U4[t] V4[t])^2-(m (U2p[0] U4[t]-U2[t] U4p[0]) (V2p[0] V4[t]-V2[t] V4p[0]))/(2 U2[t] V2[t]-2 U4[t] V4[t])^2))+(8 m veq1i veq2i (-U2p[t] U4[t]+U2[t] U4p[t]) (U2p[t] U4[t] V1[t]-U2[t] U4p[t] V1[t]-U2p[t] U3[t] V2[t]+U2[t] U3p[t] V2[t]-U3p[t] U4[t] V4[t]+U3[t] U4p[t] V4[t]) (U2[t] (-U1p[0] V2[t]+U4p[0] V3[t])+U4[t] (-U2p[0] V3[t]+U1p[0] V4[t])+U1[t] (U2p[0] V2[t]-U4p[0] V4[t])) (V2p[0] V4[t]-V2[t] V4p[0]))/((2 U2[t] V2[t]-2 U4[t] V4[t])^4 (((-U2p[0] V2[t]+U4p[0] V4[t]) (-m U2[t] V2p[0]+m U4[t] V4p[0]))/(2 U2[t] V2[t]-2 U4[t] V4[t])^2-(m (U2p[0] U4[t]-U2[t] U4p[0]) (V2p[0] V4[t]-V2[t] V4p[0]))/(2 U2[t] V2[t]-2 U4[t] V4[t])^2))+(8 m veq1i veq2i (-U2p[t] U4[t]+U2[t] U4p[t]) (U1p[t] U2[t] V2[t]-U1[t] U2p[t] V2[t]+U2p[t] U4[t] V3[t]-U2[t] U4p[t] V3[t]-U1p[t] U4[t] V4[t]+U1[t] U4p[t] V4[t]) (U2[t] (U4p[0] V1[t]-U3p[0] V2[t])+U4[t] (-U2p[0] V1[t]+U3p[0] V4[t])+U3[t] (U2p[0] V2[t]-U4p[0] V4[t])) (V2p[0] V4[t]-V2[t] V4p[0]))/((2 U2[t] V2[t]-2 U4[t] V4[t])^4 (((-U2p[0] V2[t]+U4p[0] V4[t]) (-m U2[t] V2p[0]+m U4[t] V4p[0]))/(2 U2[t] V2[t]-2 U4[t] V4[t])^2-(m (U2p[0] U4[t]-U2[t] U4p[0]) (V2p[0] V4[t]-V2[t] V4p[0]))/(2 U2[t] V2[t]-2 U4[t] V4[t])^2))+(8 m sq2q2i (-U2p[t] U4[t]+U2[t] U4p[t]) (U2p[t] U4[t] V1[t]-U2[t] U4p[t] V1[t]-U2p[t] U3[t] V2[t]+U2[t] U3p[t] V2[t]-U3p[t] U4[t] V4[t]+U3[t] U4p[t] V4[t]) (U2[t] (U4p[0] V1[t]-U3p[0] V2[t])+U4[t] (-U2p[0] V1[t]+U3p[0] V4[t])+U3[t] (U2p[0] V2[t]-U4p[0] V4[t])) (V2p[0] V4[t]-V2[t] V4p[0]))/((2 U2[t] V2[t]-2 U4[t] V4[t])^4 (((-U2p[0] V2[t]+U4p[0] V4[t]) (-m U2[t] V2p[0]+m U4[t] V4p[0]))/(2 U2[t] V2[t]-2 U4[t] V4[t])^2-(m (U2p[0] U4[t]-U2[t] U4p[0]) (V2p[0] V4[t]-V2[t] V4p[0]))/(2 U2[t] V2[t]-2 U4[t] V4[t])^2))+(8 m veq2i^2 (-U2p[t] U4[t]+U2[t] U4p[t]) (U2p[t] U4[t] V1[t]-U2[t] U4p[t] V1[t]-U2p[t] U3[t] V2[t]+U2[t] U3p[t] V2[t]-U3p[t] U4[t] V4[t]+U3[t] U4p[t] V4[t]) (U2[t] (U4p[0] V1[t]-U3p[0] V2[t])+U4[t] (-U2p[0] V1[t]+U3p[0] V4[t])+U3[t] (U2p[0] V2[t]-U4p[0] V4[t])) (V2p[0] V4[t]-V2[t] V4p[0]))/((2 U2[t] V2[t]-2 U4[t] V4[t])^4 (((-U2p[0] V2[t]+U4p[0] V4[t]) (-m U2[t] V2p[0]+m U4[t] V4p[0]))/(2 U2[t] V2[t]-2 U4[t] V4[t])^2-(m (U2p[0] U4[t]-U2[t] U4p[0]) (V2p[0] V4[t]-V2[t] V4p[0]))/(2 U2[t] V2[t]-2 U4[t] V4[t])^2))+(2 m (-U2p[t] U4[t]+U2[t] U4p[t]) (-2 u1[t] u22[t] v2[t]^2+2 u22[t] u4[t] v2[t] v3[t]+2 u1[t] u24[t] v2[t] v4[t]-u21[t] u4[t] v2[t] v4[t]+2 u1[t] u42[t] v2[t] v4[t]+2 u1[t] v2[t] v24[t] v4[t]-u24[t] u4[t] v3[t] v4[t]-u4[t] u42[t] v3[t] v4[t]-u4[t] v24[t] v3[t] v4[t]-u4[t] v2[t] v34[t] v4[t]+u14[t] u4[t] v4[t]^2+u4[t] u41[t] v4[t]^2-2 u1[t] u44[t] v4[t]^2-2 u1[t] v22[t] v4[t]^2+u4[t] v23[t] v4[t]^2+u4[t] v32[t] v4[t]^2+u12[t] v2[t] (u2[t] v2[t]-u4[t] v4[t])+2 u1[t] v2[t] v4[t] v42[t]-u4[t] v3[t] v4[t] v42[t]-u4[t] v2[t] v4[t] v43[t]+u2[t] (u21[t] v2[t]^2-u24[t] v2[t] v3[t]-u42[t] v2[t] v3[t]-v2[t] v24[t] v3[t]+v2[t]^2 v34[t]-u14[t] v2[t] v4[t]-u41[t] v2[t] v4[t]-v2[t] v23[t] v4[t]+2 u44[t] v3[t] v4[t]+2 v22[t] v3[t] v4[t]-v2[t] v32[t] v4[t]-v2[t] v3[t] v42[t]+v2[t]^2 v43[t])+2 v2[t] (-u1[t] v2[t]+u4[t] v3[t]) v44[t]) (V2p[0] V4[t]-V2[t] V4p[0]))/((u2[t] v2[t]-u4[t] v4[t])^2 (2 U2[t] V2[t]-2 U4[t] V4[t])^2 (((-U2p[0] V2[t]+U4p[0] V4[t]) (-m U2[t] V2p[0]+m U4[t] V4p[0]))/(2 U2[t] V2[t]-2 U4[t] V4[t])^2-(m (U2p[0] U4[t]-U2[t] U4p[0]) (V2p[0] V4[t]-V2[t] V4p[0]))/(2 U2[t] V2[t]-2 U4[t] V4[t])^2))+(2 m sp2p2i (U2p[0] U4[t]-U2[t] U4p[0]) (-U2p[t] U4[t]+U2[t] U4p[t])^2 (V2p[0] V4[t]-V2[t] V4p[0]))/((2 U2[t] V2[t]-2 U4[t] V4[t])^2 (-m U2[t] V2p[0]+m U4[t] V4p[0])^2 (((-U2p[0] V2[t]+U4p[0] V4[t]) (-m U2[t] V2p[0]+m U4[t] V4p[0]))/(2 U2[t] V2[t]-2 U4[t] V4[t])^2-(m (U2p[0] U4[t]-U2[t] U4p[0]) (V2p[0] V4[t]-V2[t] V4p[0]))/(2 U2[t] V2[t]-2 U4[t] V4[t])^2))+(2 m vep2i^2 (U2p[0] U4[t]-U2[t] U4p[0]) (-U2p[t] U4[t]+U2[t] U4p[t])^2 (V2p[0] V4[t]-V2[t] V4p[0]))/((2 U2[t] V2[t]-2 U4[t] V4[t])^2 (-m U2[t] V2p[0]+m U4[t] V4p[0])^2 (((-U2p[0] V2[t]+U4p[0] V4[t]) (-m U2[t] V2p[0]+m U4[t] V4p[0]))/(2 U2[t] V2[t]-2 U4[t] V4[t])^2-(m (U2p[0] U4[t]-U2[t] U4p[0]) (V2p[0] V4[t]-V2[t] V4p[0]))/(2 U2[t] V2[t]-2 U4[t] V4[t])^2))+(4 m (U2p[0] U4[t]-U2[t] U4p[0]) (-U2p[t] U4[t]+U2[t] U4p[t])^2 (u22[t] (u4[t] v1[t]-u3[t] v2[t])^2+u2[t]^2 (u44[t] v1[t]^2-v2[t] (-(u33[t]+v11[t]) v2[t]+v1[t] (u34[t]+u43[t]+v12[t]+v21[t]))+v1[t]^2 v22[t])+u24[t] u3[t] u4[t] v1[t] v4[t]-u23[t] u4[t]^2 v1[t] v4[t]-u32[t] u4[t]^2 v1[t] v4[t]+u3[t] u4[t] u42[t] v1[t] v4[t]-u4[t]^2 v1[t] v14[t] v4[t]-u24[t] u3[t]^2 v2[t] v4[t]+u23[t] u3[t] u4[t] v2[t] v4[t]+u3[t] u32[t] u4[t] v2[t] v4[t]-u3[t]^2 u42[t] v2[t] v4[t]+u3[t] u4[t] v14[t] v2[t] v4[t]+u3[t] u4[t] v1[t] v24[t] v4[t]-u3[t]^2 v2[t] v24[t] v4[t]-u3[t] u34[t] u4[t] v4[t]^2+u33[t] u4[t]^2 v4[t]^2-u3[t] u4[t] u43[t] v4[t]^2+u3[t]^2 u44[t] v4[t]^2+u4[t]^2 v11[t] v4[t]^2-u3[t] u4[t] v12[t] v4[t]^2-u3[t] u4[t] v21[t] v4[t]^2+u3[t]^2 v22[t] v4[t]^2-u4[t]^2 v1[t] v4[t] v41[t]+u3[t] u4[t] v2[t] v4[t] v41[t]+u3[t] u4[t] v1[t] v4[t] v42[t]-u3[t]^2 v2[t] v4[t] v42[t]+u2[t] (u24[t] v1[t] (-u4[t] v1[t]+u3[t] v2[t])+u4[t] (-u42[t] v1[t]^2+u23[t] v1[t] v2[t]+u32[t] v1[t] v2[t]+v1[t] v14[t] v2[t]-v1[t]^2 v24[t]+u34[t] v1[t] v4[t]+u43[t] v1[t] v4[t]+v1[t] v12[t] v4[t]-2 u33[t] v2[t] v4[t]-2 v11[t] v2[t] v4[t]+v1[t] v21[t] v4[t]+v1[t] v2[t] v41[t]-v1[t]^2 v42[t])+u3[t] (u42[t] v1[t] v2[t]-u23[t] v2[t]^2-u32[t] v2[t]^2-v14[t] v2[t]^2+v1[t] v2[t] v24[t]-2 u44[t] v1[t] v4[t]+u34[t] v2[t] v4[t]+u43[t] v2[t] v4[t]+v12[t] v2[t] v4[t]+v2[t] v21[t] v4[t]-2 v1[t] v22[t] v4[t]-v2[t]^2 v41[t]+v1[t] v2[t] v42[t]))+(u4[t] v1[t]-u3[t] v2[t])^2 v44[t]) (V2p[0] V4[t]-V2[t] V4p[0]))/((u2[t] v2[t]-u4[t] v4[t])^2 (2 U2[t] V2[t]-2 U4[t] V4[t])^2 (-m U2[t] V2p[0]+m U4[t] V4p[0])^2 (((-U2p[0] V2[t]+U4p[0] V4[t]) (-m U2[t] V2p[0]+m U4[t] V4p[0]))/(2 U2[t] V2[t]-2 U4[t] V4[t])^2-(m (U2p[0] U4[t]-U2[t] U4p[0]) (V2p[0] V4[t]-V2[t] V4p[0]))/(2 U2[t] V2[t]-2 U4[t] V4[t])^2))-(2 m vep1i vep2i (-U2p[t] U4[t]+U2[t] U4p[t])^2 (V2p[0] V4[t]-V2[t] V4p[0]))/((2 U2[t] V2[t]-2 U4[t] V4[t])^2 (-m U2[t] V2p[0]+m U4[t] V4p[0]) (((-U2p[0] V2[t]+U4p[0] V4[t]) (-m U2[t] V2p[0]+m U4[t] V4p[0]))/(2 U2[t] V2[t]-2 U4[t] V4[t])^2-(m (U2p[0] U4[t]-U2[t] U4p[0]) (V2p[0] V4[t]-V2[t] V4p[0]))/(2 U2[t] V2[t]-2 U4[t] V4[t])^2))-(4 m vep2i veq1i (U2p[0] U4[t]-U2[t] U4p[0]) (-U2p[t] U4[t]+U2[t] U4p[t]) (U1p[t] U2[t] V2[t]-U1[t] U2p[t] V2[t]+U2p[t] U4[t] V3[t]-U2[t] U4p[t] V3[t]-U1p[t] U4[t] V4[t]+U1[t] U4p[t] V4[t]) (V2p[0] V4[t]-V2[t] V4p[0]))/((2 U2[t] V2[t]-2 U4[t] V4[t])^3 (-m U2[t] V2p[0]+m U4[t] V4p[0]) (((-U2p[0] V2[t]+U4p[0] V4[t]) (-m U2[t] V2p[0]+m U4[t] V4p[0]))/(2 U2[t] V2[t]-2 U4[t] V4[t])^2-(m (U2p[0] U4[t]-U2[t] U4p[0]) (V2p[0] V4[t]-V2[t] V4p[0]))/(2 U2[t] V2[t]-2 U4[t] V4[t])^2))-(4 m sq2p2i (U2p[0] U4[t]-U2[t] U4p[0]) (-U2p[t] U4[t]+U2[t] U4p[t]) (U2p[t] U4[t] V1[t]-U2[t] U4p[t] V1[t]-U2p[t] U3[t] V2[t]+U2[t] U3p[t] V2[t]-U3p[t] U4[t] V4[t]+U3[t] U4p[t] V4[t]) (V2p[0] V4[t]-V2[t] V4p[0]))/((2 U2[t] V2[t]-2 U4[t] V4[t])^3 (-m U2[t] V2p[0]+m U4[t] V4p[0]) (((-U2p[0] V2[t]+U4p[0] V4[t]) (-m U2[t] V2p[0]+m U4[t] V4p[0]))/(2 U2[t] V2[t]-2 U4[t] V4[t])^2-(m (U2p[0] U4[t]-U2[t] U4p[0]) (V2p[0] V4[t]-V2[t] V4p[0]))/(2 U2[t] V2[t]-2 U4[t] V4[t])^2))-(4 m vep2i veq2i (U2p[0] U4[t]-U2[t] U4p[0]) (-U2p[t] U4[t]+U2[t] U4p[t]) (U2p[t] U4[t] V1[t]-U2[t] U4p[t] V1[t]-U2p[t] U3[t] V2[t]+U2[t] U3p[t] V2[t]-U3p[t] U4[t] V4[t]+U3[t] U4p[t] V4[t]) (V2p[0] V4[t]-V2[t] V4p[0]))/((2 U2[t] V2[t]-2 U4[t] V4[t])^3 (-m U2[t] V2p[0]+m U4[t] V4p[0]) (((-U2p[0] V2[t]+U4p[0] V4[t]) (-m U2[t] V2p[0]+m U4[t] V4p[0]))/(2 U2[t] V2[t]-2 U4[t] V4[t])^2-(m (U2p[0] U4[t]-U2[t] U4p[0]) (V2p[0] V4[t]-V2[t] V4p[0]))/(2 U2[t] V2[t]-2 U4[t] V4[t])^2))-(4 m vep2i veq1i (-U2p[t] U4[t]+U2[t] U4p[t])^2 (U2[t] (-U1p[0] V2[t]+U4p[0] V3[t])+U4[t] (-U2p[0] V3[t]+U1p[0] V4[t])+U1[t] (U2p[0] V2[t]-U4p[0] V4[t])) (V2p[0] V4[t]-V2[t] V4p[0]))/((2 U2[t] V2[t]-2 U4[t] V4[t])^3 (-m U2[t] V2p[0]+m U4[t] V4p[0]) (((-U2p[0] V2[t]+U4p[0] V4[t]) (-m U2[t] V2p[0]+m U4[t] V4p[0]))/(2 U2[t] V2[t]-2 U4[t] V4[t])^2-(m (U2p[0] U4[t]-U2[t] U4p[0]) (V2p[0] V4[t]-V2[t] V4p[0]))/(2 U2[t] V2[t]-2 U4[t] V4[t])^2))-(4 m sq2p2i (-U2p[t] U4[t]+U2[t] U4p[t])^2 (U2[t] (U4p[0] V1[t]-U3p[0] V2[t])+U4[t] (-U2p[0] V1[t]+U3p[0] V4[t])+U3[t] (U2p[0] V2[t]-U4p[0] V4[t])) (V2p[0] V4[t]-V2[t] V4p[0]))/((2 U2[t] V2[t]-2 U4[t] V4[t])^3 (-m U2[t] V2p[0]+m U4[t] V4p[0]) (((-U2p[0] V2[t]+U4p[0] V4[t]) (-m U2[t] V2p[0]+m U4[t] V4p[0]))/(2 U2[t] V2[t]-2 U4[t] V4[t])^2-(m (U2p[0] U4[t]-U2[t] U4p[0]) (V2p[0] V4[t]-V2[t] V4p[0]))/(2 U2[t] V2[t]-2 U4[t] V4[t])^2))-(4 m vep2i veq2i (-U2p[t] U4[t]+U2[t] U4p[t])^2 (U2[t] (U4p[0] V1[t]-U3p[0] V2[t])+U4[t] (-U2p[0] V1[t]+U3p[0] V4[t])+U3[t] (U2p[0] V2[t]-U4p[0] V4[t])) (V2p[0] V4[t]-V2[t] V4p[0]))/((2 U2[t] V2[t]-2 U4[t] V4[t])^3 (-m U2[t] V2p[0]+m U4[t] V4p[0]) (((-U2p[0] V2[t]+U4p[0] V4[t]) (-m U2[t] V2p[0]+m U4[t] V4p[0]))/(2 U2[t] V2[t]-2 U4[t] V4[t])^2-(m (U2p[0] U4[t]-U2[t] U4p[0]) (V2p[0] V4[t]-V2[t] V4p[0]))/(2 U2[t] V2[t]-2 U4[t] V4[t])^2))+(2 m (U2p[0] U4[t]-U2[t] U4p[0]) (-U2p[t] U4[t]+U2[t] U4p[t]) (2 u22[t] v2[t] (-u4[t] v1[t]+u3[t] v2[t])+u2[t] (u24[t] v1[t] v2[t]+u42[t] v1[t] v2[t]-u23[t] v2[t]^2-u32[t] v2[t]^2-v14[t] v2[t]^2+v1[t] v2[t] v24[t]-2 u44[t] v1[t] v4[t]+u34[t] v2[t] v4[t]+u43[t] v2[t] v4[t]+v12[t] v2[t] v4[t]+v2[t] v21[t] v4[t]-2 v1[t] v22[t] v4[t]-v2[t]^2 v41[t]+v1[t] v2[t] v42[t])+v4[t] (u24[t] (u4[t] v1[t]-2 u3[t] v2[t])+u4[t] (-(u34[t]+u43[t]+v12[t]+v21[t]) v4[t]+v2[t] (u23[t]+u32[t]+v14[t]+v41[t])+v1[t] (u42[t]+v24[t]+v42[t]))-2 u3[t] (-(u44[t]+v22[t]) v4[t]+v2[t] (u42[t]+v24[t]+v42[t])))+2 v2[t] (-u4[t] v1[t]+u3[t] v2[t]) v44[t]) (V2p[0] V4[t]-V2[t] V4p[0]))/((u2[t] v2[t]-u4[t] v4[t])^2 (2 U2[t] V2[t]-2 U4[t] V4[t])^2 (-m U2[t] V2p[0]+m U4[t] V4p[0]) (((-U2p[0] V2[t]+U4p[0] V4[t]) (-m U2[t] V2p[0]+m U4[t] V4p[0]))/(2 U2[t] V2[t]-2 U4[t] V4[t])^2-(m (U2p[0] U4[t]-U2[t] U4p[0]) (V2p[0] V4[t]-V2[t] V4p[0]))/(2 U2[t] V2[t]-2 U4[t] V4[t])^2))-(2 m (-U2p[t] U4[t]+U2[t] U4p[t])^2 (u12[t] u2[t] u4[t] v1[t] v2[t]+u2[t] u21[t] u4[t] v1[t] v2[t]-u2[t]^2 u41[t] v1[t] v2[t]+u13[t] u2[t]^2 v2[t]^2-u12[t] u2[t] u3[t] v2[t]^2-u2[t] u21[t] u3[t] v2[t]^2+u2[t]^2 u31[t] v2[t]^2+u2[t]^2 v13[t] v2[t]^2-u2[t]^2 v1[t] v2[t] v23[t]-2 u2[t] u24[t] u4[t] v1[t] v3[t]+2 u22[t] u4[t]^2 v1[t] v3[t]-2 u2[t] u4[t] u42[t] v1[t] v3[t]+2 u2[t]^2 u44[t] v1[t] v3[t]+u2[t] u24[t] u3[t] v2[t] v3[t]-u2[t]^2 u34[t] v2[t] v3[t]+u2[t] u23[t] u4[t] v2[t] v3[t]-2 u22[t] u3[t] u4[t] v2[t] v3[t]+u2[t] u32[t] u4[t] v2[t] v3[t]+u2[t] u3[t] u42[t] v2[t] v3[t]-u2[t]^2 u43[t] v2[t] v3[t]-u2[t]^2 v12[t] v2[t] v3[t]+u2[t] u4[t] v14[t] v2[t] v3[t]-u2[t]^2 v2[t] v21[t] v3[t]+2 u2[t]^2 v1[t] v22[t] v3[t]-2 u2[t] u4[t] v1[t] v24[t] v3[t]+u2[t] u3[t] v2[t] v24[t] v3[t]+u2[t]^2 v2[t]^2 v31[t]-u2[t]^2 v1[t] v2[t] v32[t]+u2[t] u4[t] v1[t] v2[t] v34[t]-u2[t] u3[t] v2[t]^2 v34[t]-u12[t] u4[t]^2 v1[t] v4[t]-u21[t] u4[t]^2 v1[t] v4[t]+u2[t] u4[t] u41[t] v1[t] v4[t]-2 u13[t] u2[t] u4[t] v2[t] v4[t]+u12[t] u3[t] u4[t] v2[t] v4[t]+u21[t] u3[t] u4[t] v2[t] v4[t]-2 u2[t] u31[t] u4[t] v2[t] v4[t]+u2[t] u3[t] u41[t] v2[t] v4[t]-2 u2[t] u4[t] v13[t] v2[t] v4[t]+u2[t] u4[t] v1[t] v23[t] v4[t]+u2[t] u3[t] v2[t] v23[t] v4[t]+u24[t] u3[t] u4[t] v3[t] v4[t]+u2[t] u34[t] u4[t] v3[t] v4[t]-u23[t] u4[t]^2 v3[t] v4[t]-u32[t] u4[t]^2 v3[t] v4[t]+u3[t] u4[t] u42[t] v3[t] v4[t]+u2[t] u4[t] u43[t] v3[t] v4[t]-2 u2[t] u3[t] u44[t] v3[t] v4[t]+u2[t] u4[t] v12[t] v3[t] v4[t]-u4[t]^2 v14[t] v3[t] v4[t]+u2[t] u4[t] v21[t] v3[t] v4[t]-2 u2[t] u3[t] v22[t] v3[t] v4[t]+u3[t] u4[t] v24[t] v3[t] v4[t]-2 u2[t] u4[t] v2[t] v31[t] v4[t]+u2[t] u4[t] v1[t] v32[t] v4[t]+u2[t] u3[t] v2[t] v32[t] v4[t]-u4[t]^2 v1[t] v34[t] v4[t]+u3[t] u4[t] v2[t] v34[t] v4[t]+u13[t] u4[t]^2 v4[t]^2+u31[t] u4[t]^2 v4[t]^2-u3[t] u4[t] u41[t] v4[t]^2+u4[t]^2 v13[t] v4[t]^2-u3[t] u4[t] v23[t] v4[t]^2+u4[t]^2 v31[t] v4[t]^2-u3[t] u4[t] v32[t] v4[t]^2-u14[t] (u2[t] v1[t]-u3[t] v4[t]) (u2[t] v2[t]-u4[t] v4[t])+u2[t] u4[t] v2[t] v3[t] v41[t]-u4[t]^2 v3[t] v4[t] v41[t]-2 u2[t] u4[t] v1[t] v3[t] v42[t]+u2[t] u3[t] v2[t] v3[t] v42[t]+u3[t] u4[t] v3[t] v4[t] v42[t]+u2[t] u4[t] v1[t] v2[t] v43[t]-u2[t] u3[t] v2[t]^2 v43[t]-u4[t]^2 v1[t] v4[t] v43[t]+u3[t] u4[t] v2[t] v4[t] v43[t]+2 u4[t] (u4[t] v1[t]-u3[t] v2[t]) v3[t] v44[t]+u1[t] (2 u22[t] v2[t] (-u4[t] v1[t]+u3[t] v2[t])+u2[t] (u24[t] v1[t] v2[t]+u42[t] v1[t] v2[t]-u23[t] v2[t]^2-u32[t] v2[t]^2-v14[t] v2[t]^2+v1[t] v2[t] v24[t]-2 u44[t] v1[t] v4[t]+u34[t] v2[t] v4[t]+u43[t] v2[t] v4[t]+v12[t] v2[t] v4[t]+v2[t] v21[t] v4[t]-2 v1[t] v22[t] v4[t]-v2[t]^2 v41[t]+v1[t] v2[t] v42[t])+v4[t] (u24[t] (u4[t] v1[t]-2 u3[t] v2[t])+u4[t] (-(u34[t]+u43[t]+v12[t]+v21[t]) v4[t]+v2[t] (u23[t]+u32[t]+v14[t]+v41[t])+v1[t] (u42[t]+v24[t]+v42[t]))-2 u3[t] (-(u44[t]+v22[t]) v4[t]+v2[t] (u42[t]+v24[t]+v42[t])))+2 v2[t] (-u4[t] v1[t]+u3[t] v2[t]) v44[t])) (V2p[0] V4[t]-V2[t] V4p[0]))/((u2[t] v2[t]-u4[t] v4[t])^2 (2 U2[t] V2[t]-2 U4[t] V4[t])^2 (-m U2[t] V2p[0]+m U4[t] V4p[0]) (((-U2p[0] V2[t]+U4p[0] V4[t]) (-m U2[t] V2p[0]+m U4[t] V4p[0]))/(2 U2[t] V2[t]-2 U4[t] V4[t])^2-(m (U2p[0] U4[t]-U2[t] U4p[0]) (V2p[0] V4[t]-V2[t] V4p[0]))/(2 U2[t] V2[t]-2 U4[t] V4[t])^2))+(4 sq1q1[t] (2 U2[t] V2[t]-2 U4[t] V4[t])^2 (-(((U2p[t] V2[t]-U4p[t] V4[t]) (-m U2[t] V2p[0]+m U4[t] V4p[0]))/(2 U2[t] V2[t]-2 U4[t] V4[t])^2)+(m (-U2p[t] U4[t]+U2[t] U4p[t]) (V2p[0] V4[t]-V2[t] V4p[0]))/(2 U2[t] V2[t]-2 U4[t] V4[t])^2)^2)/(-m U2[t] V2p[0]+m U4[t] V4p[0])^2+(4 m sq2p2i (-U2p[t] U4[t]+U2[t] U4p[t])^2 (U2[t] (-V1p[0] V2[t]+V1[t] V2p[0])+U4[t] (V1p[0] V4[t]-V1[t] V4p[0])+U3[t] (-V2p[0] V4[t]+V2[t] V4p[0])))/((2 U2[t] V2[t]-2 U4[t] V4[t]) (-m U2[t] V2p[0]+m U4[t] V4p[0])^2)+(4 m vep2i veq2i (-U2p[t] U4[t]+U2[t] U4p[t])^2 (U2[t] (-V1p[0] V2[t]+V1[t] V2p[0])+U4[t] (V1p[0] V4[t]-V1[t] V4p[0])+U3[t] (-V2p[0] V4[t]+V2[t] V4p[0])))/((2 U2[t] V2[t]-2 U4[t] V4[t]) (-m U2[t] V2p[0]+m U4[t] V4p[0])^2)-(8 m veq1i veq2i (-U2p[t] U4[t]+U2[t] U4p[t]) (U1p[t] U2[t] V2[t]-U1[t] U2p[t] V2[t]+U2p[t] U4[t] V3[t]-U2[t] U4p[t] V3[t]-U1p[t] U4[t] V4[t]+U1[t] U4p[t] V4[t]) (U2[t] (-V1p[0] V2[t]+V1[t] V2p[0])+U4[t] (V1p[0] V4[t]-V1[t] V4p[0])+U3[t] (-V2p[0] V4[t]+V2[t] V4p[0])))/((2 U2[t] V2[t]-2 U4[t] V4[t])^2 (-m U2[t] V2p[0]+m U4[t] V4p[0]))-(8 m sq2q2i (-U2p[t] U4[t]+U2[t] U4p[t]) (U2p[t] U4[t] V1[t]-U2[t] U4p[t] V1[t]-U2p[t] U3[t] V2[t]+U2[t] U3p[t] V2[t]-U3p[t] U4[t] V4[t]+U3[t] U4p[t] V4[t]) (U2[t] (-V1p[0] V2[t]+V1[t] V2p[0])+U4[t] (V1p[0] V4[t]-V1[t] V4p[0])+U3[t] (-V2p[0] V4[t]+V2[t] V4p[0])))/((2 U2[t] V2[t]-2 U4[t] V4[t])^2 (-m U2[t] V2p[0]+m U4[t] V4p[0]))-(8 m veq2i^2 (-U2p[t] U4[t]+U2[t] U4p[t]) (U2p[t] U4[t] V1[t]-U2[t] U4p[t] V1[t]-U2p[t] U3[t] V2[t]+U2[t] U3p[t] V2[t]-U3p[t] U4[t] V4[t]+U3[t] U4p[t] V4[t]) (U2[t] (-V1p[0] V2[t]+V1[t] V2p[0])+U4[t] (V1p[0] V4[t]-V1[t] V4p[0])+U3[t] (-V2p[0] V4[t]+V2[t] V4p[0])))/((2 U2[t] V2[t]-2 U4[t] V4[t])^2 (-m U2[t] V2p[0]+m U4[t] V4p[0]))+(4 m vep1i veq2i (-U2p[t] U4[t]+U2[t] U4p[t]) (U2p[t] V2[t]-U4p[t] V4[t]) (U2[t] (-V1p[0] V2[t]+V1[t] V2p[0])+U4[t] (V1p[0] V4[t]-V1[t] V4p[0])+U3[t] (-V2p[0] V4[t]+V2[t] V4p[0])))/((2 U2[t] V2[t]-2 U4[t] V4[t])^3 (((-U2p[0] V2[t]+U4p[0] V4[t]) (-m U2[t] V2p[0]+m U4[t] V4p[0]))/(2 U2[t] V2[t]-2 U4[t] V4[t])^2-(m (U2p[0] U4[t]-U2[t] U4p[0]) (V2p[0] V4[t]-V2[t] V4p[0]))/(2 U2[t] V2[t]-2 U4[t] V4[t])^2))+(8 m veq1i veq2i (U2p[0] U4[t]-U2[t] U4p[0]) (U2p[t] V2[t]-U4p[t] V4[t]) (U1p[t] U2[t] V2[t]-U1[t] U2p[t] V2[t]+U2p[t] U4[t] V3[t]-U2[t] U4p[t] V3[t]-U1p[t] U4[t] V4[t]+U1[t] U4p[t] V4[t]) (U2[t] (-V1p[0] V2[t]+V1[t] V2p[0])+U4[t] (V1p[0] V4[t]-V1[t] V4p[0])+U3[t] (-V2p[0] V4[t]+V2[t] V4p[0])))/((2 U2[t] V2[t]-2 U4[t] V4[t])^4 (((-U2p[0] V2[t]+U4p[0] V4[t]) (-m U2[t] V2p[0]+m U4[t] V4p[0]))/(2 U2[t] V2[t]-2 U4[t] V4[t])^2-(m (U2p[0] U4[t]-U2[t] U4p[0]) (V2p[0] V4[t]-V2[t] V4p[0]))/(2 U2[t] V2[t]-2 U4[t] V4[t])^2))+(8 m sq2q2i (U2p[0] U4[t]-U2[t] U4p[0]) (U2p[t] V2[t]-U4p[t] V4[t]) (U2p[t] U4[t] V1[t]-U2[t] U4p[t] V1[t]-U2p[t] U3[t] V2[t]+U2[t] U3p[t] V2[t]-U3p[t] U4[t] V4[t]+U3[t] U4p[t] V4[t]) (U2[t] (-V1p[0] V2[t]+V1[t] V2p[0])+U4[t] (V1p[0] V4[t]-V1[t] V4p[0])+U3[t] (-V2p[0] V4[t]+V2[t] V4p[0])))/((2 U2[t] V2[t]-2 U4[t] V4[t])^4 (((-U2p[0] V2[t]+U4p[0] V4[t]) (-m U2[t] V2p[0]+m U4[t] V4p[0]))/(2 U2[t] V2[t]-2 U4[t] V4[t])^2-(m (U2p[0] U4[t]-U2[t] U4p[0]) (V2p[0] V4[t]-V2[t] V4p[0]))/(2 U2[t] V2[t]-2 U4[t] V4[t])^2))+(8 m veq2i^2 (U2p[0] U4[t]-U2[t] U4p[0]) (U2p[t] V2[t]-U4p[t] V4[t]) (U2p[t] U4[t] V1[t]-U2[t] U4p[t] V1[t]-U2p[t] U3[t] V2[t]+U2[t] U3p[t] V2[t]-U3p[t] U4[t] V4[t]+U3[t] U4p[t] V4[t]) (U2[t] (-V1p[0] V2[t]+V1[t] V2p[0])+U4[t] (V1p[0] V4[t]-V1[t] V4p[0])+U3[t] (-V2p[0] V4[t]+V2[t] V4p[0])))/((2 U2[t] V2[t]-2 U4[t] V4[t])^4 (((-U2p[0] V2[t]+U4p[0] V4[t]) (-m U2[t] V2p[0]+m U4[t] V4p[0]))/(2 U2[t] V2[t]-2 U4[t] V4[t])^2-(m (U2p[0] U4[t]-U2[t] U4p[0]) (V2p[0] V4[t]-V2[t] V4p[0]))/(2 U2[t] V2[t]-2 U4[t] V4[t])^2))+(8 m veq1i veq2i (-U2p[t] U4[t]+U2[t] U4p[t]) (U2p[t] V2[t]-U4p[t] V4[t]) (U2[t] (-U1p[0] V2[t]+U4p[0] V3[t])+U4[t] (-U2p[0] V3[t]+U1p[0] V4[t])+U1[t] (U2p[0] V2[t]-U4p[0] V4[t])) (U2[t] (-V1p[0] V2[t]+V1[t] V2p[0])+U4[t] (V1p[0] V4[t]-V1[t] V4p[0])+U3[t] (-V2p[0] V4[t]+V2[t] V4p[0])))/((2 U2[t] V2[t]-2 U4[t] V4[t])^4 (((-U2p[0] V2[t]+U4p[0] V4[t]) (-m U2[t] V2p[0]+m U4[t] V4p[0]))/(2 U2[t] V2[t]-2 U4[t] V4[t])^2-(m (U2p[0] U4[t]-U2[t] U4p[0]) (V2p[0] V4[t]-V2[t] V4p[0]))/(2 U2[t] V2[t]-2 U4[t] V4[t])^2))+(8 m sq2q2i (-U2p[t] U4[t]+U2[t] U4p[t]) (U2p[t] V2[t]-U4p[t] V4[t]) (U2[t] (U4p[0] V1[t]-U3p[0] V2[t])+U4[t] (-U2p[0] V1[t]+U3p[0] V4[t])+U3[t] (U2p[0] V2[t]-U4p[0] V4[t])) (U2[t] (-V1p[0] V2[t]+V1[t] V2p[0])+U4[t] (V1p[0] V4[t]-V1[t] V4p[0])+U3[t] (-V2p[0] V4[t]+V2[t] V4p[0])))/((2 U2[t] V2[t]-2 U4[t] V4[t])^4 (((-U2p[0] V2[t]+U4p[0] V4[t]) (-m U2[t] V2p[0]+m U4[t] V4p[0]))/(2 U2[t] V2[t]-2 U4[t] V4[t])^2-(m (U2p[0] U4[t]-U2[t] U4p[0]) (V2p[0] V4[t]-V2[t] V4p[0]))/(2 U2[t] V2[t]-2 U4[t] V4[t])^2))+(8 m veq2i^2 (-U2p[t] U4[t]+U2[t] U4p[t]) (U2p[t] V2[t]-U4p[t] V4[t]) (U2[t] (U4p[0] V1[t]-U3p[0] V2[t])+U4[t] (-U2p[0] V1[t]+U3p[0] V4[t])+U3[t] (U2p[0] V2[t]-U4p[0] V4[t])) (U2[t] (-V1p[0] V2[t]+V1[t] V2p[0])+U4[t] (V1p[0] V4[t]-V1[t] V4p[0])+U3[t] (-V2p[0] V4[t]+V2[t] V4p[0])))/((2 U2[t] V2[t]-2 U4[t] V4[t])^4 (((-U2p[0] V2[t]+U4p[0] V4[t]) (-m U2[t] V2p[0]+m U4[t] V4p[0]))/(2 U2[t] V2[t]-2 U4[t] V4[t])^2-(m (U2p[0] U4[t]-U2[t] U4p[0]) (V2p[0] V4[t]-V2[t] V4p[0]))/(2 U2[t] V2[t]-2 U4[t] V4[t])^2))-(8 m sq2p2i (U2p[0] U4[t]-U2[t] U4p[0]) (-U2p[t] U4[t]+U2[t] U4p[t]) (U2p[t] V2[t]-U4p[t] V4[t]) (U2[t] (-V1p[0] V2[t]+V1[t] V2p[0])+U4[t] (V1p[0] V4[t]-V1[t] V4p[0])+U3[t] (-V2p[0] V4[t]+V2[t] V4p[0])))/((2 U2[t] V2[t]-2 U4[t] V4[t])^3 (-m U2[t] V2p[0]+m U4[t] V4p[0]) (((-U2p[0] V2[t]+U4p[0] V4[t]) (-m U2[t] V2p[0]+m U4[t] V4p[0]))/(2 U2[t] V2[t]-2 U4[t] V4[t])^2-(m (U2p[0] U4[t]-U2[t] U4p[0]) (V2p[0] V4[t]-V2[t] V4p[0]))/(2 U2[t] V2[t]-2 U4[t] V4[t])^2))-(8 m vep2i veq2i (U2p[0] U4[t]-U2[t] U4p[0]) (-U2p[t] U4[t]+U2[t] U4p[t]) (U2p[t] V2[t]-U4p[t] V4[t]) (U2[t] (-V1p[0] V2[t]+V1[t] V2p[0])+U4[t] (V1p[0] V4[t]-V1[t] V4p[0])+U3[t] (-V2p[0] V4[t]+V2[t] V4p[0])))/((2 U2[t] V2[t]-2 U4[t] V4[t])^3 (-m U2[t] V2p[0]+m U4[t] V4p[0]) (((-U2p[0] V2[t]+U4p[0] V4[t]) (-m U2[t] V2p[0]+m U4[t] V4p[0]))/(2 U2[t] V2[t]-2 U4[t] V4[t])^2-(m (U2p[0] U4[t]-U2[t] U4p[0]) (V2p[0] V4[t]-V2[t] V4p[0]))/(2 U2[t] V2[t]-2 U4[t] V4[t])^2))+(8 m^2 sq2p2i (U2p[0] U4[t]-U2[t] U4p[0]) (-U2p[t] U4[t]+U2[t] U4p[t])^2 (V2p[0] V4[t]-V2[t] V4p[0]) (U2[t] (-V1p[0] V2[t]+V1[t] V2p[0])+U4[t] (V1p[0] V4[t]-V1[t] V4p[0])+U3[t] (-V2p[0] V4[t]+V2[t] V4p[0])))/((2 U2[t] V2[t]-2 U4[t] V4[t])^3 (-m U2[t] V2p[0]+m U4[t] V4p[0])^2 (((-U2p[0] V2[t]+U4p[0] V4[t]) (-m U2[t] V2p[0]+m U4[t] V4p[0]))/(2 U2[t] V2[t]-2 U4[t] V4[t])^2-(m (U2p[0] U4[t]-U2[t] U4p[0]) (V2p[0] V4[t]-V2[t] V4p[0]))/(2 U2[t] V2[t]-2 U4[t] V4[t])^2))+(8 m^2 vep2i veq2i (U2p[0] U4[t]-U2[t] U4p[0]) (-U2p[t] U4[t]+U2[t] U4p[t])^2 (V2p[0] V4[t]-V2[t] V4p[0]) (U2[t] (-V1p[0] V2[t]+V1[t] V2p[0])+U4[t] (V1p[0] V4[t]-V1[t] V4p[0])+U3[t] (-V2p[0] V4[t]+V2[t] V4p[0])))/((2 U2[t] V2[t]-2 U4[t] V4[t])^3 (-m U2[t] V2p[0]+m U4[t] V4p[0])^2 (((-U2p[0] V2[t]+U4p[0] V4[t]) (-m U2[t] V2p[0]+m U4[t] V4p[0]))/(2 U2[t] V2[t]-2 U4[t] V4[t])^2-(m (U2p[0] U4[t]-U2[t] U4p[0]) (V2p[0] V4[t]-V2[t] V4p[0]))/(2 U2[t] V2[t]-2 U4[t] V4[t])^2))-(4 m^2 vep1i veq2i (-U2p[t] U4[t]+U2[t] U4p[t])^2 (V2p[0] V4[t]-V2[t] V4p[0]) (U2[t] (-V1p[0] V2[t]+V1[t] V2p[0])+U4[t] (V1p[0] V4[t]-V1[t] V4p[0])+U3[t] (-V2p[0] V4[t]+V2[t] V4p[0])))/((2 U2[t] V2[t]-2 U4[t] V4[t])^3 (-m U2[t] V2p[0]+m U4[t] V4p[0]) (((-U2p[0] V2[t]+U4p[0] V4[t]) (-m U2[t] V2p[0]+m U4[t] V4p[0]))/(2 U2[t] V2[t]-2 U4[t] V4[t])^2-(m (U2p[0] U4[t]-U2[t] U4p[0]) (V2p[0] V4[t]-V2[t] V4p[0]))/(2 U2[t] V2[t]-2 U4[t] V4[t])^2))-(8 m^2 veq1i veq2i (U2p[0] U4[t]-U2[t] U4p[0]) (-U2p[t] U4[t]+U2[t] U4p[t]) (U1p[t] U2[t] V2[t]-U1[t] U2p[t] V2[t]+U2p[t] U4[t] V3[t]-U2[t] U4p[t] V3[t]-U1p[t] U4[t] V4[t]+U1[t] U4p[t] V4[t]) (V2p[0] V4[t]-V2[t] V4p[0]) (U2[t] (-V1p[0] V2[t]+V1[t] V2p[0])+U4[t] (V1p[0] V4[t]-V1[t] V4p[0])+U3[t] (-V2p[0] V4[t]+V2[t] V4p[0])))/((2 U2[t] V2[t]-2 U4[t] V4[t])^4 (-m U2[t] V2p[0]+m U4[t] V4p[0]) (((-U2p[0] V2[t]+U4p[0] V4[t]) (-m U2[t] V2p[0]+m U4[t] V4p[0]))/(2 U2[t] V2[t]-2 U4[t] V4[t])^2-(m (U2p[0] U4[t]-U2[t] U4p[0]) (V2p[0] V4[t]-V2[t] V4p[0]))/(2 U2[t] V2[t]-2 U4[t] V4[t])^2))-(8 m^2 sq2q2i (U2p[0] U4[t]-U2[t] U4p[0]) (-U2p[t] U4[t]+U2[t] U4p[t]) (U2p[t] U4[t] V1[t]-U2[t] U4p[t] V1[t]-U2p[t] U3[t] V2[t]+U2[t] U3p[t] V2[t]-U3p[t] U4[t] V4[t]+U3[t] U4p[t] V4[t]) (V2p[0] V4[t]-V2[t] V4p[0]) (U2[t] (-V1p[0] V2[t]+V1[t] V2p[0])+U4[t] (V1p[0] V4[t]-V1[t] V4p[0])+U3[t] (-V2p[0] V4[t]+V2[t] V4p[0])))/((2 U2[t] V2[t]-2 U4[t] V4[t])^4 (-m U2[t] V2p[0]+m U4[t] V4p[0]) (((-U2p[0] V2[t]+U4p[0] V4[t]) (-m U2[t] V2p[0]+m U4[t] V4p[0]))/(2 U2[t] V2[t]-2 U4[t] V4[t])^2-(m (U2p[0] U4[t]-U2[t] U4p[0]) (V2p[0] V4[t]-V2[t] V4p[0]))/(2 U2[t] V2[t]-2 U4[t] V4[t])^2))-(8 m^2 veq2i^2 (U2p[0] U4[t]-U2[t] U4p[0]) (-U2p[t] U4[t]+U2[t] U4p[t]) (U2p[t] U4[t] V1[t]-U2[t] U4p[t] V1[t]-U2p[t] U3[t] V2[t]+U2[t] U3p[t] V2[t]-U3p[t] U4[t] V4[t]+U3[t] U4p[t] V4[t]) (V2p[0] V4[t]-V2[t] V4p[0]) (U2[t] (-V1p[0] V2[t]+V1[t] V2p[0])+U4[t] (V1p[0] V4[t]-V1[t] V4p[0])+U3[t] (-V2p[0] V4[t]+V2[t] V4p[0])))/((2 U2[t] V2[t]-2 U4[t] V4[t])^4 (-m U2[t] V2p[0]+m U4[t] V4p[0]) (((-U2p[0] V2[t]+U4p[0] V4[t]) (-m U2[t] V2p[0]+m U4[t] V4p[0]))/(2 U2[t] V2[t]-2 U4[t] V4[t])^2-(m (U2p[0] U4[t]-U2[t] U4p[0]) (V2p[0] V4[t]-V2[t] V4p[0]))/(2 U2[t] V2[t]-2 U4[t] V4[t])^2))-(8 m^2 veq1i veq2i (-U2p[t] U4[t]+U2[t] U4p[t])^2 (U2[t] (-U1p[0] V2[t]+U4p[0] V3[t])+U4[t] (-U2p[0] V3[t]+U1p[0] V4[t])+U1[t] (U2p[0] V2[t]-U4p[0] V4[t])) (V2p[0] V4[t]-V2[t] V4p[0]) (U2[t] (-V1p[0] V2[t]+V1[t] V2p[0])+U4[t] (V1p[0] V4[t]-V1[t] V4p[0])+U3[t] (-V2p[0] V4[t]+V2[t] V4p[0])))/((2 U2[t] V2[t]-2 U4[t] V4[t])^4 (-m U2[t] V2p[0]+m U4[t] V4p[0]) (((-U2p[0] V2[t]+U4p[0] V4[t]) (-m U2[t] V2p[0]+m U4[t] V4p[0]))/(2 U2[t] V2[t]-2 U4[t] V4[t])^2-(m (U2p[0] U4[t]-U2[t] U4p[0]) (V2p[0] V4[t]-V2[t] V4p[0]))/(2 U2[t] V2[t]-2 U4[t] V4[t])^2))-(8 m^2 sq2q2i (-U2p[t] U4[t]+U2[t] U4p[t])^2 (U2[t] (U4p[0] V1[t]-U3p[0] V2[t])+U4[t] (-U2p[0] V1[t]+U3p[0] V4[t])+U3[t] (U2p[0] V2[t]-U4p[0] V4[t])) (V2p[0] V4[t]-V2[t] V4p[0]) (U2[t] (-V1p[0] V2[t]+V1[t] V2p[0])+U4[t] (V1p[0] V4[t]-V1[t] V4p[0])+U3[t] (-V2p[0] V4[t]+V2[t] V4p[0])))/((2 U2[t] V2[t]-2 U4[t] V4[t])^4 (-m U2[t] V2p[0]+m U4[t] V4p[0]) (((-U2p[0] V2[t]+U4p[0] V4[t]) (-m U2[t] V2p[0]+m U4[t] V4p[0]))/(2 U2[t] V2[t]-2 U4[t] V4[t])^2-(m (U2p[0] U4[t]-U2[t] U4p[0]) (V2p[0] V4[t]-V2[t] V4p[0]))/(2 U2[t] V2[t]-2 U4[t] V4[t])^2))-(8 m^2 veq2i^2 (-U2p[t] U4[t]+U2[t] U4p[t])^2 (U2[t] (U4p[0] V1[t]-U3p[0] V2[t])+U4[t] (-U2p[0] V1[t]+U3p[0] V4[t])+U3[t] (U2p[0] V2[t]-U4p[0] V4[t])) (V2p[0] V4[t]-V2[t] V4p[0]) (U2[t] (-V1p[0] V2[t]+V1[t] V2p[0])+U4[t] (V1p[0] V4[t]-V1[t] V4p[0])+U3[t] (-V2p[0] V4[t]+V2[t] V4p[0])))/((2 U2[t] V2[t]-2 U4[t] V4[t])^4 (-m U2[t] V2p[0]+m U4[t] V4p[0]) (((-U2p[0] V2[t]+U4p[0] V4[t]) (-m U2[t] V2p[0]+m U4[t] V4p[0]))/(2 U2[t] V2[t]-2 U4[t] V4[t])^2-(m (U2p[0] U4[t]-U2[t] U4p[0]) (V2p[0] V4[t]-V2[t] V4p[0]))/(2 U2[t] V2[t]-2 U4[t] V4[t])^2))+(4 m^2 sq2q2i (-U2p[t] U4[t]+U2[t] U4p[t])^2 (U2[t] (-V1p[0] V2[t]+V1[t] V2p[0])+U4[t] (V1p[0] V4[t]-V1[t] V4p[0])+U3[t] (-V2p[0] V4[t]+V2[t] V4p[0]))^2)/((2 U2[t] V2[t]-2 U4[t] V4[t])^2 (-m U2[t] V2p[0]+m U4[t] V4p[0])^2)+(4 m^2 veq2i^2 (-U2p[t] U4[t]+U2[t] U4p[t])^2 (U2[t] (-V1p[0] V2[t]+V1[t] V2p[0])+U4[t] (V1p[0] V4[t]-V1[t] V4p[0])+U3[t] (-V2p[0] V4[t]+V2[t] V4p[0]))^2)/((2 U2[t] V2[t]-2 U4[t] V4[t])^2 (-m U2[t] V2p[0]+m U4[t] V4p[0])^2)-(8 m^2 sq2q2i (U2p[0] U4[t]-U2[t] U4p[0]) (-U2p[t] U4[t]+U2[t] U4p[t]) (U2p[t] V2[t]-U4p[t] V4[t]) (U2[t] (-V1p[0] V2[t]+V1[t] V2p[0])+U4[t] (V1p[0] V4[t]-V1[t] V4p[0])+U3[t] (-V2p[0] V4[t]+V2[t] V4p[0]))^2)/((2 U2[t] V2[t]-2 U4[t] V4[t])^4 (-m U2[t] V2p[0]+m U4[t] V4p[0]) (((-U2p[0] V2[t]+U4p[0] V4[t]) (-m U2[t] V2p[0]+m U4[t] V4p[0]))/(2 U2[t] V2[t]-2 U4[t] V4[t])^2-(m (U2p[0] U4[t]-U2[t] U4p[0]) (V2p[0] V4[t]-V2[t] V4p[0]))/(2 U2[t] V2[t]-2 U4[t] V4[t])^2))-(8 m^2 veq2i^2 (U2p[0] U4[t]-U2[t] U4p[0]) (-U2p[t] U4[t]+U2[t] U4p[t]) (U2p[t] V2[t]-U4p[t] V4[t]) (U2[t] (-V1p[0] V2[t]+V1[t] V2p[0])+U4[t] (V1p[0] V4[t]-V1[t] V4p[0])+U3[t] (-V2p[0] V4[t]+V2[t] V4p[0]))^2)/((2 U2[t] V2[t]-2 U4[t] V4[t])^4 (-m U2[t] V2p[0]+m U4[t] V4p[0]) (((-U2p[0] V2[t]+U4p[0] V4[t]) (-m U2[t] V2p[0]+m U4[t] V4p[0]))/(2 U2[t] V2[t]-2 U4[t] V4[t])^2-(m (U2p[0] U4[t]-U2[t] U4p[0]) (V2p[0] V4[t]-V2[t] V4p[0]))/(2 U2[t] V2[t]-2 U4[t] V4[t])^2))+(8 m^3 sq2q2i (U2p[0] U4[t]-U2[t] U4p[0]) (-U2p[t] U4[t]+U2[t] U4p[t])^2 (V2p[0] V4[t]-V2[t] V4p[0]) (U2[t] (-V1p[0] V2[t]+V1[t] V2p[0])+U4[t] (V1p[0] V4[t]-V1[t] V4p[0])+U3[t] (-V2p[0] V4[t]+V2[t] V4p[0]))^2)/((2 U2[t] V2[t]-2 U4[t] V4[t])^4 (-m U2[t] V2p[0]+m U4[t] V4p[0])^2 (((-U2p[0] V2[t]+U4p[0] V4[t]) (-m U2[t] V2p[0]+m U4[t] V4p[0]))/(2 U2[t] V2[t]-2 U4[t] V4[t])^2-(m (U2p[0] U4[t]-U2[t] U4p[0]) (V2p[0] V4[t]-V2[t] V4p[0]))/(2 U2[t] V2[t]-2 U4[t] V4[t])^2))+(8 m^3 veq2i^2 (U2p[0] U4[t]-U2[t] U4p[0]) (-U2p[t] U4[t]+U2[t] U4p[t])^2 (V2p[0] V4[t]-V2[t] V4p[0]) (U2[t] (-V1p[0] V2[t]+V1[t] V2p[0])+U4[t] (V1p[0] V4[t]-V1[t] V4p[0])+U3[t] (-V2p[0] V4[t]+V2[t] V4p[0]))^2)/((2 U2[t] V2[t]-2 U4[t] V4[t])^4 (-m U2[t] V2p[0]+m U4[t] V4p[0])^2 (((-U2p[0] V2[t]+U4p[0] V4[t]) (-m U2[t] V2p[0]+m U4[t] V4p[0]))/(2 U2[t] V2[t]-2 U4[t] V4[t])^2-(m (U2p[0] U4[t]-U2[t] U4p[0]) (V2p[0] V4[t]-V2[t] V4p[0]))/(2 U2[t] V2[t]-2 U4[t] V4[t])^2))+(4 m vep2i veq1i (-U2p[t] U4[t]+U2[t] U4p[t])^2 (U2[t] (V2p[0] V3[t]-V2[t] V3p[0])+U1[t] (-V2p[0] V4[t]+V2[t] V4p[0])+U4[t] (V3p[0] V4[t]-V3[t] V4p[0])))/((2 U2[t] V2[t]-2 U4[t] V4[t]) (-m U2[t] V2p[0]+m U4[t] V4p[0])^2)-(8 m sq1q1i (-U2p[t] U4[t]+U2[t] U4p[t]) (U1p[t] U2[t] V2[t]-U1[t] U2p[t] V2[t]+U2p[t] U4[t] V3[t]-U2[t] U4p[t] V3[t]-U1p[t] U4[t] V4[t]+U1[t] U4p[t] V4[t]) (U2[t] (V2p[0] V3[t]-V2[t] V3p[0])+U1[t] (-V2p[0] V4[t]+V2[t] V4p[0])+U4[t] (V3p[0] V4[t]-V3[t] V4p[0])))/((2 U2[t] V2[t]-2 U4[t] V4[t])^2 (-m U2[t] V2p[0]+m U4[t] V4p[0]))-(8 m veq1i^2 (-U2p[t] U4[t]+U2[t] U4p[t]) (U1p[t] U2[t] V2[t]-U1[t] U2p[t] V2[t]+U2p[t] U4[t] V3[t]-U2[t] U4p[t] V3[t]-U1p[t] U4[t] V4[t]+U1[t] U4p[t] V4[t]) (U2[t] (V2p[0] V3[t]-V2[t] V3p[0])+U1[t] (-V2p[0] V4[t]+V2[t] V4p[0])+U4[t] (V3p[0] V4[t]-V3[t] V4p[0])))/((2 U2[t] V2[t]-2 U4[t] V4[t])^2 (-m U2[t] V2p[0]+m U4[t] V4p[0]))-(8 m veq1i veq2i (-U2p[t] U4[t]+U2[t] U4p[t]) (U2p[t] U4[t] V1[t]-U2[t] U4p[t] V1[t]-U2p[t] U3[t] V2[t]+U2[t] U3p[t] V2[t]-U3p[t] U4[t] V4[t]+U3[t] U4p[t] V4[t]) (U2[t] (V2p[0] V3[t]-V2[t] V3p[0])+U1[t] (-V2p[0] V4[t]+V2[t] V4p[0])+U4[t] (V3p[0] V4[t]-V3[t] V4p[0])))/((2 U2[t] V2[t]-2 U4[t] V4[t])^2 (-m U2[t] V2p[0]+m U4[t] V4p[0]))+(4 m sq1p1i (-U2p[t] U4[t]+U2[t] U4p[t]) (U2p[t] V2[t]-U4p[t] V4[t]) (U2[t] (V2p[0] V3[t]-V2[t] V3p[0])+U1[t] (-V2p[0] V4[t]+V2[t] V4p[0])+U4[t] (V3p[0] V4[t]-V3[t] V4p[0])))/((2 U2[t] V2[t]-2 U4[t] V4[t])^3 (((-U2p[0] V2[t]+U4p[0] V4[t]) (-m U2[t] V2p[0]+m U4[t] V4p[0]))/(2 U2[t] V2[t]-2 U4[t] V4[t])^2-(m (U2p[0] U4[t]-U2[t] U4p[0]) (V2p[0] V4[t]-V2[t] V4p[0]))/(2 U2[t] V2[t]-2 U4[t] V4[t])^2))+(4 m vep1i veq1i (-U2p[t] U4[t]+U2[t] U4p[t]) (U2p[t] V2[t]-U4p[t] V4[t]) (U2[t] (V2p[0] V3[t]-V2[t] V3p[0])+U1[t] (-V2p[0] V4[t]+V2[t] V4p[0])+U4[t] (V3p[0] V4[t]-V3[t] V4p[0])))/((2 U2[t] V2[t]-2 U4[t] V4[t])^3 (((-U2p[0] V2[t]+U4p[0] V4[t]) (-m U2[t] V2p[0]+m U4[t] V4p[0]))/(2 U2[t] V2[t]-2 U4[t] V4[t])^2-(m (U2p[0] U4[t]-U2[t] U4p[0]) (V2p[0] V4[t]-V2[t] V4p[0]))/(2 U2[t] V2[t]-2 U4[t] V4[t])^2))+(8 m sq1q1i (U2p[0] U4[t]-U2[t] U4p[0]) (U2p[t] V2[t]-U4p[t] V4[t]) (U1p[t] U2[t] V2[t]-U1[t] U2p[t] V2[t]+U2p[t] U4[t] V3[t]-U2[t] U4p[t] V3[t]-U1p[t] U4[t] V4[t]+U1[t] U4p[t] V4[t]) (U2[t] (V2p[0] V3[t]-V2[t] V3p[0])+U1[t] (-V2p[0] V4[t]+V2[t] V4p[0])+U4[t] (V3p[0] V4[t]-V3[t] V4p[0])))/((2 U2[t] V2[t]-2 U4[t] V4[t])^4 (((-U2p[0] V2[t]+U4p[0] V4[t]) (-m U2[t] V2p[0]+m U4[t] V4p[0]))/(2 U2[t] V2[t]-2 U4[t] V4[t])^2-(m (U2p[0] U4[t]-U2[t] U4p[0]) (V2p[0] V4[t]-V2[t] V4p[0]))/(2 U2[t] V2[t]-2 U4[t] V4[t])^2))+(8 m veq1i^2 (U2p[0] U4[t]-U2[t] U4p[0]) (U2p[t] V2[t]-U4p[t] V4[t]) (U1p[t] U2[t] V2[t]-U1[t] U2p[t] V2[t]+U2p[t] U4[t] V3[t]-U2[t] U4p[t] V3[t]-U1p[t] U4[t] V4[t]+U1[t] U4p[t] V4[t]) (U2[t] (V2p[0] V3[t]-V2[t] V3p[0])+U1[t] (-V2p[0] V4[t]+V2[t] V4p[0])+U4[t] (V3p[0] V4[t]-V3[t] V4p[0])))/((2 U2[t] V2[t]-2 U4[t] V4[t])^4 (((-U2p[0] V2[t]+U4p[0] V4[t]) (-m U2[t] V2p[0]+m U4[t] V4p[0]))/(2 U2[t] V2[t]-2 U4[t] V4[t])^2-(m (U2p[0] U4[t]-U2[t] U4p[0]) (V2p[0] V4[t]-V2[t] V4p[0]))/(2 U2[t] V2[t]-2 U4[t] V4[t])^2))+(8 m veq1i veq2i (U2p[0] U4[t]-U2[t] U4p[0]) (U2p[t] V2[t]-U4p[t] V4[t]) (U2p[t] U4[t] V1[t]-U2[t] U4p[t] V1[t]-U2p[t] U3[t] V2[t]+U2[t] U3p[t] V2[t]-U3p[t] U4[t] V4[t]+U3[t] U4p[t] V4[t]) (U2[t] (V2p[0] V3[t]-V2[t] V3p[0])+U1[t] (-V2p[0] V4[t]+V2[t] V4p[0])+U4[t] (V3p[0] V4[t]-V3[t] V4p[0])))/((2 U2[t] V2[t]-2 U4[t] V4[t])^4 (((-U2p[0] V2[t]+U4p[0] V4[t]) (-m U2[t] V2p[0]+m U4[t] V4p[0]))/(2 U2[t] V2[t]-2 U4[t] V4[t])^2-(m (U2p[0] U4[t]-U2[t] U4p[0]) (V2p[0] V4[t]-V2[t] V4p[0]))/(2 U2[t] V2[t]-2 U4[t] V4[t])^2))+(8 m sq1q1i (-U2p[t] U4[t]+U2[t] U4p[t]) (U2p[t] V2[t]-U4p[t] V4[t]) (U2[t] (-U1p[0] V2[t]+U4p[0] V3[t])+U4[t] (-U2p[0] V3[t]+U1p[0] V4[t])+U1[t] (U2p[0] V2[t]-U4p[0] V4[t])) (U2[t] (V2p[0] V3[t]-V2[t] V3p[0])+U1[t] (-V2p[0] V4[t]+V2[t] V4p[0])+U4[t] (V3p[0] V4[t]-V3[t] V4p[0])))/((2 U2[t] V2[t]-2 U4[t] V4[t])^4 (((-U2p[0] V2[t]+U4p[0] V4[t]) (-m U2[t] V2p[0]+m U4[t] V4p[0]))/(2 U2[t] V2[t]-2 U4[t] V4[t])^2-(m (U2p[0] U4[t]-U2[t] U4p[0]) (V2p[0] V4[t]-V2[t] V4p[0]))/(2 U2[t] V2[t]-2 U4[t] V4[t])^2))+(8 m veq1i^2 (-U2p[t] U4[t]+U2[t] U4p[t]) (U2p[t] V2[t]-U4p[t] V4[t]) (U2[t] (-U1p[0] V2[t]+U4p[0] V3[t])+U4[t] (-U2p[0] V3[t]+U1p[0] V4[t])+U1[t] (U2p[0] V2[t]-U4p[0] V4[t])) (U2[t] (V2p[0] V3[t]-V2[t] V3p[0])+U1[t] (-V2p[0] V4[t]+V2[t] V4p[0])+U4[t] (V3p[0] V4[t]-V3[t] V4p[0])))/((2 U2[t] V2[t]-2 U4[t] V4[t])^4 (((-U2p[0] V2[t]+U4p[0] V4[t]) (-m U2[t] V2p[0]+m U4[t] V4p[0]))/(2 U2[t] V2[t]-2 U4[t] V4[t])^2-(m (U2p[0] U4[t]-U2[t] U4p[0]) (V2p[0] V4[t]-V2[t] V4p[0]))/(2 U2[t] V2[t]-2 U4[t] V4[t])^2))+(8 m veq1i veq2i (-U2p[t] U4[t]+U2[t] U4p[t]) (U2p[t] V2[t]-U4p[t] V4[t]) (U2[t] (U4p[0] V1[t]-U3p[0] V2[t])+U4[t] (-U2p[0] V1[t]+U3p[0] V4[t])+U3[t] (U2p[0] V2[t]-U4p[0] V4[t])) (U2[t] (V2p[0] V3[t]-V2[t] V3p[0])+U1[t] (-V2p[0] V4[t]+V2[t] V4p[0])+U4[t] (V3p[0] V4[t]-V3[t] V4p[0])))/((2 U2[t] V2[t]-2 U4[t] V4[t])^4 (((-U2p[0] V2[t]+U4p[0] V4[t]) (-m U2[t] V2p[0]+m U4[t] V4p[0]))/(2 U2[t] V2[t]-2 U4[t] V4[t])^2-(m (U2p[0] U4[t]-U2[t] U4p[0]) (V2p[0] V4[t]-V2[t] V4p[0]))/(2 U2[t] V2[t]-2 U4[t] V4[t])^2))-(8 m vep2i veq1i (U2p[0] U4[t]-U2[t] U4p[0]) (-U2p[t] U4[t]+U2[t] U4p[t]) (U2p[t] V2[t]-U4p[t] V4[t]) (U2[t] (V2p[0] V3[t]-V2[t] V3p[0])+U1[t] (-V2p[0] V4[t]+V2[t] V4p[0])+U4[t] (V3p[0] V4[t]-V3[t] V4p[0])))/((2 U2[t] V2[t]-2 U4[t] V4[t])^3 (-m U2[t] V2p[0]+m U4[t] V4p[0]) (((-U2p[0] V2[t]+U4p[0] V4[t]) (-m U2[t] V2p[0]+m U4[t] V4p[0]))/(2 U2[t] V2[t]-2 U4[t] V4[t])^2-(m (U2p[0] U4[t]-U2[t] U4p[0]) (V2p[0] V4[t]-V2[t] V4p[0]))/(2 U2[t] V2[t]-2 U4[t] V4[t])^2))+(8 m^2 vep2i veq1i (U2p[0] U4[t]-U2[t] U4p[0]) (-U2p[t] U4[t]+U2[t] U4p[t])^2 (V2p[0] V4[t]-V2[t] V4p[0]) (U2[t] (V2p[0] V3[t]-V2[t] V3p[0])+U1[t] (-V2p[0] V4[t]+V2[t] V4p[0])+U4[t] (V3p[0] V4[t]-V3[t] V4p[0])))/((2 U2[t] V2[t]-2 U4[t] V4[t])^3 (-m U2[t] V2p[0]+m U4[t] V4p[0])^2 (((-U2p[0] V2[t]+U4p[0] V4[t]) (-m U2[t] V2p[0]+m U4[t] V4p[0]))/(2 U2[t] V2[t]-2 U4[t] V4[t])^2-(m (U2p[0] U4[t]-U2[t] U4p[0]) (V2p[0] V4[t]-V2[t] V4p[0]))/(2 U2[t] V2[t]-2 U4[t] V4[t])^2))-(4 m^2 sq1p1i (-U2p[t] U4[t]+U2[t] U4p[t])^2 (V2p[0] V4[t]-V2[t] V4p[0]) (U2[t] (V2p[0] V3[t]-V2[t] V3p[0])+U1[t] (-V2p[0] V4[t]+V2[t] V4p[0])+U4[t] (V3p[0] V4[t]-V3[t] V4p[0])))/((2 U2[t] V2[t]-2 U4[t] V4[t])^3 (-m U2[t] V2p[0]+m U4[t] V4p[0]) (((-U2p[0] V2[t]+U4p[0] V4[t]) (-m U2[t] V2p[0]+m U4[t] V4p[0]))/(2 U2[t] V2[t]-2 U4[t] V4[t])^2-(m (U2p[0] U4[t]-U2[t] U4p[0]) (V2p[0] V4[t]-V2[t] V4p[0]))/(2 U2[t] V2[t]-2 U4[t] V4[t])^2))-(4 m^2 vep1i veq1i (-U2p[t] U4[t]+U2[t] U4p[t])^2 (V2p[0] V4[t]-V2[t] V4p[0]) (U2[t] (V2p[0] V3[t]-V2[t] V3p[0])+U1[t] (-V2p[0] V4[t]+V2[t] V4p[0])+U4[t] (V3p[0] V4[t]-V3[t] V4p[0])))/((2 U2[t] V2[t]-2 U4[t] V4[t])^3 (-m U2[t] V2p[0]+m U4[t] V4p[0]) (((-U2p[0] V2[t]+U4p[0] V4[t]) (-m U2[t] V2p[0]+m U4[t] V4p[0]))/(2 U2[t] V2[t]-2 U4[t] V4[t])^2-(m (U2p[0] U4[t]-U2[t] U4p[0]) (V2p[0] V4[t]-V2[t] V4p[0]))/(2 U2[t] V2[t]-2 U4[t] V4[t])^2))-(8 m^2 sq1q1i (U2p[0] U4[t]-U2[t] U4p[0]) (-U2p[t] U4[t]+U2[t] U4p[t]) (U1p[t] U2[t] V2[t]-U1[t] U2p[t] V2[t]+U2p[t] U4[t] V3[t]-U2[t] U4p[t] V3[t]-U1p[t] U4[t] V4[t]+U1[t] U4p[t] V4[t]) (V2p[0] V4[t]-V2[t] V4p[0]) (U2[t] (V2p[0] V3[t]-V2[t] V3p[0])+U1[t] (-V2p[0] V4[t]+V2[t] V4p[0])+U4[t] (V3p[0] V4[t]-V3[t] V4p[0])))/((2 U2[t] V2[t]-2 U4[t] V4[t])^4 (-m U2[t] V2p[0]+m U4[t] V4p[0]) (((-U2p[0] V2[t]+U4p[0] V4[t]) (-m U2[t] V2p[0]+m U4[t] V4p[0]))/(2 U2[t] V2[t]-2 U4[t] V4[t])^2-(m (U2p[0] U4[t]-U2[t] U4p[0]) (V2p[0] V4[t]-V2[t] V4p[0]))/(2 U2[t] V2[t]-2 U4[t] V4[t])^2))-(8 m^2 veq1i^2 (U2p[0] U4[t]-U2[t] U4p[0]) (-U2p[t] U4[t]+U2[t] U4p[t]) (U1p[t] U2[t] V2[t]-U1[t] U2p[t] V2[t]+U2p[t] U4[t] V3[t]-U2[t] U4p[t] V3[t]-U1p[t] U4[t] V4[t]+U1[t] U4p[t] V4[t]) (V2p[0] V4[t]-V2[t] V4p[0]) (U2[t] (V2p[0] V3[t]-V2[t] V3p[0])+U1[t] (-V2p[0] V4[t]+V2[t] V4p[0])+U4[t] (V3p[0] V4[t]-V3[t] V4p[0])))/((2 U2[t] V2[t]-2 U4[t] V4[t])^4 (-m U2[t] V2p[0]+m U4[t] V4p[0]) (((-U2p[0] V2[t]+U4p[0] V4[t]) (-m U2[t] V2p[0]+m U4[t] V4p[0]))/(2 U2[t] V2[t]-2 U4[t] V4[t])^2-(m (U2p[0] U4[t]-U2[t] U4p[0]) (V2p[0] V4[t]-V2[t] V4p[0]))/(2 U2[t] V2[t]-2 U4[t] V4[t])^2))-(8 m^2 veq1i veq2i (U2p[0] U4[t]-U2[t] U4p[0]) (-U2p[t] U4[t]+U2[t] U4p[t]) (U2p[t] U4[t] V1[t]-U2[t] U4p[t] V1[t]-U2p[t] U3[t] V2[t]+U2[t] U3p[t] V2[t]-U3p[t] U4[t] V4[t]+U3[t] U4p[t] V4[t]) (V2p[0] V4[t]-V2[t] V4p[0]) (U2[t] (V2p[0] V3[t]-V2[t] V3p[0])+U1[t] (-V2p[0] V4[t]+V2[t] V4p[0])+U4[t] (V3p[0] V4[t]-V3[t] V4p[0])))/((2 U2[t] V2[t]-2 U4[t] V4[t])^4 (-m U2[t] V2p[0]+m U4[t] V4p[0]) (((-U2p[0] V2[t]+U4p[0] V4[t]) (-m U2[t] V2p[0]+m U4[t] V4p[0]))/(2 U2[t] V2[t]-2 U4[t] V4[t])^2-(m (U2p[0] U4[t]-U2[t] U4p[0]) (V2p[0] V4[t]-V2[t] V4p[0]))/(2 U2[t] V2[t]-2 U4[t] V4[t])^2))-(8 m^2 sq1q1i (-U2p[t] U4[t]+U2[t] U4p[t])^2 (U2[t] (-U1p[0] V2[t]+U4p[0] V3[t])+U4[t] (-U2p[0] V3[t]+U1p[0] V4[t])+U1[t] (U2p[0] V2[t]-U4p[0] V4[t])) (V2p[0] V4[t]-V2[t] V4p[0]) (U2[t] (V2p[0] V3[t]-V2[t] V3p[0])+U1[t] (-V2p[0] V4[t]+V2[t] V4p[0])+U4[t] (V3p[0] V4[t]-V3[t] V4p[0])))/((2 U2[t] V2[t]-2 U4[t] V4[t])^4 (-m U2[t] V2p[0]+m U4[t] V4p[0]) (((-U2p[0] V2[t]+U4p[0] V4[t]) (-m U2[t] V2p[0]+m U4[t] V4p[0]))/(2 U2[t] V2[t]-2 U4[t] V4[t])^2-(m (U2p[0] U4[t]-U2[t] U4p[0]) (V2p[0] V4[t]-V2[t] V4p[0]))/(2 U2[t] V2[t]-2 U4[t] V4[t])^2))-(8 m^2 veq1i^2 (-U2p[t] U4[t]+U2[t] U4p[t])^2 (U2[t] (-U1p[0] V2[t]+U4p[0] V3[t])+U4[t] (-U2p[0] V3[t]+U1p[0] V4[t])+U1[t] (U2p[0] V2[t]-U4p[0] V4[t])) (V2p[0] V4[t]-V2[t] V4p[0]) (U2[t] (V2p[0] V3[t]-V2[t] V3p[0])+U1[t] (-V2p[0] V4[t]+V2[t] V4p[0])+U4[t] (V3p[0] V4[t]-V3[t] V4p[0])))/((2 U2[t] V2[t]-2 U4[t] V4[t])^4 (-m U2[t] V2p[0]+m U4[t] V4p[0]) (((-U2p[0] V2[t]+U4p[0] V4[t]) (-m U2[t] V2p[0]+m U4[t] V4p[0]))/(2 U2[t] V2[t]-2 U4[t] V4[t])^2-(m (U2p[0] U4[t]-U2[t] U4p[0]) (V2p[0] V4[t]-V2[t] V4p[0]))/(2 U2[t] V2[t]-2 U4[t] V4[t])^2))-(8 m^2 veq1i veq2i (-U2p[t] U4[t]+U2[t] U4p[t])^2 (U2[t] (U4p[0] V1[t]-U3p[0] V2[t])+U4[t] (-U2p[0] V1[t]+U3p[0] V4[t])+U3[t] (U2p[0] V2[t]-U4p[0] V4[t])) (V2p[0] V4[t]-V2[t] V4p[0]) (U2[t] (V2p[0] V3[t]-V2[t] V3p[0])+U1[t] (-V2p[0] V4[t]+V2[t] V4p[0])+U4[t] (V3p[0] V4[t]-V3[t] V4p[0])))/((2 U2[t] V2[t]-2 U4[t] V4[t])^4 (-m U2[t] V2p[0]+m U4[t] V4p[0]) (((-U2p[0] V2[t]+U4p[0] V4[t]) (-m U2[t] V2p[0]+m U4[t] V4p[0]))/(2 U2[t] V2[t]-2 U4[t] V4[t])^2-(m (U2p[0] U4[t]-U2[t] U4p[0]) (V2p[0] V4[t]-V2[t] V4p[0]))/(2 U2[t] V2[t]-2 U4[t] V4[t])^2))+(8 m^2 veq1i veq2i (-U2p[t] U4[t]+U2[t] U4p[t])^2 (U2[t] (-V1p[0] V2[t]+V1[t] V2p[0])+U4[t] (V1p[0] V4[t]-V1[t] V4p[0])+U3[t] (-V2p[0] V4[t]+V2[t] V4p[0])) (U2[t] (V2p[0] V3[t]-V2[t] V3p[0])+U1[t] (-V2p[0] V4[t]+V2[t] V4p[0])+U4[t] (V3p[0] V4[t]-V3[t] V4p[0])))/((2 U2[t] V2[t]-2 U4[t] V4[t])^2 (-m U2[t] V2p[0]+m U4[t] V4p[0])^2)-(16 m^2 veq1i veq2i (U2p[0] U4[t]-U2[t] U4p[0]) (-U2p[t] U4[t]+U2[t] U4p[t]) (U2p[t] V2[t]-U4p[t] V4[t]) (U2[t] (-V1p[0] V2[t]+V1[t] V2p[0])+U4[t] (V1p[0] V4[t]-V1[t] V4p[0])+U3[t] (-V2p[0] V4[t]+V2[t] V4p[0])) (U2[t] (V2p[0] V3[t]-V2[t] V3p[0])+U1[t] (-V2p[0] V4[t]+V2[t] V4p[0])+U4[t] (V3p[0] V4[t]-V3[t] V4p[0])))/((2 U2[t] V2[t]-2 U4[t] V4[t])^4 (-m U2[t] V2p[0]+m U4[t] V4p[0]) (((-U2p[0] V2[t]+U4p[0] V4[t]) (-m U2[t] V2p[0]+m U4[t] V4p[0]))/(2 U2[t] V2[t]-2 U4[t] V4[t])^2-(m (U2p[0] U4[t]-U2[t] U4p[0]) (V2p[0] V4[t]-V2[t] V4p[0]))/(2 U2[t] V2[t]-2 U4[t] V4[t])^2))+(16 m^3 veq1i veq2i (U2p[0] U4[t]-U2[t] U4p[0]) (-U2p[t] U4[t]+U2[t] U4p[t])^2 (V2p[0] V4[t]-V2[t] V4p[0]) (U2[t] (-V1p[0] V2[t]+V1[t] V2p[0])+U4[t] (V1p[0] V4[t]-V1[t] V4p[0])+U3[t] (-V2p[0] V4[t]+V2[t] V4p[0])) (U2[t] (V2p[0] V3[t]-V2[t] V3p[0])+U1[t] (-V2p[0] V4[t]+V2[t] V4p[0])+U4[t] (V3p[0] V4[t]-V3[t] V4p[0])))/((2 U2[t] V2[t]-2 U4[t] V4[t])^4 (-m U2[t] V2p[0]+m U4[t] V4p[0])^2 (((-U2p[0] V2[t]+U4p[0] V4[t]) (-m U2[t] V2p[0]+m U4[t] V4p[0]))/(2 U2[t] V2[t]-2 U4[t] V4[t])^2-(m (U2p[0] U4[t]-U2[t] U4p[0]) (V2p[0] V4[t]-V2[t] V4p[0]))/(2 U2[t] V2[t]-2 U4[t] V4[t])^2))+(4 m^2 sq1q1i (-U2p[t] U4[t]+U2[t] U4p[t])^2 (U2[t] (V2p[0] V3[t]-V2[t] V3p[0])+U1[t] (-V2p[0] V4[t]+V2[t] V4p[0])+U4[t] (V3p[0] V4[t]-V3[t] V4p[0]))^2)/((2 U2[t] V2[t]-2 U4[t] V4[t])^2 (-m U2[t] V2p[0]+m U4[t] V4p[0])^2)+(4 m^2 veq1i^2 (-U2p[t] U4[t]+U2[t] U4p[t])^2 (U2[t] (V2p[0] V3[t]-V2[t] V3p[0])+U1[t] (-V2p[0] V4[t]+V2[t] V4p[0])+U4[t] (V3p[0] V4[t]-V3[t] V4p[0]))^2)/((2 U2[t] V2[t]-2 U4[t] V4[t])^2 (-m U2[t] V2p[0]+m U4[t] V4p[0])^2)-(8 m^2 sq1q1i (U2p[0] U4[t]-U2[t] U4p[0]) (-U2p[t] U4[t]+U2[t] U4p[t]) (U2p[t] V2[t]-U4p[t] V4[t]) (U2[t] (V2p[0] V3[t]-V2[t] V3p[0])+U1[t] (-V2p[0] V4[t]+V2[t] V4p[0])+U4[t] (V3p[0] V4[t]-V3[t] V4p[0]))^2)/((2 U2[t] V2[t]-2 U4[t] V4[t])^4 (-m U2[t] V2p[0]+m U4[t] V4p[0]) (((-U2p[0] V2[t]+U4p[0] V4[t]) (-m U2[t] V2p[0]+m U4[t] V4p[0]))/(2 U2[t] V2[t]-2 U4[t] V4[t])^2-(m (U2p[0] U4[t]-U2[t] U4p[0]) (V2p[0] V4[t]-V2[t] V4p[0]))/(2 U2[t] V2[t]-2 U4[t] V4[t])^2))-(8 m^2 veq1i^2 (U2p[0] U4[t]-U2[t] U4p[0]) (-U2p[t] U4[t]+U2[t] U4p[t]) (U2p[t] V2[t]-U4p[t] V4[t]) (U2[t] (V2p[0] V3[t]-V2[t] V3p[0])+U1[t] (-V2p[0] V4[t]+V2[t] V4p[0])+U4[t] (V3p[0] V4[t]-V3[t] V4p[0]))^2)/((2 U2[t] V2[t]-2 U4[t] V4[t])^4 (-m U2[t] V2p[0]+m U4[t] V4p[0]) (((-U2p[0] V2[t]+U4p[0] V4[t]) (-m U2[t] V2p[0]+m U4[t] V4p[0]))/(2 U2[t] V2[t]-2 U4[t] V4[t])^2-(m (U2p[0] U4[t]-U2[t] U4p[0]) (V2p[0] V4[t]-V2[t] V4p[0]))/(2 U2[t] V2[t]-2 U4[t] V4[t])^2))+(8 m^3 sq1q1i (U2p[0] U4[t]-U2[t] U4p[0]) (-U2p[t] U4[t]+U2[t] U4p[t])^2 (V2p[0] V4[t]-V2[t] V4p[0]) (U2[t] (V2p[0] V3[t]-V2[t] V3p[0])+U1[t] (-V2p[0] V4[t]+V2[t] V4p[0])+U4[t] (V3p[0] V4[t]-V3[t] V4p[0]))^2)/((2 U2[t] V2[t]-2 U4[t] V4[t])^4 (-m U2[t] V2p[0]+m U4[t] V4p[0])^2 (((-U2p[0] V2[t]+U4p[0] V4[t]) (-m U2[t] V2p[0]+m U4[t] V4p[0]))/(2 U2[t] V2[t]-2 U4[t] V4[t])^2-(m (U2p[0] U4[t]-U2[t] U4p[0]) (V2p[0] V4[t]-V2[t] V4p[0]))/(2 U2[t] V2[t]-2 U4[t] V4[t])^2))+(8 m^3 veq1i^2 (U2p[0] U4[t]-U2[t] U4p[0]) (-U2p[t] U4[t]+U2[t] U4p[t])^2 (V2p[0] V4[t]-V2[t] V4p[0]) (U2[t] (V2p[0] V3[t]-V2[t] V3p[0])+U1[t] (-V2p[0] V4[t]+V2[t] V4p[0])+U4[t] (V3p[0] V4[t]-V3[t] V4p[0]))^2)/((2 U2[t] V2[t]-2 U4[t] V4[t])^4 (-m U2[t] V2p[0]+m U4[t] V4p[0])^2 (((-U2p[0] V2[t]+U4p[0] V4[t]) (-m U2[t] V2p[0]+m U4[t] V4p[0]))/(2 U2[t] V2[t]-2 U4[t] V4[t])^2-(m (U2p[0] U4[t]-U2[t] U4p[0]) (V2p[0] V4[t]-V2[t] V4p[0]))/(2 U2[t] V2[t]-2 U4[t] V4[t])^2));
\end{spverbatim}

\section{Initial conditions for auxiliary equations in the optimization}

The initial (``final'') conditions for the auxiliary equations in Eq. (\ref{eq:auxequ}) are given by
\begin{equation}
\mathbf{p}^{(i)}(t_f)=\frac{\partial h}{\partial\mathbf{x}}(\mathbf{x}^{(i)}(t_f)),
\end{equation}
therefore, the initial conditions are
\begin{equation}
p_{U1}(t_f)=\frac{1}{2}\left( \frac{\partial\langle \hat{p}^{2}\rangle}{\partial U1} + \frac{\partial\langle \hat{q}^{2}\rangle}{\partial U1} \right)
\end{equation}
\begin{equation}
p_{U2}(t_f)=\frac{1}{2}\left( \frac{\partial\langle \hat{p}^{2}\rangle}{\partial U2} + \frac{\partial\langle \hat{q}^{2}\rangle}{\partial U2} \right)
\end{equation}
\begin{equation}
p_{U3}(t_f)=\frac{1}{2}\left( \frac{\partial\langle \hat{p}^{2}\rangle}{\partial U3} + \frac{\partial\langle \hat{q}^{2}\rangle}{\partial U3} \right)
\end{equation}
\begin{equation}
p_{U4}(t_f)=\frac{1}{2}\left( \frac{\partial\langle \hat{p}^{2}\rangle}{\partial U4} + \frac{\partial\langle \hat{q}^{2}\rangle}{\partial U4} \right)
\end{equation}
\begin{equation}
p_{V1}(t_f)=\frac{1}{2}\left( \frac{\partial\langle \hat{p}^{2}\rangle}{\partial V1} + \frac{\partial\langle \hat{q}^{2}\rangle}{\partial V1} \right)
\end{equation}
\begin{equation}
p_{V2}(t_f)=\frac{1}{2}\left( \frac{\partial\langle \hat{p}^{2}\rangle}{\partial V2} + \frac{\partial\langle \hat{q}^{2}\rangle}{\partial V2} \right)
\end{equation}
\begin{equation}
p_{V3}(t_f)=\frac{1}{2}\left( \frac{\partial\langle \hat{p}^{2}\rangle}{\partial V3} + \frac{\partial\langle \hat{q}^{2}\rangle}{\partial V3} \right)
\end{equation}
\begin{equation}
p_{V4}(t_f)=\frac{1}{2}\left( \frac{\partial\langle \hat{p}^{2}\rangle}{\partial V4} + \frac{\partial\langle \hat{q}^{2}\rangle}{\partial V4} \right)
\end{equation}
\begin{equation}
p_{U1p}(t_f)=\frac{1}{2}\left( \frac{\partial\langle \hat{p}^{2}\rangle}{\partial U1p} + \frac{\partial\langle \hat{q}^{2}\rangle}{\partial U1p} \right)
\end{equation}
\begin{equation}
p_{U2p}(t_f)=\frac{1}{2}\left( \frac{\partial\langle \hat{p}^{2}\rangle}{\partial U2p} + \frac{\partial\langle \hat{q}^{2}\rangle}{\partial U2p} \right)
\end{equation}
\begin{equation}
p_{U3p}(t_f)=\frac{1}{2}\left( \frac{\partial\langle \hat{p}^{2}\rangle}{\partial U3p} + \frac{\partial\langle \hat{q}^{2}\rangle}{\partial U3p} \right)
\end{equation}
\begin{equation}
p_{U4p}(t_f)=\frac{1}{2}\left( \frac{\partial\langle \hat{p}^{2}\rangle}{\partial U4p} + \frac{\partial\langle \hat{q}^{2}\rangle}{\partial U4p} \right)
\end{equation}
\begin{equation}
p_{V1p}(t_f)=\frac{1}{2}\left( \frac{\partial\langle \hat{p}^{2}\rangle}{\partial V1p} + \frac{\partial\langle \hat{q}^{2}\rangle}{\partial V1p} \right)
\end{equation}
\begin{equation}
p_{V2p}(t_f)=\frac{1}{2}\left( \frac{\partial\langle \hat{p}^{2}\rangle}{\partial V2p} + \frac{\partial\langle \hat{q}^{2}\rangle}{\partial V2p} \right)
\end{equation}
\begin{equation}
p_{V3p}(t_f)=\frac{1}{2}\left( \frac{\partial\langle \hat{p}^{2}\rangle}{\partial V3p} + \frac{\partial\langle \hat{q}^{2}\rangle}{\partial V3p} \right)
\end{equation}
\begin{equation}
p_{V4p}(t_f)=\frac{1}{2}\left( \frac{\partial\langle \hat{p}^{2}\rangle}{\partial V4p} + \frac{\partial\langle \hat{q}^{2}\rangle}{\partial V4p} \right)
\end{equation}

\begin{equation}
p_{u1}(t_f)=\frac{1}{2}\left[ \frac{\partial\langle \hat{p}^{2}\rangle}{\partial u1} + \frac{\partial\langle \hat{q}^{2}\rangle}{\partial u1} + \sum_{i=1}^{4}\left( \frac{\partial\langle \hat{p}^{2}\rangle}{\partial u1i} + \frac{\partial\langle \hat{q}^{2}\rangle}{\partial u1i} \right)ui_{ini} + \sum_{i=1,i \neq 1}^{4}\left( \frac{\partial\langle \hat{p}^{2}\rangle}{\partial ui1} + \frac{\partial\langle \hat{q}^{2}\rangle}{\partial ui1} \right)ui_{ini} \right]
\end{equation}
\begin{equation}
p_{u2}(t_f)=\frac{1}{2}\left[ \frac{\partial\langle \hat{p}^{2}\rangle}{\partial u2} + \frac{\partial\langle \hat{q}^{2}\rangle}{\partial u2} + \sum_{i=1}^{4}\left( \frac{\partial\langle \hat{p}^{2}\rangle}{\partial u2i} + \frac{\partial\langle \hat{q}^{2}\rangle}{\partial u2i} \right)ui_{ini} + \sum_{i=1,i \neq 2}^{4}\left( \frac{\partial\langle \hat{p}^{2}\rangle}{\partial ui2} + \frac{\partial\langle \hat{q}^{2}\rangle}{\partial ui2} \right)ui_{ini} \right]
\end{equation}
\begin{equation}
p_{u3}(t_f)=\frac{1}{2}\left[ \frac{\partial\langle \hat{p}^{2}\rangle}{\partial u3} + \frac{\partial\langle \hat{q}^{2}\rangle}{\partial u3} + \sum_{i=1}^{4}\left( \frac{\partial\langle \hat{p}^{2}\rangle}{\partial u3i} + \frac{\partial\langle \hat{q}^{2}\rangle}{\partial u3i} \right)ui_{ini} + \sum_{i=1,i \neq 3}^{4}\left( \frac{\partial\langle \hat{p}^{2}\rangle}{\partial ui3} + \frac{\partial\langle \hat{q}^{2}\rangle}{\partial ui3} \right)ui_{ini} \right]
\end{equation}
\begin{equation}
p_{u4}(t_f)=\frac{1}{2}\left[ \frac{\partial\langle \hat{p}^{2}\rangle}{\partial u4} + \frac{\partial\langle \hat{q}^{2}\rangle}{\partial u4} + \sum_{i=1}^{4}\left( \frac{\partial\langle \hat{p}^{2}\rangle}{\partial u4i} + \frac{\partial\langle \hat{q}^{2}\rangle}{\partial u4i} \right)ui_{ini} + \sum_{i=1,i \neq 4}^{4}\left( \frac{\partial\langle \hat{p}^{2}\rangle}{\partial ui4} + \frac{\partial\langle \hat{q}^{2}\rangle}{\partial ui4} \right)ui_{ini} \right]
\end{equation}
\begin{equation}
p_{v1}(t_f)=\frac{1}{2}\left[ \frac{\partial\langle \hat{p}^{2}\rangle}{\partial v1} + \frac{\partial\langle \hat{q}^{2}\rangle}{\partial v1} + \sum_{i=1}^{4}\left( \frac{\partial\langle \hat{p}^{2}\rangle}{\partial v1i} + \frac{\partial\langle \hat{q}^{2}\rangle}{\partial v1i} \right)vi_{ini} + \sum_{i=1,i \neq 1}^{4}\left( \frac{\partial\langle \hat{p}^{2}\rangle}{\partial vi1} + \frac{\partial\langle \hat{q}^{2}\rangle}{\partial vi1} \right)vi_{ini} \right]
\end{equation}
\begin{equation}
p_{v2}(t_f)=\frac{1}{2}\left[ \frac{\partial\langle \hat{p}^{2}\rangle}{\partial v2} + \frac{\partial\langle \hat{q}^{2}\rangle}{\partial v2} + \sum_{i=1}^{4}\left( \frac{\partial\langle \hat{p}^{2}\rangle}{\partial v2i} + \frac{\partial\langle \hat{q}^{2}\rangle}{\partial v2i} \right)vi_{ini} + \sum_{i=1,i \neq 2}^{4}\left( \frac{\partial\langle \hat{p}^{2}\rangle}{\partial vi2} + \frac{\partial\langle \hat{q}^{2}\rangle}{\partial vi2} \right)vi_{ini} \right]
\end{equation}
\begin{equation}
p_{v3}(t_f)=\frac{1}{2}\left[ \frac{\partial\langle \hat{p}^{2}\rangle}{\partial v3} + \frac{\partial\langle \hat{q}^{2}\rangle}{\partial v3} + \sum_{i=1}^{4}\left( \frac{\partial\langle \hat{p}^{2}\rangle}{\partial v3i} + \frac{\partial\langle \hat{q}^{2}\rangle}{\partial v3i} \right)vi_{ini} + \sum_{i=1,i \neq 3}^{4}\left( \frac{\partial\langle \hat{p}^{2}\rangle}{\partial vi3} + \frac{\partial\langle \hat{q}^{2}\rangle}{\partial vi3} \right)vi_{ini} \right]
\end{equation}
\begin{equation}
p_{v4}(t_f)=\frac{1}{2}\left[ \frac{\partial\langle \hat{p}^{2}\rangle}{\partial v4} + \frac{\partial\langle \hat{q}^{2}\rangle}{\partial v4} + \sum_{i=1}^{4}\left( \frac{\partial\langle \hat{p}^{2}\rangle}{\partial v4i} + \frac{\partial\langle \hat{q}^{2}\rangle}{\partial v4i} \right)vi_{ini} + \sum_{i=1,i \neq 4}^{4}\left( \frac{\partial\langle \hat{p}^{2}\rangle}{\partial vi4} + \frac{\partial\langle \hat{q}^{2}\rangle}{\partial vi4} \right)vi_{ini} \right]
\end{equation}

\cleardoublepage
\manualmark
\markboth{\spacedlowsmallcaps{\bibname}}{\spacedlowsmallcaps{\bibname}} 
\refstepcounter{dummy}
\addtocontents{toc}{\protect\vspace{\beforebibskip}} 
\addcontentsline{toc}{chapter}{\tocEntry{\bibname}}
\bibliographystyle{ieeetr}
\label{app:bibliography} 
{\footnotesize
\bibliography{Bibliography}}

\end{document}